\documentclass[12pt]{report}
\usepackage[english]{babel}
\usepackage{graphicx}
\usepackage{tikz}
\usepackage{subcaption}
\usepackage{framed}
\usepackage[normalem]{ulem}
\usepackage{amsmath}
\usepackage{bbm}
\usepackage{amsthm}
\usepackage{amssymb}
\usepackage{amsfonts}
\usepackage{enumerate}
\usepackage{enumitem}
\usepackage[utf8]{inputenc}
\usepackage[top=1 in,bottom=1in, left=1 in, right=1 in]{geometry}
\usepackage{bbm}
\usepackage{tensor}
\usepackage{dsfont}

\theoremstyle{definition}

\usepackage[english]{babel}
\usepackage{graphicx}
\usepackage{braket}
\usepackage{tikz}
\usepackage{arydshln}
\usepackage{subcaption}
\usepackage{manfnt}
\usepackage{framed}
\usepackage[normalem]{ulem}
\usepackage{amsmath}
\usepackage{bbm}
\usepackage{amsthm}
\usepackage{amssymb}
\usepackage{amsfonts}
\usepackage{bm}
\usepackage{enumerate}
\usepackage{enumitem}
\usepackage[utf8]{inputenc}
\usepackage[top=1 in,bottom=1in, left=1 in, right=1 in]{geometry}
\usepackage{bbm}
\usepackage{tensor}
\usepackage{dsfont}

\theoremstyle{definition}
\newcommand{\comment}[1]{}

\def\be{\begin{equation}}
\def\ee{\end{equation}}

\def\rp{\textcolor{red}{\bullet}}
\def\gp{\textcolor{green}{\bullet}}

\def\confa{\tikz[baseline=0cm]{
		\begin{scope}[scale = 0.5]
			\draw[very thick] [red] (-2,0) -- (-2,2);
			\draw[thick,dashed] (0,0) -- (2,0) -- (2,2) -- (0,2) -- (0,0);
			\draw[very thick] [green] (0,0) -- (1,1);
			\draw[very thick] [green] (1,1) -- (2,2);
			\draw[thick,very thick] [red] (2,0) -- (1,1);
			\draw[thick,very thick] [red] (1,1) -- (0,2);
			\draw (2,1) node[anchor=west] {$$};
			
			\draw (0,2) node[anchor=south] {$z_{2}$};
			\draw (2,2) node[anchor=south] {$z_{3}$};
			\draw (-2,2) node[anchor=south] {$z_{1}$};
			
			\draw(-2,-2) node[anchor=north] {$z_{3}$} ;
			\draw(2,-2) node[anchor=north] {$z_{2}$} ;
			\draw (0,-2) node[anchor=north] {$z_{1}$} ;
			
			\draw[green,fill=green] (2,2) circle (.5ex) ;
			\draw[red,fill=red] (0,2) circle (.5ex);
			\draw[red,fill=red] (-2,2) circle (.5ex);
			
			\draw[green,fill=green] (-2,-2) circle (.5ex);
			\draw[red,fill=red] (2,-2) circle (.5ex);
			\draw[red,fill=red] (0,-2) circle (.5ex);
			
			\begin{scope}[shift=({-2,-2})]
				
				\draw[very thick] [red] (4,0) -- (4,2);
				
				\draw[thick,dashed] (0,0) -- (2,0) -- (2,2) -- (0,2) -- (0,0);
				\draw[very thick] [green] (0,0) -- (1,1);
				\draw[very thick] [green] (1,1) -- (2,2);
				\draw[thick,very thick] [red] (2,0) -- (1,1);
				\draw[thick,very thick] [red] (1,1) -- (0,2);
				\draw (2,1) node[anchor=west] {$$};
				\draw (0,2) node[anchor=south] {$$};
				\draw (2,2) node[anchor=south] {$$};
				
			\end{scope}
			
		\end{scope}

	}
}

\def\confb{\tikz[baseline=-0.5cm]{
		\begin{scope}[scale = 0.5]
			\draw[very thick] [red] (-2,0) -- (-2,2);
			
			\draw[thick,dashed] (0,0) -- (2,0) -- (2,2) -- (0,2) -- (0,0);
			\draw[very thick] [green] (0,0) -- (1,1);
			\draw[very thick] [green] (1,1) -- (2,2);
			\draw[thick,very thick] [red] (2,0) -- (1,1);
			\draw[thick,very thick] [red] (1,1) -- (0,2);
			\draw (2,1) node[anchor=west] {$$};
			
			\draw (0,2) node[anchor=south] {$z_{2}$};
			\draw (2,2) node[anchor=south] {$z_{3}$};
			\draw (-2,2) node[anchor=south] {$z_{1}$};
			
			\draw(-2,-4) node[anchor=north] {$z_{3}$} ;
			\draw(2,-4) node[anchor=north] {$z_{1}$} ;
			\draw (0,-4) node[anchor=north] {$z_{2}$} ;
			
			\draw[green,fill=green] (2,2) circle (.5ex) ;
			\draw[red,fill=red] (0,2) circle (.5ex);
			\draw[red,fill=red] (-2,2) circle (.5ex);
			
			\draw[green,fill=green] (-2,-4) circle (.5ex);
			\draw[red,fill=red] (2,-4) circle (.5ex);
			\draw[red,fill=red] (0,-4) circle (.5ex);
			
			\begin{scope}[shift=({-2,-2})]
				
				\draw[very thick] [red] (4,0) -- (4,2);
				
				\draw[thick,dashed] (0,0) -- (2,0) -- (2,2) -- (0,2) -- (0,0);
				\draw[very thick] [green] (0,0) -- (1,1);
				\draw[very thick] [green] (1,1) -- (2,2);
				\draw[thick,very thick] [red] (2,0) -- (1,1);
				\draw[thick,very thick] [red] (1,1) -- (0,2);
				
			\end{scope}

			\begin{scope}[shift=({0,-4})]
				\draw[very thick] [green] (-2,0) -- (-2,2);
				
				\draw[thick,dashed] (0,0) -- (2,0) -- (2,2) -- (0,2) -- (0,0);
				\draw[very thick] [red] (0,0) -- (1,1);
				\draw[very thick] [red] (1,1) -- (2,2);
				\draw[thick,very thick] [red] (2,0) -- (1,1);
				\draw[thick,very thick] [red] (1,1) -- (0,2);
				\draw (2,1) node[anchor=west] {$$};

			\end{scope}
			
		\end{scope}

	}
}

\def\confc{\tikz[baseline=0.5cm]{
		\begin{scope}[scale = 0.5]
			\draw[very thick] [red] (-2,0) -- (-2,2);
			
			\draw[thick,dashed] (0,0) -- (2,0) -- (2,2) -- (0,2) -- (0,0);
			\draw[very thick] [green] (0,0) -- (1,1);
			\draw[very thick] [green] (1,1) -- (2,2);
			\draw[thick,very thick] [red] (2,0) -- (1,1);
			\draw[thick,very thick] [red] (1,1) -- (0,2);
			\draw (2,1) node[anchor=west] {$$};
			
			\draw (0,2) node[anchor=south] {$z_{2}$};
			\draw (2,2) node[anchor=south] {$z_{3}$};
			\draw (-2,2) node[anchor=south] {$z_{1}$};
			
			\draw(-2,0) node[anchor=north] {$z_{1}$} ;
			\draw(2,0) node[anchor=north] {$z_{2}$} ;
			\draw (0,0) node[anchor=north] {$z_{3}$} ;

			\draw[green,fill=green] (2,2) circle (.5ex) ;
			\draw[red,fill=red] (0,2) circle (.5ex);
			\draw[red,fill=red] (-2,2) circle (.5ex);
			
			\draw[green,fill=green] (0,0) circle (.5ex);
			\draw[red,fill=red] (2,0) circle (.5ex);
			\draw[red,fill=red] (-2,0) circle (.5ex);
		\end{scope}

	}
}

\def\confd{\tikz[baseline=0cm]{
		\begin{scope}[scale = 0.5]
			\draw[very thick] [red] (-2,0) -- (-2,2);
			
			\draw[thick,dashed] (0,0) -- (2,0) -- (2,2) -- (0,2) -- (0,0);
			\draw[very thick] [green] (0,0) -- (1,1);
			\draw[very thick] [green] (1,1) -- (2,2);
			\draw[thick,very thick] [red] (2,0) -- (1,1);
			\draw[thick,very thick] [red] (1,1) -- (0,2);
			\draw (2,1) node[anchor=west] {$$};
			
			\draw (0,2) node[anchor=south] {$z_{2}$};
			\draw (2,2) node[anchor=south] {$z_{3}$};
			\draw (-2,2) node[anchor=south] {$z_{1}$};
			
			\draw(-2,-2) node[anchor=north] {$z_{3}$} ;
			\draw(2,-2) node[anchor=north] {$z_{2}$} ;
			\draw (0,-2) node[anchor=north] {$z_{1}$} ;
			
			\draw[green,fill=green] (2,2) circle (.5ex) ;
			\draw[red,fill=red] (0,2) circle (.5ex);
			\draw[red,fill=red] (-2,2) circle (.5ex);
			
			\draw[green,fill=green] (0,-2) circle (.5ex);
			\draw[red,fill=red] (2,-2) circle (.5ex);
			\draw[red,fill=red] (-2,-2) circle (.5ex);
			
			\begin{scope}[shift=({-2,-2})]
				
				\draw[very thick] [red] (4,0) -- (4,2);
				
				\draw[thick,dashed] (0,0) -- (2,0) -- (2,2) -- (0,2) -- (0,0);
				\draw[very thick] [red] (0,0) -- (1,1);
				\draw[very thick] [green] (1,1) -- (2,2);
				\draw[thick,very thick] [green] (2,0) -- (1,1);
				\draw[thick,very thick] [red] (1,1) -- (0,2);
				\draw (2,1) node[anchor=west] {$$};
				\draw (0,2) node[anchor=south] {$$};
				\draw (2,2) node[anchor=south] {$$};
				
			\end{scope}
			
		\end{scope}

	}
}

\def\confe{\tikz[baseline=0cm]{
		\begin{scope}[scale = 0.5]
			\begin{scope}[shift=({0,-2})]
				
				\draw[very thick] [red] (-2,0) -- (-2,2);
				
				\draw[thick,dashed] (0,0) -- (2,0) -- (2,2) -- (0,2) -- (0,0);
				\draw[very thick] [green] (0,0) -- (1,1);
				\draw[very thick] [green] (1,1) -- (2,2);
				\draw[thick,very thick] [red] (2,0) -- (1,1);
				\draw[thick,very thick] [red] (1,1) -- (0,2);
				\draw (2,1) node[anchor=west] {$$};
				
			\end{scope}

			\draw (0,2) node[anchor=south] {$z_{2}$};
			\draw (2,2) node[anchor=south] {$z_{3}$};
			\draw (-2,2) node[anchor=south] {$z_{1}$};
			
			\draw(-2,-2) node[anchor=north] {$z_{2}$} ;
			\draw(2,-2) node[anchor=north] {$z_{1}$} ;
			\draw (0,-2) node[anchor=north] {$z_{3}$} ;
			
			\draw[green,fill=green] (2,2) circle (.5ex) ;
			\draw[red,fill=red] (0,2) circle (.5ex);
			\draw[red,fill=red] (-2,2) circle (.5ex);
			
			\draw[green,fill=green] (0,-2) circle (.5ex);
			\draw[red,fill=red] (-2,-2) circle (.5ex);
			\draw[red,fill=red] (2,-2) circle (.5ex);
			
			\begin{scope}[shift=({-2,0})]
				
				\draw[very thick] [green] (4,0) -- (4,2);
				
				\draw[thick,dashed] (0,0) -- (2,0) -- (2,2) -- (0,2) -- (0,0);
				\draw[very thick] [red] (0,0) -- (1,1);
				\draw[very thick] [red] (1,1) -- (2,2);
				\draw[thick,very thick] [red] (2,0) -- (1,1);
				\draw[thick,very thick] [red] (1,1) -- (0,2);
				\draw (2,1) node[anchor=west] {$$};
				\draw (0,2) node[anchor=south] {$$};
				\draw (2,2) node[anchor=south] {$$};
				
			\end{scope}
			
		\end{scope}

	}
}

\def\conff{\tikz[baseline=-0.5cm]{
		\begin{scope}[scale = 0.5]
			\draw[very thick] [red] (-2,0) -- (-2,2);
			
			\draw[thick,dashed] (0,0) -- (2,0) -- (2,2) -- (0,2) -- (0,0);
			\draw[very thick] [red] (0,0) -- (1,1);
			\draw[very thick] [green] (1,1) -- (2,2);
			\draw[thick,very thick] [green] (2,0) -- (1,1);
			\draw[thick,very thick] [red] (1,1) -- (0,2);
			\draw (2,1) node[anchor=west] {$$};
			
			\draw (0,2) node[anchor=south] {$z_{2}$};
			\draw (2,2) node[anchor=south] {$z_{3}$};
			\draw (-2,2) node[anchor=south] {$z_{1}$};
			
			\draw(-2,-4) node[anchor=north] {$z_{3}$} ;
			\draw(2,-4) node[anchor=north] {$z_{1}$} ;
			\draw (0,-4) node[anchor=north] {$z_{2}$} ;
			
			\draw[green,fill=green] (2,2) circle (.5ex) ;
			\draw[red,fill=red] (0,2) circle (.5ex);
			\draw[red,fill=red] (-2,2) circle (.5ex);
			
			\draw[green,fill=green] (0,-4) circle (.5ex);
			\draw[red,fill=red] (2,-4) circle (.5ex);
			\draw[red,fill=red] (-2,-4) circle (.5ex);
			
			\begin{scope}[shift=({-2,-2})]
				
				\draw[very thick] [green] (4,0) -- (4,2);
				
				\draw[thick,dashed] (0,0) -- (2,0) -- (2,2) -- (0,2) -- (0,0);
				\draw[very thick] [red] (0,0) -- (1,1);
				\draw[very thick] [red] (1,1) -- (2,2);
				\draw[thick,very thick] [red] (2,0) -- (1,1);
				\draw[thick,very thick] [red] (1,1) -- (0,2);
				
			\end{scope}

			\begin{scope}[shift=({0,-4})]
				\draw[very thick] [red] (-2,0) -- (-2,2);
				
				\draw[thick,dashed] (0,0) -- (2,0) -- (2,2) -- (0,2) -- (0,0);
				\draw[very thick] [green] (0,0) -- (1,1);
				\draw[very thick] [green] (1,1) -- (2,2);
				\draw[thick,very thick] [red] (2,0) -- (1,1);
				\draw[thick,very thick] [red] (1,1) -- (0,2);
				\draw (2,1) node[anchor=west] {$$};

			\end{scope}
		\end{scope}

	}
}

\def\confg{\tikz[baseline=-0.5cm]{
		\begin{scope}[scale = 0.5]
			\draw[very thick] [red] (-2,0) -- (-2,2);
			
			\draw[thick,dashed] (0,0) -- (2,0) -- (2,2) -- (0,2) -- (0,0);
			\draw[very thick] [green] (0,0) -- (1,1);
			\draw[very thick] [green] (1,1) -- (2,2);
			\draw[thick,very thick] [red] (2,0) -- (1,1);
			\draw[thick,very thick] [red] (1,1) -- (0,2);
			\draw (2,1) node[anchor=west] {$$};
			
			\draw (0,2) node[anchor=south] {$z_{2}$};
			\draw (2,2) node[anchor=south] {$z_{3}$};
			\draw (-2,2) node[anchor=south] {$z_{1}$};
			
			\draw(-2,-4) node[anchor=north] {$z_{3}$} ;
			\draw(2,-4) node[anchor=north] {$z_{1}$} ;
			\draw (0,-4) node[anchor=north] {$z_{2}$} ;
			
			\draw[green,fill=green] (2,2) circle (.5ex) ;
			\draw[red,fill=red] (0,2) circle (.5ex);
			\draw[red,fill=red] (-2,2) circle (.5ex);
			
			\draw[green,fill=green] (0,-4) circle (.5ex);
			\draw[red,fill=red] (2,-4) circle (.5ex);
			\draw[red,fill=red] (-2,-4) circle (.5ex);
			
			\begin{scope}[shift=({-2,-2})]
				
				\draw[very thick] [red] (4,0) -- (4,2);
				
				\draw[thick,dashed] (0,0) -- (2,0) -- (2,2) -- (0,2) -- (0,0);
				\draw[very thick] [red] (0,0) -- (1,1);
				\draw[very thick] [green] (1,1) -- (2,2);
				\draw[thick,very thick] [green] (2,0) -- (1,1);
				\draw[thick,very thick] [red] (1,1) -- (0,2);
				
			\end{scope}

			\begin{scope}[shift=({0,-4})]
				\draw[very thick] [red] (-2,0) -- (-2,2);
				
				\draw[thick,dashed] (0,0) -- (2,0) -- (2,2) -- (0,2) -- (0,0);
				\draw[very thick] [green] (0,0) -- (1,1);
				\draw[very thick] [red] (1,1) -- (2,2);
				\draw[thick,very thick] [red] (2,0) -- (1,1);
				\draw[thick,very thick] [green] (1,1) -- (0,2);
				\draw (2,1) node[anchor=west] {$$};

			\end{scope}
		\end{scope}

	}
}

\def\rp{\textcolor{red}{\bullet}}
\def\gp{\textcolor{green}{\bullet}}

\usepackage{amsthm}
\theoremstyle{plain}
\newtheorem{thm}{Theorem}[section]
\newtheorem{lem}[thm]{Lemma}
\newtheorem{conj}[thm]{Conjecture}

\newtheorem{coro}[thm]{Corollary}

\expandafter\let\csname equation*\endcsname\relax
\expandafter\let\csname endequation*\endcsname\relax
\usepackage{graphicx}
\usepackage{amssymb}
\usepackage{amsmath}
\usepackage{amsthm}
\usepackage{cite}
\usepackage{hyperref}
\usepackage{epstopdf}
\usepackage{tikz}
\usepackage{caption}
\usepackage{subcaption}
\usepackage{hyperref}
\usepackage{float}
\usepackage{enumitem}
\usepackage{calc}

\usepackage{lipsum}
\usepackage[Sonny]{fncychap}

\usepackage{fancybox}

\usepackage{tcolorbox}
\newtcolorbox{mybox}{colback=red!5!white,colframe=red!75!black}

\theoremstyle{remark}

\usepackage[font=small, labelfont=bf, width=0.8\textwidth]{caption}

\def\be{\begin{equation}}
\def\ee{\end{equation}}

\def\bfz{{\bf z}}

\def\bfJ{{\bf J}}
\def\bfrho{\boldsymbol{\rho}}

\makeatletter
\newcommand*{\textoverline}[1]{$\overline{\hbox{#1}}\m@th$}
\makeatother

\begin{document}

\thispagestyle{empty} 
\vspace{-2cm}

\voffset-10pt

\begin{figure}
\centering  
\begin{tabular}{ccc}                          
	\includegraphics[scale=0.4]{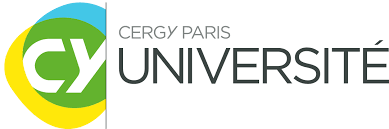}&
	\hfill
	& 
	\hspace*{4cm} 
	\includegraphics[scale=0.39]{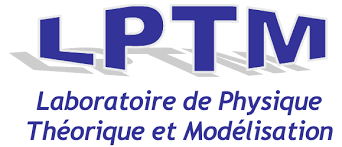}                      
\end{tabular}
\end{figure}

\vspace*{-0.5cm}

\begin{center}
{\bf \small \'Ecole Doctorale EM2PSI (ED 405) } \\
 \vspace*{0.4cm}
\end{center}


\begin{center}
 {\Huge \bf TH\`ESE DE DOCTORAT }\\
 \vspace*{0.1cm}
{  \sc  sp\'ecialit\'e : physique th\'eorique}\\
 \vspace{0.2cm}
\end{center}

\begin{center}
{\bf Soutenue le 29 Novembre 2022}\\
\vspace{0.4cm}
{\Large\bf  Ali Zahra }
\end{center}

\vspace{5mm}

\noindent \rule{1\linewidth}{0.4mm}
\vspace{0.01cm}
\begin{center}
{\Large\bf Multi-Species Generalization of the Totally Asymmetric Simple Exclusion Process}

\end{center}

\begin{center}
	{\color{darkgray} \large Integrability and Hydrodynamic Aspects }
\end{center}
\noindent \rule{1\linewidth}{0.2mm}

\vspace{1cm}

\begin{center}
{\bf Pr\'esent\'ee en vue de l'obtention du grade de DOCTEUR }\\
\vspace{0.4cm}
 {{  Dirig\'ee par : } \bf Luigi Cantini}\\
\end{center} \vspace{-0.2cm}


\vspace{1mm}

\begin{center}
\noindent \rule{1\linewidth}{0.2mm}
{\small  Jury  de soutenance}
\noindent \rule{1\linewidth}{0.2mm}
\end{center}

\begin{tabular}{llll}
	
{\sc  \footnotesize Kirone Mallick} & {\footnotesize Directeur de recherche} & {\footnotesize CEA \scriptsize (IPhT-Saclay) } & {\footnotesize Rapporteur }\vspace{1mm}\\

{\sc  \footnotesize Gunter M. Schütz} & {\footnotesize Professeur} & {\footnotesize IST \scriptsize (Universidade de Lisboa)} & {\footnotesize Rapporteur}\vspace{1mm}\\

{\sc  \footnotesize Flora Koukiou} & {\footnotesize Professeur} & {\footnotesize CNRS \scriptsize (CY Universit\'e) } & {\footnotesize Examinateur }\vspace{1mm}\\

{\sc  \footnotesize Sylvain Prolhac} & {\footnotesize Maitre de conférence} & {\footnotesize IRSAMC \scriptsize (Université Paul Sabatier) } & {\footnotesize Examinateur }\vspace{1mm}\\

{\sc  \footnotesize Filippo Colomo} & {\footnotesize Chargé de recherche} & {\footnotesize INFN \scriptsize (Sezione di Firenze) } & {\footnotesize Examinateur }\vspace{1mm}\\

{\sc  \footnotesize Jean Avan} & {\footnotesize Directeur de recherche} & {\footnotesize CNRS \scriptsize (CY Universit\'e) } & {\footnotesize Examinateur }\vspace{1mm}\\

{\sc  \footnotesize Luigi Cantini} & {\footnotesize Maitre de conférence} & {\footnotesize CNRS \scriptsize (CY Universit\'e) } & {\footnotesize Directeur de th\`ese}\vspace{1mm}\\

\end{tabular}

\vfill
\noindent

\newpage

\begin{abstract}
Exclusion processes in one dimension first appeared in the 70s and have since dragged much attention from communities in different domains: stochastic processes, out of equilibriums statistical physics, and more recently integrable systems. While the state of the art for a single species totally asymmetric simple exclusion process (TASEP) can be described, from different aspects as mature, much less is known when multiple interacting species are present. Using tools from integrable systems and hydrodynamics in the first place and stochastic processes in the second place, this work attempts to study the behavior of a novel version of the model with different species of particles having hierarchical dynamics that depend on arbitrary parameters.
While Burger's equation famously represents the hydrodynamic limit of TASEP with a single species, we present a counterpart coupled system of PDE representing the hydrodynamic limit for a model with two species. The solutions of these PDEs display a rich phenomenology of solutions best characterized through the underlying normal modes. We discuss the associated Riemann problem and validate our results with numerical simulations.
This system with two species can be used as a toy model for studying driven diffusive systems with open boundaries. Using heuristics, we present results suggesting a general principle governing the boundary induced phase diagram of systems with multiple coupled driven conserved quantities, generalizing thus the extremal current principle known for the case of a single driven quantity.
The integrability side of our study is mainly concerned with developing a formalism allowing the computation of the finite-time probability distribution of particle positions on the 1D lattice, generalizing therefore known results for TASEP and other multi-species models.
We finally study the behavior and the impact of a single second class impurity initially located at the interface separating two regions of different densities of first class particles. Different limit shapes are deduced and observed. Using tools from probability theory, we generalize the asymptotic speed properties of the impurity for a regime of the hopping parameters.
\end{abstract}

\pagebreak
\hspace{0pt}
\vfill
\begin{center}
	\textbf{Résumé}
\end{center}
Les processus d'exclusion à une dimension sont apparus pour la première fois dans les années 70 et ont depuis attiré beaucoup d'attention de la part des communautés dans différents domaines : processus stochastiques, physique statistique hors équilibre, et plus récemment systèmes intégrables. Alors que l'état de l'art pour un processus d'exclusion simple totalement asymétrique (TASEP) d'une seule espèce peut être décrit, sous différents aspects comme mature, on en sait beaucoup moins lorsque plusieurs espèces en interaction sont présentes. En utilisant des outils issus des systèmes intégrables et de l'hydrodynamique en premier lieu et des processus stochastiques en second lieu, ce travail tente d'étudier le comportement d'une nouvelle version du modèle avec différentes espèces de particules ayant une dynamique hiérarchique qui dépend de paramètres arbitraires.
Alors que l'équation de Burger représente la limite hydrodynamique de TASEP avec une seule espèce, nous présentons un système couplé d'EDP représentant la limite hydrodynamique pour un modèle avec deux espèces. Les solutions de ces EDP présentent une riche phénoménologie de solutions mieux caractérisée par les modes normaux sous-jacents. Nous discutons du problème de Riemann associé et validons nos résultats par des simulations numériques.
Ce système à deux espèces peut être utilisé comme un modèle jouet pour étudier les systèmes diffusifs pilotés avec des bords ouverts. En utilisant des heuristiques, nous présentons des résultats suggérant un principe général régissant le diagramme de phase induit par les frontières des systèmes avec de multiples quantités conservées couplées, généralisant ainsi le principe du courant extrémal connu pour le cas d'une seule quantité entraînée.
L'aspect intégrabilité de notre étude concerne principalement le développement d'un formalisme permettant le calcul de la distribution de probabilité en temps fini des positions des particules sur le réseau à 1D, généralisant ainsi les résultats connus pour TASEP et d'autres modèles multi-espèces.
Nous étudions enfin le comportement et l'impact d'une seule impureté de seconde classe initialement située à l'interface séparant deux régions de densités différentes de particules de première classe. Différentes formes limites sont déduites et observées. En utilisant des outils de la théorie des probabilités, nous généralisons les propriétés de vitesse asymptotique de l'impureté pour un régime des taux.
\vfill
\hspace{0pt}
\pagebreak

\pagebreak
\hspace{0pt}
\vfill
\begin{center}
	\textbf{Acknowledgment}
	
\end{center}

First and foremost, I would like to thank my supervisor Luigi Cantini. Working with him has been simply great. Our meetings have always been a source of inspiration to me. I can't be grateful enough to him for what I learned during my Ph.D. Besides all of the scientific side, his support and kindness are exceptional. 

I am thankful to all the members of jury for having kindly accepted to evaluate this work. Most of them had to make a long trip to physically attend my defense. I would like to thank the two reporters who put a remarkable effort into reading my manuscript and writing the reports. Their comments and suggestions have been very useful for improving the quality of this dissertation. I want to thank all the members of LPTM, who made this lab such a great environment both on the professional and social levels. The lunch breaks with Jean Avan and Genvieve Rollet are always rich in culture and humor. I'm indebted to both of them for the generous support they offered to me on multiple occasions.
A particular thanks go to Andreas Honecker who has been always very kind and helpful starting from my Master year and throughout the following years. I had plenty of pleasure sharing teaching duties with Guy Trambly de Laissardière, Jean Philippe Kownacki, Claire Pinette, Geneviève Rollet and Andreas Honecker. They were always generous with their insightful pedagogical hints. 
I would like to thank the administrator of our lab Sylvie Villemin who is always there to help with a big smile.  I want to thank my friends and family for their continuous encouragement and support. 
This text has been linguistically checked and refined thanks to the effort of my friends Laurence Verges, Dovile Jankauskaite, Ibrahim Saideh and Marta Pedrosa García-Moreno. 
Finally, I want to thank the doctoral school EM2PSI for their financial support through the doctoral contract.

\vfill
\hspace{0pt}
\pagebreak

\tableofcontents

\setcounter{chapter}{-1}

\chapter{Introduction}

Statistical physics at equilibrium is one of the most impressive success stories in physics, it allows us to explain the properties of matter surrounding us. Its mathematical foundations are well established too \cite{landau2013statistical} \cite{sethna2021statistical}.  Given a Hamiltonian system, one can find the probability distribution of microscopic states as the one that maximizes the entropy, so for a system coupled to a reservoir, this probability is given by the Boltzmann-Gibbs ensemble
\begin{equation}
	P_{eq}(C) = \frac{1}{Z} e^{-H(C)/k_{B}T}
\end{equation}
This allows in principle to compute all physical quantities such as free energy and correlation functions. To perform these computations, one often needs approximation methods such as the mean-field approach, the renormalization group, and series expansion. Analytical exact expressions are possible only for a minor number of models giving them a major role in the theory \cite{baxter2016exactly}. A prominent pioneer example is the Ising model in 2D, that was solved by Lars Onsager in 1944 \cite{onsager1944crystal} and for which critical exponents were computed exactly for the first time and were different from the mean-field ones. This had a major impact in understanding the the critical behavior around phase transition in equilibrium statistical physics and paved the way for the emergence of the idea of universality where this behavior depends in many situations only on the dimensionality and the symmetries of interactions \cite{baxter2016exactly}. However, most of the collective phenomena going on in nature are out of equilibrium. If one defines systems out of equilibrium as merely the complementary set of systems at equilibrium, then this set is so vast that it is not reasonable to expect it to have some common theoretical features. So we are usually led to work within a particular setting. For instance, in closed quantum systems, we typically consider particular schemes of a time-dependent Hamiltonian that makes the problem tractable \cite{eisert2015quantum} common examples include periodic driving, where the Hamiltonian has a time periodicity $H(t) = H(t + \tau)$ \cite{berdanier2020universality}. Another one is quantum quenching, for which a system is prepared at the ground state of a time-independent Hamiltonian, and at some instant, we suddenly turn on a perturbing Hamiltonian and observe the dynamic evolution of the system till thermalization \cite{mitra2018quantum}. These systems are often hard to analyze analytically, and even numerically using only a classical computer, they rather require quantum simulators such as the ones based on ultracold atoms in optical lattices \cite{gross2017quantum}. Most importantly, quantum systems in nature are usually not closed but rather coupled to an environment, they thus exhibit decoherence and their effective behavior collapses in many situations to non-Hamiltonian stochastic evolution, see chapter 8 of \cite{joos2013decoherence} for details. This brings us to the focus of this dissertation, which is the stochastic systems out of equilibrium. In particular, we will be considering Markovian systems, which are as well adapted to classical Hamiltonian systems at the mesoscopic time scale. In such systems the stochastic evolution depends only on the current configuration of the system, and not on its history, in other words, these systems don't have intrinsic memory. Mathematically, the relevant information about the system is reduced to the set of transition rates between the different microscopic configurations, denote $w_{C' \rightarrow C}$ the hopping rate from the configuration $C'$ to the configuration $C$. Once these rates are known, one can write the evolution equation of the probability distribution over the configuration space \footnote{Assume this space is countable. The more general framework is briefly mentioned in chapter 5}

$$\frac{dP(C)}{dt} =
\sum_{C' \neq C} \underbrace{w_{C' \rightarrow C} P(C')}_{\text{gain}}
-
\sum_{C' \neq C} \underbrace{w_{C \rightarrow C'} P(C) }_{\text{loss}} $$
This is called the master equation. It can be written in a compact form: $\frac{dP}{dt} = M P$, where $M$ is called the Markov matrix.  $M_{C,C^{'}} := w_{C' \rightarrow C}$ for the off-diagonal elements, and $M_{C,C} = - \sum_{C' \neq C} w_{C \rightarrow C'} $ for the diagonal ones.
Starting form some initial probability distribution over the configurations, and evolving in time with a Markov operator, the system relaxes in time to a stationary state for which the probability weights are static.
\footnote{For an infinite system, these weights might not be normalisable, and it's more accurate to speak about an invariant measure, as it will be explained in chapter 3}
If $\pi$ is the this stationary distribution, it should verify $M \pi = 0$. In other words, this is the eigenvector corresponding to the zero eigenvalue. The stochastic structure of the Markov matrix makes it so that all the other eigenvalues have a negative real part, so they correspond to decaying modes. The eigenvalue with the largest non zero real part provides an estimation (through its inverse) of the typical relaxation time of the system, which is a relevant physical observable quantity. The Markovian framework is adapted for both equilibrium and out equilibrium systems. In the equilibrium case, although the Boltzmann-Gibbs doesn't tell us about the transition rates, it is always possible to assume detailed balance, meaning that there is no net probability current between any two configurations at equilibrium, this is expressed as:

$$ P_{eq}(C)w_{C \rightarrow C'} = P_{eq}(C')w_{C' \rightarrow C} $$

Given a system satisfying detailed balance, if we record its time evolution, we can't tell in which direction the film is played. So, this is equivalent to time reversibility.
The existence of a distribution verifying the detailed balance can be expressed as a restriction on the elements of the Markov matrix, known as the Kolmogorov criteria
\footnote{
	Although Kolmogorov implies the existence of detailed balance, the opposite implication is valid only for irreducible Markov chain, i.e. chains for which any state is accessible from any other state (not necessarily directly).
},which states that around any closed cycle of states, there is no net flow of probability, for example, for any three configurations, we should have: $w_{1\rightarrow2}w_{2\rightarrow3}w_{3\rightarrow1}=w_{1\rightarrow3}w_{3\rightarrow2}w_{2\rightarrow1}$, \cite{kelly2011reversibility}. The simplest way to create a system out of equilibrium is to take a system in equilibrium and to perturb it slightly so that it is driven out of its equilibrium distribution, which breaks the detailed balance. Linear response theory is adapted to deal with this situation \cite{kubo1957statistical}.It applies typically to a system with a small gradient of thermodynamic variables inducing purely diffusive currents. Reciprocity relations over the elements of the diffusion matrix were revealed by Onsager based on the local microscopic time reversibility of the interactions and were the origin of his Nobel Prize in chemistry in 1968. Another basic situation to be out of equilibrium  is to have a configuration $C$ such that $w_{C \rightarrow C'}=0$ for all $C{'}$ and  $w_{C' \rightarrow C}\neq 0$ for some $C^{'}$, the state $C$ is called an absorbing state, once the system reaches it, it cannot get out of it, 
this creates a uni-directional probability currents towards the absorbing state and ensures being out of equilibrium. An example for these systems is a model of the spread of an epidemic, a recovered population would be an absorbing state. Although, universal behaviors have been observed for absorbing state phase transitions, most notably the direct percolation universality class where universal critical exponents were observed \cite{janssen1981nonequilibrium} \cite{hinrichsen2006non}. However, this model is not exactly solvable, so the critical exponents are known only approximately through numerical means. A remarkable setting where we both have exactly solvable models with non-equilibrium steady state(NESS) \cite{zhang2012stochastic}, is the driven diffusive systems,  they can be thought of,for instance, as a gas of charged particles with a driving electromagnetic force breaking the space isotropy and inducing a permanent current even when the system is homogeneous \cite{zia2000contrasts}. They can be of course defined on the continuous space and for an arbitrary dimension. However, we will only consider lattice gas models defined on the lattice in 1D, this choice is justified by the availability of an exactly solvable toy model, that is as well relevant for applications. This model is the Asymmetric Simple Exclusion Process (ASEP), which is considered a paradigmatic model for driven diffusive systems in general and transport models in 1D in particles. It's often as well described as the Ising model for out-of-equilibrium statistical physics\footnote{The same claim is made for the directed percolation, however, integrability makes the analogy to the Ising model more relevant for ASEP}. Let's provide a definition and review briefly its most elementary properties. These properties can be found in details in few classical reviews on the topic, for instance: \cite{blythe2007nonequilibrium} \cite{chou2011non}.  

\subsection*{The Asymmetric Simple Exclusion Process}
This model is defined as a gas of identical particles on the $\mathbb{Z}$ lattice. Each particle is a random walker in continuous time, it hops forward at a rate $p$, but only if the site in front of it is empty, and can hop backward at a rate $q$, only if the site behind is empty. So there is a hardcore exclusion between the particles that leads to a maximum number of one particle per site. The initial motivation and context of the appearance of the model will be mentioned latter in this introduction. A simple particular case is when  $p=q$ then the model is called SSEP (Symmetric Simple Exclusion Process), this model is not out of equilibrium, it's easy to understand that there is no average current and that detailed balance is conserved, however, this particular case is still relevant either from a mathematical point of view where it has been historically the first case to be solved exactly by mapping it to spin chains, or sometimes it can be seen as a critical system where we have a transition between non-equilibrium and equilibrium. Another particular case that is interesting for many reasons is when the particles move only in one direction, take for instance $q=0$, we speak here about the Totally Asymmetric Simple Exclusion Process (TASEP). We can set $p=1$ by a change of the scale of time. This particular case allows, in many cases, for exact computations that are much harder for the general ASEP, so it provided the simplest out-of-equilibrium exactly solvable model.

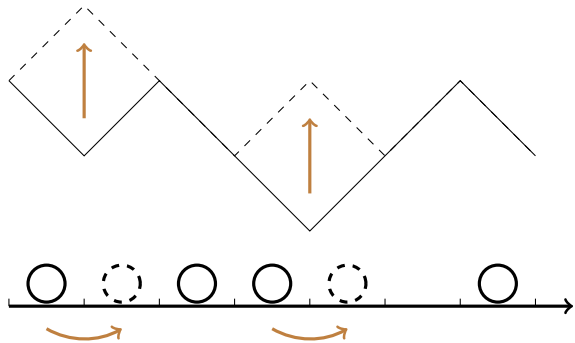
\begin{figure}[h!]
	\centering
	
	\begin{tikzpicture}
	\draw [very thick, ->] (0,0) -- (7.5,0);
	
	\foreach \i in {0,...,7}
	{
		\draw (\i,0) -- (\i,0.1) ;
	}
	
	\foreach \i in {0,2,3,6}
	{
		\node at (\i+0.5,0.5){};
		\draw [very thick] (\i+0.5,0.3) circle (7pt);
	}
	
	\foreach \i in {1,4}
	{
		\node at (\i+0.5,0.5){};
		\draw [very thick, dashed] (\i+0.5,0.3) circle (7pt);
	}
	
	\begin{scope}[shift=({0.5,-0.3})]
	\draw  [very thick, ->,brown]  (0,0) arc (60:120:-1) ;
	\end{scope}
	
	\begin{scope}[shift=({3.5,-0.3})]
	\draw  [very thick, ->,brown]  (0,0) arc (60:120:-1) ;
	\end{scope}
	
	\draw [very thick, ->,brown] (1,2.5) -- (1,3.5);
	\draw [very thick, ->,brown] (4,1.5) -- (4,2.5);
	
	\begin{scope}[shift=({0,3})]
	\draw (0,0) -- (1,-1)--(2,0)--(3,-1)--(4,-2)--(5,-1)--(6,0)--(7,-1);
	
	\draw [dashed] (0,0) -- (1,1)--(2,0)--(3,-1)--(4,0)--(5,-1)--(6,0)--(7,-1);
	
	\end{scope}
	
	\end{tikzpicture}
	
	\caption{The corner growth process as an interpretation of TASEP}
	\label{growth process}
\end{figure}

What adds to the interest of TASEP is that it has different interpretations. For instance, it can be mapped to a surface growth model. This was first pointed out by Rost \cite{rost1981non}. Let a 2D surface with an upper boundary represented by an affine continuous function $h(x,t)$ defined up to an additive constant and verifying:
\begin{equation}
h(j+\frac{1}{2},t) - h(j-\frac{1}{2},t) = \begin{cases}
-1 \quad &\text{if the site at $j$ is occupied} \\
1 \quad &\text{if the site at $j$ is empty} \\
\end{cases}
\end{equation}

One can understand that the time evolution of the TASEP corresponds to a random growth process of the surface, figure \ref{growth process}.

\begin{figure}[h!]
	\centering
	\begin{tikzpicture}[scale=0.7]
	
	\draw (0.6,-6.7) node [below] {$(a)$} ;
	
	\draw [very thick] (0.5,-3) circle (3);
	
	\foreach \j in {0,20,...,360}
	{
		\begin{scope} [rotate around={\j:(0.5,-3)}]
		
		\foreach \i in {0}
		{
			\draw (\i,0) -- (\i,0.1) ;
		}
		
		\end{scope}
	}
	
	\foreach \j in {0,1,4,5,7,8,10,13,15,17}
	{
		\begin{scope} [rotate around={20*\j:(0.5,-3)}]

		\foreach \i in {0}
		{
			\node at (\i+0.5,0.5){};
			\draw [very thick,fill=red] (\i+0.5,0.3) circle (7pt);
		}
		
		\end{scope}
	}
	
	\begin{scope}[shift=({6,-1})]
	\begin{scope}[shift=({0.4+3,0.75}),scale=0.8]
	\draw  [very thick, ->,brown]  (0,0) arc (60:120:1) ;
	\end{scope}
	\draw (3,1.5) node [below] {$q$} ;
	
	\begin{scope}[shift=({0.6+3,0.75}),scale=0.8]
	\draw  [very thick, ->,brown]  (0,0) arc (120:60:1) ;
	\end{scope}
	\draw (4,1.5) node [below] {$p$} ;

	\begin{scope}[shift=({0.6+5,0.75}),scale=0.8]
	\draw  [very thick, ->,brown]  (0,0) arc (120:60:1) ;
	\end{scope}
	\draw (6,1.5) node [below] {$\times$} ;
	
	\begin{scope}[shift=({0.4+9,0.75}),scale=0.8]
	\draw  [very thick, ->,brown]  (0,0) arc (60:120:1) ;
	\end{scope}
	\draw (9,1.5) node [below] {$\times$} ;

	\foreach \i in {1,...,11}
	{
		\draw (\i,0) -- (\i,0.1) ;
	}
	
	\foreach \i in {0,3,5,6,8,9}
	{
		\node at (\i+0.5,0.5){};
		\draw [very thick,fill=red] (\i+0.5,0.3) circle (7pt);
	}
	
	\draw [very thick, <->] (0,0) -- (12,0);
	\draw (12.5,0.1) node {$$} [anchor=north] ;

	\draw (6,-0.5) node [below] {$(b)$} ;

	\end{scope}

	\begin{scope}[shift=({7,-5})]
	\begin{scope}[shift=({-0.5,0.75}),scale=0.8]
	\draw  [very thick, ->,brown]  (0,0) arc (120:60:1) ;
	\end{scope}
	\draw (0,1.5) node [below] {$\alpha$} ;
	\begin{scope}[shift=({-0.5,-0.3}),scale=0.8]
	\draw  [very thick, <-,brown]  (0,0) arc (60:120:-1) ;
	\end{scope}
	\draw (0,-0.5) node [below] {$\delta$} ;
	
	\draw (-1.5,0.9) node [below] {$Left$} ;
	\draw (-1.5,0.3) node [below] {$Res.$} ;
	
	\begin{scope}[shift=({-0.5+12,0.75}),scale=0.8]
	\draw  [very thick, ->,brown]  (0,0) arc (120:60:1) ;
	\end{scope}
	\draw (0+12,1.6) node [below] {$\beta$} ;
	\begin{scope}[shift=({-0.5+12,-0.3}),scale=0.8]
	\draw  [very thick, <-,brown]  (0,0) arc (60:120:-1) ;
	\end{scope}
	\draw (0+12,-0.5) node [below] {$\gamma$} ;
	
	\draw (-1.5+15,0.9) node [below] {$Right$} ;
	\draw (-1.5+15,0.3) node [below] {$Res.$} ;

	\draw (6,-1) node [below] {$(c)$} ;

	\foreach \i in {0,...,12}
	{
		\draw (\i,0) -- (\i,0.1) ;
	}
	
	\foreach \i in {0,3,5,6,8,9}
	{
		\node at (\i+0.5,0.5){};
		\draw [very thick,fill=red] (\i+0.5,0.3) circle (7pt);
	}
	
	\draw [very thick] (0,0) -- (12,0);
	\draw (12.5,0.1) node {$$} [anchor=north] ;
	\end{scope}

	\end{tikzpicture}
	
	\caption{Different boundary conditions for the exclusion process. (a) Periodic boundary condition where the sites are identifiable to $\mathbb{Z}/L\mathbb{Z}$, $L$ being the number of sites. (b) ASEP on the line; the lattice is $\mathbb{Z}$. (c) Open boundary conditions, where we have a finite number of site. The system is coupled to a reservoir on the left where particles can hop inside at a rate $\alpha$ if the first site is empty. If it is not, then the occupying particle can escape the system at a rate $\delta$. Similar mechanism occurs on the right but with different rates.}
	\label{boundaries}
\end{figure}
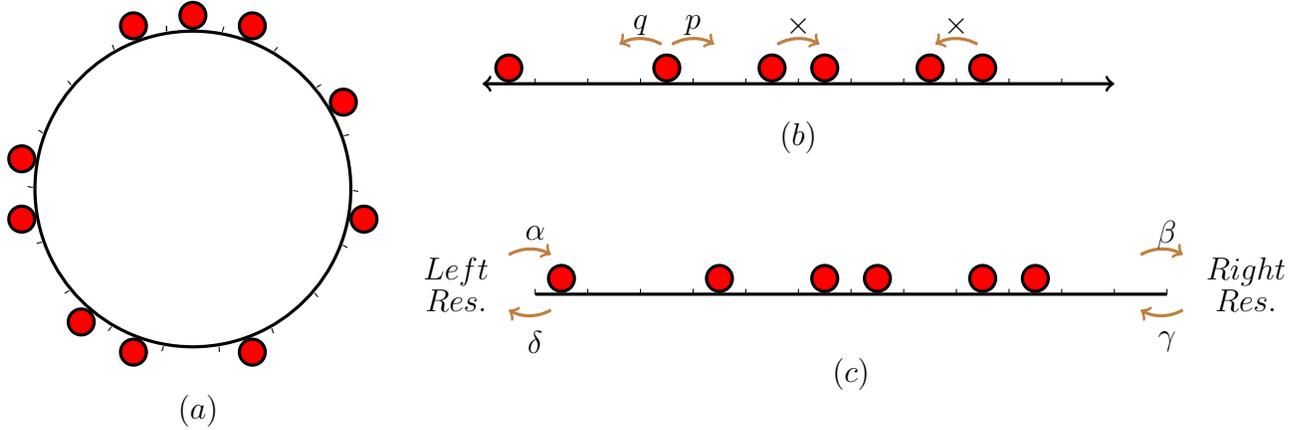

Another interesting interpretation of TASEP is in terms of queuing theory: The particles can be though of as servers and the voids as clients, so each server has a queue of clients in front of it. When a particle jumps, a client is served, this client will queue up in the queue belonging to the following server, and waits again for its turn. This image was exploited by \cite{spitzer1970interaction} who found the invariant measure for TASEP. More details are in the introduction of chapter 5.

Different types of boundary conditions are possible for ASEP, each has its own interest. They are illustrated in figure \ref{boundaries}. Let's look at the most classical properties ASEP on the ring, and TASEP with open boundary conditions.

\subsubsection*{ASEP on the ring}
The periodic boundary condition is the simplest. it's almost a trivial, yet pedagogical exercise to find the stationary state for ASEP with periodic boundary conditions. Consider a configuration composed of $l$ blocks of particles, where a block is a set of adjacent particles surrounded by voids. The system can leave or join the configuration only by the front or the backs of a block:
\begin{center}
	\begin{tikzpicture}

	\draw (2.4,0) node {$ \circ \bullet \bullet \bullet \circ $} [anchor=north] ;
	
	\draw [->] (3,-0.3) -- (3.5,-1.2);
	\draw (3.25,-0.75) node [left] {$q$} ;
	
	\draw [<-] (3.2,-0.3) -- (3.7,-1.2);
	\draw (3.45,-0.75) node [right] {$p$} ;

	\draw (1,-1.5) node {$ \circ \bullet \bullet \circ \bullet $} [anchor=east] ;
	
	\draw (4,-1.5) node {$ \bullet \circ \bullet \bullet \circ $} [anchor=north] ;

\draw [->] (3-1 - 0.2,-0.3) -- (2.5-1- 0.2,-1.2);
\draw (3.25-1.5- 0.2,-0.75) node [left] {$p$} ;

\draw [<-] (3.2-1- 0.2,-0.3) -- (2.7-1- 0.2,-1.2);
\draw (3.45-1.5- 0.2,-0.75) node [right] {$q$} ;
	
	\end{tikzpicture}
\end{center}
Now it's easy to understand that if we chose a uniform probability distribution for the configurations (each configuration has a probability $P_{eq}$), then each of the escaping rate and the entering rate will be equal to $l(p+q)P_{eq}$ which leads to a stationary system. If there are $M$ particles and $N$ sites, each configuration will have the probability: $P_{eq}= 1/\binom{N}{M}$. Now the current can be found exactly by choosing a site and counting the number of configurations such that this site is occupied and followed by a void or the other way around:

$$J = p E(  \bullet  \circ) - q E( \circ \bullet ) = (p-q) \dfrac{\binom{N-2}{M-1}}{\binom{N}{M}}
= (p-q) \frac{M}{N}(\frac{M-N}{M-1}) 
\rightarrow (p-q) \rho(1-\rho) $$
where the limit is taken for infinite $N$ and $M$ and keeping a fixed ratio $\rho := \frac{M}{N}$ which is the average density. We notice that this expression in the limit of a large system is the same as one obtained by a mean field. We will briefly see in chapter 5 that this is due to the fact that the product measure is invariant for ASEP in an infinite system.

\subsubsection*{Hydrodynamic behaviour of TASEP}
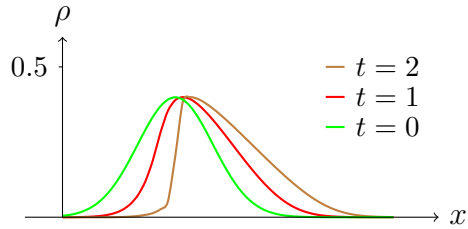
\begin{figure}[h!]
	\centering
\begin{tikzpicture}[xscale=0.5, yscale=4]
\draw [red, thick] plot [,smooth] coordinates {
	(-2.999999999999958,-4.804405523179467E-5)   
	(-2.7999999999999576,9.116331538308122E-5)   
	(-2.5999999999999552,3.81355108467048E-4)   
	(-2.3999999999999533,9.588328072964121E-4)   
	(-2.199999999999953,0.0020650605972153675)   
	(-1.999999999999953,0.004112266200276614)   
	(-1.799999999999951,0.0077842073976911045)   
	(-1.599999999999949,0.014188463029671805)   
	(-1.3999999999999488,0.025075645197769232)   
	(-1.1999999999999469,0.0431259782060273)   
	(-0.9999999999999449,0.07228241705764889)   
	(-0.7999999999999439,0.1181770670217098)   
	(-0.5999999999999428,0.1883066135589916)   
	(-0.39999999999994174,0.27934496530875236)   
	(-0.19999999999994067,0.3530070408066131)   
	(6.217248937900877E-14,0.3895891468250185)   
	(0.20000000000006501,0.3993345374170186)   
	(0.4000000000000661,0.39253366051276967)   
	(0.6000000000000689,0.3753761539876034)   
	(0.8000000000000718,0.3516422455661045)   
	(1.0000000000000746,0.32371454419203494)   
	(1.2000000000000774,0.29315281295209356)   
	(1.4000000000000785,0.2611000676817944)   
	(1.6000000000000814,0.22848089221165405)   
	(1.8000000000000842,0.1960945707017437)   
	(2.0000000000000857,0.16467875436634383)   
	(2.200000000000088,0.13494176397275345)   
	(2.400000000000091,0.1075541322498946)   
	(2.600000000000094,0.08310420147624793)   
	(2.800000000000096,0.06203562504225658)   
	(3.0000000000000977,0.04458905215336793)   
	(3.2000000000001005,0.030766612973016055)   
	(3.4000000000001034,0.020331647508529803)   
	(3.600000000000105,0.012849297560498926)   
	(3.8000000000001073,0.007762666810924639)   
	(4.00000000000011,0.00448501936817513)  
	(4.200000000000113,0.0024808402807798913)   
	(4.400000000000116,0.0013155023735260066)   
	(4.600000000000117,6.696069693837132E-4)   
	(4.80000000000012,3.275593022178478E-4)   
	(5.000000000000123,1.5413812627950867E-4)   
	(5.200000000000124,6.982235432465106E-5)   
	(5.4000000000001265,3.0463558028926056E-5)   
	(5.600000000000129,1.280709554189528E-5)   
	(5.800000000000132,5.189799253279892E-6) };

\draw [brown, thick] plot [smooth] coordinates {
	(-2.999999999999958,-1.508348970519064E-4) 
	(-2.7999999999999576,-1.8755518573105196E-4)  
	(-2.5999999999999552,-2.2711933936328044E-4)  
	(-2.3999999999999533,-2.617711036017571E-4)  
	(-2.199999999999953,-2.759717742313783E-4)  
	(-1.999999999999953,-2.4127890779174728E-4)
	(-1.799999999999951,-1.0840073105727276E-4)
	(-1.599999999999949,2.0456449334522634E-4) 
	(-1.3999999999999488,8.296129652081002E-4) 
	(-1.1999999999999469,0.001979144853100037) 
	(-0.9999999999999449,0.004013799332930589) 
	(-0.7999999999999439,0.00757895208925731) 
	(-0.5999999999999428,0.013416712737186169) 
	(-0.39999999999994174,0.025950293942266948) 
	(-0.19999999999994067,0.051177764495953584) 
	(6.217248937900877E-14,0.20431882520540692)  
	(0.20000000000006501,0.38447409372663066) 
	(0.4000000000000661,0.3989799412310865) 
	(0.6000000000000689,0.39323573148077207)  
	(0.8000000000000718,0.3794709395183348) 
	(1.0000000000000746,0.36132891075739887) 
	(1.2000000000000774,0.3406054068376967) 
	(1.4000000000000785,0.31820640643158893) 
	(1.6000000000000814,0.2947143594681802) 
	(1.8000000000000842,0.27056778752642424)  
	(2.0000000000000857,0.24610237161846565)  
	(2.200000000000088,0.2215859107055851) 
	(2.400000000000091,0.19724709858781148) 
	(2.600000000000094,0.17330102896079053)  
	(2.800000000000096,0.14997118093151338) 
	(3.0000000000000977,0.1275053090603495) 
	(3.2000000000001005,0.10618273796279147) 
	(3.4000000000001034,0.08631164979240417)  
	(3.600000000000105,0.06821477716133154)  
	(3.8000000000001073,0.05220058627153642)  
	(4.00000000000011,0.038518619327417805)  
	(4.200000000000113,0.027305889328415812)  
	(4.400000000000116,0.01854265753869263) 
	(4.600000000000117,0.012039974021496305) 
	(4.80000000000012,0.007470187744498799) 
	(5.000000000000123,0.004430585702299693)  
	(5.200000000000124,0.0025149386793250805)  
	(5.4000000000001265,0.0013684958726963231)  
	(5.600000000000129,7.151418245243229E-4) 
	(5.800000000000132,3.595253199349741E-4) } ;

\draw [green, thick] plot [smooth] coordinates {
	(-2.999999999999958,0.004431796766043165)   
	(-2.7999999999999576,0.007915365735690613)   
	(-2.5999999999999552,0.013582833500930466)   
	(-2.3999999999999533,0.022394326766328326)   
	(-2.199999999999953,0.035474304510619614)   
	(-1.999999999999953,0.05399058240865261)   
	(-1.799999999999951,0.07894967990556752)   
	(-1.599999999999949,0.11092028139967558)   
	(-1.3999999999999488,0.14972687620740827)   
	(-1.1999999999999469,0.19418548198819746)   
	(-0.9999999999999449,0.24197022182808786)   
	(-0.7999999999999439,0.2896911599355672)   
	(-0.5999999999999428,0.3332243339517981)   
	(-0.39999999999994174,0.3682699837242945)   
	(-0.19999999999994067,0.39104262520343314)   
	(6.217248937900877E-14,0.398942280401252)   
	(0.20000000000006501,0.39104276274695354)   
	(0.4000000000000661,0.36827029688131896)   
	(0.6000000000000689,0.3332248718299748)   
	(0.8000000000000718,0.28969194558478967)   
	(1.0000000000000746,0.2419712272072691)   
	(1.2000000000000774,0.1941866279757425)   
	(1.4000000000000785,0.14972805506272774)   
	(1.6000000000000814,0.11092138795929667)   
	(1.8000000000000842,0.07895063669750885)   
	(2.0000000000000857,0.053991350619756594)   
	(2.200000000000088,0.035474881184094316)   
	(2.400000000000091,0.02239473382543012)   
	(2.600000000000094,0.013583104968126185)   
	(2.800000000000096,0.007915537431517104)   
	(3.0000000000000977,0.004431900058689275)   
	(3.2000000000001005,0.002384117821648202)   
	(3.4000000000001034,0.0012322353892572514)   
	(3.600000000000105,6.11910421798866E-4)   
	(3.8000000000001073,2.9195117882098014E-4)   
	(4.00000000000011,1.3383226477647217E-4)   
	(4.200000000000113,5.8944003841670956E-5)   
	(4.400000000000116,2.494288290383882E-5)   
	(4.600000000000117,1.0141025450032537E-5)   
	(4.80000000000012,3.961369061608097E-6)   
	(5.000000000000123,1.486746568036669E-6)   
	(5.200000000000124,5.361135556374396E-7)   
	(5.4000000000001265,1.8573974061118973E-7)   
	(5.600000000000129,6.182741380088322E-8)   
	(5.800000000000132,1.9773589916727864E-8)} ;

\draw [ ->] (-4,0) -- (7,0) node[right] {$x$};

\draw [ ->] (-3,0) -- (-3,0.6) node[above] {$\rho$};

\draw (-3,0.5) -- (-3.1,0.5) node[left] {\small 0.5};


\draw [brown, thick] (4,0.5) -- (4.5,0.5) node[right, black] {\small $t=2$};
\draw [red, thick] (4,0.5-0.1) -- (4.5,0.5-0.1) node[right, black] {\small $t=1$};
\draw [green, thick] (4,0.5-0.2) -- (4.5,0.5-0.2) node[right, black] {\small $t=0$};
\end{tikzpicture}
\caption{Illustation of the formation of the shocks in Burgers equation as a result of the group velocity $v = 1 - 2 \rho$ }
\label{fig:shockformation}
\end{figure}

Now imagine a system with an average local coarse-grained density changing over space and time $\rho(x,t)$, regardless of the boundaries,consider the TASEP case, we can write a conservation equation associated with the previous expression of the current,
$$\partial_{t} \rho + (1-2 \rho) \partial_{x} \rho = 0$$
This equation is called the non-viscous Burgers equation. Its more precise meaning will be given later in this introduction. However, note that $1-2 \rho$ expresses the speed of the front wave around the density $\rho$. Note that it is a decreasing function of the density. If we have an increasing initial profile of density over space, then the upper parts will move faster than the lower parts, creating an even steeper profile, till we finally reach a discontinuous profile that is called a shock, cite \ref{fig:shockformation}. This shock is not static, if the density on its left is $\rho^{L}$ and on its right is $\rho^{R}$ then the speed of the shock is given $1-\rho^{L} - \rho^{R}$, as we will see in chapter 2. On the other hand, if the initial profile is decreasing as a function of the space, then its slope will get even smaller, and the solution will stay regular, more details will be provided in chapter 2.

\subsubsection*{TASEP with open boundaries}

\begin{figure}[h!]
	\centering
	\begin{tikzpicture}[thick, scale=0.8]
	\draw[thick,->] (0,0) -- (7,0) node[anchor=north] {$\alpha$};
	\draw[thick,->] (0,0) -- (0,7) node[anchor=south ]{$\beta$};	
	
	\draw (3 cm,1pt) -- (3 cm,-1pt) node[anchor=north] {$\frac{1}{2}$};
	
	\draw (6 cm,1pt) -- (6 cm,-1pt) node[anchor=north] {$1$};

	\draw (1pt,6 cm) -- (-1pt,6 cm) node[anchor=east] {$1$};
	
	\draw [very thick, orange]  (3,3) -- (6,3);
	\draw  (0,6) -- (6,6);
	\draw [very thick, orange] (3,6) -- (3,3);
	\draw [very thick, blue] (0,0) -- (3,3);
	\draw (6,0) -- (6,6);
	
	\draw (5.7,1) node[anchor=east] {};
	
	\draw (4.5,5) node[anchor=north] {MC};
	
	\draw (1.5,4) node[anchor=north] {LD};
	
	\draw (4,1.9) node[anchor=north] {HD};

	\end{tikzpicture}
	\caption{Boundary induced phase diagram of TASEP. LD: Low-density phase. HD: high-density phase. MC: Maximal current phase. The blue line represents boundaries where a phase order phase transition occurs. The orange line represents a second-order phase transition. The bulk density can be regarded as the order parameter}
	\label{fig:TASEPopenboundaries}
\end{figure}
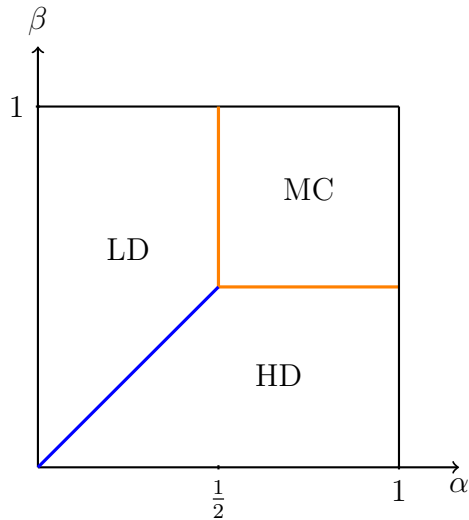

Consider TASEP with open boundaries where particles can hop inside the system from the left at rate $\alpha$ if the first site is empty, and can leave the system from the right at rate $\beta$. The left boundary behaves as if has a density $\rho^{R} = \alpha$, and the right boundary behaves as if it has a density $\rho^{L} = 1- \beta$. Now, it is possible to sketch most of the behavior of the system using a heuristic hydrodynamic approach based on the front wave speed $ v  = 1-2\rho$. Note first that if $\alpha<\frac{1}{2}$ then $v^{L}  = 1-2\alpha>0$, so there is a kinetic wave at density $\alpha$ trying to penetrate the system from the left. If $\beta<\frac{1}{2}$ then $v^{R}  = 1-2(1-\beta) = 2\beta -1 <0$, so now the kinetic wave of density $1-\beta$ is trying to enter from the right. Now, we can distinguish four cases, figure \ref{fig:TASEPopenboundaries}

\begin{itemize}
\item $\alpha<\frac{1}{2}$ and $\beta>\frac{1}{2}$, only the wave from the left is entering the system, and it will reach the bulk, so we have a system dominated by a density $\alpha<\frac{1}{2}$, this phase is called a low-density phase (LD)
\item $\alpha>\frac{1}{2}$ and $\beta<\frac{1}{2}$, the opposite of the previous situation, the bulk density will be $1-\beta>\frac{1}{2}$. This is called a high-density phase (HD)
\item $\alpha<\frac{1}{2}$ and $\beta<\frac{1}{2}$, both of kinetic waves are entering the system, so will create a shock that moves at a speed $1- \rho^{L} - \rho^{R} = \beta-\alpha$. If this speed is positive then the left boundary dominates the bulk, extending the low-density phase. Otherwise, the right boundary wins and the system is in the high-density phase.
\item $\alpha>\frac{1}{2}$ and $\beta>\frac{1}{2}$, then both of the waves are leaving the system, creating a phase where the current is maximal (MC) and the bulk density is $\frac{1}{2}$
\end{itemize}
This qualitative hydrodynamic approach has been confirmed by an exact solution \cite{schutz1993phase} by solving recursion relations of the probability profile on the size of the system.
The boundary induced phase transition for TASEP is one of the simplest for driven diffusive systems, and it paved the way for developing a more general principle describing the boundary-induced phase transitions of any system with a single driven quantity \cite{krug1991boundary}  \cite{popkov1999steady}, \cite{hager2001minimal}, \cite{katz_nonequilibrium_1984}. We will see how it will be generalized in chapter 4 for systems with multiple coupled driven quantities. 

\subsubsection{A brief historical perspective:}
Let's now stop at the most prominent stations during the lifetime of the model that was an inspiration to our work:

\begin{enumerate}
	
	\item In 1968,  MacDonald, Gibbs, and Pipkin first proposed ASEP \cite{macdonald1968kinetics}in the context of transport in biology modeling the situation of multiple enzymes copying sequentially from the same DNA template. They actually introduced a more general version of ASEP where the exclusion rule extends over $L$ neighboring sites rather than just one, so particles can't have a distance less than $L$ \footnote{Or equivalently, as the original formulation, they are not particles, but segments of length $L$}. They found using a mean-field analysis, the expression of the current for a uniform $\rho$ density system. In addition, they sketched the main features of the behavior of ASEP with open boundaries using the hydrodynamic approach.
	
	\item In 1978, Alexander and Holstein \cite{alexander1978lattice} mapped the master equation for SSEP to the Heisenberg spin chain. Since this spin chain was diagonalized exactly by Bethe in 1931, this sparked the interest of the integrability community in particle systems. The mapping was later extended to various other one-dimensional reaction-diffusion processes. Check \cite{alcaraz1994reaction} for an early review which highlighted the underlying Heck algebra that is common among the evolution operators of that family of models. In 1992 Gwa and Spohn \cite{gwa1992bethe} mapped ASEP to XXZ spin chain, which allowed them to diagonalize the Markov matrix using Bethe Ansatz and to estimate the relaxation time. More details will be provided in chapter 3.

	\item In 1981, Rost \cite{rost1981non} noticed that if time and space are scaled in the same way (in other words, you compress the space and accelerate the time with the same large factor), the corresponding density profile converges to a deterministic limit shape given by Burgers equation.  The limit shape is properly defined for the height function through a hydrodynamic scaling:
	\begin{equation}
		\rho(x,t) = \partial_{x} \lim_{\epsilon \rightarrow 0} \epsilon h(x \epsilon^{-1},t\epsilon^{-1}),
	\end{equation}where the macroscopic density is the physical solution of Burgers equation.
\footnote{ As we will see in the next chapter, a weak form of Burgers equation admits unstable solutions that we refer to as non-physical}
Although this result is true for any initial condition, Rost proved it with a step initial profile, all negative sites are occupied, and all positive sites are empty. This initial profile plays a role comparable to that of quench in quantum out-of-equilibrium systems. The limit shape with this initial condition is:
	\begin{equation}
\rho(x,t) =
		\begin{cases}
		1 \quad \text{if} \quad x < -t \\
		\frac{1}{2} (1-\frac{x}{t}) \quad \text{if} \quad -t < x < t \\
		0 \quad \text{if} \quad x > t
		\end{cases}
	\end{equation}
And the height will have a limit shape:
\begin{equation}
\lim_{t \rightarrow \infty} \frac{h(vt,t)}{t} =
\begin{cases}
|v| \quad \quad \quad \quad \text{if} \quad \quad |v| < 1 \\
\frac{1}{2} (v^{2}+1) \quad \text{if} \quad -1 < v < 1 \\
\end{cases}
\end{equation}
Although the Burgers equation existed much earlier, it was the first time the exclusion process was proposed as a microscopic description of the Burgers equation.

\item In 1991, P.A. Ferrari, while studying the shocks fluctuation of ASEP \cite{ferrari1992shocks}, introduces a second class particle, this particle jumps as a normal particle when the following site is empty: $20 \rightarrow 02$, but the normal particle (named first class) see it as void, and can thus swap with it: $12 \rightarrow 21$. The second-class particle can't overtake the first-class particle, hence the terminology. This particle was introduced as a means to identify microscopically the shock. It was inspired by the basic coupling technique introduced by Ligget \cite{liggett1976coupling} for TASEP. In the same year Ferrari, Kipnis and Saada proved that if a second-class particle is added to the origin of a rarefaction fan, it will choose a random asymptotic speed within the available ones with a uniform measure. Soon the second-class particle attracts the attention of a wider audience. In 1993, Derrida, Janowsky, Lebowitz, and Speer \cite{derrida1993exact2}  determine the shock profile as seen from the perspective of a second-class particle. Not much later, the second-class particle acquires an interest on its own besides its role as a theoretical mean. In 1996, Derrida \cite{derrida1996statphys} and Mallick \cite{mallick1996shocks} generalized this concept into a defect or impurity, which is a second-class particle that can jump with an arbitrary hopping rate and can be taken over with another arbitrary rate. This was the birth of a new model: the two species TASEP.

\item In 1997 Schütz \cite{schutz1997exact}
obtained for TASEP on the infinite line, an exact expression for the conditional probability $ P(x_{1},...,x_{N};t|y_{1},...,y_{N};0)$ of $N$ particles being at positions $\{x_{1},...,x_{N}\}$ at time $t$ given their initial positions $\{y_{1},...,y_{N}\}$ at time zero. The result is obtained using Bethe Ansatz and was expressed as an $N \times N$ determinant. This was an early result connecting TASEP to the domain of integrable probability. Tracy and Widom generalized it to ASEP  \cite{tracy2008integral} \cite{tracy2008fredholm}. Latter Tasep with second class particles was treated by
Chatterjee  and Schütz \cite{chatterjee2010determinant}. This will be reviewed and extended in chapter 3.
\item In 1993, the exact phase diagram for of TASEP with open boundaries was derived in two independant papers: Schütz and Domany \cite{schutz1993phase} solved recursion relations on the size of the system for the stationary state, allowing its explicit expression, and discussed the phase diagram in terms of the dynamics a domain wall. 
The second paper is \cite{derrida1993exact} where Derrida, Evans, Hakim and Pasquier determined this stationary state using a Matrix Product Ansatz (MPA) formulation. 
This opened the door for a series of cases where a non-equilibrium steady state is expressed in a matrix product form. For a pedagogical review check \cite{blythe2007nonequilibrium}. A brief explanation of MPA will be provided in chapter 5.
	
\item In 1999, Johansson \cite{johansson2000shape} revealed a connection to Random Matrix Theory (RMT) that triggered an impressive quantity of subsequent investigations. This is important to out-of-equilibrium statistical physics in particular because RMT has been a gold mine for universal behaviors. Let's state the main result:

\begin{equation}\label{TW}
	\lim_{t \rightarrow \infty}
	Prob
	\big(
	h(v t, t) > 
	\underbrace{\frac{1+ v^{2}}{2} t}_\text{Limit shape} 
	- s
	\underbrace{\frac{(1- v^{2})^{2/3}}{2^{1/3}}
	t^{1/3}}_\text{Fluctuations} 
	\big)
	= F_{GUE}(s)
\end{equation}

where $F_{GUE}$ is the cumulative Tracy–Widom distribution, precisely, it is the distribution of the rescaled largest eigenvalue $\lambda_{max}$ of random matrix sampled from the Unitary Gaussian Ensemble. If $n\times n$ is the size of the matrix, this eigenvalue grows as $\sqrt{2n}$ and fluctuates with a standard deviation of $n^{-\frac{1}{6}}$. Then we have: $ F_{GUE}(s) := \lim_{n \rightarrow \infty} Prob( (\lambda_{max} -\sqrt{2n}) \sqrt{2} n^{\frac{1}{6}} < s )  $.
The term $\frac{1+ v^{2}}{2}$ represents the speed of the growth. The most important information in that equation is the exponent in $t^{1/3}$. It has been previously conjectured that this growth model belongs to the KPZ universality class, and thus fluctuates as $t^{1/3}$ however, it was the first time this was proven and the only model for which it was proven rigorously. To make sure that \ref{TW} is appreciated correctly, one can compare to the central limit theorem, where $t^{1/3}$ plays the role of $t^{1/2}$ and $F_{GUE}$ plays the role of the integral of a Gaussian. The poof of the previous result is based on combinatorics, there is a correspondence with the problem of the distribution of the length of the longest
increasing subsequence in a random permutation that has the same limiting distribution \cite{baik1999distribution}.

\end{enumerate}

\subsubsection*{From single-species to multi-species TASEP}
\begin{figure}[h!]
	\centering
	\includegraphics[scale=0.5]{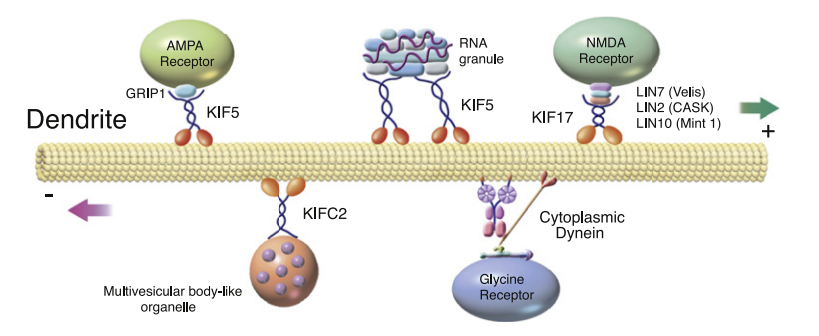} 
	\caption{Molecular transport on an axon, the main nerve fiber of a neuron, featuring different types of molecular motors,  adapted from  \cite{hirokawa2010molecular} }
	\label{fig:multi}
\end{figure}

The exclusion process, and TASEP in particular, is far from being only a mathematical model in 1D that theoretical physicists get excited about. It is used as a mesoscopic vehicle model in the field of traffic flow, for instance, the phase diagram for TASEP with open boundaries is celebrated in the traffic literature \cite{heibig1994existence}   \cite{chowdhury2000statistical}. However, this model is too idealized to be applied to real systems, It's rather only suited for one-lane, one-direction identical vehicles on a homogeneous freeway with no accidents. We all know how is it in daily situations.
A similar narrative can be made regarding intracellular transport in biology where the exclusion process is still relevant, for instance in molecular motor proteins moving along micro-tubules filaments \cite{parmeggiani2009non} \cite{chowdhury2005physics}, but again, real transport in cellular biology involves complex phenomena not counted by TASEP, such as the existence of multiple types of molecules transported on the same filament. See figure  \ref{fig:multi} for an example of transport molecular motors in neurons, studying this transport phenomenon has been relevant for the understanding of brain function, development, and disease \cite{hirokawa2010molecular}.

Hence the need for a model taking into account the presence of different types of particles that have different rates and that can swap between each other. In this model the exclusion rule is still valid as well as the local update. Different species swap with arbitrary entra-species rates: $$(\bullet \textcolor{red}{\bullet} \rightarrow \textcolor{red}{\bullet}  \bullet ) \; \text{with rate } \: \tau_{\bullet \textcolor{red}{\bullet} }  $$

For this model to be exactly solvable, some restrictions have to be obeyed by the rates as we will see in chapter 3. In the particular case of two species + void, it’s enough to assume hierarchy for the model to be exactly solvable, meaning that if we denote the particles as $0,1,2$, the only possible swaps are: $10 \rightarrow 01$, $20 \rightarrow 02$, $12 \rightarrow 21$ with arbitrary rates. This model was first introduced by Derrida \cite{derrida1996statphys} and Mallick \cite{mallick1996shocks} for a single second-class particle. Cantini later found the currents for an arbitrary number of defects \cite{cantini2008algebraic}.
This model, besides its applications, represents a much richer spectrum of phenomenology compared to TASEP, even when the simplest questions are asked. The objective of this dissertation is to be a building block for the bulk of knowledge for the 2-species exclusion process.

\subsubsection*{Novelties of this work:}
\begin{itemize}
\item Addressing the hydrodynamic behavior of two species TASEP. Although the rigorous convergence to the limit shape is a mathematically subtle question, we will rather make use of the integrability of the model that provides the currents and solves coupled conservation partial differential equations. The solutions are substantially more complex and rich than the TASEP one. This is the subject of our publication \cite{cantini2022hydrodynamic} which is included as in chapter 2.

\item We provide in chapter 3 a framework allowing the calculations of finite time conditional probability for the position of a finite number of particles of multiple species. The formalism can be though of as  a stochastic vertex model and leads explicit formulas in particular situations, generalizing the work of Schütz \cite{schutz1997exact} and Schütz et al. \cite{chatterjee2010determinant}

\item  We investigate in chapter 4 a method that allows to determine the steady state of a driven diffusive system with multiple driven coupled quantities, generalizing thus the extremum current principle proposed by Krug \cite{krug1991boundary}, Schütz and others \cite{popkov1999steady}, \cite{hager2001minimal}, \cite{katz_nonequilibrium_1984}. This method is operational even for models where the stationary measure is not a product measure, completing thus other method proposed in \cite{popkov2011hierarchy} \cite{popkov2004infinite} \cite{popkov2004hydrodynamic}. We apply this formalism to multiple particle models, 2-TASEP being one of them.   

\item In Chapter 5, we treat the question of the interaction between a defect particle and a density field for TASEP on the line with a Riemann initial condition. Besides the different phenomenology encountered, we expand the proof of the uniform density for the asymptotic speed for the case of a step initial profile.
\end{itemize}

For the unfamiliar reader, the first chapter is dedicated to providing all the necessary tools from the domain of conservation laws. This is required for the second chapter as well as the fourth one.

Each of chapters 3,4 and 5 will be the core of a future separate publication.

\chapter{Introduction to conservation laws}
\label{con}
In 1757, Leonhard Euler wrote in his memoir "Principes généraux du mouvement des fluides" an equation for the conservation of momentum and another for the mass. These equations were among the first partial differential equations ever written \cite{christodoulou2007euler,euler1757principes} and raised the initial problems that led later to the development of the domain of conservation laws with widespread applications in physics and chemistry. From a mathematical point of view, they are often qualified as hyperbolic due to their wavelike solutions. Yet, they are famous for having shocks singular solutions, requiring mostly an ad-hoc mathematical framework and placing them often in the last chapter of PDE textbooks. Despite being an old subject, research is still active in the domain \cite{bressan2011open}. Although the space multi-dimensional conservation laws are nowadays an exciting frontier of research, we restrict our presentation to 1D space, focusing mainly on the aspects related to the needs of the other chapters.

This essay starts with a discussion of scalar conservation laws in section \ref{Scalar conservation laws}, with an emphasis on the techniques that are generalizable to  non-scalar systems with multiple coupled conserved quantities. In particular, the stability conditions for the weak solution are discussed in details.  Burgers equation is used as a toy example for the scalar laws, the version used here is $u_{t} + u u_{x} = 0$ which is slightly simpler than the TASEP one but completely equivalent. A flavor of the vanishing viscosity method is given in section \ref{Hopf's treatment of the Burgers equation} during Hopf's treatment of the Burgers equation. This method will be relevant to chapter \ref{open} when dealing with multiple conservation laws in a system with open boundaries.
In section \ref{Hyperbolic Systems of Conservation Laws}, we review the most classical features of systems of conservation laws, this provides the background necessary for chapter \ref{hydro} where we solved a system with two conserved quantities resulting from a scaled two species TASEP model. 
We finish this section with a brief discussion of a particular family of conservation laws known as the Temple class, which has a curious connection with integrable models that we briefly investigate on the hydrodynamic level.

Despite that in section \ref{Relation to Hamilton-Jacobi equation}, we present the Hopf-Lax formula that allows formally to treat a wide class of initial conditions, the focus is later given only to the Riemann initial condition. The relevance of the Riemann problem can be compared to quenching in a quantum system; it's a popular procedure that provides insights into the dynamical behavior of the system and has been used in chapter \ref{hydro} for the 2-species TASEP.

This chapter is largely mathematical and mostly based on classical texts  \cite{evans2010partial, serre1999systems1, serre1999systems2}, \cite{serre1999systems1},
\cite{serre1999systems2}, \cite{lax2006hyperbolic},
\cite{dafermos2005hyperbolic},
\cite{bressan2013hyperbolic},
\cite{spinolonotes},
\cite{menonpde}.

\section{Scalar conservation laws}
\label{Scalar conservation laws}
\subsection{Introduction}

In this part we consider a single unknown function: $u(x,t): \mathbb{R \times R^{+}} \rightarrow \mathbb{R}  $ that represents a density satisfying a conservation laws with initial data at $t=0$:
\begin{equation}\label{key}
\begin{split}
& u_{t} + (f(u))_{x} = 0 \\
& u(x,0) = u_{0}(x)
\end{split}
\end{equation}
with $f$ in the class $\mathcal{C}^{1}$, representing the flux of $u$. We are concerned here in investigating the solutions of this problem in the most general setting. To give an initial flavor, although not representative of general solutions, let's start with the trivial case of a linear flux.
\subsubsection*{A linear flux:}
The most simple case is when $f(u) = c u $. The equation becomes: 
\begin{equation}\label{key}
u_{t} + c u_{x} = 0 
\end{equation}
This means that the directional derivative of $u$ in the direction $(1,c)$ is zero, so $u$ is constant over the lines $x = ct + x_{0}$:
$$u(ct+x_{0}, t) = u(x_{0}, 0) = u_{0}(x_{0})$$
With a change of variable we have:
\begin{equation}\label{key}
u(x,t) = u_{0}(x-ct)
\end{equation}
So the initial profile will be just moving at a constant speed  $c$.
Note that if we write the general equation in the form: $u_{t} + f'(u)u_{x} = 0 $ and consider an initial profile that is uniform with an infinitesimal perturbation around $\tilde{u}$ $u_{0}(x) \approx \tilde{u}$ then it will evolve translating with the speed $f'(\tilde{u})$, we call this speed, the speed of perturbations.

If the flux is non-linear, the differential equation is said to be quasi-linear (A fully non-linear equation requires a non-linearity of the highest derivative: i.e. $u_{x}$ or $u_{t}$). In the next paragraph, we remind a general method that is used not only for conservation laws but for a wider class of non-linear first order PDE.
\subsection{Method of characteristics}
The method of characteristics consists of partitioning the variables’ space into a family of curves where the PDE transforms into a system of ODEs (Ordinary differential equation) on the curves.
It is adapted for the general class of non-linear first order equations, i.e. equations of the form: $H(Du,u,\textbf{x})=0$ defined on an open domain $\textbf{x} \in U \subset \mathbb{R}^{n}$ and subject to a boundary condition $u = g$ on a curve $\Gamma \subset \partial U$. The characteristics are the three functions of a real parameter:
\begin{equation}\label{}
\begin{split}
& \textbf{x}(s) \\
&  z(s) :=  u(\textbf{x}(s)) \\
& \textbf{p}(s) := Du(\textbf{x}(s))
\end{split}
\end{equation}
Deriving $F$ with respect to $x_{i}$ gives:

\begin{equation}
\sum_{j} \frac{\partial H}{\partial p_{j}} u_{x_{j}x_{i}} + \frac{\partial H}{\partial z} u_{x_{i}} + \frac{\partial H}{\partial x_{i}} = 0
\end{equation}
We can identify the first term with $\dot{p}_{i}(s) = \sum_{j} u_{x_{i}x_{j}}\dot{x}_{j}(s)$ providing that we identify $\frac{\partial H}{\partial p_{j}}$ with $\dot{x}_{j}$. So finally, we have this system of ODE:

\begin{equation}\label{chara}
\begin{split}
& \dot{\textbf{x}} = D_{p}H \\
&\dot{ \textbf{p}} = - D_{x}H - \textbf{p} D_{z}H \\
&  \dot{z} =  \textbf{p}D_{p}H 
\end{split}
\end{equation}
(Note that if we forget about the third equation and the second term of the second equation, we get Hamilton-Jacobi equations. We will come back to this later). This equivalence between the PDE and the set of ODEs is formally valid for regular solutions $u \in \mathcal{C}^{2}$. Under this condition, the Cauchy problem of the ODE has a unique solution for sufficiently regular $H$ (Lipschitz) providing that the boundary condition is compatible with the characteristics.


\subsubsection{Application to the scalar conservation law}
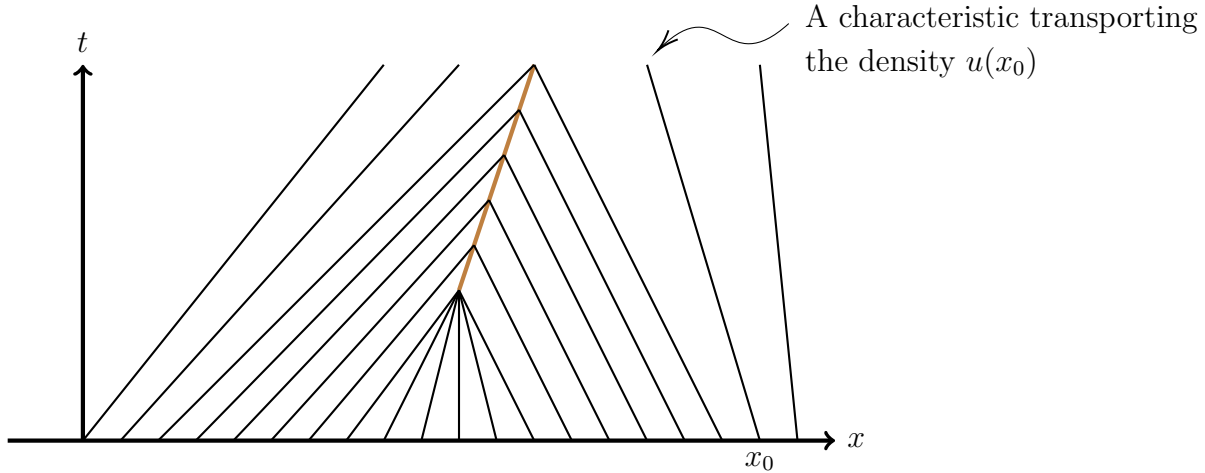
\begin{figure}[h!]
	\centering
	\begin{tikzpicture}
	
	\draw[->,ultra thick] (-1,0)--(10,0) node[right]{$x$};
	\draw[->,ultra thick] (0,0)--(0,5) node[above]{$t$};
	
	\draw[color = brown, ultra thick] (5,2)--(6,5);
	\draw[thick] (0,0)--(4,5);
	\draw[thick] (0.5,0)--(5,5);
	\draw[thick] (1,0)--(6,5);
	\draw[thick] (1.5,0)--(5.8,4.4);
	\draw[thick] (2,0)--(5.6,3.8);
	\draw[thick] (2.5,0)--(5.4,3.2);
	\draw[thick] (3,0)--(5.2,2.6);
	\draw[thick] (3.5,0)--(5,2);
	\draw[thick] (4,0)--(5,2);
	\draw[thick] (4.5,0)--(5,2);
	\draw[thick] (5,0)--(5,2);
	\draw[thick] (5.5,0)--(5,2);
	\draw[thick] (6,0)--(5,2);
	\draw[thick] (6.5,0)--(5.2,2.6);
	\draw[thick] (7,0)--(5.4,3.2);
	\draw[thick] (7.5,0)--(5.6,3.8);
	\draw[thick] (8,0)--(5.8,4.4);
	\draw[thick] (8.5,0)--(6,5);
	\draw[thick] (9,0)node[below]{$x_{0}$}--(7.5,5) ;
	\draw[thick] (9.5,0)--(9,5);
	
	\begin{scope}[x=0.75pt,y=0.75pt,yscale=-1,xscale=1, shift={(60,-300)}]
	
	\draw    (296,89.5) .. controls (263.5,119.05) and (262.04,71.95) .. (231.42,101.59) ;
	\draw [shift={(230,103)}, rotate = 314.56] [color={rgb, 255:red, 0; green, 0; blue, 0 }  ][line width=0.75]    (10.93,-3.29) .. controls (6.95,-1.4) and (3.31,-0.3) .. (0,0) .. controls (3.31,0.3) and (6.95,1.4) .. (10.93,3.29)   ;
	
	\draw (303,79) node [anchor=north west][inner sep=0.75pt]   [align=left] {A characteristic transporting
	};
	\draw (303,100) node [anchor=north west][inner sep=0.75pt]   [align=left] {the density $u(x_{0})$};
	
	\end{scope}
	
	\end{tikzpicture}
	
	\caption{Illustration of singularity formation through characteristic intersections}
	\label{shock formation}
\end{figure}

For our purpose, it's quite simple: $\textbf{x} =(x,t)$, $\textbf{p} = (u_{x}, u_{t})$, $H(u_{t}, u_{x},u,x,t) = u_{t} + f'(u)u_{x}$
The characteristics are:
\begin{equation}\label{key}
\begin{split}
& \dot{t} = 0 \\
&\dot{x} = f'(z) \\
&  \dot{z} =  0
\end{split}
\end{equation}
They form a closed system. We can obviously use the time as a parameter:
\begin{equation}\label{key}
\begin{split}
&\frac{dx}{dt}(t) = f'(u(x,t)) \\
&\frac{d}{dt}u(x(t),t) = 0
\end{split}
\end{equation}
So $u$ is constant all over the characteristics, and they are simply straight lines:
\begin{equation}
\begin{split}
& x(t) = f'(u_{0}(x_{0}))t + x_{0} \\
& u(x(t),t) = u_{0}(x_{0})
\end{split}
\end{equation}
If the initial condition is smooth then the solution at time $t>0$ is still smooth as far as the characteristics don't intersect, so we have a classical $\mathcal{C}^{1}$ solution for $0 \leq t <T $. If $f'$ is $\kappa_{1}$-Lipschitz and $u_{0}$ is $\kappa_{2}$-Lipschitz then the first intersection of characteristics will appear at $T = \kappa_{1}\kappa_{2}$.
At the points $(x,t)$ of the intersection of characteristics, the value of the solution is not defined since different values carried from different characteristics contradict. The limit of $u(x,t)$ at the point of intersection of characteristics depends on the path, so the solution forms a finite discontinuity, figure \ref{shock formation}.

\subsection{Weak solution, Rankine-Hugoniot condition}
Clearly, we need a weaker interpretation of the equation that takes into account discontinuous solutions. A possible way is to consider the equation in the distribution sense, so $u$ would be a distribution acting on a Schwartz space (a space of test functions $\phi(x,t)$ of the class $\mathcal{C}^{\infty}$ with compact support)

\begin{equation}
\int_{0}^{\infty} \int_{-\infty}^{\infty} (u_{t}(x,t) + f(u(x,t))_{x} ) \phi(x,t) dxdt = 0 
\end{equation}
The relevance of this writing is that it allows performing the integration by part:

\begin{equation}\label{int equa}
\int_{0}^{\infty} \int_{-\infty}^{\infty} u(x,t)\phi_{t}(x,t) + f(u(x,t))\phi_{x}(x,t)  dxdt = 0 
\end{equation}
This equation may be called the integral form of the conservation law. It doesn't impose a regularity restriction on the solutions. Functions verifying this equation are called weak solutions.

Let's consider now a situation where we have a solution $u(x,t)$ that is regular all over $\mathbb{R} \times \mathbb{R}^{+}$ except on some continuous path $\Gamma$ parameterized by   $x = s(t)$ (so we assume that there is a single finite discontinuity at each instant). We are interested in describing the behavior of this path. Let's assume as well the following limits exist:

\begin{equation}\label{key}
\begin{split}
& \lim_{x \xrightarrow{<} s(t)}u(x,t)  := u^{L}(t)  \\
&  \lim_{x \xrightarrow{>} s(t)}u(x,t)  := u^{R}(t) 
\end{split}
\end{equation}

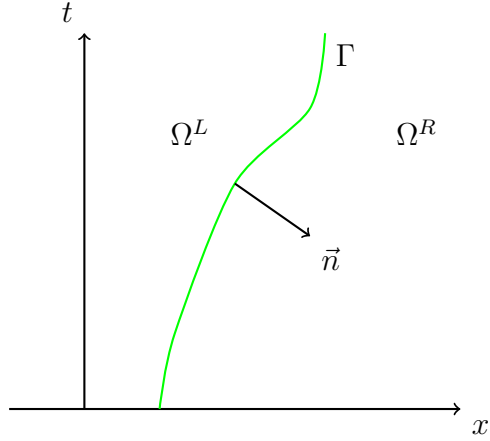
\begin{figure}[h!]
	\centering
	\begin{tikzpicture}
	\draw[thick,->] (0,0) -- (6,0) node[anchor=north west] {$x$};
	\draw[thick,->] (1,0) -- (1,5) node[anchor=south east] {$t$};
	\draw [thick,green] plot [smooth, tension=0.5] coordinates { (2,0) (2.2,1) (3,3) (4,4) (4.2,5)};
	\draw[thick,->] (3,3) -- (4,2.3) node[anchor=north west] {$\vec{n}$};
	\draw (2,4) node[anchor=north west] {$\Omega^{L}$};
	\draw (5,4) node[anchor=north west] {$\Omega^{R}$};
	\draw (4.2,5) node[anchor=north west] {$\Gamma$};
	\end{tikzpicture}
	\caption{A singular path propagating in the $(x,t)$ domain}
	\label{Rankin}
\end{figure}

The path divides the domain $\mathbb{R} \times \mathbb{R}^{+}$ into two subdomains, one located on its left $\Omega^{L}$ and another on its right $\Omega^{R}$. 
figure \ref{Rankin}

We can decompose eq. \eqref{int equa} into :

\begin{equation}
\int_{\mathbb{R} \times \mathbb{R}^{+}} u\phi_{t} + f(u)\phi_{x}  dxdt = \int_{\Omega^{R}} u\phi_{t} + f(u)\phi_{x}  dxdt + \int_{\Omega^{L}} u\phi_{t} + f(u)\phi_{x}  dxdt = 0
\end{equation}

Let's integrate by part the first term:

\begin{equation}
\int_{\Omega^{L}} u\phi_{t} + f(u)\phi_{x}  dxdt = \int_{\Omega^{L}} u_{t}\phi + f(u)_{x}\phi  dxdt - \int_{\partial \Omega^{L}} u\phi n^{t} + f(u)\phi n^{x}  dxdt \\
\end{equation}

\begin{equation}
= - \int_{\Gamma} (u^{L} n^{t} + f(u^{L})  n^{x} ) \phi dxdt
\end{equation}

Where $ (n^{x}, n^{t}) $ is a unit vector normal to the boundaries

We can treat the term on the right in a similar fashion except that we will have a minus sign from $ (n^{x}, n^{t}) $ since we will use the same normal vector as previously. so finally, we get:
\begin{equation}
\int_{\Gamma} ((u^{L} - u^{R}) n^{t} + (f(u^{L}) - f(u^{R}))  n^{x} ) \phi dxdt = 0
\end{equation}
Which means: 
\begin{equation}
(u^{L} - u^{R}) n^{t} + (f(u^{L}) - f(u^{R}))  n^{x} = 0
\end{equation}
knowing $ (n^{x}, n^{t}) $ allows us to have the tangent to $\Gamma$, which gives the derivative:
\begin{equation}
\frac{ds}{dt} = \frac{- n^{t}}{n^{x}}
\end{equation}
This is nothing but the speed of the shock, let's note it $\sigma$. So we reach the famous Rankine-Hugoniot formula:
\begin{equation}
(u^{L}(t) - u^{R}(t)) \sigma(t) = f(u^{L}(t)) - f(u^{R}(t))
\end{equation}
There is a simple way to grasp this identity, simply by imagining the shock as a level of water in a 2D tank that has a source and a sink. Each side represents the rate of filling of the tank expressed in two different ways. The problem with weak solutions is that they are not always unique as we will see in what follows.
\subsection{Non-unicity of weak solutions}
While the strong form of the conservation equation doesn't always have a solution, the weak form might have more than one solution with the same initial data. Let's give the Burgers equation as an example:
\begin{equation}
u_{t} + u u_{x} = 0
\end{equation}
With the initial condition:
\begin{equation}
u_{0}(x) = \mathds{1}_{x>0}(x)
\end{equation}
The flux for this equation is $f(u) = u^{2}/2$. It admits two weak solutions: a regular one, called a rarefaction fan:
\begin{equation}
u(x,t) = \frac{x}{t} \mathds{1}_{0<x<t}(x) + \mathds{1}_{x>t}(x)
\end{equation}
And a shock with a speed $\frac{1}{2}$
\begin{equation}\label{bad sh}
u(x,t) =  \mathds{1}_{x>\frac{t}{2}}(x)
\end{equation}
One can argue that the second solution is not stable. Consider for instance this small (in the sens of $\mathcal{L}^{1}$) perturbation of the initial profile:

\begin{equation}
u_{\epsilon}(x) = \frac{x}{\epsilon} \mathds{1}_{0<x<\epsilon}(x) + \mathds{1}_{x>\epsilon}(x)
\end{equation}
It's clear by looking at the characteristics that this profile will evolve in time like the first solution (the fan) and will thus divert from the shock solution corresponding to $\epsilon = 0$. We will see later more formal notions of stability of solutions.

If we now change the initial condition to this one:
\begin{equation}
u_{0}(x) = \mathds{1}_{x<0}(x)
\end{equation}
Then we have this weak solution:
\begin{equation}\label{good sh}
u(x,t) =  \mathds{1}_{x<\frac{t}{2}}(x)
\end{equation}
However, this is a stable one: if we apply a similar perturbation to the initial data as previously:
\begin{equation}
u_{\epsilon}(x) = \frac{\epsilon-x}{\epsilon} \mathds{1}_{0<x<\epsilon}(x) + \mathds{1}_{x<0}(x)
\end{equation}
then it is easy to understand that the effect of the perturbation vanishes quickly and the evolution will continue as a shock. One can describe this solution as physical in contrast to the non-physical previous one.
Formal criteria allowing to classify solutions according to their admissibility is the subject of much literature, often called admissibility conditions, or entropy conditions, we will revise some of them later.
\subsection{Hopf's treatment of the Burgers equation}
\label{Hopf's treatment of the Burgers equation}
The Burgers equation originates from fluid mechanics. It is the simplest non-trivial example of a scalar conservation law. A more realistic version of it includes the viscosity:
\begin{equation}\label{Bur}
u_{t} + u u_{x} = \nu u_{xx}
\end{equation}
Assume the initial data $u_{0}$ to be bounded. The viscous term expresses a dependence of the flux on the gradient of $u$ in addition to its dependence on $u$. In other words, it takes into account the diffusion. What Eberhard Hopf showed in his paper \cite{hopf1950partial} is that unlike the non-viscous Burgers equation, the viscous one does not suffer from a lack of unicity of weak solutions, and it's only when we set $\nu = 0$ where we can encounter this issue of multiple solutions. Among these multiple solutions, only one is actually resulting from taking the limit $\nu \rightarrow 0$. This sets a reasonable admissibility condition for the non-viscous Burgers equation. Solutions verifying this condition are often called viscous solutions, and the method is called: the vanishing viscosity method. 
Because its mathematical and physical importance, we will be synthesizing Hopf's paper.

Let first $U$ be an integral of $u$ (sometimes called the hight function as a reference to the growth process):
\begin{equation}
U(x,t) = \int_{0}^{x} u(y,t) dy
\end{equation}
Then:
\begin{equation}
U_{x} = u
\end{equation}
We can express Burgers equation in terms of $U$:
\begin{equation}
U_{t} = \int_{0}^{x} u_{t} dy = \int_{0}^{x} u u_{x} - \nu u_{xx} dy = \frac{1}{2} u^{2} - \nu u_{x}
\end{equation}
\begin{equation}\label{HJ}
U_{t} +  \frac{1}{2}U_{x}^{2} = \nu U_{xx}
\end{equation}
We will see how this equation can be understood as the Hamilton-Jacobi equation.
Let's consider the change of variable:
\begin{equation}
\phi = e^{-\frac{U}{2\nu}}
\end{equation}
With elementary operations, we can show that \eqref{HJ} get reduced to the heat equation:
\begin{equation}
\phi_{t} = \nu \phi_{xx}
\end{equation}
The solution of this equation is unique and obtained by a convolution of the initial condition: $\phi(x,0) = e^{\frac{-U(x,0)}{2 \nu}}$ with the solution of a Dicac initial condition:
\begin{equation}
\begin{split}
\phi(x,t) = (K_{0}(.,t) \ast \phi(.,0)) (x) & = \frac{1}{2 \sqrt{\pi \nu t}} \int_{\mathbb{R}} \exp (-\frac{(x-y)^{2}}{{4\nu t}}) \phi(y,0) dy \\ & =  \int_{\mathbb{R}} \exp (\frac{-G(x,y,t)}{2 \nu}) dy
\end{split}
\end{equation}
where:
\begin{equation}
K_{0}(x,t) = \frac{1}{2\sqrt{\pi \nu t }} e^{-\frac{x^{2}}{4 \nu t}}
\end{equation}
and:
\begin{equation}
G(x,y,t) = \frac{(x-y)^{2}}{2 t} + U_{0}(y)
\end{equation}
We are interested in solving equation \eqref{Bur}
\begin{equation}
u^{\nu} = -2 \nu \frac{\phi_{x}}{\phi} = \frac{\int_{\mathbb{R}} \frac{(x-y)}{{t}} \exp (\frac{-G}{2 \nu}) dy}{\int_{\mathbb{R}} \exp (\frac{-G}{2 \nu}) dy}
\end{equation}
\begin{equation}
= \frac{x}{t} - \frac{1}{t}\frac{\int_{\mathbb{R}} y  \exp (\frac{-G}{2 \nu}) dy}{\int_{\mathbb{R}} \exp (\frac{-G}{2 \nu}) dy}
\end{equation}
We are looking for the limit $  \lim_{\nu \rightarrow 0} u^{\nu}(x,t) $
If $G$ is regular ($\mathcal{C}^{2}$) with a unique minimum, we can use the saddle point method, by 
expanding $G$ to the second order in the neighborhood of the minimum. We get:
$$   \lim_{\nu \rightarrow 0} u^{\nu}(x,t) = \frac{x-y^{0}(x,t)}{t}  $$
Where $y^{0}(x,t)$ is the point where $G(x,y,t)$ reaches its minimum for the $y$ variable.

In case we have multiple points for which $G$ reaches its minimum then this limit is not defined, and we need a couple of extra tools.  Let's denote $y^{*}$ as the maximum and $y_{*}$ as the minimum of the set of points where $G$ reaches its minimum. One can show the following:
\begin{enumerate}
	\item $ x_{1}< x_{2} \implies y^{*}(x_{1},t) \leq y_{*}(x_{2},t) $
	\item $y_{*}(x-0,t) = y_{*}(x,t)$ and $y^{*}(x+0,t) = y^{*}(x,t)$
	\item  $y_{*}(\infty,t) = \infty$ and $y^{*}(-\infty,t) = -\infty$
\end{enumerate}

These properties can be proved rigorously, but we will instead contend with an intuitive understanding. 
First let's notice that $G(x,y,t)$ attains its minimum at the same values as the function :
\begin{equation}\label{G}
G(0,y,t) - \frac{x}{t} y
\end{equation}

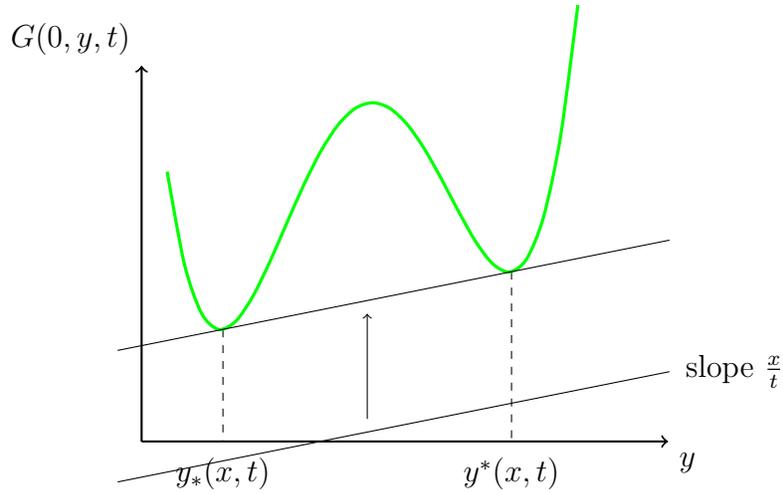
\begin{figure}[h!]
	\centering
	
	\begin{tikzpicture}
	
	\draw[thick,->] (-3,-1) -- (4,-1) node[anchor=north west] {$y$};
	\draw[thick,->] (-3,-1) -- (-3,4) node[anchor=south east] {$G(0,y,t)$};
	
	\draw [dashed,scale=0.7]  (-2.74 , 0.702004) -- (-2.74 , -1/0.7)  node[below=2pt] {$y_{*}(x,t)$};

	\draw [dashed,scale=0.7]  (2.74, 1.798) -- (2.74, -1/0.7)  node[below=2pt] {$y^{*}(x,t)$};

	\draw[scale=0.7, domain=-3.8:4, smooth, variable=\x, very thick ,green] plot ({\x}, { -(-\x)^2 + (-\x)^4/15 + \x/5 + 5});

	\draw[scale=0.7] (-2.74 - 2 , 0.702004 - 2*0.2) -- (2.74+3, 1.798+0.2*3);
	
	\draw[scale=0.7] (-2.74 - 2 , 0.702004 - 2*0.2 -2.5) -- (2.74+3, 1.798+0.2*3 - 2.5) node[right=2pt] {slope $\frac{x}{t}$};
	
	\draw[scale=0.7, ->] (0,-1) -- (0,1);
	\end{tikzpicture}
	
	\caption{Illustration of the behavior of the set of minimums of $G$. }
	
	\label{minG}
\end{figure}

Now we can understand the previous properties with the help of figure \ref{minG}

\paragraph{Theorem}
\begin{equation}\label{key}
\frac{x-y^{*}(x,t)}{t} = \limsup_{\nu \rightarrow 0} u^{\nu}(x,t) \leq \liminf_{\nu \rightarrow 0} u^{\nu}(x,t) =  \frac{x-y_{*}(x,t)}{t}
\end{equation}

\paragraph{Remarks}

\begin{itemize}
	\item This formula is valid even when G is not smooth.
	\item For every $x \in \mathbb{R}$ except on a countable set (forming the shocks) we have
	\begin{equation}\label{key}
	\lim_{\nu \rightarrow 0} u^{\nu}(x,t)  =  \frac{x-y^{*}(x,t)}{t} = \frac{x-y_{*}(x,t)}{t}
	\end{equation}
	
	\item If $G(0,y,t)$ is convex then $u(.,t)$ is continuous.

\end{itemize}

\paragraph{A first example}

Let's apply this formalism to find the solution of The Burgers equation for in initial step function:
\begin{equation}
u_{0}(x) = \mathds{1}_{x>0}(x)
\end{equation}
So we need to find where \eqref{G} attains its minimum:

\begin{equation}
G(0,y,t) - \frac{x}{t} y = \frac{y^{2}}{2t} + y\mathds{1}_{y>0}(y) - \frac{x}{t} y
\end{equation}

It is smooth everywhere except at $y=0$.
If $x<0$ then the minimum is at $y=x$.
If $x > t$ then the minimum is $x-t$.
\begin{equation}\label{key}
y^{*}(x,t) = y_{*}(x,t) =
\begin{cases}
x \quad if \quad x <0 \\
x-t \quad if \quad x > t \\
0 \quad if \quad 0< x <t \\
\end{cases}
\end{equation}
So the solution $ \frac{x-y^{*}}{t} $ becomes
\begin{equation}\label{key}
u(x,t) =
\begin{cases}
1 \quad if \quad x <0 \\
x-t \quad if \quad x > t \\
\frac{x}{t} \quad if \quad 0< x <t \\
\end{cases}
\end{equation}
Which is the expected solution.
\paragraph{A second example}
Let's consider the other classical initial profile:
\begin{equation}
u_{0}(x) = \mathds{1}_{x<0}(x)
\end{equation}

In this case we have:

\begin{equation}
G(0,y,t) = \frac{y^{2}}{2t} + y\mathds{1}_{y<0}(y)
\end{equation}

We can see from the figure that the locus of the minimum of $G(0,y,y) - x\frac{y}{t}$ will not be continuous with respect to $x$. The discontinuity can be found simply: $x = \frac{t}{2}$
\begin{equation}\label{key}
\begin{split}
&y^{*}(\frac{1}{2},t) = 1 \\
&y_{*}(\frac{1}{2},t) = 0
\end{split}
\end{equation}
\begin{equation}
y^{*}(x,t) = y_{*}(x,t) =
\begin{cases}
x \quad   \quad if  \quad x <\frac{t}{2} \\
x-t \quad if \quad x > \frac{t}{2} \\
\end{cases}
\end{equation}
And finally the expected solution
\begin{equation}
u(x,t) =
\begin{cases}
1 \quad if \quad x <\frac{t}{2} \\
0 \quad if \quad x > \frac{t}{2} \\
\end{cases}
\end{equation}
We notice that this formalism identifies and select naturally the stable solution of Burgers equation among the weak ones. We will be visiting in what follows more formal admissibility conditions that apply more generally.
\subsection{Kružkov entropy condition}
The work of Kružkov in the 70's \cite{kruvzkov1970first} represents a major development in the understanding of conservation laws.
Let $u$ be a smooth solution (so in the strong sense) to the problem:
\begin{equation}
\begin{split}
& u_{t} + (f(u))_{x} = 0 \\
& u(x,0) = u_{0}(x)
\end{split}
\end{equation}
with $f\in \mathcal{C}^{1}$, Let $\eta$ be a positive convex $\mathcal{C}^{1}$ function. The claim is that $\eta(u)$ will be a conserved quantity under the evolution of $u$. This is quite easy to show:
\begin{equation}
\eta(u)_{t} = \eta'(u) u_{t} = \eta'(u)(-(f(u))_{x}) = -\eta'(u)f^{'}(u) u_{x}
\end{equation}
So if we define $q$ so that:
\begin{equation}
q^{'}(u) =  \eta'(u)f^{'}(u)
\end{equation}
We get:
\begin{equation}
\eta(u)_{t} + q_{x} = 0
\end{equation}
$\eta(.)$ is called an \textbf{entropy function} and $q$ is the associated \textbf{entroy flux}.

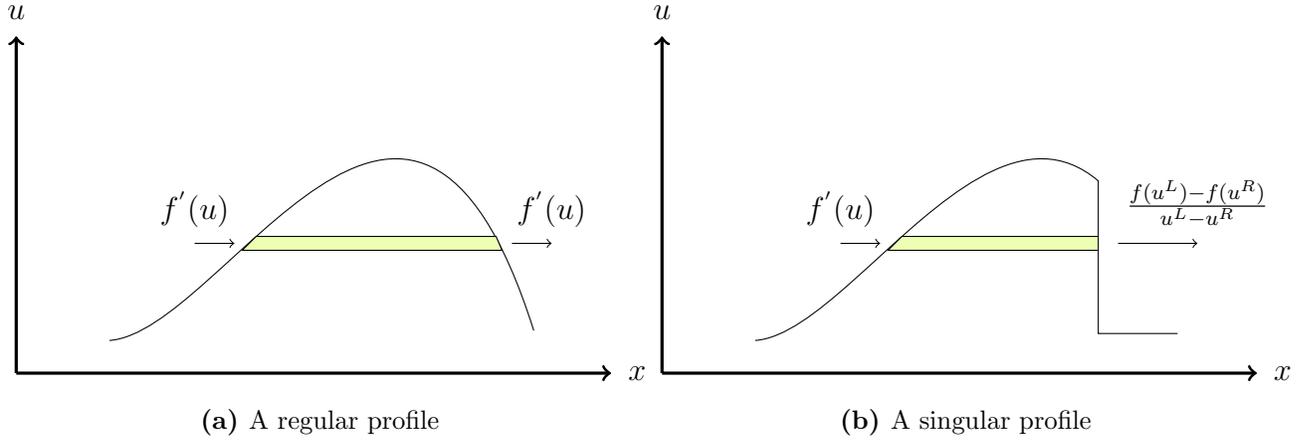
\begin{figure}[h!]
	\centering
	\begin{subfigure}{.52\textwidth}
		\centering
		
		\begin{tikzpicture}[x=0.75pt,y=0.75pt,yscale=-1,xscale=1]

		
		
		\draw [very thick, ->] (100,220)--(100, 50) node[above=2pt] {$u$};
		\draw [very thick, ->] (100,220)--(400,220) node[right=2pt] {$x$};
		
		\draw    (147,203.5) .. controls (205,199.5) and (299,2.5) .. (361,198.5) ;
		
		\draw [->]   (350,154.5) -- (370,154.5) node[above=2pt] {$f^{'}(u)$}; 
		
		\draw [->]   (190,154.5) node[above=2pt] {$f^{'}(u)$} -- (210,154.5)  ;
		
		\draw [fill = lime!30] (221,151) -- (342,151) --  (345,158) -- (214,158) -- (221,151) ;
		
		\end{tikzpicture}
		
		\caption{A regular profile}
		\label{fig:sub1}
	\end{subfigure}%
	\begin{subfigure}{.52\textwidth}
		\centering

		\begin{tikzpicture}[x=0.75pt,y=0.75pt,yscale=-1,xscale=1]
		
		\draw [very thick, ->] (100,220)--(100, 50) node[above=2pt] {$u$};
		\draw [very thick, ->] (100,220)--(400,220) node[right=2pt] {$x$};	
		\draw    (147,203.5) .. controls (205,199.5) and (299,2.5) .. (361,198.5) ;
		
		\draw [->]   (350,154.5) -- (370,154.5) node[above=2pt] {$f^{'}(u)$}; 
		
		\draw [->]   (190,154.5) node[above=2pt] {$f^{'}(u)$} -- (210,154.5)  ;
		
		\draw [fill = lime!30] (221,151) -- (342,151) --  (345,158) -- (214,158) -- (221,151) ;
		
		\draw [color = white, fill = white] (320,100) -- (400,100) --  (400,200) -- (320,200) -- (320,100) ;
		
		\draw (320,122.5) -- (320,200) -- (360,200)  ;
		
		\draw [->]   (330,154.5) -- (370,154.5) node[above=2pt] {$\frac{f(u^{L}) - f(u^{R})}{u^{L}- u^{R}}$};
		
		\end{tikzpicture}

		\caption{A singular profile}
		\label{fig:surrrrr}
	\end{subfigure}
	\caption{For a regular profile(on the left), the front of an infinitesimal slice moves at the same speed as its back. This property is conserved by the application of $\eta$. On the other side, for a weak solution, the "mass" moves between different slices, this is responsible for the violation of the conservation of $\eta(u)$}

	\label{fig:entropy understanding}
\end{figure}

This property seems quite counter-intuitive at a first glance, especially because there are few assumptions on $\eta$, but the key point is that it is true only for the "strong" evolution, and can actually be understood geometrically with the help of the figure \ref{fig:entropy understanding}. This property allows us to distinguish a strong solution from a weak solution, and thus will allow us to establish admissibility criteria for selecting the viscous solution among the weak ones, as we will see in what follows.

\subsubsection{Entropic Admissibility condition}
$u$ is said to be an entropy solution if for all entropy functions $\eta$ with the corresponding flux $q$, this inequality is verified:

\begin{equation}\label{entropy inequality}
\eta(u)_{t} + q_{x} \leq 0
\end{equation}

It's fairly easy to show that this is a necessary condition for any viscous solution: Consider a viscous solution $u = \lim_{\epsilon \rightarrow 0} u^{\epsilon}$, where $u^{\epsilon}$ verifies:

\begin{equation}
u^{\epsilon}_{t} + f(u^{\epsilon})_{x} = \epsilon u^{\epsilon}_{xx}
\end{equation}
Then by multiplying both sides by $\eta^{'}(u)$  we have:

\begin{equation}
\eta^{'}(u^{\epsilon}) u^{\epsilon}_{t} + \eta^{'}(u^{\epsilon})f'(u^{\epsilon})u^{\epsilon}_{x} = \epsilon  \eta^{'}(u^{\epsilon})  u^{\epsilon}_{xx}
\end{equation}

\begin{equation}
\eta(u^{\epsilon})_{t} + q'(u^{\epsilon})u^{\epsilon}_{x} = \epsilon ( \eta(u^{\epsilon})_{xx} - \eta^{''}(u^{\epsilon}) (u^{\epsilon}_{x})^{2})
\end{equation}

Since $ \eta^{''}(u^{\epsilon}) (u^{\epsilon}_{x})^{2} \geq 0$

\begin{equation}
\eta(u^{\epsilon})_{t} + q'(u^{\epsilon})u^{\epsilon}_{x} \leq \epsilon  \eta(u^{\epsilon})_{xx} 
\end{equation}
so in the limit $\epsilon \rightarrow 0$ we get:

\begin{equation}\label{entropy inequality}
\eta(u)_{t} + q_{x} \leq 0
\end{equation}

This inequality has to be understood in the weak sens:

\begin{equation}\label{entropy inequality}
\int_{ \mathbb{R \times R^{+}}} (\eta(u)_{t} \phi(x,t) + q_{x} \phi(x,t) )dx dt\leq 0
\end{equation}
For a positive test function $\phi$. Then by Green's formula:
\begin{equation}\label{entropy inequality}
\int_{ \mathbb{R \times R^{+}}} ( \eta(u) \phi_{t}(x,t) + q(u) \phi_{x}(x,t) ) dx dt - \int_{\mathbb{R^{+}}} \eta(u) \phi(0,t) dt  \geq 0
\end{equation}

\paragraph{Remarks}

\begin{enumerate}
	
	\item Suppose the previous inequality is verified for $(\eta_{1},q_{1})$ and for $(\eta_{2},q_{2})$ then it is verified for $\eta = \eta_{1} + \eta_{2}$ associated with the flux $q = q_{1} + q_{2}$. Note however that the sum might not be convex.

	\item One needs not to check the previous inequality for all convex continuous $\eta$, it's enough to check it for a this special family: $\{ \eta_{k}(u) = |u-k| \quad k \in \mathbb{R}  \}$. The associated flux of this family is $q_{k}(u) = (f(u) - f(k))sgn(u-k)$, with $sgn$ is the sign function. The entropic inequality then becomes:
	
	\begin{equation}\label{entropy inequality}
	\int_{ \mathbb{R \times R^{+}}} (|u-k| \phi_{t} + (f(u)-f(k))sgn(u-k)\phi_{x} ) dx dt - \int_{\mathbb{R^{+}}} |u_{0}(x)-k| \phi(x,0) dx  \geq 0
	\end{equation}
	
	\textbf{Proof:} It's actually possible to establish a sequence of piece-wise affine functions that converge to any continuous convex function such that each element of the sequence is of the form:
	
	\begin{equation}
	\eta^{n} = a^{n} + b^{n} u + \sum_{k} c^{n}_{k} |u-k|
	\end{equation}
	Each term of the sequence verifies the inequality thanks to the previous remark, so is the limit.

	\item If $f$ is convex then it's enough to check for one $\eta$

	\item \textbf{Oleinik’s entropy condition} If we consider a solution that is smooth everywhere except for a discontinuity at $s(t)$ at time $t$, then we can apply on $\eta(u)$ the same calculations as for the Hugnoiot-Rankine condition, and it would lead to the inequality:

	\begin{equation}
	(\eta(u^{L}) - \eta(u^{R})) n^{t} + (q(u^{L}) - q(u^{R}))  n^{x} \geq 0
	\end{equation}

	if we apply this to the family $\eta_{k}(u) = |u-k| = (u-k)sgn(u-k)$, we get:
	
	\begin{equation}
	\underset{u_R}{\overset{u^{L}}{[}} |u - k |] n^{t} + \underset{u_R}{\overset{u^{L}}{[}} (f(u) - f(k))sgn(u-k) ]  n^{x} \geq 0
	\end{equation}

	Let's choose $k$ in between $u^{L}$ and $u^{R}$: $k = \lambda u^{L} + (1-\lambda) u^{R} $ with $\lambda \in [0,1]$

	\begin{equation}
	((u^{L} + u^{R} - 2k)n^{t} + (f(u^{L}) + f(u^{R}) - 2 f(k))n^{x}   ) 	sgn(u^{L} - u^{R}) \geq 0
	\end{equation}
	
	Now using the Hugoniot-Rankine relation:
	\begin{equation}
	(     \frac{f(u^{L}) - f(u^{R})}{u^{L} - u^{R}}(u^{L} + u^{R} - 2k) +(f(u^{L}) + f(u^{R}) - 2 f(k))   ) 	sgn(u^{L} - u^{R}) \geq 0
	\end{equation}
	noticing that $u^{L} + u^{R} -2k = (2\lambda -1)(u^{L} - u^{R}) $ we get:
	
	\begin{equation}
	(  (\lambda f(u^{L}) + (1-\lambda) f(u^{R}) -  f(k))   ) 	sgn(u^{L} - u^{R}) \geq 0
	\end{equation}
	Which means:
	
	\begin{itemize}
		\item If $u^{L} > u^{R}$, then $f$ restricted to the $[u^{R}, u^{L}]$ is under its chord, figure ~\ref{fig:sub2}
		
		\item If $u^{L} < u^{R}$, then $f$ restricted to the $[u^{L}, u^{R}]$ is above its chord , figure ~\ref{fig:sub1}
		
	\end{itemize}
	
	This constitutes \textbf{Oleinik’s entropy condition} for the admissibility of discontinuous shocks.

	\item \textbf{The stability interpretation}:
	
	One can understand Oleinik’s entropy condition intuitively in terms of the stability of the shock:
	assume $u^{R} < u^{*} < u^{L}$, then the condition can be written as: (figure ~\ref{fig:oleinik}
	)
	
	\begin{equation}\label{stability}
	\frac{f(u^{*}) - f(u^{R})   }{u^{*} - u^{R} } \leq \frac{f(u^{L}) - f(u^{*})}{u^{L} - u^{*} } 
	\end{equation}
	
	This means that if the shock between $u^{L}$ and $u^{R}$ is split (due to a small perturbation) into two shocks: one between $u^{L}$ and an intermediate value $u^{*}$ followed by one between $u^{*}$ and $u^{R}$, in order for the two shocks to unite again, the first shock has to be faster than the second, which is given by \eqref{stability}.See figure  ~\ref{fig:shock stab} for illustration. This condition is referred to sometimes as \textbf{Liu entropy condition}.

	\begin{figure}
		\centering
		\begin{subfigure}{.5\textwidth}
			\centering
			
			\begin{tikzpicture}
			
			\draw[thick,->] (0,0) -- (6,0) node[anchor=north west] {$u$};
			\draw[thick,->] (0,0) -- (0,5) node[anchor=south east] {$f(u)$};
			
			\draw [very thick ,green] plot [smooth, tension=0.7] coordinates { (1,1) (1.1, 1.8) (1.6,3) (3,3) (4,4) (4.6 , 2.5) (5,0.5)};

			\draw (1.1,1.8) -- (3.5,3.5) --(4.6 , 2.5) -- cycle node[anchor=south east] {};
			
			\draw [dashed] (1.1,1.8) -- (1.1, 0)  node[below=2pt] {$u^{L}$};
			
			\draw [dashed] (4.6 , 2.5) -- (4.6, 0) node[below=2pt] {$u^{R}$};
			
			\draw [dashed] (3.5,3.5)  -- (3.5,0)  node[below=2pt] {$u^{*}$};

			\end{tikzpicture}

			\caption{}
			\label{fig:sub1}
		\end{subfigure}%
		\begin{subfigure}{.5\textwidth}
			\centering
			
			\begin{tikzpicture}
			
			\draw[thick,->] (0,0) -- (6,0) node[anchor=north west] {$u$};
			\draw[thick,->] (0,0) -- (0,5) node[anchor=south east] {$f(u)$};
			
			\draw [very thick ,green] plot [smooth, tension=0.7] coordinates { (5-1,5-1) (5-1.1, 5-1.8) (5-1.6,5-3) (5-3,5-3) (5-4,5-4) (5-4.6 , 5-2.5) (5-5,5-0.5)};

			\draw (5-1.1,5-1.8) -- (5-3.5,5-3.5) --(5-4.6 , 5-2.5) -- cycle node[anchor=south east] {};
			
			\draw [dashed] (5-1.1,5-1.8) -- (5-1.1, 0)  node[below=2pt] {$u^{L}$};
			
			\draw [dashed] (5-4.6 , 5-2.5) -- (5-4.6, 0) node[below=2pt] {$u^{R}$};
			
			\draw [dashed] (5-3.5,5-3.5)  -- (5-3.5,0)  node[below=2pt] {$u^{*}$};

			\end{tikzpicture}
			
			\caption{}
			\label{fig:sub2}
		\end{subfigure}
		\caption{The situations where the Oleinik condition is verified}
		\label{fig:oleinik}
	\end{figure}
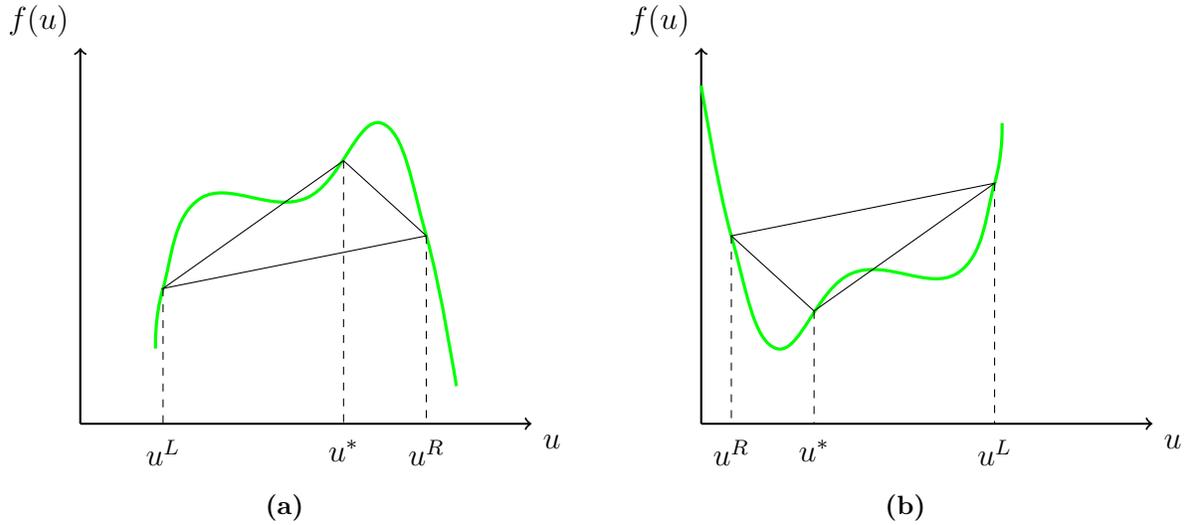

	\begin{figure}[h!]
		
		\centering
		
		\begin{tikzpicture}
		
		\draw[thick,->] (0,0) -- (6,0) node[anchor=north west] {$x$};
		\draw[thick,->] (0,0) -- (0,3) node[anchor=south east] {$u$};
		
		\draw[thick] (1,2) -- (3,2) -- (3,1.2) -- (3.2,1.2) -- (3.2,0.5) -- (5.5,0.5) node[anchor=south east] {};

		\draw[dashed] (2,2) -- (0,2)  node[anchor= east] {$ u^{L} $};
		
		\draw[thick,->,brown] (3.2,0.8) -- (3.7,0.8) node[anchor=north west] {};
		
		\draw[thick,->,brown] (3,1.7) -- (4,1.7) node[anchor=north west] {};

		\draw[dashed] (3,1.2) -- (0,1.2) node[anchor= east] {$u^{*}$};

		\draw[dashed] (3.2,0.5) -- (0,0.5) node[anchor= east] {$u^{R}$};
		
		\end{tikzpicture}
		
		\caption{Illustration of the stability condition of a decreasing shock: the upper sub shock should have a higher speed than the lower sub shock for any intermediate split value $u^{*}$}
		\label{fig:shock stab}
	\end{figure}
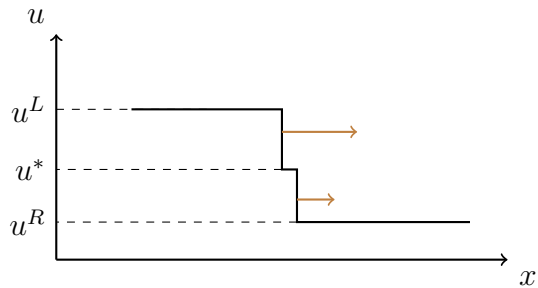

	\item If $f$ is convex, then the Oleinik’s entropy condition becomes particularly simple: only decreasing shocks are admissible:
	\begin{equation}
	u^{L} > u^{R}
	\end{equation}
	If $f$ is concave, the admissible shocks are the increasing ones.

	\item  We can rewrite \ref{stability} slightly differently:
	
	\begin{equation}
	\frac{f(u^{L}) - f(u^{R})   }{u^{L} - u^{R} } \leq \frac{f(u^{L}) - f(u^{*})}{u^{L} - u^{*} } 
	\end{equation}
	One can understand the equivalence between the two inequality easily by contemplating figure \ref{fig:oleinik}.
	the relevance of this form is its adaptability to a generalization to the non-scalar case, as we will see later.

\end{enumerate}

We saw that the entropy condition is a necessary condition for a viscous solution. It is possible to show that an entropic solution is unique. so, if we admit the existence of a viscous solution, the entropic condition is sufficient for selecting it.
This result is known as \textbf{Kružkov uniqueness theorem}. The proof relies on an $ L^{1} $ contraction property of entropic solutions. It can be found in chapter 3 of \cite{evans2010partial}

\subsection{Relation to Hamilton-Jacobi equation}
\label{Relation to Hamilton-Jacobi equation}
The objective here is to state one of the most classical formulas for scalar conservation laws: Lax-Okeinik formula that describes an entropic solution for arbitrary bounded initial condition. We choose to arrive from a path familiar to most physicist: the Hamilton-Jacobi equation.

Let's start by recalling the context of the HJE. Consider a system endowed with a Lagrangian: $\mathcal{L}(q,\dot{q},t)$. We define the action as:
\begin{equation}\label{Action}
S_{q_{0},t_{0}}(q,t) := min \int_{t_{0}, q_{0}}^{t,q} \mathcal{L}(q,\dot{q},\tau) d\tau
\end{equation}

Where the minimum is taken over all the trajectories $q(.)$ such that $q(t_{0}) = q_{0}$ and $q(t_{1}) = q_{1}$. Note that this definition is slightly technically different from the usual definition of the action that is a functional of the trajectories and not a function of $(q,t) \in \mathbb{R}^{n} \times \mathbb{R}^{+}$. For what follows we set $n=1$ for simplicity. Many properties can be generalized trivially to higher dimensions.

Let's notice that the action verifies the following property: consider an infinite path $\gamma$ that divides the the $(q,t)$ space into two parts, one containing $ (q_{0},t_{0}) $ and the other $ (q,t) $ then:

\begin{equation}
S_{q_{0},t_{0}}(q,t) := \min_{(\tilde{q},\tilde{t})\in \Gamma} (S_{q_{0},t_{0}}(\tilde{q},\tilde{t}) + S_{\tilde{q},\tilde{t}}(q,t) )
\end{equation}

This suggest a different way to initialize the action. Consider that we know the action over a path (for simplicity $\mathbb{R} \times \{0\}$ ):
\begin{equation}
S(q,0) = S_{0}(q)
\end{equation}

then we can define the action all over the space by:

\begin{equation}
S(q,t) = \min_{\tilde{q}} (S_{0}(\tilde{q}) + S_{\tilde{q},0}(q,t) )
\end{equation}
We will see a bit later the relevance of this definition. Let's continue first our development of the HJE. elementary calculus of variations of \eqref{Action} allows to establish:
\begin{equation}\label{p}
\frac{\partial S}{\partial q} = \frac{\partial \mathcal{L} }{\partial \dot{q}} := p
\end{equation}
And equally:
\begin{equation}
\frac{\partial S}{\partial t} = -\dot{q}p + \mathcal{L} := -H
\end{equation}
Note that $q$ and $t$ are only spectators regarding the Legendre transformation between $H$ and $\mathcal{L}$. After writing $\dot{q}$ in terms of $p$,$q$ and $t$ and then $H$ in terms of $(q,p,t)$, one can use the previous two equations to have a first order PDE for $S$:
\begin{equation}\label{HJE}
\frac{\partial S}{\partial t} + H(q,\frac{\partial S}{\partial q},t) = 0
\end{equation}
This is the Hamilton Jacobi equation. Once $S$ is found, one can find the trajectories in a similar fashion to finding the light rays in geometrical optics out of the wavefronts. ( When $q$ is scalar, the image of the trajectories is trivial, one needs only to find $\dot{q}$ as a function of $q$, which can be done directly from \eqref{p} ), for more details: \cite{houchmandzadeh2020hamilton}. Actually, the characteristics \eqref{chara} of this equation are Hamilton's equations. At this stage, it's easy to make sense of the famous Hopf-Lax formula
\paragraph{Hopf Lax formula:}
If we consider the HJE with the initial condition:

\begin{equation}
S(q,t) = S_{0}(q)
\end{equation}

From the previous discussion, we can write $S$ as:

\begin{equation}
S(x,t) = \inf\{ \int_{0}^{t} L(\dot{q}(s))ds + S_{0}(y) \quad | \quad  q_{0} = y, q(t) = x  \}
\end{equation}

Let's now assume that $H$ is convex and superlinear $\lim_{|p| \rightarrow \infty} \frac{H(p)}{p} = \infty $, then the Hopfs-Lax formula tells us that we don't actually need to minimize over all the trajectories starting from the path and reaching the point$(x,t)$, it's enough to minimize among straight lines with the constant speed $\frac{x-y}{t}$ :
\begin{equation}\label{key}
S(x,t) = \min_{y \in \mathbb{R}} \{ t \mathcal{L}(\frac{x-y}{t} ) + S_{0}(y) \}
\end{equation}
\begin{proof}
	If we choose a straight trajectory from point $y\in \mathbb{R}$ to $x$ with a constant speed $ \dot{q} =  \frac{x-y}{t} $
	It's obvious that $$  S(x,t) \leq \min_{y \in \mathbb{R}} \{ t \mathcal{L}(\frac{x-y}{t} ) + S_{0}(y) \} $$
	to show the other direction inequality, we use the convexity of $\mathcal{L}$ by applying Jensen's inequality:
	\begin{equation}
	\mathcal{L} (\int_{0}^{t} \frac{1}{t} \dot{q}(s)ds) \leq \frac{1}{t} \int_{0}^{t} \mathcal{L}(\dot{q}(s))ds
	\end{equation}
	So that makes:
	\begin{equation}
	t \mathcal{L} ( \frac{x-y}{t} ) \leq  \int_{0}^{t} \mathcal{L}(\dot{q}(s))ds
	\end{equation}
	Which completes the proof.
\end{proof}
\paragraph{Regularity and uniqueness of the solution}
The Hopf-Lax formula provides a weak form solution for the HJE as it doesn't require for $S$ to be differentiable. Let's be more precise: assumes $S_{0}$ to be Lipschitz with $Lip(S_{0}) \leq M$ then $S$ defined by Hopf-Lax solves the HJE a.e., and it is Lipschitz with $Lip(S_{t}) \leq M$ and differentiable almost everywhere.


Now we reach the stage where we can show the link to the conservation law. If we derive \eqref{HJE} with respect to $q$, and ignore the dependence of $H$ on $q$ and $t$  we get

\begin{equation}
\frac{\partial }{\partial t} \frac{\partial S }{\partial q}+ \frac{\partial H}{\partial q}(\frac{\partial S}{\partial q}) = 0
\end{equation}

Now we can identify $u$ with $\frac{\partial S }{\partial q}$ and $ H $ with $f$ and we find our conservation law.



\paragraph{Lax-Oleinik formula}
The solution $S$ defined by Lax-Hopf is differentiable almost everywhere (Rademacher's theorem). We would like to work out its derivative:

\begin{equation}\label{key}
u(x,t) :=  \frac{\partial }{\partial x} \min_{y \in \mathbb{R}} \{ t \mathcal{L}(\frac{x-y}{t} ) + S_{0}(y) \}
\end{equation}
Under some assumptions:
\begin{itemize}
	\item $f(0) = 0$
	\item $f$ is uniformly convex
	\item $f$ is smooth
	\item $u_{0}$ is bounded	
\end{itemize}
And let $G(x) = (f')^{-1}(x)$, then
\begin{itemize}
	\item There exists for almost all values of $x$, a unique $y(x,t)$ such that the minimum is attained:
	\begin{equation}\label{key}
	\min_{y \in \mathbb{R}} \{ t \mathcal{L}(\frac{x-y}{t} ) + S_{0}(y) \} = t \mathcal{L}(\frac{x-y(x,t)}{t} ) + S_{0}(y(x,t))
	\end{equation}
	
	\item $x \rightarrow y(x,t)$ is non decreasing 
	\item Almost everywhere for $x$ the previous derivative is:
	\begin{equation}
	u(x,t) = G(\frac{x-y(x,t)}{t})
	\end{equation}
	
\end{itemize}
This generalizes the Hopfs treatment of Burgers equation previously encountered.

\subsection{Riemann problem }
The Riemann problem is a conservation system with constant initial data except at zero:
\begin{equation}
u_{0}(x) =
\begin{cases}
u^{L}& \text{if  $ x<0 $} \\
u^{R}& \text{if $ x>0 $}
\end{cases}
\end{equation}
The advantage of this initial condition is that it allows for a solution which is invariant under the resealing:
\begin{equation}
u(x,t) = u(\lambda x, \lambda t )
\end{equation}
In other words, the solution is a function of $\frac{x}{t} := \xi $
\begin{equation}
u(x,t) = u(\frac{x}{t}) := u(\xi)
\end{equation}
We can write the conservation equation as:
\begin{equation}
u^{'}(\xi)(f^{'}(u(\xi)) - \xi) = 0
\end{equation}

This equation in the strong sense can give us insight into the regular solutions: they can be either constants or of the form: $u(\xi) = (f^{'})^{-1}(\xi)$. This requires $f^{'}$ to be invertible. We need to take into account the shocks and the initial condition and to treat the case where $f^{'}$ is not invertible. It's convenient to start the discussion with a convex (or concave), then move to the general form of flux.
\subsubsection{Convex flux}
If $u^{L} > u^{R}$ then there is a simple solution which is a shock at a speed $\frac{f(u^{L}) - f(u^{R})}{u^{L} - u^{R}} $ and it is an entropic solution since it verifies the Oleinik condition.

If $u^{L} < u^{R}$, then we can have a continuous entropic solution: 

\begin{equation}
u(\xi) =
\begin{cases}
u^{L}& \text{if  $ \xi< f^{'}(u^{L}) $} \\
(f^{'})^{-1}(\xi)& \text{if  $f^{'}(u^{L}) <\xi< f^{'}(u^{R})$}\\
u^{R}& \text{if $ \xi > f^{'}(u^{R})  $}
\end{cases}
\end{equation}
Thanks to the convexity, $f^{'}$ is increasing and thus invertible.
The part the solution on the interval $[f^{'}(u^{L}) ,  f^{'}(u^{R})]$ is called a  \textbf{rarefaction fan}

Note that if $f$ is not differentiable (but only continuous) at some point  $\tilde{u}$ then the solution is constant on the interval $[f^{'}( \tilde{u} - 0^{+}  ) , f^{'}( \tilde{u} + 0^{+} ) ]$, so we can  have multiples rarefaction fans.
For a convex flux, one cannot observe at the same time a shock and a rarefaction fan.

The case of a concave flux is treated in exactly similar fashion except that the shock will appear now when $u^{L} > u^{R}$ while the rarefaction for $u^{L} < u^{R}$. We have actually the symmetry $f(u) \leftrightarrow -f(-u)$ that allows to passe from one case to the other. 

\subsubsection{Non convex flux}

The general non-convex, non-concave flux case has been treated by S.Osheri in 1983 \cite{osher1983riemann} According to him the solution is:

if $u^{L} < u^{R} $:

\begin{equation}
\xi u(\xi) - f(u(\xi)) = \max_{v \in [u^{L}, u^{R}]} \{ \xi v - f(v) \}
\end{equation}

if $u^{L} > u^{R} $:

\begin{equation}
\xi u(\xi) - f(u(\xi)) = \min_{v \in [u^{L}, u^{R}]} \{ \xi v - f(v) \}
\end{equation}

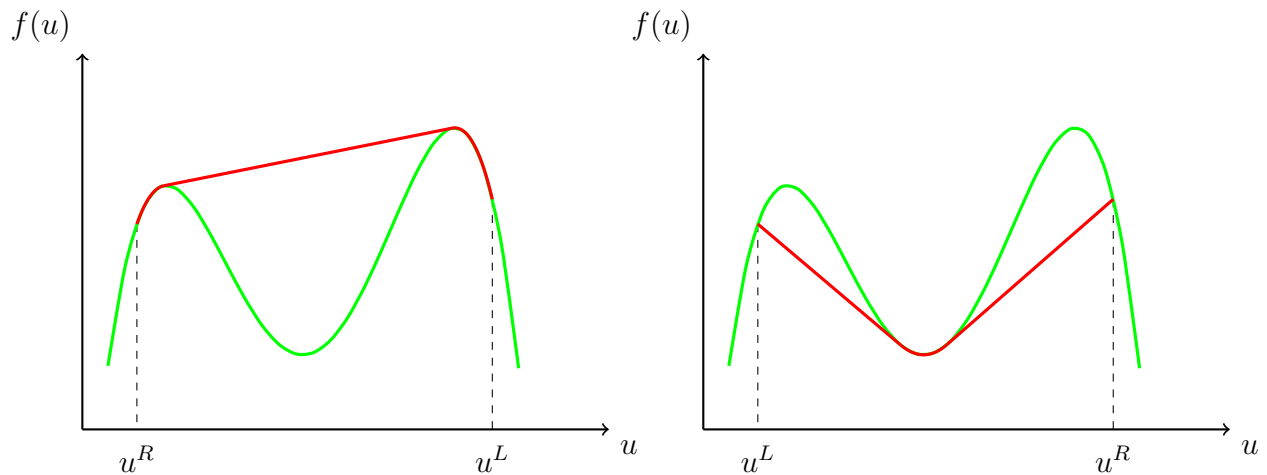
\begin{figure}[h!]
	\centering
	\begin{subfigure}{.5\textwidth}
		\centering
		
		\begin{tikzpicture}
		
		\draw[thick,->] (-3,-1) -- (4,-1) node[anchor=north west] {$u$};
		\draw[thick,->] (-3,-1) -- (-3,4) node[anchor=south east] {$f(u)$};

		\draw [dashed]  (0.7*7/2, 0.7*707/240) -- (0.7*7/2,-1)  node[below=2pt] {$u^{L}$};
		
		\draw [dashed]  (-0.7*13/4, 0.7*9503/3840) -- (-0.7*13/4,-1)  node[below=2pt] {$u^{R}$};

		\draw[scale=0.7, domain=-3.8:4, smooth, variable=\x, very thick ,green] plot ({\x}, { (\x)^2 - (\x)^4/15 + \x/5});

		\draw[scale=0.7, domain=-sqrt(15/2):sqrt(15/2), smooth, variable=\x, very thick ,red] plot ({\x}, { (3*(\x - sqrt(15/2)) + 225/4 + 3*sqrt(15/2))/15 });
		
		\draw[scale=0.7, domain=-13/4:-sqrt(15/2), smooth, variable=\x,  very thick ,red] plot ({\x}, { (\x)^2 - (\x)^4/15 + \x/5});
		
		\draw[scale=0.7, domain=sqrt(15/2):7/2, smooth, variable=\x,  very thick ,red] plot ({\x}, { (\x)^2 - (\x)^4/15 + \x/5});

		\end{tikzpicture}

		\caption{The concave hull (in red) of flux (in green), needed when $u^{R} < u^{L}$}
		\label{fig:concave}
	\end{subfigure}%
	\begin{subfigure}{.5\textwidth}
		\centering

		\begin{tikzpicture}
		
		\draw[thick,->] (-3,-1) -- (4,-1) node[anchor=north west] {$u$};
		\draw[thick,->] (-3,-1) -- (-3,4) node[anchor=south east] {$f(u)$};
		\draw [dashed]  (0.7*7/2, 0.7*707/240) -- (0.7*7/2,-1)  node[below=2pt] {$u^{R}$};
		
		\draw [dashed]  (-0.7*13/4, 0.7*9503/3840) -- (-0.7*13/4,-1)  node[below=2pt] {$u^{L}$};

		\draw[scale=0.7, domain=-3.8:4, smooth, variable=\x, very thick ,green] plot ({\x}, { (\x)^2 - (\x)^4/15 + \x/5});

		\draw[scale=0.7, domain=(1/12)*(13 - sqrt(382)):(1/6)*(-7 + sqrt(82)), smooth, variable=\x,  very thick ,red] plot ({\x}, { (\x)^2 - (\x)^4/15 + \x/5});

		\draw[scale=0.7, domain=(1/6)*(-7 + sqrt(82)):7/2, smooth, variable=\x, very thick ,red] plot ({\x}, { ((3 + 5*(-7 + sqrt(82)]) - (1/54)*(-7 + sqrt(82))^3)*(\x - 7/2) + 707/16)/15 });

		\draw[scale=0.7, domain=-13/4:(1/12)*(13 - sqrt(382)), smooth, variable=\x, very thick ,red] plot ({\x}, { ((3 + (5/2)*(13 - sqrt(382)) - (1/432)*(13 - sqrt(382))^3)*(\x + 13/4) + 9503/256)/
			15 });
		
		\end{tikzpicture}

		\caption{The convex hull (in red) of flux (in green), needed when $u^{R} > u^{L}$}
		\label{fig:convex}
	\end{subfigure}
	\caption{Non convex/concave flux treatment, equivalent of Osheri's solution.  }
	\label{fig:osheri}
\end{figure}

There is an equivalent very simple formation, figure \ref{fig:osheri} (that I astonishingly haven't encountered it in the literature):

if $u^{L} < u^{R} $, we replace $f$ by its convex hull:

\begin{equation}
\check{f} = \max\{ g \in \mathcal{C}^{0} ; g \leq f \}
\end{equation}

if $u^{L} > u^{R} $, we replace $f$ by its concave hull:

\begin{equation}
\hat{f} = \min\{ g \in \mathcal{C}^{0} ; g \geq f \}
\end{equation}

One can understand the equivalence of the two formulations with the help of some elementary geometrical constructions.

\paragraph{Remarks}

\begin{itemize}
	\item if part of the flux is linear then this part corresponds to a discontinuity moving at a speed equal to the slope of straight line, which means that the Oleinik condition is a particular case of this formulation since the shock for convex $f$ and $u^{L} > u^{R}$ can be as well seen as the solution of a linear flux in the interval $[u^{R} , u^{L}]$. This flux is called "contact flux" in the literature. 
	
	\item If the initial condition is "Riemann like", in the sense that it's uniform on the left and on the right, except on some bounded interval, then the re-scaled solution converges to the limit shape of the corresponding Riemann problem.
	
\end{itemize}

\section{Hyperbolic Systems of Conservation Laws}
\label{Hyperbolic Systems of Conservation Laws}
In the first part we treated the case of a single conserved quantity, the problem becomes significantly more complex with $n$ conserved quantities $\mathbf{u} = (u_{1},...,u_{n})^{t}$ with a flux for the quantity $i$ that depends on all of the quantities: $\mathbf{f} : \mathbb{R}^{n} \rightarrow \mathbb{R}^{n}$. The conservation law becomes a system of $n$ coupled PDE's:
\begin{equation}
\begin{split}
& \mathbf{u}_{t} + (\mathbf{f}(\mathbf{u}))_{x} = 0 \\
& \mathbf{u}(x,0) = \mathbf{u}_{0}(x)
\end{split}
\end{equation}
This system is said to be strictly hyperbolic when the differential of the flux $A(u) = D_{u}f$ is diagonalisable in $\mathbb{R}$ and its eigenvalues are distinct for all $\mathbf{u}$:
\begin{equation}
\lambda_{1} < \lambda_{2} < .. < \lambda_{n} 
\end{equation}
This allows to choose the left and the right eigenvectors ($ \mathbf{l}_{i} $ and $ \mathbf{r}_{i} $ respectively) such that:
\begin{equation}
\mathbf{l}_{i}.\mathbf{r}_{j} = \delta_{i,j}
\end{equation}
The proof of this is elementary:
\begin{equation}
\mathbf{l}_{i}^{t}A\mathbf{r}_{j} =   \mathbf{l}_{i}^{t}\lambda_{i}\mathbf{r}_{j} =  \mathbf{l}_{i}^{t}\lambda_{j}\mathbf{r}_{j}
\end{equation}
Finally:
\be
(\lambda_{i} - \lambda_{j} )\mathbf{l}_{i}.\mathbf{r}_{j} = 0
\ee
Before treating the general non-linear system, let's have a look at the simple case of a linear one:
\subsection{A linear system}
A simple situation when is $\mathbf{f}$ is linear: $\mathbf{f}(\mathbf{u}) = A \mathbf{u} $
The conservation system is:
\begin{equation}\label{sys}
\mathbf{u}_{t} + A \mathbf{u}_{x} = 0
\end{equation}
We can write the vector $ \mathbf{u} $ as :
\begin{equation}
\mathbf{u} = \sum_{i} (\mathbf{u}.\mathbf{l_{i}})\mathbf{r_{i}}
\end{equation}
Lets define $n$ new quantities: $ \mathbf{\tilde{u}} = (\tilde{u}_{i})_{1\leq i \leq n} $
\begin{equation}
\tilde{u}_{i} = \mathbf{u}.\mathbf{l_{i}}
\end{equation}
Which are the densities in the base of the right eigenvectors.
Then we can realize by multiplying both sides of equation \eqref{sys} by $ \mathbf{l_{i}} $ that each of these quantities verifies a scalar conservation law:
\begin{equation}\label{}
(\tilde{u}_{i})_{t} + \lambda_{i} (\tilde{u}_{i})_{x} = 0
\end{equation}
Where $\lambda_{i}$ is the eigenvalue associated with $\mathbf{l_{i}}$. So, the new quantities evolve independently, and the solution of the original system is simply:
\begin{equation}
\mathbf{u}(x,t) = \sum_{i} (\mathbf{u}(x - \lambda_{i}t,0).\mathbf{l_{i}})\mathbf{r_{i}}
\end{equation}
The situation becomes significantly more complex when the matrix $A$ is a function of $\mathbf{u}$. The different waves can now interact with each other. We will consider only the situation when the system can be written in a conservative form, i.e. when $A$ is the differential of a flux function. We assume as well the strict hyperbolic condition as previously.
\subsection{Weak solutions, the Rankine-Hugoniot condition }

Similarly to the scalar case, one has to interpret the conservation equation in the sense of distributions. This allows for discontinuous solutions to exist. A straightforward generalization of the Rankine-Hugoniot condition is possible: The discontinuities should verify:

\begin{equation}\label{RHC}
(\mathbf{u}^{L} - \mathbf{u}^{R}) \sigma = (\mathbf{f}(\mathbf{u}^{L}) - \mathbf{f}(\mathbf{u}^{R}))
\end{equation}

Unlike its scalar counterpart, this condition doesn't only provide the shock speed, but also constraint the densities between which one can have a shock solution, namely for all $1 \leq i,j \leq n$

\be
Det \begin{pmatrix}
	u_{i}^{L} - u_{i}^{R} & f_{i}(\mathbf{u}^{L}) - f_{i}(\mathbf{u}^{R})   \\
	u_{j}^{L} - u_{j}^{R} & f_{j}(\mathbf{u}^{L}) - f_{j}(\mathbf{u}^{R})
\end{pmatrix} = 0
\ee

\subsection{The shock curves}

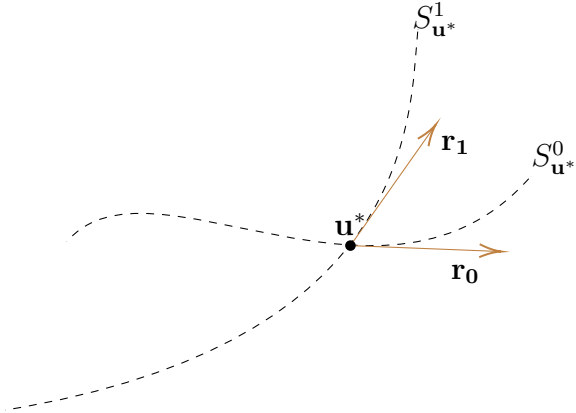
\begin{figure}[h!]
	\centering
	
	\begin{tikzpicture}[x=0.75pt,y=0.75pt,yscale=-1,xscale=1]
	\draw    (204,148.5) .. controls (246,99.5) and (365,199.5) .. (438,115.5) ;
	
	\draw  [dashed, color = white, very thick]  (204,148.5) .. controls (246,99.5) and (365,199.5) .. (438,115.5) ;

	\draw  [color=brown]  (347,150.5) -- (421,153.42) ;
	\draw [shift={(423,153.5)}, rotate = 182.26] [color=brown ][line width=0.75]    (10.93,-3.29) .. controls (6.95,-1.4) and (3.31,-0.3) .. (0,0) .. controls (3.31,0.3) and (6.95,1.4) .. (10.93,3.29)   ;
	\draw    (173,233.5) .. controls (381,199.5) and (377,84) .. (381,41.5) ;
	
	\draw  [dashed, color = white, very thick]   (173,233.5) .. controls (381,199.5) and (377,84) .. (381,41.5) ;

	\draw [color=brown]   (347,150.5) -- (388.85,91.13) ;
	\draw [shift={(390,89.5)}, rotate = 125.18] [color=brown][line width=0.75]    (10.93,-3.29) .. controls (6.95,-1.4) and (3.31,-0.3) .. (0,0) .. controls (3.31,0.3) and (6.95,1.4) .. (10.93,3.29)   ;
	
	\fill (347,150.5) circle(2pt) node[above]{$ \mathbf{u^{*}} $};
	
	\fill (405,175) node[above] {$ \mathbf{r_{0}} $};
	\fill (400,110) node[above] {$ \mathbf{r_{1}} $};
	\fill (390,50) node[above] {$ S^{1}_{\mathbf{u^{*}}} $};
	\fill (450,120) node[above] {$ S^{0}_{\mathbf{u^{*}}} $};
	\end{tikzpicture}
	
	\caption{Shock curves in a 2d density space}
	
	\label{sh curves}
\end{figure}

Let's fix a point in the density space $\mathbf{u^{*}}$ and search for all the points that can be connected to $\mathbf{u^{*}}$ through a shock.
One can see this set as parameterized by $\sigma$, so it is expected to form a 1d manifold, i.e a curve, but actually, it is composed of $n$ curves that passes by $\mathbf{u^{*}}$. To see this, one can linearize \ref{RHC} in the neighborhood of $\mathbf{u^{*}}$:
\be
\mathbf{u} = \mathbf{u^{*}} + \frac{1}{\sigma}A(\mathbf{u^{*}})(\mathbf{u} - \mathbf{u^{*}})
\ee
Since $A$ has $n$ real distinct eigenvalues, we can conclude that this equation admits $n$ independent 1d eigenspaces that would represent the tangents of $n$ shock curves. We can parameterize each by its speed: $S^{i}_{\mathbf{u^{*}}}(\sigma)$, figure \ref{sh curves}. Obviously, the small perturbations in the i-shocks propagate at a speed $\lambda_{i}$:

\be
\lim_{\sigma \rightarrow \lambda_{i} }S^{i}_{\mathbf{u^{*}}}(\sigma) = \mathbf{u^{*}}
\ee

Note that the $i$-shock curve emerging from one point does not coincide in general with the $i$-shock curve emerging from another point located at the former. 

\subsection{Admissibility conditions}
Since weak solutions are not necessarily unique, we need to select among them the "physical" ones. Conceptually, we can add an infinitesimal diffusion term to the conservation equation:
\begin{equation}
\mathbf{u}_{t} + (\mathbf{f}(\mathbf{u}))_{x} + \epsilon \mathbf{u}_{xx}  = 0 
\end{equation}
And search for a solution that are a limit in $L^{^{1}}_{loc}$ when $\epsilon \rightarrow 0$.

Although, it was possible for the Burgers equation to be treated in this manner as Hopf did, this approach doesn't provide a practical procedure allowing to eliminate non-physical solutions. One has to look for alternative approaches.

\subsubsection{Entropy condition}

The notion of Entropy can be extended to the multi-dimensional case; however, its existence is no longer guaranteed. Let's recall that entropy is a smooth convex scalar function, associated with a flux that verifies:

\begin{equation}
D_{\mathbf{u}} q  =  D_{\mathbf{u}} \eta A(\mathbf{u})
\end{equation}

This is a system of $n$ first-order PDE with two scalar variables. For $n>2$ the system is over-determined and doesn't in general have a solution. If it does, then it allows for classifying solutions within three categories:
\begin{itemize}
	\item regular solutions (in the sense $C^{1}$) conserve the entropy under time evolution.
	\item physical singular solutions consume entropy
	\item non-physical solutions produce entropy.
\end{itemize}
The second category means that a $\mathbf{u^{L}}-\mathbf{u^{R}}$ discontinuity, traveling at a speed $\lambda$, should verify:

\begin{equation}
(\eta(\mathbf{u^{L}}) - \eta(\mathbf{u^{R}})) \lambda   \leq q(\mathbf{u^{L}}) - q(\mathbf{u^{R}})  
\end{equation}

It's not possible to extend the Oleinik condition in a straightforward way even if we restrict ourselves to a particular i-shock curve.
For this to happen, the stability condition has to be formulated in the sense of Liu.

\subsubsection{Liu Condition}
Consider an i-shock curve that originates at $\mathbf{u^{L}}$ and let $\mathbf{u^{R}}$ be a point that belongs to that curve: $\mathbf{u^{R}} = S^{i}_{\mathbf{u^{L}}}(\sigma) $ and let $\mathbf{u^{*}}$ be an intermediate point on the curve between $\mathbf{u^{L}}$ and $\mathbf{u^{R}}$: $\mathbf{u^{*}} = S^{i}_{\mathbf{u^{L}}}(s) $. The Liu stability condition states that:
\be
\sigma(\mathbf{u^{L}},\mathbf{u^{R}}) \leq \sigma(\mathbf{u^{L}},\mathbf{u^{*}})
\ee

This obviously means that the intermediate perturbation shock will not form a separate shock but rather join back with the mother shock. This follows the same logic as the scalar case, figure \ref{fig:shock stab}, restricted on one i-shock curve. This condition was developed by Liu in his paper: \cite{liu1976entropy}. Another handy directly applicable condition is the Lax condition, introduced in the next paragraph.

\subsubsection{Lax condition}
\begin{figure}[h!]
	\centering
	\begin{subfigure}{.5\textwidth}
		\centering

		\begin{tikzpicture}
		
		\draw[->,ultra thick] (1,2)--(8.5,2) node[right]{$x$};
		\draw[->,ultra thick] (2,1.5)--(2,6) node[above]{$t$};
		
		\draw[color = brown, ultra thick] (5,2)--(6,5);
		
		\draw[thick] (2.5,2)--(6,5);
		\draw[thick] (3,2)--(5.8,4.4);
		\draw[thick] (3.5,2)--(5.6,3.8);
		\draw[thick] (4,2)--(5.4,3.2);
		\draw[thick] (4.5,2)--(5.2,2.6);
		
		\draw[thick] (5.5,2)--(5.2,2.6);
		\draw[thick] (6,2)--(5.4,3.2);
		\draw[thick] (6.5,2)--(5.6,3.8);
		\draw[thick] (7,2)--(5.8,4.4);
		\draw[thick] (7.5,2)--(6,5);

		\draw (5,5.5) node[right=2pt] {Admissible};
		
		\end{tikzpicture}

		\label{}
		\caption{Characteristics are joining the shock}
	\end{subfigure}%
	\begin{subfigure}{.5\textwidth}
		\centering
		
		\begin{tikzpicture}
		
		\draw[->,ultra thick] (1,2)--(8.5,2) node[right]{$x$};
		\draw[->,ultra thick] (2,1.5)--(2,6) node[above]{$t$};
		
		\begin{scope}[shift={(10.5,7)}, rotate=180]

		\draw[color = brown, ultra thick] (5,2)--(6,5);
		
		\draw[thick] (2.5,2)--(6,5);
		\draw[thick] (3,2)--(5.8,4.4);
		\draw[thick] (3.5,2)--(5.6,3.8);
		\draw[thick] (4,2)--(5.4,3.2);
		\draw[thick] (4.5,2)--(5.2,2.6);
		
		\draw[thick] (5.5,2)--(5.2,2.6);
		\draw[thick] (6,2)--(5.4,3.2);
		\draw[thick] (6.5,2)--(5.6,3.8);
		\draw[thick] (7,2)--(5.8,4.4);
		\draw[thick] (7.5,2)--(6,5);
		
		\end{scope}
		
		\draw (4,5.5) node[right=2pt] {Not admissible};
		
		\end{tikzpicture}

		\caption{Characteristics are leaving the shock}
		\label{}
	\end{subfigure}
	\caption{Illustration of Lax condition in terms of the behavior of of characteristic around the shock represented by the brown segment}
	\label{fig:lax}
\end{figure}
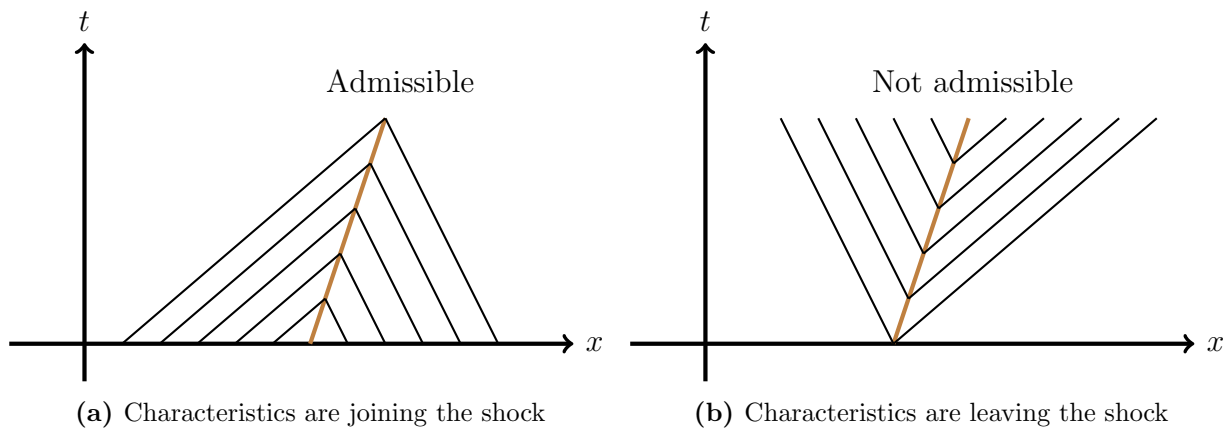

Let $\lambda_{i}(\mathbf{u^{L}})$,$\lambda_{i}(\mathbf{u^{R}})$ be the i-eigenvalues at $\mathbf{u^{L}}$, $\mathbf{u^{R}}$ respectively, then Lax stability condition can be expressed as:	
\be
\lambda_{i}(\mathbf{u^{L}}) \geq \sigma(\mathbf{u^{L}},\mathbf{u^{R}}) \geq \lambda_{i}(\mathbf{u^{R}})
\ee
This means that the small perturbations on the left and on the right of the shock should move towards the shock. In terms of characteristics: In the neighborhood of the i-shock, the neighboring i-characteristics should be entering the shock and not leaving it, figure \ref{fig:lax}.
In a sense, this condition represents the time-irreversible character of the singular solution. The information of the initial data is lost at the shocks. mathematical solutions that inverse this process are not physical.
\subsubsection{Conclusion}
Concretely, the admissibility conditions will eliminate for each shock curve originating from $\mathbf{u^{*}}$, one of the two parts of the shock curve separated by $\mathbf{u^{*}}$. Let's denote $S^{i+}_{\mathbf{u^{*}}} $ the non-admissible part and $S^{i-}_{\mathbf{u^{*}}} $ the admissible one.

\begin{figure}[h!]
	\centering
	
	\begin{tikzpicture}[x=0.75pt,y=0.75pt,yscale=-1,xscale=1]
	
	\tikzset{
		bicolor/.style 2 args={
			dashed,dash pattern=on 150pt off 90pt,#1,
			postaction={draw,dashed,dash pattern=on 90pt off 150pt,#2,dash phase=130pt}
		},
	}

	\draw  [bicolor={green}{red}]  (204,148.5) .. controls (246,99.5) and (365,199.5) .. (438,115.5) ;
	
	\draw [dashed, color = white, very thick]  (204,148.5) .. controls (246,99.5) and (365,199.5) .. (438,115.5) ;
	
	\draw  [color=brown]  (347,150.5) -- (421,153.42) ;
	
	\draw [shift={(423,153.5)}, rotate = 182.26] [color=brown ][line width=0.75]
	(10.93,-3.29) .. controls (6.95,-1.4) and (3.31,-0.3) .. (0,0) .. controls (3.31,0.3) and (6.95,1.4) .. (10.93,3.29)   ;

	\tikzset{
		bicolor/.style 2 args={
			dashed,dash pattern=on 150pt off 90pt,#1,
			postaction={draw,dashed,dash pattern=on 90pt off 150pt,#2,dash phase=90pt}
		},
	}
	
	\draw [bicolor={green}{red}]   (173,233.5) .. controls (381,199.5) and (377,84) .. (381,41.5) ;
	
	\draw [dashed, color = white, very thick]  (173,233.5) .. controls (381,199.5) and (377,84) .. (381,41.5) ;

	\draw [color=brown]   (347,150.5) -- (388.85,91.13) ;
	\draw [shift={(390,89.5)}, rotate = 125.18] [color=brown][line width=0.75]   (10.93,-3.29) .. controls (6.95,-1.4) and (3.31,-0.3) .. (0,0) .. controls (3.31,0.3) and (6.95,1.4) .. (10.93,3.29)   ;
	
	\fill (347,150.5) circle(2pt) node[above]{$ \mathbf{u^{*}} $};
	
	\fill (405,175) node[above] {$ \mathbf{r_{0}} $};
	\fill (400,110) node[above] {$ \mathbf{r_{1}} $};
	\fill (390,50) node[above] {$ S^{1}_{\mathbf{u^{*}}} $};
	\fill (450,120) node[above] {$ S^{0}_{\mathbf{u^{*}}} $};
	\end{tikzpicture}
	
	\caption{Admissible sections of shock curves(green). Non-admissible ones are colored(red)}

	\label{fig:admis}
\end{figure}
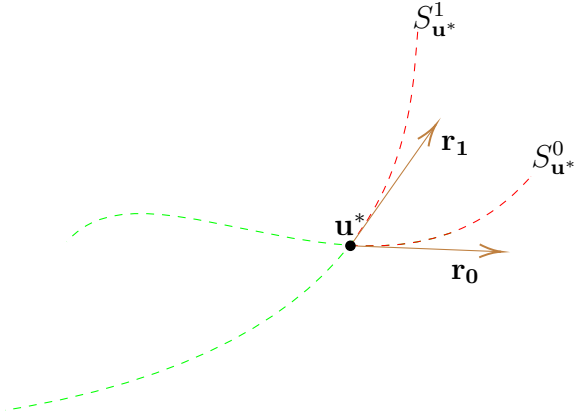

We orientate $ \mathbf{r}_{i} $ to the non-admissible part, figure \ref{fig:admis}. This orientation will be compatible with further developments.

Recall that in the case of a scalar system, the convexity of the current allowed a simplification of the analysis in particular because the mapping from the density to the speed of perturbations becomes monotonous, and thus one to one. In higher dimensions, we need this character to be conserved on the integral curves the eigenvectors of the Jacobian matrix as we will see in what follows.

\subsubsection*{A simplifying hypothesis}
A typical hypothesis in the textbooks that has its origins to Lax 1957 \cite{lax1957hyperbolic} is to assume that each of the eigenvectors' fields falls into one of the two categories:

\begin{itemize}
	\item$  D_{\mathbf{u}} \lambda_{i}. \mathbf{r}_{i}(\mathbf{u}) > 0  $ for all $\mathbf{u}$ and in this case it's said to be genuinely non linear
	\item $  D_{\mathbf{u}} \lambda_{i}. \mathbf{r}_{i}(\mathbf{u}) = 0  $ for all $\mathbf{u}$ and is said to be linearly degenerate. 
\end{itemize} 
The first case means that the directional derivative of $\lambda_{i}$ in the direction of $\mathbf{r}_{i}$ is positive. Obviously, the sign is irrelevant as far as it doesn't change. However, for later convenience, we choose the direction of $\mathbf{r}_{i}$ so that this sign is positive. This means that $\lambda_{i}$ is an increasing function along the directed integral curves of the i-field. The second case simply means that $\lambda_{i}$ is constant along these curves.
\subsection{Rarefaction Curves}
For a genuinely non-linear field, we obviously can parameterize an integral curve by the corresponding eigenvalue field. So i-curve passing by a density point $\mathbf{u^{*}}$ can be described by:
\be
\lambda_{i} \rightarrow R_{\mathbf{u^{*}}}^{i}(\lambda_{i})
\ee
In other words, this is the solution (and the flow for the parameter $ \mathbf{u^{*}} $) of the ODE:
\be
\frac{d \mathbf{u} }{ds } = \frac{\mathbf{r}_{i}(\mathbf{u}) }{D_{\mathbf{u}} \lambda_{i}. \mathbf{r}_{i}(\mathbf{u}) }
\ee
This curve is called the i-rarefaction curve.

\begin{figure}[h!]
	\centering
	
	\begin{tikzpicture}[x=0.75pt,y=0.75pt,yscale=-1,xscale=1]
	
	\draw     (204,148.5) .. controls (246,99.5) and (365,199.5) .. (438,115.5) ;
	
	\draw  [dashed, color = white, very thick]   (204,148.5) .. controls (246,99.5) and (365,199.5) .. (438,115.5) ;

	\draw    (347,150.5) -- (421,153.42) ;
	\draw [shift={(423,153.5)}, rotate = 182.26] [color={rgb, 255:red, 0; green, 0; blue, 0 }  ][line width=0.75]    (10.93,-3.29) .. controls (6.95,-1.4) and (3.31,-0.3) .. (0,0) .. controls (3.31,0.3) and (6.95,1.4) .. (10.93,3.29)   ;
	\draw [->]   (175,104.5) .. controls (399,176.5) and (382,161.5) .. (446,72.5) ;

	\fill (347,150.5) circle(2pt) node[above]{$ \mathbf{u^{*}} $};
	
	\fill (405,175) node[above] {$ \mathbf{r_{0}} $};

	\fill (450,120) node[above] {$ S^{0}_{\mathbf{u^{*}}} $};
	
	\fill (460,85) node[above] {$ R^{0}_{\mathbf{u^{*}}} $};
	\end{tikzpicture}
	
	\caption{A rarefaction curve with a shock curve}
	
	\label{rarefaction}
\end{figure}
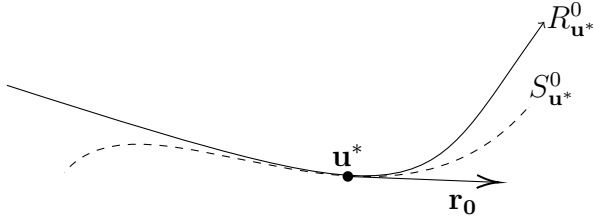

The i-rarefaction curve that passes by $ \mathbf{u^{*}} $  is tangent to the i-shock curve that originates at $ \mathbf{u^{*}} $. It is even possible to show that they have the same curvature, figure \ref{rarefaction}. In a class of conservation laws known as the Temple class, the two curves are identical for all the fields.

$\mathbf{u^{*}} $ divides the rarefaction curve passing by it into two parts. We denote $ R^{i+}_{\mathbf{u^{*}}} $ the part verifying $\lambda_{i}(\mathbf{u}) > \lambda_{i}(\mathbf{u^{*}}) $, and by $ R^{i-}_{\mathbf{u^{*}}} $, the other part.
\subsection{T-curves}
It will soon become meaningful to define a new curve by sticking $ R^{i+}_{\mathbf{u^{*}}} $ to $ S^{i-}_{\mathbf{u^{*}}} $. This curve is called a T-curve.

\begin{figure}[h!]
	\centering
	\begin{tikzpicture}[x=0.75pt,y=0.75pt,yscale=-1,xscale=1]
	
	
	\draw [dashed]   (183,180.5) .. controls (235,146.5) and (276,154.5) .. (329,149.5) ;
	
	\fill (183,180.5) node[below] {$ S^{i-}_{\mathbf{u^{*}}} $};

	\draw  [color = brown]  (329,149.5) -- (408.03,135.84) ;
	
	\draw [shift={(410,135.5)}, rotate = 170.19] [color= brown][line width=0.75]    (10.93,-3.29) .. controls (6.95,-1.4) and (3.31,-0.3) .. (0,0) .. controls (3.31,0.3) and (6.95,1.4) .. (10.93,3.29)   ;
	
	\fill (415,135.84) node[below] {$ \mathbf{r_{i}} $};

	\fill (329,149.5) circle(2pt) node[above]{$ \mathbf{u^{*}} $};

	\draw    (329,149.5) .. controls (392,140.5) and (433,115.5) .. (446,72.5) ;
	
	\fill (446,72.5) node[right] {$ R^{i+}_{\mathbf{u^{*}}} $};
	
	\end{tikzpicture}
	\caption{The structure of a T-curve}
	\label{key}
\end{figure}
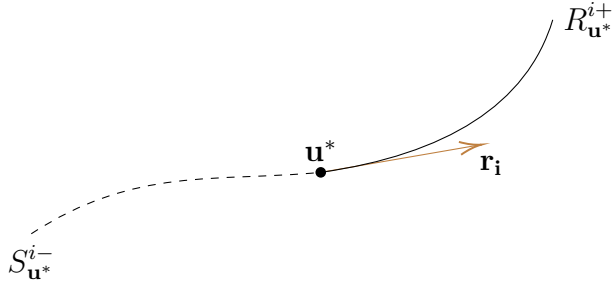

In the next paragraph, we will be considering the solutions of the system of conservation laws for the particular Riemann initial condition. 

\subsection{The Riemann Problem}

We consider the system of conservation laws associated with an initial condition:
\be
\mathbf{u}(x,0) = \mathbf{u^{L}} \mathds{1}_{x<0} + \mathbf{u^{R}} \mathds{1}_{x>0}
\ee
It is easy to verify that the system is invariant under the transformation $(x,t) \rightarrow (\mu x, \mu t)$, so one can express the solution in terms of the variable $ \xi = \frac{x}{t} $ and we can convert the system into an equation of $\mathbf{u}(\xi)$:

\be \label{con compact}
\mathbf{u}^{'}(\xi)(\xi  - A) = 0 
\ee

Besides trivial constant solutions we can identify the following ones:

\subsubsection{Elementary solutions}
We can distinguish the following simple solutions composed of a single type of waves: 

\begin{itemize}
	
	\item
	If $ \mathbf{u^{L}} $ and  $ \mathbf{u^{R}} $ belong to the same i- shock curve and verify: $ \lambda_{i}(\mathbf{u^{L}}) \geq  \lambda_{i}(\mathbf{u^{R}}) $ then the solution of the Riemann problem is a simple shock:
	\be
	\mathbf{u}(\xi) = \begin{cases}
		\mathbf{u^{L}} & \text{if } \xi < \sigma(\mathbf{u^{L}},\mathbf{u^{R}}) \\
		\mathbf{u^{R}}  & \text{if } \xi > \sigma(\mathbf{u^{L}},\mathbf{u^{R}})
	\end{cases}
	\ee
	
	\item
	If $ \mathbf{u^{L}} $ and  $ \mathbf{u^{R}} $ belong to the same genuinely non-linear i-rarefaction curve and verify: $ \lambda_{i}(\mathbf{u^{L}}) <  \lambda_{i}(\mathbf{u^{R}}) $ then the solution of the Riemann problem is a simple Rarefaction wave:
	\be
	\mathbf{u}(\xi) = \begin{cases}
		\mathbf{u^{L}} & \text{if } \xi < \lambda_{i}(\mathbf{u^{L}}) \\

		R_{\mathbf{u^{L}}}^{i}(\xi) & \text{if } \lambda_{i}(\mathbf{u^{L}}) < \xi < \lambda_{i}(\mathbf{u^{R}}) \\ 
		
		\mathbf{u^{R}}  & \text{if } \xi > \lambda_{i}(\mathbf{u^{R}})
	\end{cases}
	\ee
	
	To verify why the branch in the middle holds, it's enough to multiply the equation \ref{con compact} from the right by $\mathbf{r_{i}}$, we get a factor with the eigenvalue equation that is solved by this branch.

	\item If $ \mathbf{u^{L}} $ and  $ \mathbf{u^{R}} $ belong to the same linearly degenerate i- rarefaction curve with a constant eigenvalue $\lambda_{i}$ then the solution of the Riemann problem is a shock:
	\be
	\mathbf{u}(\xi) = \begin{cases}
		\mathbf{u^{L}} & \text{if } \xi < \lambda_{i}  \\
		\mathbf{u^{R}}  & \text{if } \xi > \lambda_{i}
	\end{cases}
	\ee
	This type of shock is sometimes referred to as contact discontinuity. Unlike the two previous cases, the linearly degenerate curve is bi-directional, it's at the same time a rarefaction and a shock curve.  
\end{itemize}

In conclusion, to have a physical elementary solution, $\mathbf{u^{R}}$ has to belong to one of the T-curves generated by $\mathbf{u^{L}}$.

\subsubsection{Combined solutions}

If $ \mathbf{u^{L}} $ and  $ \mathbf{u^{R}} $ don't belong to the same curve, one has to combine different T-curves to connect the two. It's possible to show the existence and unicity of this combination of curves for $ \mathbf{u^{L}} $ and  $ \mathbf{u^{R}} $ sufficiently close.

\subsection{Riemann Variables}
Let's contemplate the vector field of the right eigenvectors:
\be
\mathbf{l_{i}} A =  \lambda_{i} \mathbf{l_{i}}
\ee
Let's assume that up to a multiplicative scalar field $ \mathbf{l_{i}} $ can be derived from a scalar potential  $z_{i}(\mathbf{u})$. i.e: $ \mathbf{l_{i}} \propto \nabla z_{i} $. We can choose the norm of $ \mathbf{l_{i}} $ so that we have equality:
\be
\mathbf{l_{i}} = \nabla z_{i}
\ee 
We call these scalar fields, the Riemann variables. They always exist for $n=2$. For $n>2$ they don't exist in general.
Riemann variables simplify the analysis of the conservation laws. They allow to partially decouple the system. More precisely We have: 
\be
\begin{split}
	\partial_{t} z_{i}(\mathbf{u}) + \lambda_{i} \partial_{x} z_{i}(\mathbf{u}) & = 
	\nabla  z_{i} \partial_{t}(\mathbf{u}) + \lambda_{i}\nabla  z_{i} \partial_{x}(\mathbf{u}) \\
	&= \nabla  z_{i}( \partial_{t}(\mathbf{u}) + \lambda_{i}\partial_{x}(\mathbf{u})) \\
	&= \mathbf{l_{i}} ( \partial_{t}(\mathbf{u}) + \lambda_{i}\partial_{x}(\mathbf{u})) \\
	&= \mathbf{l_{i}} ( \partial_{t}(\mathbf{u}) + A\partial_{x}(\mathbf{u})) = 0
\end{split}
\ee
This means that if we express the system in terms of the Riemann variables, the coupling between the equations appears only in the velocity coefficient $\lambda_{i}(\mathbf{z})$.

The sets where $z_{i}$ is constant form a foliation of manifolds of dimension $n-1$ in the density space, that are perpendicular to the integral curves of $\mathbf{l_{i}}$ and since we have, $\mathbf{l_{i}}.\mathbf{r_{j}} = \delta_{i,j}$, This means that $z_{i}$ is constant over all the j-rarefaction curves such that $j \neq i$. For this reason, they are sometimes called the Riemann invariants.

One can show that these speeds can be obtained by:
\be
\lambda_{i}(\boldsymbol{z}) = \frac{\partial J_{k}}{\partial z_{i}} / \frac{\partial u_{k}}{\partial z_{i}}
\ee
This is true for any $k$ and it implies that:

\be
\frac{\partial J_{n}}{\partial z_{i}} \frac{\partial u_{m}}{\partial z_{i}} = \frac{\partial J_{m}}{\partial z_{i}} \frac{\partial u_{n}}{\partial z_{i}}
\ee

\subsubsection*{Remark}
The condition for the existence of Riemann variables is provided Frobenius theorem. It's more convenient to express it in the language of differentiable forms. Let's see $ \mathbf{l_{i}} $ as a 1-form. If they exist, then we have a scalar field $f$ such that  $ \mathbf{l_{i}} = f d z_{i} $. We have: $ d \mathbf{l_{i}} = d f \wedge d z_{i} $. Now since $ f d z_{i} \wedge df \wedge d z_{i} = 0 $, we have the Frobenius condition:
$$ \mathbf{l_{i}} \wedge d \mathbf{l_{i}} = 0 $$
It's easy to see that in 2D, this is always verified

\subsection{Temple Class Systems}
A particular case of conservation laws for which explicit calculations are typically possible and simpler is the Temple class. It was first noticed and studied by Blake Temple 1982 \cite{temple1983systems}. I will be simplifying the main ideas of this paper.

\subsubsection{Definition}
We say that a system is of temple class if for all $i$ the i-rarefaction curve coincides with the i-shock curve.
This of course happens trivially for an i-field in the case of a contact discontinuity, which means that $\lambda_{i}$ is constant over each of the integral curves. i.e the i-field is linearly degenerate. We will assume that none of the fields in this situation for what follows.

\subsubsection{Important Property}
A system is of Temple class if and only if all the rarefaction curves and the shock. curves are affine.

One of the directions of the implications is trivial. Since we know that the i-rarefaction and the i-shock curve originated from a point are tangent, being affine implies coinciding. For the other direction, I will not provide rigorous proof, it can be found in the paper of Temple, but I can provide an intuitive understanding: Since the shock curve coincides with the rarefaction curve, it means that the shock curve in this case does not depend on the point where it is originated from along the rarefaction curve. so if we take three distinct points on this curve: $\mathbf{u_{1}}$, $\mathbf{u_{2}}$, $\mathbf{u_{3}}$. we can write three Hugoniot conditions:
\begin{equation}
f(\mathbf{u_{1}}) - f(\mathbf{u_{2}}) = \sigma_{1} (\mathbf{u_{1}} - \mathbf{u_{2}})
\end{equation}
\begin{equation}
f(\mathbf{u_{2}}) - f(\mathbf{u_{3}}) = \sigma_{2} (\mathbf{u_{2}} - \mathbf{u_{3}})
\end{equation}
\begin{equation}
f(\mathbf{u_{1}}) - f(\mathbf{u_{3}}) = \sigma_{3} (\mathbf{u_{1}} - \mathbf{u_{3}})
\end{equation}
If the curve is linearly degenerate, then $\sigma_{1} = \sigma_{2} = \sigma_{3}$. However, we excluded this possibility. so the sigmas are not all identical (except potentially at a subset of the curve of Lebesgue measure zero, at which we can prolong the arguments by continuity). Obviously, the left side of the third equation is the sum of the first two, so the same must hold for the right side:
\begin{equation}\label{key}
\sigma_{1} (\mathbf{u_{1}} - \mathbf{u_{2}}) + \sigma_{2} (\mathbf{u_{2}} - \mathbf{u_{3}}) = \sigma_{3} (\mathbf{u_{1}} - \mathbf{u_{3}})
\end{equation}
It is easy to see now that if the sigmas are not all equal, then they are all different. This implies a linear relationship between the three points and means that they are aligned.

\subsubsection*{Remarks}

\begin{itemize}
	\item If the initial data of our problem belong to a single rarefaction-shock curve, then it will evolve staying on this curve for any time, so this curve is sometimes called an invariant manifold of dimension one. In other words, it is possible to restrict the system of conservation laws to the manifold, and the restriction will be a scalar conservation law:
	\be
	\partial_{t} \rho + \lambda(\phi(\rho)) \partial_{x} \rho = 0
	\ee
	where $\phi: \mathbb{R} \rightarrow \mathbb{R}^{n}  $ is a parameterization  of the curve. This similarity between Temple class and scalar conservation laws allows extending some of the general theories that were proven only for scalar systems to Temple classes. For instance, the existence and uniqueness of physical solutions were proven in  \cite{heibig1994existence}

	\item
	Riemann variables are build-in within Temple systems: Let $\mathbf{l}_{i}(\mathbf{u}^{*})$ be the left eigenvector at $\mathbf{u}^{*}$. The j-rarefaction lines for all $j\neq i$ form a hyperplane perpendicular to $\mathbf{l}_{i}(\mathbf{u}) $ will be perpendicular to this hyperplane for all $\mathbf{u}$ belonging to it. If we parameterize the family of hyperplanes associated with the i-left field by $z_{i}$ then this will constitute obviously a Riemann variable: $ \mathbf{l_{i}} \propto \nabla z_{i} $. For a more rigorous treatment of the existence of Riemann variables, one has to make use of the Frobenius theorem.

	\item A conservation system can of course be partially of Temple class, in the sense that the properties of Temple apply only to some of the characteristic fields. These fields would form an invariant manifold where initial data stay on it and do not leave it. The restriction of the conservation system to this manifold would be a Temple system.
	
\end{itemize}

\subsubsection{Temple class for a system of two conservation laws.}

Consider two conservation laws corresponding to non-degenerate fields.

\begin{equation}
\begin{split}
\partial_{t} u + \partial_{x}f(u,v) = 0 \\
\partial_{t} u + \partial_{x}g(u,v) = 0 
\end{split}
\end{equation}

We assume that the rarefaction lines are not parallel for either of the fields. If one of them is, then this one decouples trivially from the system. We can parameterize the i-rarefaction field by the slope of its lines, say $z_{i}(u,v)$. These variables would obviously be Riemann invariant. We are interested here in determining the class of currents that can generate such systems. 
One can show that:

\begin{equation}
\begin{split}
g = f z_{0} + H_{1}(z_{0}) \\
g = f z_{1} + H_{2}(z_{1})
\end{split}
\end{equation}
This means that if we know the currents on the boundaries of a temple class system, we can determine the currents on the bulk.

\chapter{Hydrodynamic behavior of the  two--TASEP}
\label{hydro}

\begin{tcolorbox}
The content of this chapter is based on the article Cantini, Zahra (2022) that is published in Journal of Physics A: Mathematical and Theoretical 55.30 (2022) https://iopscience.iop.org/article/10.1088/1751-8121/ac79e3/meta
\end{tcolorbox}

\section*{Abstract}
We address the question of the hydrodynamic
behavior of a 2-species generalization of the TASEP, called 2--TASEP, introduced by Derrida \cite{derrida1996statphys} and Mallick \cite{mallick1996shocks}.
We find that the auxiliary variables, introduced previously in the literature to express the density dependence of particle currents,  
turn out to be the Riemann variables of the 
conservation equations. This allows us to work out quite explicitly the rarefaction and shock solutions and to completely solve the associated Riemann problem. Our theoretical results are confirmed by Monte Carlo simulations.



\section{Introduction}

The asymmetric simple exclusion process (ASEP) is a minimal model of transport in (quasi) one--dimensional systems. It consists of particles which occupy the sites of a one dimensional lattice with only one particle allowed on each lattice site.
These particles hop under the effect of an external driving force which breaks detailed balance and creates a stationary current. 
This model was introduced in the late 60s in biology to model translation in
protein synthesis \cite{macdonald1968kinetics} and independently in probability \cite{spitzer1970interaction} and afterwards it has found a wide spectrum of applications, ranging from theoretical and experimental studies of biophysical transport \cite{chou2011non} to 
%
%
%
%
modeling  traffic  flow  \cite{chowdhury2000statistical,evans1996bose}. 
As soon one considers models which are more suited for physical/biological systems, one will encounter variants of ASEP containing localized or mobile defects and several species of particles, which have different behaviors.  As a result, typically these models are not exactly solvable and even 
for some of the most basic questions, like the study of large scale behavior of the system (which in the case of ASEP is known to be described by the Burgers equation \cite{rost1981non,benassi1987hydrodynamical,
	rezakhanlou1991hydrodynamic,kipnis1998scaling}), approximations schemes like mean--field are  necessary. 

In this paper we address the question of the large scale or hydrodynamic 
behavior of an exactly solvable multispecies generalization of ASEP, consisting of two kinds of particles, $\bullet$--particles and $\circ$--particles, moving in opposite directions. One can think of them as opposite charged particles moving under the influence of an external electric field or as cars moving on two opposite lanes.
Each site of a one--dimensional lattice is either empty or occupied by one of the two kinds of particles. For convenience, empty sites can
be  treated as a third species of particles, the $\ast$--particles.  
%
%
In continuous time, a $\bullet$--particle jumps forward on empty sites  with rate $\beta$, while a white $\circ$--particle jumps backward on empty sites with rate $\alpha$. On top of this, an adjacent pair $\bullet\circ$ swaps to $\circ\bullet$ with rate $1$. 
\be
\begin{split}
	\bullet\,\ast& ~ \rightarrow ~ \ast\,\bullet\quad\text{rate}\quad \beta\\
	\ast\,\circ& ~ \rightarrow ~ \circ\,\ast\quad\text{rate}\quad \alpha\\
	\bullet\,\circ& ~ \rightarrow ~ \circ\,\bullet\quad\text{rate}\quad 1
\end{split}
\ee
This model has appeared in the literature under different names. It has been first considered in \cite{derrida1996statphys,mallick1996shocks}, where the stationary measure on a finite periodic lattice was written in a matrix product ansatz form \cite{derrida1993exact,blythe2007nonequilibrium}. It is also a particular case ($q=0$) of the so called AHR model 
\cite{arndt1998spontaneous,arndt1999spontaneous,rajewsky2000spatial},  in which the swap $\circ\bullet\rightarrow \bullet\circ$ is allowed with rate $q$.
Being  a natural 2--species generalization of TASEP we shall call this model $2$--TASEP.
It turns out that the 2--TASEP is Yang--Baxter integrable, this was first proven in \cite{popkov2002sufficient} in a particular case with the constraint $\alpha + \beta = 1$, which happens to be the same condition for the system to have a product invariant measure. For arbitrary values of $\alpha$ and $\beta$, the Yang-Baxter integrability was proven in \cite{cantini2008algebraic}.
It belongs indeed to a larger family of integrable multispecies exclusion processes introduced in \cite{cantini2016inhomogenous}. Bethe ansatz techniques can be used to solve  exactly 
for the long time limit behavior of the generating function of the  
currents \cite{derrida1999bethe,cantini2008algebraic}. 
More recently, 
in the case $\alpha+\beta=1$, the transition probabilities as well as the 
joint current distribution for some specific initial distribution of a 
finite number of $\bullet$ and $\circ$--particles have been obtained 
\cite{chen2018exact,chen2021limiting}, and an asymptotic analysis of 
these results has allowed to prove that the joint current distribution is 
given by a product of a Gaussian and a GUE Tracy-Widom distribution in the 
long time limit, as predicted by non--linear fluctuating hydrodynamics
\cite{van2012exact,spohn2014nonlinear,ferrari2013coupled}.

When $\alpha+\beta=1$ the stationary measure factorizes and the currents 
have a simple expression as function of the densities.
In \cite{fritz2004derivation,toth2005perturbation} the 
hydrodynamic limit of the 2--TASEP for $\alpha=\beta=\frac{1}{2}$ has 
been 
studied and proven to converge to the classical Leroux system of 
conservation laws \cite{leroux1978analyse,serre1988existence}.
The Leroux system is a notable example of a \emph{Temple class system} 
i.e. a 2--components conservation law whose shock and rarefaction 
curves coincide  \cite{temple1983systems}. 
The theory of Temple class systems shares several common features with the theory of single component conservation laws \cite{serre1999systems}, in particular 
well-posedness results for Temple systems are available for
a much larger class of initial data compared to general systems of conservation laws.

For arbitrary $\alpha$ and $\beta$ only numerical results based on  mean field approximation are available 
\cite{arndt2002spontaneous}.
In the present paper we study the exact hydrodynamic equations of the 2--TASEP and show that they are Temple class for arbitrary $\alpha$ and $\beta$. This allows us to compute their rarefaction and shock solution, as well as to solve completely the Riemann problem, which consists in determining the density profile starting from a domain-wall initial data.

%

\vspace{.3cm}

The paper is organized as follows. In Section \ref{sect:curr} we review and expand on results about the 2--TASEP currents obtained in \cite{cantini2008algebraic}. The core of the paper is Section \ref{sect:conserv} where the conservation laws are studied. We derive the rarefaction waves as well as the shock solutions and finally we solve the full Riemann problem. In Section \ref{sec:montecarlo} we compare the prediction of the hydrodynamic equations with Monte Carlo simulations.
Conclusions and some outlooks for further works are discussed in Section \ref{section:conclusion}.

\section{Currents}\label{sect:curr}

In this section we reproduce and expand the results of the analysis in \cite{cantini2008algebraic} in a convenient way, which makes manifest the symmetries of the model. 
In order to compute the particle currents as functions of 
the local densities, we consider our model on a periodic ring with a fixed 
number of particles of each species. Call $M_i$ the number of particles of 
species $i$, they are related to $N$, the length of the ring, by $N=M_\bullet+M_\circ+M_\ast$.
Let the system evolve starting at time $t=0$ from an arbitrary fixed configuration and call $n_{i,j}(t)$ the number of swaps of consecutive ordered pairs of particles of type $i,j$ up to time $t$. This number increases by $+1$ each time two consecutive ordered  particles of species $i,j$ exchange their position $i,j\rightarrow j,i$. 
The average rate of swaps  $i,j\rightarrow j,i$ in the steady state is just given by $\lim_{t\rightarrow+\infty}\frac{1}{t}\mathbb{E}\left[  n_{i,j}(t)\right]$, irrespectively of the initial state. The particle currents in the steady state are hence given by 
\begin{equation}
J_i= \lim_{t\rightarrow+\infty}\frac{1}{Nt}\mathbb{E}\left[ \sum_{j\ne i} n_{i,j}(t)-n_{j,i}(t)\right]
\end{equation}
In our case, it is convenient to introduce the following quantity
\be 
\Phi(\nu_{\bullet,\circ},\nu_{\bullet,\ast},\nu_{\ast,\circ})=\lim_{t\rightarrow+\infty}\frac{1}{t}\mathbb{E}\left[\nu_{\bullet,\circ}\,  n_{\bullet,\circ}(t)  + \nu_{\bullet,\ast}\, 
n_{\bullet,\ast}(t) 
+\nu_{\ast,\circ}\, n_{\ast,\circ}(t) \right].
\ee
The currents are obtained as specialization of $\Phi(\nu_{\bullet,\circ},\nu_{\bullet,\ast},\nu_{\ast,\circ})$
\be 
J_\bullet=\frac{\Phi(1,1,0)}{N},\quad J_\circ=	\frac{ \Phi(-1,0,-1)}{N},\quad J_\ast=\frac{\Phi(0,-1,1)}{N}.
\ee
In \cite{cantini2008algebraic} the function $\Phi(\nu_{\bullet,\circ},\nu_{\bullet,\ast},\nu_{\ast,\circ})$  was shown to be given by the solution of the following equation
\be\label{det-phi} 
\det G(\Phi(\nu_{\bullet,\circ},\nu_{\bullet,\ast},\nu_{\ast,\circ}),\nu_{\bullet,\circ},\nu_{\bullet,\ast},\nu_{\ast,\circ})=0.
\ee
where the matrix $G(\Phi,\nu_{\bullet,\circ},\nu_{\bullet,\ast},\nu_{\ast,\circ})$ is given by
\be 
G(\Phi,\nu_{\bullet,\circ},\nu_{\bullet,\ast},\nu_{\ast,\circ})=
\scalebox{.75}{$\left(
	\begin{array}{ccc}
	\Phi& F_\alpha[M_\circ,M_\bullet,M_\ast] & F_\beta[M_\bullet,M_\circ,M_\ast]\\
	\nu_{\bullet,\circ}M_\bullet+\nu_{\ast,\circ}M_\ast& F_\alpha[M_\circ+1,M_\bullet,M_\ast] & -F_\beta[M_\bullet,M_\circ+1,M_\ast]\\
	\nu_{\bullet,\circ}M_\circ+\nu_{\bullet,\ast}M_\ast &
	-F_\alpha[M_\circ,M_\bullet+1,M_\ast]& F_\beta[M_\bullet+1,M_\circ,M_\ast]
	\end{array}
	\right)$}
\ee
with
\be
F_\gamma[a,b,c]:= \oint_0 \frac{dz}{2\pi i}
\frac{1}{z^{a}(z-1)^{b}(z-\gamma)^{c}}.
\ee
When one of the particle species
is \emph{strictly} absent (i.e. when one among $M_\bullet,M_\circ,M_\ast$ vanishes) the model reduces to a single species TASEP and it is not difficult to see that one of the currents vanishes, while the others boil down to the usual TASEP  current. On the other hand in the following we shall assume that at least one particle per species is present ($M_i\neq 0$) and we shall be mainly interested in the thermodynamic limit of these quantities as $N\rightarrow \infty$, with $\lim_{N\rightarrow\infty}\frac{M_i}{N}=\rho_i$. 
We shall see, as already found in \cite{mallick1996shocks,derrida1996statphys}, that the presence of even a single particle of a given species (i.e. an infinitesimally vanishing but not strictly zero density) can affect the macroscopic behavior of the system. 
With this in mind, we consider the limit 
$a\longrightarrow \infty$, with $b/a$  and $c/a$ fixed, of the function
$F_\gamma[a,b,c]$,
that behaves like\footnote{Here we are supposing $a,b,c,\gamma>0$. } 
$$
F_\gamma[a,b,c]\sim \frac{1}{z_\gamma^a(z_\gamma-1)^{b}(z_\gamma-\gamma)^{c}},
$$
where $z_\gamma$ is the zero  of the saddle point equation
$
\frac{a}{z}+\frac{b}{z-1}+\frac{c}{z-\gamma}=0
$,  belonging to the interval $[0,\min[1,\gamma]]$.
Applying this expression in eq.(\ref{det-phi})
we get in the thermodynamic limit  
\be 
\lim_{N\rightarrow \infty}\frac{\Phi(\nu_{\bullet,\circ},\nu_{\bullet,\ast},\nu_{\ast,\circ})}{N}=(\nu_{\bullet,\circ}\rho_\bullet+\nu_{\ast,\circ}\rho_\ast)z_\alpha(1-z_\beta)+(\nu_{\bullet,\circ}\rho_\circ+\nu_{\bullet,\ast}\rho_\ast)z_\beta(1-z_\alpha)
\ee
where with $z_\alpha \in [0,\min(1,\alpha)]$ and $z_\beta \in [0,\min(1,\beta)]$ are the solution of the saddle point equations
\begin{gather}\label{chang-var1}
\frac{\rho_\circ}{z_\alpha}+\frac{\rho_\bullet}{z_\alpha-1}+\frac{1-\rho_\circ-\rho_\bullet}{z_\alpha-\alpha}=0\\ \label{chang-var2}
\frac{\rho_\bullet}{z_\beta}+\frac{\rho_\circ}{z_\beta-1}+\frac{1-\rho_\circ-\rho_\bullet}{z_\beta-\beta}=0.
\end{gather}
The result for the currents then reads
\begin{gather}\label{0Jrz} 
J_\circ= z_\alpha(z_\beta-1)+\rho_\circ(z_\alpha-z_\beta)\\
\label{1Jrz}
J_\bullet= z_\beta(1-z_\alpha)+\rho_\bullet(z_\alpha-z_\beta)\\
\label{2Jrz}
J_\ast= \rho_\ast(z_\alpha-z_\beta)
\end{gather}
Notice that eqs.(\ref{chang-var1},\ref{chang-var2}) are invariant 
under exchange $\rho_\circ\leftrightarrow \rho_\bullet$, $\alpha
\leftrightarrow \beta$ and $z_\alpha\leftrightarrow z_\beta$. This 
implies as expected, that under 
exchange $\rho_\circ\leftrightarrow \rho_\bullet$ and $\alpha\leftrightarrow
\beta$   we have  
$J_\circ\leftrightarrow-J_\bullet$.
Let us finish this section by showing how some known results fit in the analysis above.   
\begin{itemize}[leftmargin=0.5cm]
	\item [$\Diamond$] $\beta=1$. In this case $\bullet$--particles 
	do not distinguish $\circ$--particles from $\ast$--particles and so they behave just as particles in a single species TASEP. 
	This is reflected in eq.(\ref{chang-var2}), where $\rho_\circ$ disappears 
	and one finds $z_\beta=\rho_\bullet$, which replaced in eq.(\ref{1Jrz}) 
	gives $J_\bullet=\rho_\bullet(1-\rho_\bullet)$. The case $\alpha=1$ is 
	completely analogous: $z_\alpha=\rho_\circ$ and $J_\circ=\rho_\circ(\rho_\circ-1)$.  
	\item [$\Diamond$] $\alpha+\beta=1$. In this case it is known  that the stationary measure takes a factorized form  \cite{rajewsky2000spatial}. At the level of the currents, we have indeed $J_\circ=-\rho_\circ(\rho_\bullet+\alpha \rho_\ast)$ and $J_\bullet=\rho_\bullet(\rho_\circ+\beta \rho_\ast)$.
\end{itemize}

\subsection{The $z$ variables}

In our analysis the variables $\bfz=(z_\alpha, z_\beta)$ will play a prominent role, it is therefore important to work out their domain of definition $\mathcal{D}_z(\alpha,\beta)$ corresponding to the physical domain 
$\mathcal{D}$ in the variables $\bfrho=(\rho_\circ,\rho_\bullet)$, $\rho_\circ,\rho_\bullet\geq 0, \rho_\circ+\rho_\bullet\leq 1$.
%
%
%

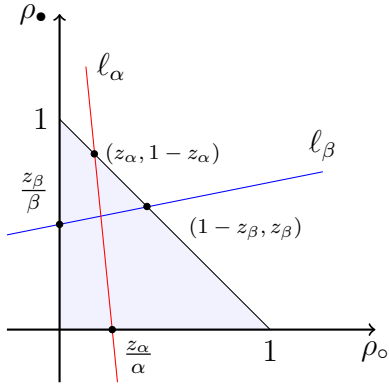
\begin{figure}[h!]
	\centering
	\begin{tikzpicture}[scale=1.4]
	\draw [fill=white!95!blue] (0,0)--(2,0) node [below] {$1$}--(0,2)node [left] {$1$}--cycle;
	\draw [thick,->] (-.5,0)--(3,0) node[below] {$\rho_\circ$};
	\draw [thick,->] (0,-.5)--(0,3) node[left] {$\rho_\bullet$};
	\draw[blue,domain=-.5:2.5,smooth,variable=\t]plot (\t,0.2*\t+1);
	\draw (2.5,0.2*2.5+1) node[above] {$\ell_\beta$};
	\draw[fill=black] (0,1) circle  (.03) node[above left] {$\frac{z_\beta}{\beta}$};
	\draw[fill=black] (0.83,1.17) circle  (.03) node[below right] {\quad \scriptsize $(1-z_\beta,z_\beta)$};
	\draw[red,domain=-.5:2.5,smooth,variable=\t]plot (-0.1*\t+0.5,\t);
	\draw (-0.1*2.5+0.5,2.5) node[right] {$\ell_\alpha$};
	\draw[fill=black] (0.5,0) circle  (.03) node[below right] {$\frac{z_\alpha}{\alpha}$};
	\draw[fill=black] (0.33,1.67) circle  (.03) node[right] {\scriptsize $(z_\alpha,1-z_\alpha)$};
	\end{tikzpicture}
	\caption{The shaded triangle is the physical region $\mathcal{D}$.
		Given $z_\alpha \in [0,\min(1,\alpha)]$, the  line $\ell_\alpha$ (red) intersects the boundary of $\mathcal{D}$ at the bottom side and at the diagonal side at the point $C_\alpha$ of coordinates $\rho_\circ=z_\alpha,\rho_\bullet=1-z_\alpha$. Similarly given $z_\beta \in [0,\min(1,\beta)]$, the  line $\ell_\beta$ (blue) intersects the boundary of $\mathcal{D}$ at the left side and at the diagonal side at the point $C_\beta$ of coordinates $\rho_\circ=1-z_\beta,\rho_\bullet=z_\beta$. For the two lines to cross we need the point $C_\alpha$ to lie above the point $C_\beta$, which is the case iff $z_\alpha+z_\beta\leq 1$. }\label{cross-lines}
\end{figure}

First of all we have already seen above that $\bfz$ has to satisfy  $z_\alpha \in [0,\min(1,\alpha)]$ and $z_\beta \in [0,\min(1,\beta)]$. 
At fixed $\bfz$, the system of equations (\ref{chang-var1},\ref{chang-var2}) is just the crossing of two lines in the $\bfrho$ plane: $\ell_\alpha$ coming from eq.(\ref{chang-var1}) and $\ell_\beta$ coming from eq.(\ref{chang-var2}). 
In Fig. \ref{cross-lines} we show by a simple geometrical argument that these lines cross inside $\mathcal{D}$ whenever $z_\alpha+z_\beta\leq 1$. 
So in conclusion the domain $\mathcal{D}_z(\alpha,\beta)$ is given by $z_\alpha \in [0,\min(1,\alpha)]$, $z_\beta \in [0,\min(1,\beta)]$ and $z_\alpha+z_\beta\leq 1$.   
The geometrical reasoning explained in Fig. \ref{cross-lines} allows also to conclude that at fixed $z_
\alpha$, $\rho_\bullet$ is an increasing function of $z_\beta$, while at fixed 
$z_\beta$, $\rho_\circ$ is an increasing function of $z_\alpha$.

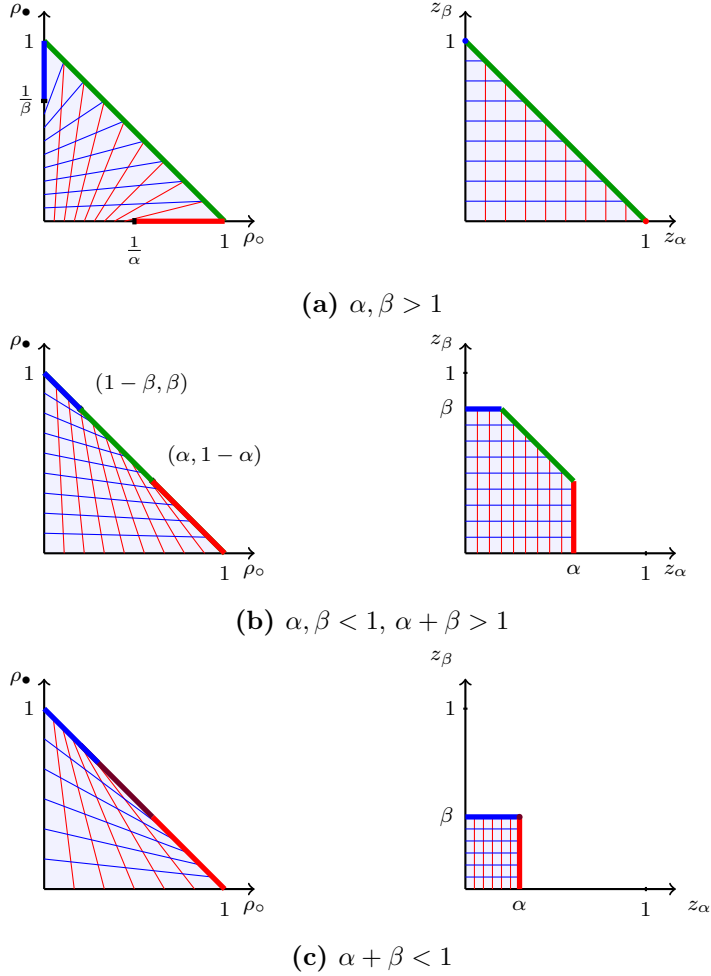
\begin{figure}[h!]
	\centering
	\begin{subfigure}{.6\linewidth}
		\begin{tikzpicture}
		\begin{scope}[scale = .4]
		\fill [fill=white!95!blue] (0,0)--(0,6)--(6,0)--cycle;
		\draw[thick,->] (0,0) -- (7,0) node[below] {\scriptsize $\rho_\circ$};
		\draw[thick,->] (0,0) -- (0,7) node[left] {\scriptsize $\rho_\bullet$};
		\foreach \x in {1,2,...,8}
		\draw [red] (\x/3,0) -- (\x*2/3, 6- \x*2/3);
		
		\foreach \y in {1,2,...,8}
		\draw [blue] (0,4*\y/9) -- (6-\y*2/3, \y*2/3);
		
		\draw [line width=0.7 mm, black!40!green] (0,6) -- (6,0);
		\draw [line width=0.7 mm, red ]  (3,0) -- (6,0);
		\draw [line width=0.7 mm, blue ] (0,4) -- (0,6);
		
		\draw [ultra thick](-3pt,4 cm) -- (3pt,4 cm) node[left] {\scriptsize $\frac{1}{\beta}$};
		\draw [ultra thick] (3 cm,3pt) -- (3 cm,-3pt) node[below] {\scriptsize $\frac{1}{\alpha}$};
		\draw [thick](2pt,6 cm) node[left] {\scriptsize $1$};
		\draw [thick](6 cm,-2pt) node[below] {\scriptsize $1$};
		\end{scope}
		\begin{scope}[scale = .4, xshift=14cm]
		\fill [fill=white!95!blue] (0,0)--(0,6)--(6,0)--cycle;
		\draw[thick,->] (0,0) -- (7,0) node[below] {\scriptsize $z_{\alpha}$};
		\draw[thick,->] (0,0) -- (0,7) node[left] {\scriptsize $z_{\beta}$};
		
		\foreach \x in {1,2,...,8}
		\draw [red] (2*\x/3,0) -- (2*\x/3, 6- 2*\x/3);
		
		\foreach \y in {1,2,...,8}
		\draw [blue] (0,2*\y/3) -- (6-2*\y/3, 2*\y/3);

		\draw [line width=0.7 mm, black!40!green] (0,6) -- (6,0);
		
		\fill [blue]  (0,6) circle (1mm) ;
		\fill [red]  (6,0) circle (1mm) ;
		
		\draw [thick](2pt,6 cm) node[anchor=east] {\scriptsize $1$};
		\draw [thick](6 cm,-2pt) node[anchor=north] {\scriptsize $1$};
		
		\end{scope}
		\end{tikzpicture}
		\caption{$\alpha,\beta>1$}\label{domains-maps-a}
	\end{subfigure}
	\begin{subfigure}[b]{.6\linewidth}
		\begin{tikzpicture}
		\begin{scope}[scale = .4]
		\fill [fill=white!95!blue] (0,0)--(0,6)--(6,0)--cycle;
		
		\draw[thick,->] (0,0) -- (7,0) node[below] {\scriptsize $\rho_\circ$};
		\draw[thick,->] (0,0) -- (0,7) node[left] {\scriptsize $\rho_\bullet$};

		\draw [ultra thick] (6*0.2,6-6*0.2)++(45:-3pt)--++(45:6pt) node [above right] {\scriptsize $(1-\beta,\beta)$};
		
		\draw [ultra thick] (6-0.4*6,0.4*6)++(45:-3pt)--++(45:6pt) node [above right] {\scriptsize $(\alpha,1-\alpha)$};
		
		\foreach \x in {1,2,...,8}
		\draw [red] (\x*2/3,0) -- (0.6*\x*2/3, 6- 0.6*\x*2/3);
		
		\foreach \y in {1,2,...,8}
		\draw [blue] (0,\y*2/3) -- (6-0.8*\y*2/3, 0.8*\y*2/3);
		
		\draw [line width=0.7 mm, black!40!green] (0,6) -- (6,0);
		
		\draw[line width=0.7 mm, blue] (6*0.2,6-6*0.2) -- (0,6);
		\draw [line width=0.7 mm, red] (6,0) -- (6-0.4*6,0.4*6);
		
		\draw [thick] (2pt,6 cm) node[left] {\scriptsize $1$};
		\draw [thick] (6 cm,-2pt) node[below] {\scriptsize $1$};
		\end{scope}

		\begin{scope}[scale = .4, xshift=14cm]
		\fill [fill=white!95!blue] (0,0)--(0,0.8*6) -- (6-0.8*6,0.8*6) --(0.6*6,6-0.6*6)--(0.6*6,0)--cycle;

		\draw[thick,->] (0,0) -- (7,0) node[below] {\scriptsize $z_{\alpha}$};
		\draw[thick,->] (0,0) -- (0,7) node[left] {\scriptsize $z_{\beta}$};

		\foreach \x in {3,4,...,8}
		\draw [red] (0.6*\x*2/3,0) -- (0.6*\x*2/3, 6- 0.6*\x*2/3);
		
		\foreach \x in {1,2}
		\draw [red] (0.6*\x*2/3,0) -- (0.6*\x*2/3, 6*0.8);
		
		\foreach \y in {1,2,3,4}
		\draw [blue] (0,0.8*\y*2/3) -- (6*0.6, 0.8*\y*2/3);
		\foreach \y in {5,6,7,8}
		\draw [blue] (0,0.8*\y*2/3) -- (6-0.8*\y*2/3, 0.8*\y*2/3);

		\draw [line width=0.7 mm, red] (0.6*6,0) -- (0.6*6,6-0.6*6);
		\draw [line width=0.7 mm, blue] (0,0.8*6) -- (6-0.8*6,0.8*6);
		
		\draw [line width=0.7 mm, black!40!green] (6-0.8*6,0.8*6) -- (0.6*6,6-0.6*6);
		
		\draw (6*0.6 cm,-1pt) node[anchor=north] {\scriptsize $\alpha$};
		
		\draw (-1pt,6*0.8 cm) node[anchor=east] {\scriptsize $\beta$};
		
		\draw [thick](-2pt,6 cm) -- (2pt,6 cm) node[anchor=east] {\scriptsize $1$};
		\draw [thick] (6 cm,2pt) -- (6 cm,-2pt) node[anchor=north] {\scriptsize $1$};
		
		\end{scope}

		\end{tikzpicture}
		\caption{$\alpha,\beta <1$, $\alpha+\beta>1$}\label{domains-maps-b}
		
	\end{subfigure}
	\begin{subfigure}[b]{.6\linewidth}
		\begin{tikzpicture}
		\begin{scope}[scale = .4]
		\fill [fill=white!95!blue] (0,0)--(0,6)--(6,0)--cycle;
		\draw[thick,->] (0,0) -- (7,0) node[below] {\scriptsize $\rho_\circ$};
		\draw[thick,->] (0,0) -- (0,7) node[left] {\scriptsize $\rho_\bullet$};

		\foreach \x in {1,2,3,4,5}
		\draw [red] (\x,0) -- (0.3*\x, 6- 0.3*\x);
		
		\foreach \y in {1,2,3,4,5}
		\draw [blue] (0,\y) -- (6-0.4*\y, 0.4*\y);

		\draw [line width=0.7 mm, black!40!purple] (6*0.2,6-6*0.2) -- (6-0.4*6,0.4*6);
		\draw[line width=0.7 mm, blue] (6*0.3,6-6*0.3) -- (0,6);
		\draw [line width=0.7 mm, red] (6,0) -- (6-0.4*6,0.4*6);
		
		\draw [thick](2pt,6 cm) node[left] {\scriptsize $1$};
		\draw [thick](6 cm,-2pt) node[below] {\scriptsize $1$};
		\end{scope}

		\begin{scope}[scale = .4, xshift=14cm]
		\fill [fill=white!95!blue] (0,0)--(0,6*0.4) -- (6*0.3,6*0.4)--(6*0.3,0)--cycle;
		\draw[thick,->] (0,0) -- (7,0) node[anchor=north west] {\scriptsize $z_{\alpha}$};
		\draw[thick,->] (0,0) -- (0,7) node[anchor=south east] {\scriptsize $z_{\beta}$};
		
		\foreach \x in {1,2,3,4,5}
		\draw [red] (0.3*\x,0) -- (0.3*\x, 6*0.4);
		
		\foreach \y in {1,2,3,4,5}
		\draw [blue] (0,0.4*\y) -- (0.3*6, 0.4*\y);

		\draw [line width=0.7 mm, blue]  (0,6*0.4) -- (6*0.3,6*0.4);
		\draw [line width=0.7 mm, red]  (6*0.3,0) -- (6*0.3,6*0.4);

		\fill [black!40!purple]  (6*0.3,6*0.4) circle (1mm) ;
		
		\draw  (6*0.3 cm,-1pt) node[anchor=north] {\scriptsize $\alpha$};
		
		\draw (-1pt,6*0.4 cm) node[anchor=east] {\scriptsize $\beta$};
		
		\draw [thick](-2pt,6 cm) -- (2pt,6 cm) node[anchor=east] {\scriptsize $1$};
		\draw [thick] (6 cm,2pt) -- (6 cm,-2pt) node[anchor=north] {\scriptsize $1$};
		
		\end{scope}
\end{tikzpicture}
		\caption{$\alpha+\beta<1$}\label{domains-maps-c}
\end{subfigure}
\caption{On the left the physical domain $\mathcal{D}$ in the densities plane, on the right the corresponding domain $\mathcal{D}_z$ in the $\bfz$ variables plane. On both sides we have drawn in red the lines $\ell_\alpha$ at constant $z_\alpha$ and in blue the lines $\ell_\beta$ at constant $z_\beta$. The lines $\ell_\alpha$ (red) have slope $\frac{\alpha(1-z_\alpha)}{(\alpha-1)z_\alpha}$ and in 
particular they have positive slope for $\alpha>1$ as in 
figure {\bf (a)} and negative slope for $\alpha<1$ as in 
figures {\bf (b)} and {\bf (c)}.  Similarly, the lines $\ell_\beta$  (blue) have slope $\frac{(\beta-1)z_\beta}{\beta(1-z_\beta)}$,  they have positive slope for $\beta>1$ as in 
figure {\bf (a)} and negative slope for $\beta<1$ as in figures {\bf (b)} and {\bf (c)}.
}\label{domains-maps}

\end{figure}
In Figs. \ref{domains-maps}  we have reported 
on the left the physical domain $\mathcal{D}$ in the densities plane and 
on the right the corresponding domain $\mathcal{D}_z$ in the $\bfz$ 
variables plane for the cases $\alpha,\beta>1$ (\ref{domains-maps-a}) , $\alpha,\beta<1$ and $\alpha+\beta>1$ (\ref{domains-maps-b}) and $\alpha+\beta<1$ (\ref{domains-maps-c}). 
Notice that in Fig. \ref{domains-maps-a} the thick 
red segment on the $\rho_\circ$ axis is mapped to the point $(z_
\alpha=1,z_\beta=0)$ and the thick blue segment on the $\rho_\bullet$ 
axis is mapped to the point  $(z_\alpha=0,z_\beta=1)$. 
In Fig. \ref{domains-maps-c}: the overlap of the red and blue segments on the boundary $\rho_\ast=0$ is mapped to the point  $(z_\alpha=\alpha,z_\beta=\beta)$.
This indicates that the mapping $\bfrho\rightarrow \bfz$ can be singular on the boundary of the physical domain $\mathcal{D}$, where at least one of the densities vanishes.
Let's analyze the different possibilities and work out the portion of the boundary where the mapping is singular. 
\begin{itemize}[leftmargin=0cm] 
	\item []\fbox{$\rho_\circ\rightarrow 0$} In this case from 
	eq.(\ref{chang-var1}) we deduce that $z_\alpha= 0$ while the two 
	solutions of eq.(\ref{chang-var2}) are $z_\beta=1,z_\beta=
	\rho_\bullet\beta$ and we have to retain the smallest one. If $\beta\leq 
	1$ then on the $\rho_\circ=0$ axis the map is $1$-to-$1$ and there are no 
	singularities, on the other hand, if $\beta>1$ then all the points $\rho_
	\bullet\geq \beta^{-1}$ are mapped to the same point $(z_\alpha=0,z_
	\beta=1)$.
	\item []\fbox{$\rho_\bullet\rightarrow 0$} This case is treated similarly to the previous one: we have $z_\beta= 0$ and the two solutions of eq.(\ref{chang-var1}) are $z_\alpha=1,z_\alpha=\rho_\circ\alpha$. If $\alpha \leq 1$ then on the $\rho_\bullet=0$ axis the map is $1$-to-$1$ and there are no singularities, on the other hand, if $\alpha>1$ then all the points $\rho_\circ\geq \alpha^{-1}$ are mapped to the same point $(z_\alpha=1,z_\beta=0)$.
	\item []\fbox{$\rho_\ast\rightarrow 0$} Eqs.(\ref{chang-var1},
	\ref{chang-var2}) have solutions, $z_\alpha=\alpha, z_\alpha=\rho_\circ$, 
	$z_\beta=\beta, z_\beta=1-\rho_\circ$. Whenever $\alpha+\beta\leq 1$, all the points on the line 
	$\rho_\circ+\rho_\bullet=1$ such that $\alpha\leq \rho_\circ\leq 1-\beta$ are 
	mapped to the single point $z_\alpha=\alpha, z_\beta=\beta$. 
\end{itemize}

\subsection{Behavior at the boundary of the physical domain}
The singularities of the mapping $\bfrho\rightarrow \bfz$ reflect some important features of the model. Let's consider  the currents
of non zero density particles at the boundary of $\mathcal{D}$

\noindent
\fbox{$\rho_\circ\rightarrow 0$} We have for the current 
\be\label{singul1}
J_\bullet(\rho_\bullet) =  \left\{
\begin{array}{ll}
	\beta \rho_\bullet(1-\rho_\bullet) & 0\leq \rho_\bullet\leq \beta^{-1}\\
	(1-\rho_\bullet) & \beta^{-1}\leq \rho_\bullet\leq 1.
\end{array}
\right.
\ee

\noindent
\fbox{$\rho_\bullet\rightarrow 0$} We have for the current
\be\label{singul0}
J_\circ(\rho_\circ) =  \left\{
\begin{array}{ll}
	-\alpha \rho_\circ(1-\rho_\circ) & 0\leq \rho_\circ\leq \alpha^{-1}\\
	-(1-\rho_\circ) & \alpha^{-1}\leq \rho_\circ\leq 1.
\end{array}
\right.
\ee

\noindent
\fbox{$\rho_\ast\rightarrow 0$} We have for the current
\be\label{singul10}
J_\bullet(\rho_\bullet) =  \left\{
\begin{array}{ll}
	\rho_\bullet(1-\rho_\bullet) & 0\leq \rho_\bullet\leq \beta, 1-\alpha\leq \rho_\bullet \leq 1 \\
	\beta(1-\alpha)+(\alpha-\beta)\rho_\bullet & \beta\leq \rho_\bullet\leq 1-\alpha.
\end{array}
\right.
\ee
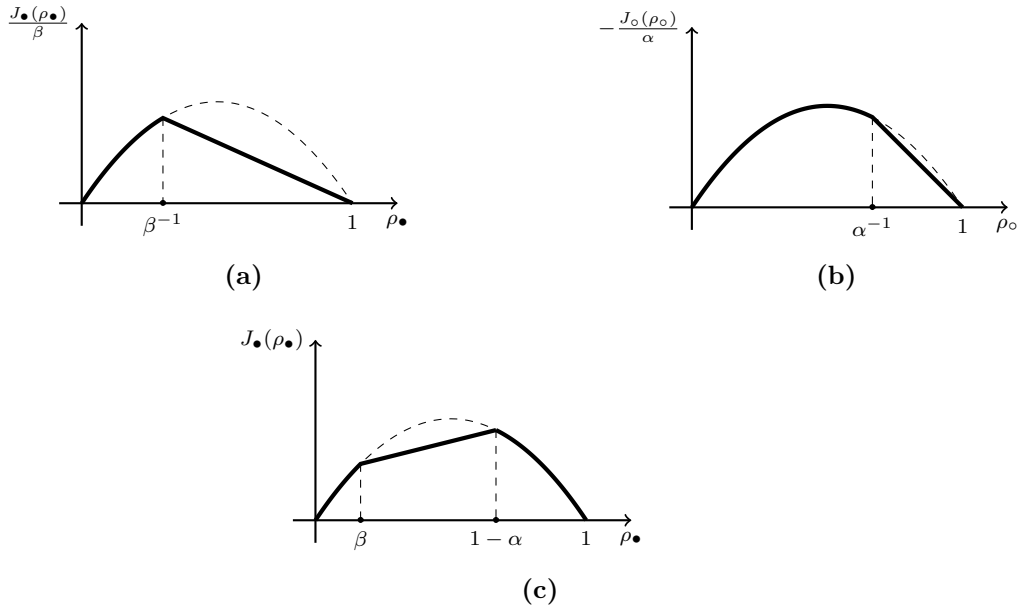
\begin{figure}[h]
	\begin{center}
		\begin{subfigure}[b]{0.4\linewidth}
			\begin{tikzpicture}[scale=1.2]
			\draw [thick,->] (-.25,0)--(3.5,0)node [below] {\scriptsize $\rho_\bullet$};
			\draw [thick,->] (0,-.25)--(0,2)node [left] {\scriptsize $\frac{J_\bullet(\rho_\bullet)}{\beta}$};
			\draw[dashed,domain=0:3,smooth,variable=\t]plot (\t,{\t*(3-\t)/2});
			\def \n {0.9}
			\draw[ultra thick,domain=0:\n,smooth,variable=\t]plot (\t,{\t*(3-\t)/2})--(3,0) node[below] {\scriptsize $1$};
			\draw[dashed,fill=black] (\n,{\n*(3-\n)/2})--(\n,0)circle  (.03) node[below] {\scriptsize $\beta^{-1}$};
			\end{tikzpicture}
			\caption{}
		\end{subfigure}
		\hspace{1cm}
		\begin{subfigure}[b]{0.4\linewidth}
			\begin{tikzpicture}[scale=1.2]
			\draw [thick,->] (-.25,0)--(3.5,0)node [below] {\scriptsize $\rho_\circ$};
			\draw [thick,->] (0,-.25)--(0,2)node [left] {\scriptsize $-\frac{J_\circ(\rho_\circ)}{\alpha}$};
			\def \r {1/2}
			\def \n {2}
			\draw[dashed,domain=0:3,smooth,variable=\t]plot (\t,{\t*(3-\t)*\r});
			\draw[ultra thick,domain=0:{\n},smooth,variable=\t]plot (\t,{\t*(3-\t)*\r})--(3,0) node[below] {\scriptsize $1$};
			\draw[dashed,fill=black] (\n,{\n*(3-\n)*\r})--(\n,0)circle  (.03) node[below] {\scriptsize $\alpha^{-1}$};
			\end{tikzpicture}
			\caption{}
		\end{subfigure}
		
		\vspace{.3cm}
		\begin{subfigure}[b]{0.5\linewidth}
			\begin{tikzpicture}[scale=1.2]
			\draw [thick,->] (-.25,0)--(3.5,0)node [below] {\scriptsize $\rho_\bullet$};
			\draw [thick,->] (0,-.25)--(0,2)node [left] {\scriptsize $J_\bullet(\rho_\bullet)$};
			\def \r {1/2}
			\def \n {.5}
			\def \f {2}
			\draw[dashed,domain=0:3,smooth,variable=\t]plot (\t,{\t*(3-\t)*\r})  node[below] {\scriptsize $1$};
			\draw[ultra thick,domain=0:{\n},smooth,variable=\t]plot (\t,{\t*(3-\t)*\r})-- (\f,{\f*(3-\f)*\r});
			\draw[ultra thick,domain={\f}:3,smooth,variable=\t]plot (\t,{\t*(3-\t)*\r});
			\draw[dashed,fill=black] (\n,{\n*(3-\n)*\r})--(\n,0)circle  (.03) node[below] {\scriptsize $\beta$};
			\draw[dashed,fill=black] (\f,{\f*(3-\f)*\r})--(\f,0)circle  (.03) node[below] {\scriptsize $1-\alpha$};
			\end{tikzpicture}
			\caption{}
		\end{subfigure}
	\end{center}
	\caption{{\bf (a)} Current $J_\bullet$ for $\beta>1$ and $\rho_\circ=0$. {\bf (b)} Current $J_\circ$ for $\alpha>1$ and $\rho_\bullet=0$. 
		{\bf (c)} Current $J_\bullet$ for $\alpha+\beta<1$ and $\rho_\ast=0$. 
		The dashed lines correspond to the current for a strict absence of $\circ$--particles {\bf (a)}, $\bullet$--particles {\bf (b)}, $\ast$--particles {\bf (c)}.}
\end{figure}
These results have to be compared to the situation in which we have \emph{strict} absence of a species of particles (and not just vanishing density).
Consider for example a system without $\circ$--particles. Such a system is effectively a single species TASEP with jump rates equal 
to $\beta$ and hence with current just $J_\bullet(\rho_\bullet)=\beta \rho_\bullet(1-\rho_\bullet)$. Comparing this with eq.(\ref{singul1}) we see that for $\beta>1$ this behavior holds only for $0\leq \rho_\bullet\leq \beta^{-1}$, while for $\rho_\bullet > \beta^{-1}$
the presence of even a single $\circ$--particle affects the macroscopic behavior of the system, giving rise to a modified current.

At the boundary of $\mathcal{D}$, the average speed of the zero density particles displays also an interesting behavior.
\begin{itemize}[leftmargin=0cm] 
	\item [] \fbox{$\rho_\circ\rightarrow 0$} Speed of $\circ$--particles 
	\be\label{vel-singul1}
	v_\circ(\rho_\bullet) =  \left\{
	\begin{array}{ll}
		-\frac{\alpha+\beta(1-\alpha)\rho_\bullet(1-\rho_\bullet)}{1+(\alpha-1)\rho_\bullet}-\beta \rho_\bullet & 0\leq \rho_\bullet\leq \beta^{-1}\\
		-1 & \beta^{-1}\leq \rho_\bullet\leq 1,
	\end{array}
	\right.
	\ee
	\item [] \fbox{$\rho_\bullet \rightarrow 0$} Speed of $\bullet$--particles 
	\be\label{vel-singul0}
	v_\bullet(\rho_\circ) =  \left\{
	\begin{array}{ll}
		\frac{\beta+\alpha(1-\beta)\rho_\circ(1-\rho_\circ)}{1+(\beta-1)\rho_\circ}-\alpha \rho_\circ & 0\leq \rho_\circ\leq \alpha^{-1}\\
		1 & \alpha^{-1}\leq \rho_\circ\leq 1,
	\end{array}
	\right.
	\ee
	\item [] \fbox{$\rho_\ast \rightarrow 0$} Speed of $\ast$--particles 
	\be\label{vel-singul10}
	v_\ast(\rho_\bullet) =  z_\alpha-z_\beta
	\ee
	with
	\be
	z_\alpha = \left\{
	\begin{array}{cc}
		\alpha & \rho_\bullet\leq 1-\alpha\\ 
		1-\rho_\bullet & \rho_\bullet \geq 1-\alpha
	\end{array}
	\right.,\qquad
	z_\beta = \left\{
	\begin{array}{cc}
		\beta & \rho_\bullet\geq \beta\\ 
		\rho_\bullet & \rho_\bullet \leq \beta
	\end{array}
	\right.
	\ee
\end{itemize} 
The results in eqs.(\ref{vel-singul1}--\ref{vel-singul10}) have been obtained in the literature by considering systems with a single particle of either species: the cases $\rho_\ast\rightarrow 0$,  eqs.(\ref{singul10},\ref{vel-singul10}) first appeared  in  
\cite{mallick1996shocks,derrida1999bethe}, the cases $\rho_\circ \rightarrow 0$ or $\rho_\bullet \rightarrow  0$,  eqs.(\ref{singul1},\ref{vel-singul1}) and eqs.(\ref{singul0},\ref{vel-singul0}) first appeared in \cite{lee1997two}.

\section{Conservations laws}\label{sect:conserv}

Under Euler--scaling (where site position and time scale as $\epsilon^{-1}n,\epsilon^{-1}t$ for $\epsilon\rightarrow 0$) the density profiles are expected to evolve deterministically as solutions of a system of conservation laws.
Consider initial data $\rho_i^{(0)}(x)$ 
and to such data associate a family of initial conditions of  the 2-TASEP of product Bernoulli form, with local probability at site $n$ given by 
$$  
\mathbb{E}[\chi_i^{\epsilon}(n,t=0)] =  \rho_i^{(0)}(\epsilon n),  
$$
where $\chi_i^{\epsilon}(n,t)$ is the $i$--th species indicator function at time $t$ and site $n$.
We expect  that the random variable $\chi_i^{\epsilon}(\lfloor \epsilon^{-1}x \rfloor,\epsilon^{-1}t,)$ converges for $\epsilon \rightarrow 0$ to a deterministic density profile. More precisely we expect that
\be \label{converg}
\lim_{\epsilon \longrightarrow 0} \sum_{n:a\leq \epsilon n \leq b}\epsilon \,\chi_i^{\epsilon}(n,\epsilon^{-1}t) = \int_{a}^{b}\rho_i(x,t)dx, \qquad a.s.    
\ee
where   $\bfrho=(\rho_\circ,\rho_\bullet)$  is the solutions of a system of conservation laws 
\be \label{conserv-rho-j}
\partial_t \bfrho +\partial_x {\bf J}=0.
\ee
with initial condition $\bfrho(t=0)=\bfrho^{(0)}=(\rho^{(0)}_\circ,\rho^{(0)}_\bullet)$.
By making the usual hypothesis of \emph{local stationarity} we identify the local currents with the stationary currents at density ${\bfrho}$, given by eqs.(\ref{0Jrz},\ref{1Jrz}).
A more precise statement and proof of this result for the case $\alpha=\beta=\frac{1}{2}$ can be found in \cite{fritz2004derivation}. While the approach developed in \cite{fritz2004derivation} can be extended to the full $\alpha+\beta=1$ line (for which the stationary measure is product), 
it is not expected to work for arbitrary values of $\alpha$ and $\beta$ \cite{tothprivate}.
In the present paper we take  eqs.(\ref{converg},\ref{conserv-rho-j}) as a working hypothesis.
Eqs.(\ref{conserv-rho-j}) form a system of coupled conservation laws, whose non-linearity is known to be at the origin of characteristic phenomena such as shocks formation in finite time, and rarefaction waves, i.e. self-similar solutions, which  present regions expanding in time at constant speed where the densities interpolate between two boundary values.
In the following we shall analyze in detail eqs.(\ref{conserv-rho-j}). We shall show that the variables $\bfz$ are Riemann variables for this system. On general grounds we know that the rarefaction fans can be expressed in an implicit form involving the Riemann variables. What is more surprising is that for our system also the shock solutions are explicitly written in terms of the Riemann variables: they correspond to a discontinuity of only one of the two Riemann variables (the other one being continuous). Putting together fans and shock we can explicitly solve the Riemann problem. 
In the Section \ref{sec:montecarlo} these theoretical results are compared to Monte Carlo simulations of the $2$--TASEP.

\subsection{The cases $\alpha=\beta=1$ and $\alpha+\beta=1$}

Before discussing the system of equation \eqref{conserv-rho-j} in full generality let us start with some comments on two particular cases:  $\alpha=\beta=1$ and $\alpha+\beta=1$

\vspace{.3cm}
\noindent
\emph{ $\bullet\quad \alpha=\beta=1$}.

\noindent
When $\beta=1$, the $\bullet$--particles don't distinguish 
$\circ$--particles from $\ast$--particles. This means that the 
$\bullet$--particles evolve as in a single species TASEP. At the level of currents, for  $\beta=1$ we have indeed $J_\bullet=\rho_\bullet(1-\rho_\bullet)$. In this case the conservation law for $\rho_\bullet$ completely decouples from that of $\rho_\circ$ and takes the usual form of the non-viscous Burgers equation
$$
\partial_t \rho_\bullet +(1-2\rho_\bullet) \partial_x \rho_\bullet=0.
$$  
Analogously, for $\alpha =1$, the conservation law for $\rho_\circ$ completely decouples from that of $\rho_\bullet$ and takes the form
$$
\partial_t \rho_\circ -(1-2\rho_\circ) \partial_x \rho_\circ=0.
$$  
So for $\alpha=\beta=1$, system \eqref{conserv-rho-j} just decouples completely into two Burgers equations.

\vspace{.3cm}
\noindent
\emph{ $\bullet\quad \alpha+\beta=1$}.

\noindent
In this case, thanks to the factorization of the stationary measure,  the currents can be explicitly written as functions of the densities
$$
J_\circ=-\rho_\circ(\rho_\bullet+\beta(1-\rho_\circ-\rho_\bullet)),\qquad J_\bullet=\rho_\bullet(\rho_\circ+\alpha(1-\rho_\circ-\rho_\bullet)).
$$ 
One can consider conserved quantities $\rho$ and $v$, defined by 
\begin{equation}
\left(
\begin{array}{c}
\rho \\
v
\end{array}
\right) =
\left(\begin{array}{cc}
-\frac{\alpha(\alpha+2\beta)}{3}& -\frac{\beta(\alpha+2\beta)}{3}  \\
\alpha & -\beta
\end{array}
\right)\left(
\begin{array}{c}
\rho_\circ \\
\rho_\bullet
\end{array}
\right)+\left(
\begin{array}{c}
\frac{(2\alpha+\beta)(2\beta+\alpha)}{9} \\
\frac{\beta-\alpha}{3}
\end{array}
\right)
\end{equation}
The associated currents are (up to irrelevant additive constants)
$$
J_\rho= \rho v,\qquad J_v= \rho+v^2.
$$
These are the currents of the Leroux system \cite{leroux1978analyse,serre1988existence} (the particular case $\alpha=\beta=\frac{1}{2}$ is the one considered in \cite{fritz2004derivation}), which is known to be a Temple class system. 

\subsection{The general case: Riemann variables}

For the problem under investigation 
one could expect that in addition to the generic complexity of the analysis of a coupled system of conservation equations, one has to face the further complication due to the implicit 
dependence of the currents on the densities, which goes through the auxiliary variables $\bfz$.
%
%
Actually, quite unexpectedly, what seems a drawback of the equations turns out to be the main feature which allows to solve them.
Indeed the variables $\bfz$ happen to be Riemann variables for our conservation laws, i.e. they diagonalize the system of eqs.(\ref{conserv-rho-j}) and simplify substantially their analysis.
From now on we want to think of both $\bfrho$ and $\bfJ$ as functions of  
$\bfz$ .

We first notice that, solving eqs.(\ref{0Jrz},\ref{1Jrz}) for the densities $\bfrho$ in terms of 
$\bfz$ and $\bfJ$ and replacing them into eqs.(\ref{chang-var1},\ref{chang-var2}), the currents are the solution 
of the following linear system of equations
\begin{gather}\label{Jz0}
\frac{J_\circ}{z_\alpha}+\frac{J_\bullet}{z_\alpha-1}-\frac{J_\circ+J_\bullet}{z_\alpha-\alpha}+1=0\\\label{Jz1}
\frac{J_\bullet}{z_\beta}+\frac{J_\circ}{z_\beta-1}-\frac{J_\circ+J_\bullet}{z_\beta-\beta}+1=0.
\end{gather}
Now differentiate the l.h.s. of eq.(\ref{chang-var1}) with respect to $t$, differentiate the l.h.s. of eq.(\ref{Jz0}) with respect to $x$ and sum the obtained results.
%
Thanks to the conservation laws eqs.(\ref{conserv-rho-j}) the derivatives $\partial_t \bfrho$ and $\partial_x \bfJ$ cancel and 
one remains with
\begin{equation}\label{conserv-za}
\partial_t z_\alpha +v_\alpha(\bfz) \partial_x z_\alpha=0,\qquad v_\alpha(\bfz)=\frac{\left(\frac{J_\circ}{z_\alpha^2}+\frac{J_\bullet}{(z_\alpha-1)^2}-\frac{J_\circ+J_\bullet}{(z_\alpha-\alpha)^2}\right)}{\left(\frac{\rho_\circ}{z_\alpha^2}+\frac{\rho_\bullet}{(z_\alpha-1)^2}+\frac{1-\rho_\circ-\rho_\bullet}{(z_\alpha-\alpha)^2}\right)}.
\end{equation}
In the same way one obtains the equation for $z_\beta$
\begin{equation}\label{conserv-zb}
\partial_t z_\beta +v_\beta(\bfz) \partial_x z_\beta=0,\qquad v_\beta(\bfz)=\frac{\left(\frac{J_\bullet}{z_\beta^2}+\frac{J_\circ}{(z_\beta-1)^2}-\frac{J_\circ+J_\bullet}{(z_\beta-\beta)^2}\right)}{\left(\frac{\rho_\bullet}{z_\beta^2}+\frac{\rho_\circ}{(z_\beta-1)^2}+\frac{1-\rho_\circ-\rho_\bullet}{(z_\beta-\beta)^2}\right)}.
\end{equation}
The speeds $v_\alpha$ and $v_\beta$ are the eigenvalues of the linearization matrix $\partial_{\rho_j}J_i$, and on general grounds they can also be written as
\begin{equation}
v_\alpha= \partial_{\rho_i}J_i|_{z_\beta}=\frac{\partial_{z_\alpha}J_i(\bfz)}{\partial_{z_\alpha}\rho_i(\bfz)},\qquad v_\beta= \partial_{\rho_i}J_i|_{z_\alpha}=\frac{\partial_{z_\beta}J_i(\bfz)}{\partial_{z_\beta}\rho_i(\bfz)}.
\end{equation}
A close inspection of their expression allows to conclude that 
\begin{equation}\label{ineq-v}
v_\beta(\bfz)\geq v_\alpha(\bfz),
\end{equation}
with the equality holding for $1-z_\alpha-z_\beta=0$, which is a non empty set only for $\alpha+\beta\geq 1$. We conclude that for $\alpha+\beta<1$, the system in \eqref{conserv-rho-j} is strictly hyperbolic on the whole physical domain $\mathcal{D}$, whereas for for $\alpha+\beta\geq 1$ it is degenerate hyperbolic on the locus $1-z_\alpha-z_\beta=0$, i.e. $\rho_\ast=0,\rho_\circ\leq \alpha, \rho_\bullet\leq \beta$  (green segments in Figs. \ref{domains-maps-a},\ref{domains-maps-b}) and strictly hyperbolic on the rest of the physical domain.

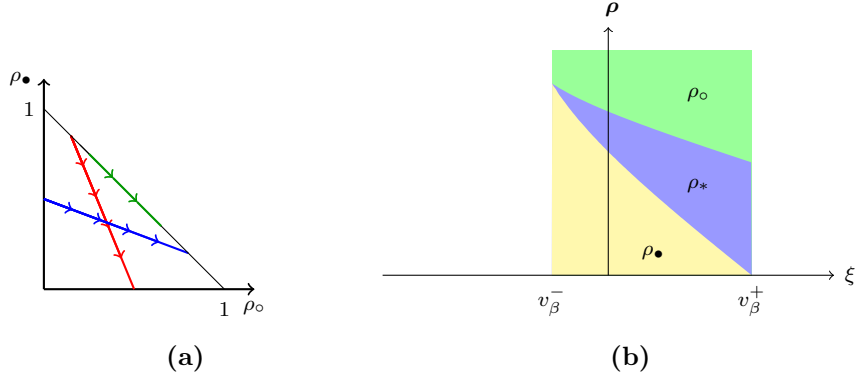
\begin{figure}[h!!]
	\begin{center}
		\begin{subfigure}[b]{0.3\linewidth}
			\begin{tikzpicture}[scale=.4]
			\draw[thick,->] (0,0) -- (7,0) node[below] {\scriptsize $\rho_\circ$};
			\draw[thick,->] (0,0) -- (0,7) node[left] {\scriptsize $\rho_\bullet$};
			
			\foreach \n in {0.2,0.4,0.6,0.8}
			\foreach \x in {3}
			\draw [thick, red, ->]  (0.3*\x, 6- 0.3*\x)-- (\x + 0.3*\x*\n - \x*\n, 6*\n- 0.3*\x*\n) ;
			
			\foreach \x in {3}
			\draw [thick, red] (\x,0) -- (0.3*\x, 6- 0.3*\x);

			\foreach \n in {0.2,0.4,0.6,0.8}
			\foreach \y in {3}
			\draw [thick, blue ,->] (0,\y) -- (6*\n-0.4*\y*\n, \y + 0.4*\y*\n - \y*\n);
			\foreach \y in {3}
			\draw [thick, blue] (0,\y) -- (6-0.4*\y,  0.4*\y );

			\draw  (6*0.2,6-6*0.2) -- (6-0.4*6,0.4*6);
			\draw (6*0.3,6-6*0.3) -- (0,6);
			\draw (6,0) -- (6-0.4*6,0.4*6);
			
			\draw [thick](2pt,6 cm) node[left] {\scriptsize $1$};
			\draw [thick](6 cm,-2pt) node[below] {\scriptsize $1$};
			\draw [thick, black!40!green ,->] (1.5,4.5) -- (2.3,6-2.3);
			\draw [thick, black!40!green ,->] (2.3,6-2.3)--(3.1,6-3.1);
			\draw [thick, black!40!green] (3.1,6-3.1)--(3.9,6-3.9);
			
			\end{tikzpicture}
			\caption{}\label{fig-fan-zeta}
		\end{subfigure}
		\begin{subfigure}[b]{0.4\linewidth}
			\begin{tikzpicture}[scale=3]


			\fill [fill=green!40!white] (0.635714,1) -- (-0.25,1) --  (-0.25,0) node[below] {\scriptsize $v_\beta^-$} -- (0.635714,0) node[below] {\scriptsize $v_\beta^+$} ;
			
			%
			
			\fill [thick,fill=blue!40!white,variable=\y,domain=0.4:0,samples=100]
			
			plot({(-0.04539 + 0.2856*\y - 0.537*\y^2 + 0.33*\y^3)/(-0.0714 + 0.204*\y - 
				0.165*\y^2)},{1 + (0.15*(-0.34 + 1.04*\y - 0.7*\y^2))/(0.102 - 0.165*\y)}) -- (0.635714,0) -- (-0.25,0) ;
			
			\fill [thick, fill=yellow!40!white,variable=\y,domain=0.4:0,samples=100]
			
			plot({(-0.04539 + 0.2856*\y - 0.537*\y^2 + 0.33*\y^3)/(-0.0714 + 0.204*\y - 
				0.165*\y^2)},{-((0.85*(0.21 - 0.3*\y)*\y)/(-0.102 + 0.165*\y))}) -- (0.635714,0) -- (-0.25,0)
			
			(0.2,0.1) node {\scriptsize $\rho_\bullet$}
			(0.4,0.4) node {\scriptsize $\rho_\ast$}
			(0.4,0.8) node {\scriptsize $\rho_\circ$} ;

			\draw[->] (0,0) -- (0,1.1) node[above] 
			{\scriptsize { $\boldsymbol \rho$}};
			
			\draw[->] (-1,0) -- (1,0) node[right] {\scriptsize $\xi$};
			\end{tikzpicture}
			\caption{}\label{fig-fan-rho}
		\end{subfigure}
	\end{center}
	\caption{
		(a) Projection in the $\boldsymbol \rho$--plane of the three types of rarefactions fans in the case $\alpha,\beta<1$: in green a TASEP-like rarefaction fan, in blue an $\alpha$--fan, in red a $\beta$--fan. (b) A plot of a $\beta$--fan. The 
		local densities correspond to the width of the corresponding colored regions. In particular the line separating the yellow region from the violet region represents the plot of the density $\rho_\bullet(\xi)$. The extremes of the fan are given by
		: $v_\beta^-=v_\beta(z_\alpha,1-z_\alpha), v_\beta^+=v_\beta(z_\alpha,0)$.
	}\label{fig:fans}
\end{figure}

Using eqs.(\ref{conserv-za},\ref{conserv-zb}) we can easily work 
out  the rarefaction fans. These are continuous solutions of eqs.(\ref{conserv-rho-j}), 
which depend only on the self-similarity variable $\xi=x/t$, 
$\bfz(x,t)=\bfz(\xi=x/t)$. They are given by the solutions of the equations 
\begin{equation}
\begin{split}
(v_\alpha-\xi)\partial_\xi z_\alpha&=0\\
(v_\beta-\xi)\partial_\xi z_\beta&=0.
\end{split}
\end{equation}
Locally we have four possibilities. 
\begin{enumerate}
	\item The trivial solution, namely both $z_\alpha$ and $z_\beta$ are constant.
	\item Both $\partial_\xi z_\alpha\neq 0,\partial_\xi z_\beta\neq 0$.
	In this case we must have  $v_\alpha-\xi=v_\beta-\xi=0$, and in particular $v_\alpha=v_\beta$. As mentioned above, this is possible only if $1-z_\alpha-z_\beta=0$. In this case we get a
	\emph{rarefaction fan} of equation
	\begin{equation}\label{single-fan}
	\rho_\circ(\xi)=z_\alpha(\xi)=\frac{1+\xi}{2},\qquad \rho_\bullet(\xi)=z_\bullet(\xi)=\frac{1-\xi}{2}.
	\end{equation}
	Notice that the condition $1-z_\alpha-z_\beta=0$ implies absence of $\ast$-particles, and the solution \eqref{single-fan} corresponds to the fan solution of the single species TASEP (upon identification of $\circ$-particles with empty sites).  
	\item $z_\alpha$ constant, $\partial_\xi z_\beta\neq 0$. In this case the rarefaction fan ($\beta$--fan) is given by 
	\be
	v_\beta(z_\alpha,z_\beta(\xi,z_\alpha))=\xi.
	\ee
	This equation, combined with the expression $\bfrho(z_\alpha,z_\beta)$, allows to write a $\beta$-fan in parametrized form
	($z_\alpha$ is kept constant while $z_\beta$ varies), see Fig. \ref{fig-fan-rho}. 
	One can show that (at fixed $z_\alpha$)  $v_\beta$ is a decreasing function of $z_\beta$, while $\rho_\bullet$ is an increasing function of $z_\beta$ . This means that $z_\beta(\xi,z_\alpha)$ and $\rho_\bullet(\xi,z_\alpha)$ are decreasing functions of $\xi$.
	\item $z_\beta$ constant, $\partial_\xi z_\alpha\neq 0$. This case is similar to the previous one, but the role of $\alpha$ and $\beta$ variables is exchanged. So we speak of an $\alpha$--fan, given by 
	\be
	v_\alpha(z_\alpha(\xi,z_\beta),z_\beta)=\xi.
	\ee
	One can show that (at fixed $z_\beta$) $v_\alpha$ and $\rho_\circ$ are increasing function of $z_\alpha$. This means that $z_\alpha(\xi,z_\beta)$ and $\rho_\circ(\xi,z_\beta)$ are increasing functions of $\xi$.
\end{enumerate}
Projection in the $\boldsymbol \rho$--plane of the three types of fans as well as an example of a $\beta$--fan are represented in Fig. \ref{fig:fans}.

\subsection{Shocks}

It is well known that a smooth solution of a nonlinear conservation 
laws like eqs.(\ref{conserv-rho-j}) may develop a shock discontinuity in finite time. One needs therefore to admit the notion of \emph{weak solution}, i.e. solution in the sense of distributions, which need not even be continuous. 
The discontinuity associated to a shock with trajectory $x_s(t)$, has to satisfy the \emph{Rankine-Hugoniot jump relations }. If we denote by $[{\bfrho}]$ and $[{\bf J}]$ respectively the discontinuities of the densities and of the currents across the shock, i.e.
$$
[{\bfrho}]=\bfrho(x_s^+)-\bfrho(x_s^-),\qquad [{\bf J}]={\bf J}(x_s^+)-{\bf J}(x_s^-)
$$
then the Rankine-Hugoniot jump relations read 
\be\label{rankine-hugoniot}
[{\bfrho} ] -v_s [{\bf J}]=0,\qquad v_s=\text{shock's speed}.
\ee
The Rankine-Hugoniot jump relations, allow to express the speed of the shock in terms of the discontinuity and at the same time put a constraint on the admissible discontinuities, the Hugoniot condition:
\be\label{hugoniot-condition}
\det 
\left(
\begin{array}{cc}
	[\rho_\circ ] &  [J_\circ]\\[5pt]
	[\rho_\bullet]  & [J_\bullet]
\end{array}
\right)=0.
\ee
In order to analyze the Hugoniot condition, we consider  $\bfrho(x_s^-)=\bfrho^-$ as fixed.
For strictly hyperbolic systems of $N$ conservation laws, it is known that the set of $\bfrho^+=\bfrho(x_s^+)$  satisfying the Hugoniot condition passes through $\bfrho^-$ and locally decompose around $\bfrho^-$  in $N$ different branches \cite{lefloch2002hyperbolic}, each one called a \emph{shock curve}. In our case we expect two shock curves passing through any point $\rho^-$, which correspond to two different kinds of shocks.
Using eq.(\ref{chang-var1}) and eq.(\ref{Jz0}) we see that if $z_\alpha^{+}=z_\alpha^{-}=z_\alpha$, then we have 
\begin{equation}
\begin{split}
\frac{[\rho_\circ]}{z_\alpha}+\frac{[\rho_\bullet]}{z_\alpha-1}-\frac{[\rho_\circ]+[\rho_\bullet]}{z_\alpha-\alpha}=0\\
\frac{[J_\circ]}{z_\alpha}+\frac{[J_\bullet]}{z_\alpha-1}-\frac{[J_\circ]+[J_\bullet]}{z_\alpha-\alpha}=0.
\end{split}
\end{equation}
This means that the matrix in the  eq.\eqref{hugoniot-condition} 
has the left null vector $(\frac{1}{z_\alpha}-\frac{1}{z_\alpha-\alpha},\frac{1}{z_\alpha-1}-\frac{1}{z_\alpha-\alpha}) $ and therefore its determinant vanishes. We conclude that a straight line
at $z_\alpha$ constant is a shock curves. In the same way we find 
that the lines at $z_\beta$ constant are also shock curves.
In conclusion we have found that we have two kind of admissible shocks
\begin{itemize}
	\item $\beta$-shocks: $z_\alpha(\bfrho^-)=z_\alpha(\bfrho^+)=z_\alpha$ with speed $v_{s,\beta}(z_\alpha;z_\beta^-,z_\beta^+)$;
	\item $\alpha$-shocks: $z_\beta(\bfrho^-)=z_\beta(\bfrho^+)=z_\beta$, with speed $v_{s,\alpha}(z_\beta;z_\alpha^-,z_\alpha^+)$.
\end{itemize}
The corresponding shocks speeds $v_{s,\beta}(z_\alpha;z_\beta^-,z_\beta^+)$ and $v_{s,\alpha}(z_\beta;z_\alpha^-,z_\alpha^+)$ do not have particularly transparent expressions except for some particular cases. 
In the case $\alpha=\beta=1$, a $\beta$--shock is a discontinuity of the density $\rho_\bullet$, with $\rho_\circ$ constant across the discontinuity, while an $\alpha$--shock is the other way round, i.e. a discontinuity of the density $\rho_\circ$, with $\rho_\bullet$ constant across the discontinuity.
Shock curves coincide with rarefaction curves so this is an example of a conservation system of Temple class \cite{temple1983systems}.

A further analysis of the Hugoniot condition allows to conclude that in the bulk of the physical domain $\mathcal{D}$, the only possible shocks are $\alpha$ and $\beta$--shocks.  
There exists however one more class of shocks when both sides of the discontinuity lie on the boundary line $\rho_\ast=0$. These shocks have speed  
$v_s=\frac{[J_\bullet]}{[\rho_\bullet]}$,
where the current $J_\bullet$ is given by eq.\eqref{singul10}.

\begin{figure}[h!!]
	\begin{center}
		\begin{tikzpicture}[xscale=3,yscale=5]
		\def\a{4};
		\def\b{5};
		\def\za{.2};
		\draw [blue, thick] plot[smooth,domain=0:{1-\za}, samples=80] ({-( (-1 + \za)*\x*(-\a + \za - \b*\za + \a*\x))/(-\a*\b + \a*\b*\za + \a*\b*\x + \za*\x - \a*\za*\x - \b*\za*\x)},
		{
			(
			(-1 +\za)*\x*
			(\a*\b + \a*\za - \a*\b*\za - (\za)^2 + \b*(\za)^2 - 
			\a*\x - \a*\b*\x + \a*(\x)^2)
			)/(-\a*\b 
			+ \a*\b*\za + \a*\b*\x + \za*\x - \a*\za*\x - \b*\za*\x)});
		\def\a{4};
		\def\b{1};
		\def\za{.2};
		\draw [black!25!green, thick] plot[smooth,domain=0:{1-\za}, samples=80] ({-( (-1 + \za)*\x*(-\a + \za - \b*\za + \a*\x))/(-\a*\b + \a*\b*\za + \a*\b*\x + \za*\x - \a*\za*\x - \b*\za*\x)},
		{
			(
			(-1 +\za)*\x*
			(\a*\b + \a*\za - \a*\b*\za - (\za)^2 + \b*(\za)^2 - 
			\a*\x - \a*\b*\x + \a*(\x)^2)
			)/(-\a*\b 
			+ \a*\b*\za + \a*\b*\x + \za*\x - \a*\za*\x - \b*\za*\x)});
		\def\a{4};
		\def\b{.3};
		\def\za{.2};
		\draw [red, thick] plot[smooth,domain=0:{\b}, samples=80] ({-( (-1 + \za)*\x*(-\a + \za - \b*\za + \a*\x))/(-\a*\b + \a*\b*\za + \a*\b*\x + \za*\x - \a*\za*\x - \b*\za*\x)},
		{
			(
			(-1 +\za)*\x*
			(\a*\b + \a*\za - \a*\b*\za - (\za)^2 + \b*(\za)^2 - 
			\a*\x - \a*\b*\x + \a*(\x)^2)
			)/(-\a*\b 
			+ \a*\b*\za + \a*\b*\x + \za*\x - \a*\za*\x - \b*\za*\x)});
		\draw [->] (0,0)-- (1.2,0) node[below] {\tiny $\rho_\bullet$};
		\draw [->] (0,0)-- (0,.6) node [left] {\tiny $J_\bullet$};
		
		\foreach \y in {0.2,0.4,0.6,0.8,1} \draw ({\y},-.01) node[below]{\tiny \y}--({\y},+.01);
		\foreach \y in {0.1,0.2,0.3,0.4,0.5} \draw (-.01,{\y}) node[left]{\tiny \y}--(+.01,{\y});
		\draw (0.6,-.2) node {\bf(a)};
		\begin{scope}[xshift = 2cm,yscale=2]
		\def\a{3};
		\def\b{.3};
		\def\za{.2};
		\draw [blue, thick] plot[smooth,domain=0:{\b}, samples=80] ({-( (-1 + \za)*\x*(-\a + \za - \b*\za + \a*\x))/(-\a*\b + \a*\b*\za + \a*\b*\x + \za*\x - \a*\za*\x - \b*\za*\x)},
		{
			(
			(-1 +\za)*\x*
			(\a*\b + \a*\za - \a*\b*\za - (\za)^2 + \b*(\za)^2 - 
			\a*\x - \a*\b*\x + \a*(\x)^2)
			)/(-\a*\b 
			+ \a*\b*\za + \a*\b*\x + \za*\x - \a*\za*\x - \b*\za*\x)});
		\def\a{1};
		\def\b{.3};
		\def\za{.2};
		\draw [black!25!green, thick] plot[smooth,domain=0:{\b}, samples=80] ({-( (-1 + \za)*\x*(-\a + \za - \b*\za + \a*\x))/(-\a*\b + \a*\b*\za + \a*\b*\x + \za*\x - \a*\za*\x - \b*\za*\x)},
		{
			(
			(-1 +\za)*\x*
			(\a*\b + \a*\za - \a*\b*\za - (\za)^2 + \b*(\za)^2 - 
			\a*\x - \a*\b*\x + \a*(\x)^2)
			)/(-\a*\b 
			+ \a*\b*\za + \a*\b*\x + \za*\x - \a*\za*\x - \b*\za*\x)});
		\def\a{.3};
		\def\b{.3};
		\def\za{.2};
		\draw [red, thick] plot[smooth,domain=0:{\b}, samples=80] ({-( (-1 + \za)*\x*(-\a + \za - \b*\za + \a*\x))/(-\a*\b + \a*\b*\za + \a*\b*\x + \za*\x - \a*\za*\x - \b*\za*\x)},
		{
			(
			(-1 +\za)*\x*
			(\a*\b + \a*\za - \a*\b*\za - (\za)^2 + \b*(\za)^2 - 
			\a*\x - \a*\b*\x + \a*(\x)^2)
			)/(-\a*\b 
			+ \a*\b*\za + \a*\b*\x + \za*\x - \a*\za*\x - \b*\za*\x)});
		\draw [->] (0,0)-- (1.2,0) node[below] {\tiny $\rho_\bullet$};
		\draw [->] (0,0)-- (0,.3) node [left] {\tiny $J_\bullet$};
		
		\foreach \y in {0.2,0.4,0.6,0.8,1} \draw ({\y},-.01) node[below]{\tiny \y}--({\y},+.01);
		\foreach \y in {0.05,0.1,0.15,0.2,0.25} \draw (-.01,{\y}) node[left]{\tiny \y}--(+.01,{\y});
		\draw (0.6,-.1) node {\bf (b)};
		\end{scope}
		\end{tikzpicture}
	\end{center}
	\caption{Current $J_\bullet$ as function of $\rho_\bullet$ at constant $z_\alpha=.2$. (a) Fixed $\alpha=4$ and decreasing $\beta$ (blue line $\beta=5$, green line $\beta=1$, red line $\beta=0.3$). (b) Fixed $\beta=0.3$ and decreasing $\alpha$ (blue line $\alpha=3$, green line $\alpha=1$, red line $\alpha=0.3$).  
	}\label{fig:current-za-const}
\end{figure}
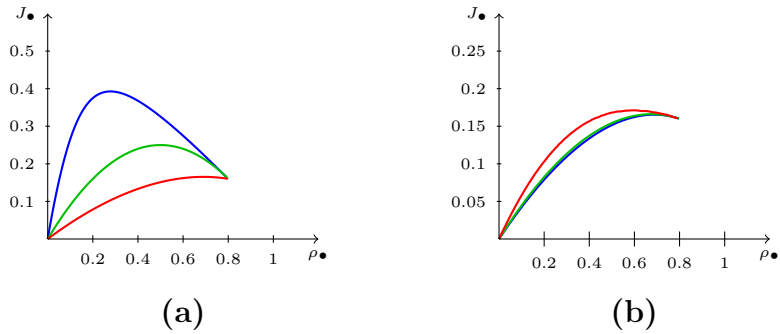

It is possible to show that at fixed $z_\alpha$ the current $J_\bullet$ is a concave function of the density $\rho_\bullet$ (see fig. \ref{fig:current-za-const} ). 
This implies that, for a fixed value of the
densities on one side of the shock, say $\bfrho^-$, the speed of a $\beta$--shock is a decreasing function of $\rho_\bullet^+$.
This property can be conveniently reformulated in terms of the $\bfz$ variables. 
At constant $z_\alpha$, $\rho_\bullet$ is an increasing function of $z_\beta$, hence at fixed $z_\alpha$ and $z_\beta^-$, 
the speed of $\beta$--shock $v_{s,\beta}(z_\alpha;z_\beta^-,z^+_\beta)$ is a decreasing function of $z^+_\beta$. In particular it takes its minimum for the largest $z_\beta$ allowed, i.e. 
$z_{\beta,\textrm{max}}=1-z_\alpha$
\begin{equation}
v_{s,\beta}(z_\alpha;z_\beta,z^+_\beta)\geq v_{s,\beta}(z_\alpha;z_\beta,1-z_\alpha) \qquad 0\leq z^+_\beta \leq 1-z_\alpha.
\end{equation}
The current $J_\circ$ is a convex function of $\rho_\circ$ at fixed $z_\beta$. A similar reasoning as the one presented above allows to  
conclude 
\begin{equation}
v_{s,\alpha}(z_\beta;z_\alpha,z^+_\alpha)\leq v_{s,\alpha}(z_\beta;z_\alpha,1-z_\beta)\qquad 0\leq  z^+_\alpha\leq  1-z_\beta.
\end{equation}
Now, from an explicit computation, we notice that 
$$
v_{s,\alpha}(z_\beta;z_\alpha,1-z_\beta)=v_{s,\beta}(z_\alpha;z_\beta,1-z_\alpha)=z_\alpha-z_\beta.
$$
This allows to conclude that 
\be\label{ineq-shock-speeds}
v_{s,\alpha}(z_\beta;z_\alpha,\widetilde{z_\alpha})\leq  v_{s,\beta}(z_\alpha;z_\beta,\widetilde{z_\beta}).
\ee
In words this means that, for a fixed value of the densities on one side of the shock, the speed of any $\beta$--shock is larger than the speed of any $\alpha$--shock.

\noindent
Since $\lim_{\widetilde{z_\alpha}\rightarrow z_\alpha} v_{s,\alpha}(z_\beta;z_\alpha,\widetilde{z_\alpha})=v_{\alpha}(z_\alpha,z_\beta)$ and 
$\lim_{\widetilde{z_\beta}\rightarrow z_\beta}v_{s,\beta}(z_\alpha;z_\beta,\widetilde{z_\beta})=v_{\beta}(z_\alpha,z_\beta)$, we get also
\begin{gather}\label{ineq-v-vshA}
v_{\alpha}(z_\alpha,z_\beta)\leq  v_{s,\beta}(z_\alpha;z_\beta,\widetilde{z_\beta}) \\\label{ineq-v-vshB}
v_{\beta}(z_\alpha,z_\beta)\geq v_{s,\alpha}(z_\beta;z_\alpha,\widetilde{z_\alpha}).
\end{gather}

The Hugoniot condition is not sufficient to select the physical shocks. Indeed eqs.\eqref{conserv-rho-j} has to be thought of as the zero viscosity limit of a set of conservation laws which contains a diffusive term, a term which comes from the  microscopic corrections to the currents and depends on the derivatives of the densities.
Inviscid limits of viscous solutions are typically characterized by entropy conditions, which for shocks take the form of the Liu entropy criterion \cite{liu1974riemann,lefloch2002hyperbolic}. Let the right densities $\bfrho^+$ lie on a Hugoniot curve emanating from $\bfrho^-$, then for all densities  
$\widetilde \bfrho$ lying on the same Hugoniot curve in between $\bfrho^-$ and $\bfrho^+$ the Liu condition states that 
\be\label{liu-condition-general}
v_s(\bfrho^-,\bfrho^+)\leq v_s(\bfrho^-,\widetilde \bfrho).
\ee
Physically this condition can be understood as a stability condition: if under a perturbation the shock were to split by inserting an intermediate state $\widetilde \bfrho$, then a violation of  condition  \eqref{liu-condition-general} would imply that the shock between $\bfrho^-$ and $\widetilde \bfrho$  would move away from the original shock between $\bfrho^-$ and $\bfrho^+$.
In the case of $\beta$--shocks the Liu condition reads
\be\label{liu-cond-b}
v_{s,\beta}(z_\alpha;z_\beta^-,z_\beta^+)  \leq v_{s,\beta}(z_\alpha;z_\beta^-,\widetilde{z_\beta}),\qquad \forall \widetilde{z_\beta} \in [\min(z_\beta^-,z_\beta^+),\max(z_\beta^-,z_\beta^+)]
\ee
Since, as mentioned above, at fixed $z_\alpha$ the current $J_\beta$ is a concave function of the density $\rho_\bullet$, $
\partial_{\rho_\bullet}^2 J_\beta|_{z_\alpha} <0
$, we conclude that the Liu constraint means 
$\rho_\bullet^-<\rho_\bullet^+$. This can be also formulated in terms of the $z$ variables. Indeed, since $\partial_{z_\beta} \rho_\bullet|_{z_\alpha}>0$, we must have  $z_\beta^-<z_\beta^+$.
A similar analysis can be performed for $\alpha$--shocks. 

\vspace*{.2cm}
\noindent
\begin{minipage}{0.7 \textwidth}
	Here is a summary of our results. They are schematized in the figure on the right, where  we have reported the direction of the possible shock--discontinuities.
	\begin{itemize}
		\item For an $\alpha$--shocks (blue oriented line), we need $\rho_\circ^->\rho_\circ^+$ or equivalently $z_\alpha^{-}>z_\alpha^{+}$. 
		\item For a $\beta$--shocks (red oriented line), we need $\rho_\bullet^-<\rho_\bullet^+$ or equivalently $z_\beta^{-}<z_\beta^{+}$. 
	\end{itemize}
\end{minipage}
\begin{minipage}{0.3 \textwidth}
	
	\begin{center}
		\begin{tikzpicture}[scale=0.4]
		\draw[thick,->] (0,0) -- (7,0) node[below] {\scriptsize $\rho_\circ$};
		\draw[thick,->] (0,0) -- (0,7) node[left] {\scriptsize  $\rho_\bullet$};
		
		\foreach \n in {0.2,0.4,0.6,0.8}
		\foreach \x in {3}
		\foreach \x in {3}
		\draw [thick, red, ->, dashed] (\x,0) -- (\x + 0.3*\x*\n - \x*\n,  6*\n- 0.3*\x*\n );
		
		\foreach \x in {3}
		\draw [thick, red, dashed] (\x,0) -- (0.3*\x, 6- 0.3*\x);

		\foreach \n in {0.2,0.4,0.6,0.8}
		\foreach \y in {3}
		\draw [thick, blue ,->, dashed] (6-0.4*\y, 0.4*\y ) -- (6*\n-0.4*\y*\n, \y + 0.4*\y*\n - \y*\n);
		
		\foreach \y in {3}
		\draw [thick, blue, dashed] (0,\y) -- (6-0.4*\y,  0.4*\y );

		\draw  (6*0.2,6-6*0.2) -- (6-0.4*6,0.4*6);
		\draw (6*0.3,6-6*0.3) -- (0,6);
		\draw (6,0) -- (6-0.4*6,0.4*6);
		
		\draw [thick](2pt,6 cm) node[left] {\scriptsize $1$};
		\draw [thick](6 cm,-2pt) node[below] {\scriptsize $1$};
		
		\end{tikzpicture}
	\end{center}

\end{minipage}

\noindent

\subsection{Riemann's problem}\label{sect-riemann}

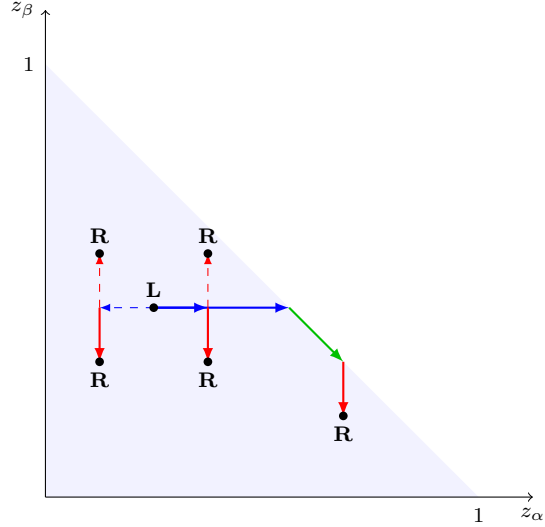
\begin{figure}[h]
	\begin{center}
		\begin{tikzpicture}[scale= 1.2]
		\begin{scope}[scale = .6, xshift=10cm]
		\fill [fill=white!95!blue] (-1,-1)--(-1,7) node [left] {\scriptsize $1$}--(7,-1)node [below] {\scriptsize $1$}--cycle;
		\draw[->] (-1,-1) -- (8,-1) node[below] {\scriptsize $z_{\alpha}$};
		\draw[->] (-1,-1) -- (-1,8) node[left] {\scriptsize  $z_{\beta}$};
		\begin{scope}[shift={(-1,.5)}]
		\draw [->,>=latex,dashed,blue](2,2) --(1,2);
		\draw [->,>=latex,dashed,red](1,2)--(1,3);
		\draw [->,>=latex,thick,red](1,2)--(1,1);
		\draw [->,>=latex,thick,blue](2,2) --(3,2);
		\draw [->,>=latex,thick,blue](2,2) --(4.5,2);
		\draw [->,>=latex,thick,black!25!green](4.5,2) --(5.5,1);
		\draw [->,>=latex,thick,red](5.5,1)--(5.5,0);
		\draw [->,>=latex,thick,red](3,2)--(3,1);
		\draw [->,>=latex,dashed,red](3,2)--(3,3);
		\fill (2,2) circle (.08) node[above] {\scriptsize \bf L};
		\fill (1,3) circle (.08) node[above] {\scriptsize \bf R};
		\fill (1,1) circle (.08) node[below] {\scriptsize \bf R};
		\fill (3,1) circle (.08) node[below] {\scriptsize \bf R};
		\fill (5.5,0) circle (.08) node[below] {\scriptsize \bf R};
		\fill (3,3) circle (.08) node[above] {\scriptsize \bf R};
		\end{scope}
		
		\end{scope}
		\end{tikzpicture}
	\end{center}
	\caption{Projection in the $\bfz$--plane of the different type of solutions of the Riemann problem. The left values of the $\bfz$  variables are represented by the point $L$. The points $R$ represent the different distinct possibilities for the right values of the $\bfz$  variables. Continuous lines represent fans, while dashed lines represent shocks. In green is indicated the possible TASEP-like fan, in red either an $\alpha$--shock or an $\alpha$--fan and in blue either a $\beta$--shock or a $\beta$--fan.}\label{fig:riemann}
\end{figure}

With the result for the rarefaction curves and the shock curves at our disposal, it is rather simple to describe the general solution of the Riemann problem, i.e. the
solution of eqs.\eqref{conserv-rho-j}  with domain wall initial conditions
\be
\bfrho(x,t=0)=\left\{
\begin{array}{l}
	\bfrho^{L}=(\rho_\circ^L,\rho_\bullet^L)\qquad x<0\\
	\bfrho^R=(\rho_\circ^R,\rho_\bullet^R)\qquad x>0.
\end{array}
\right.
\ee
By uniqueness, the solution of the Riemann problem has to take the form $\bfrho(x,t)=\bfrho(\xi)$ with $\xi=x/t$ and is given by a sequence of rarefaction waves and/or shocks.
It is best described in terms of the variables $\bfz^L=(z_\alpha^L,z_\beta^L), \bfz^R=(z_\alpha^R,z_\beta^R)$.
We have four possible situations (these are schematically summarized in figure \ref{fig:riemann}).

\vspace{.3cm}
\noindent
$\bullet$ $z^L_\alpha>z^R_\alpha, z^L_\beta<z^R_\beta$. The solution is composed of two shocks: an $\alpha$-shock with $\bfz^-=(z^L_\alpha,z^L_\beta)$ and $\bfz^+=(z^R_\alpha,z^L_\beta)$  at position $\xi_\alpha= v_{s,\alpha}(z^L_\beta;z^L_\alpha,z^R_\alpha)$, 
followed by a $\beta$--shock with $\bfz^-=(z^R_\alpha,z^L_\beta)$ and $\bfz^+=(z^R_\alpha, z^R_\beta)$ at position $\xi_\beta=v_{s,\beta}(z^R_\alpha;z^L_\beta,z^R_\beta)$. 
This result follows from the inequality \eqref{ineq-shock-speeds}
$$
\xi_\alpha= v_{s,\alpha}(z^L_\beta;z^L_\alpha,z^R_\alpha)\leq v_{s,\beta}(z^R_\alpha;z^L_\beta,z^R_\beta) =\xi_\beta.
$$

\vspace{.3cm}
\noindent
$\bullet$ $z^L_\alpha>z^R_\alpha, z^L_\beta>z^R_\beta$. The solution is composed of an $\alpha$-shock and a $\beta$--fan.
The $\alpha$--shock has $\bfz^-=(z^L_\alpha,z^L_\beta)$ and $\bfz^+=(z^R_\alpha,z^L_\beta)$    and it is located at position $\xi_\alpha= v_{s,\alpha}(z^L_\beta;z^L_\alpha,z^R_\alpha)$.
The $\beta$--fan starts with value $(z^R_\alpha,z^L_\beta)$ at $\xi_{\beta,1}= v_{\beta}(z^R_\alpha,z^L_\beta)$ and ends  with value $(z^R_\alpha,z^R_\beta)$ at $\xi_{\beta,2}= v_{\beta}(z^R_\alpha,z^R_\beta)$. 
This result follows from the inequality 
$$
\xi_\alpha= v_{s,\alpha}(z^L_\beta;z^L_\alpha,z^R_\alpha)\leq v_{\beta}(z^R_\alpha,z^L_\beta) =\xi_{\beta,1} \leq v_{\beta}(z^R_\alpha,z^R_\beta)=\xi_{\beta,2}.
$$
The first inequality is just inequality \eqref{ineq-v-vshB}, while the second one follows from the fact that  $v_{\beta}$ is a decreasing function of $z_\beta$.

\vspace{.2cm}
\noindent
$\bullet$ $z^L_\alpha<z^R_\alpha, z^L_\beta<z^R_\beta$. The solution is composed of an $\alpha$--fan and a $\beta$--shock.
The $\alpha$--fan starts with value $(z^L_\alpha,z^L_\beta)$ at $\xi_{\alpha,1}=v_{\alpha}(z^L_\alpha,z^L_\beta)$ and ends with value $(z^R_\alpha,z^L_\beta)$ at $\xi_{\alpha,2}=v_{\alpha}(z^R_\alpha,z^L_\beta)$. 
The $\beta$ shock has $\bfz^-=(z^R_\alpha,z^L_\beta)$ and $\bfz^+=(z^R_\alpha, z^R_\beta)$ and is located at position $\xi_\beta=v_{s,\beta}(z^R_\alpha;z^L_\beta,z^R_\beta)$.

\vspace{.2cm}
\noindent
$\bullet$ $z^L_\alpha<z^R_\alpha, z^L_\beta>z^R_\beta$. The solution is composed of fans. One has to distinguish two cases depending whether $z_\alpha^R+z_\beta^L$ is larger or smaller than $1$. 
If $z_\alpha^R+z_\beta^L<1$, the solution consists of an $\alpha$--fan starting with value $(z_\alpha^L,z_\beta^L)$ at $\xi_{\alpha,1}=v_{\alpha}(z^L_\alpha,z^L_\beta)$
and ending with value $(z_\alpha^R,z_\beta^L)$ at $\xi_{\alpha,2}=v_{\alpha}(z^R_\alpha,z^L_\beta)$ followed by  a $\beta$--fan starting at $\xi_{\beta,1}=v_{\beta}(z^R_\alpha,z^L_\beta)$ and ending at $\xi_{\beta,2}=v_{\beta}(z^R_\alpha,z^R_\beta)$.
If $\alpha+\beta>1$ then it is possible to have $z_\alpha^R+z_\beta^L>1$. In this case the $\alpha$-fan cannot reach the value $(z_\alpha^R,z_\beta^L)$, which lies outside the physical domain. It ends at the value $(1-z_\beta^L,z_\beta^L)$, followed by a degenerate fan till $(z_\alpha^R,1-z_\alpha^R)$ and then by a $\beta$--fan till  $(z_\alpha^R,z_\beta^R)$.   
%
%

\section{Monte Carlo simulations}\label{sec:montecarlo}

We come back to the original microscopic stochastic model and compare the predictions of the hydrodynamic equations with numerical simulations. 
We have simulated our model on a finite lattice of integer coordinates, running on the interval $[-L,L]$, with $L=2100$.
The system is initialized in a random configuration sampled from a product measure of local densities $\bfrho^{L}$ on sites of coordinate $i<0$ and $\bfrho^{R}$ on sites of coordinates $i\geq 0$.
At the left and right boundaries the particles are chosen neither to leave nor to enter the system. 
This means that we expect to find three distinct regions: two kinetic waves coming from the boundaries  and a kinetic wave originating from the discontinuity at the origin.
Whereas we make no prediction on the boundary waves, we expect that as long as they don't meet the bulk one, they do not influence the latter.

%
Let us introduce the height function, which is defined up to an arbitrary  additive constant by
%
%
%
\begin{equation}\label{integrated-density}
h_{i}\left(\frac{n}{t},t\right)- h_{i}\left(-1,t\right) := \frac{1}{t} \sum_{-t < k \leq n} \chi_i\left(k,t\right)\quad \quad  - t < n < t,
\end{equation}
From our assumption eq.(\ref{converg}), it follows that at large time and for each sample, the height function 
should converge to the deterministic shape
%
\begin{equation}
h_{i}(\xi)-h_{i}(-1) = \int_{-1}^{\xi} \rho_{i}(\xi') d \xi',
\end{equation}
where  $\rho_{i}(\xi)$ is the solution of the Riemann problem found in Section \ref{sect-riemann}.

\begin{figure}[h!]
	\begin{subfigure}{.45\textwidth}
		\centering
		\begin{tikzpicture}
		\node (0,0) {\includegraphics[width=6cm]{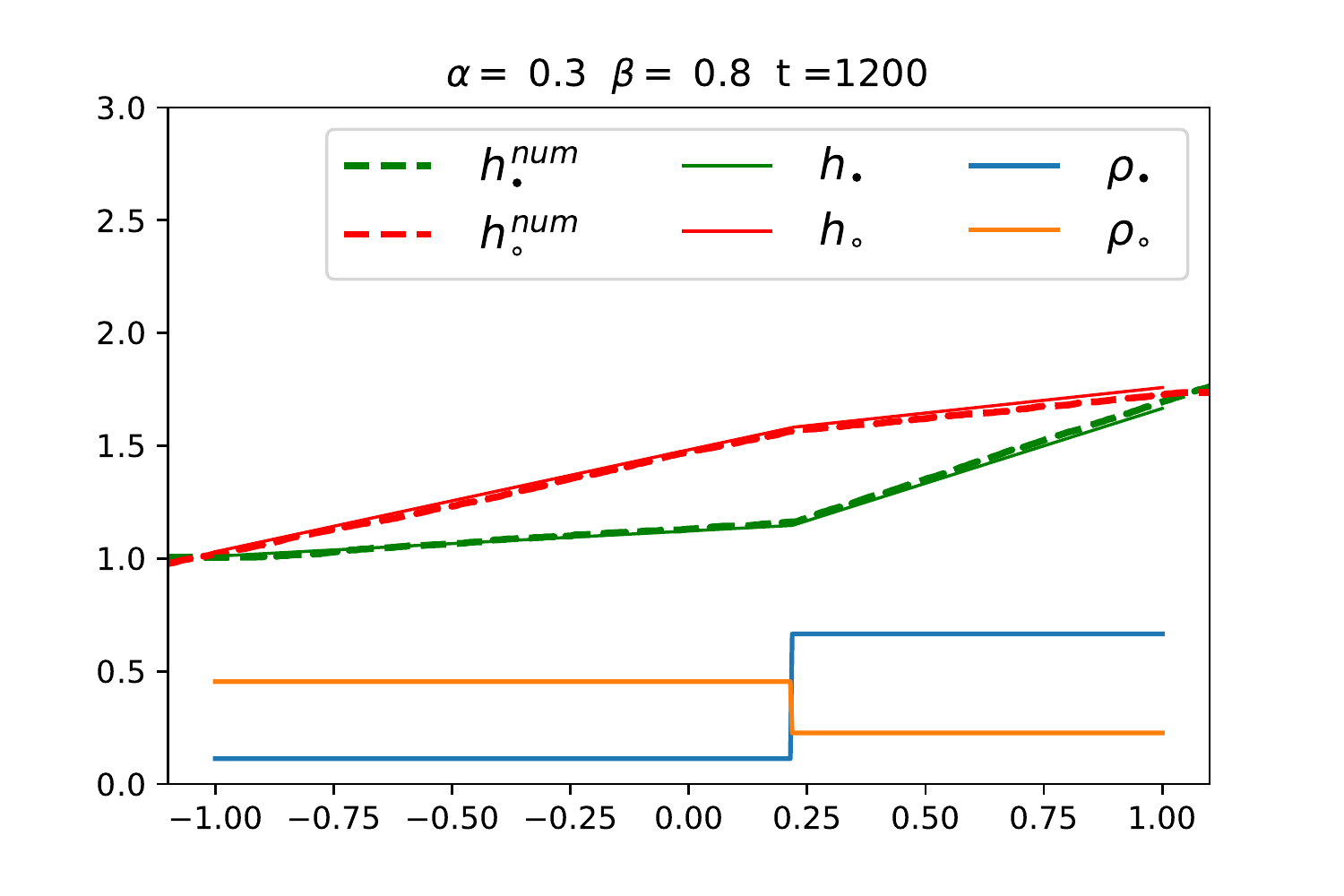}};
		\end{tikzpicture}
		\caption{\scriptsize $(\rho_\circ^L=0.45,\rho_\bullet^L=0.11)$, $(\rho_\circ^R=0.23,\rho_\bullet^R=0.67)$}
		\label{fig:sub1}
	\end{subfigure}
	\begin{subfigure}{.45\textwidth}
		\centering
		\begin{tikzpicture}
		\node (0,0) {\includegraphics[width=6cm]{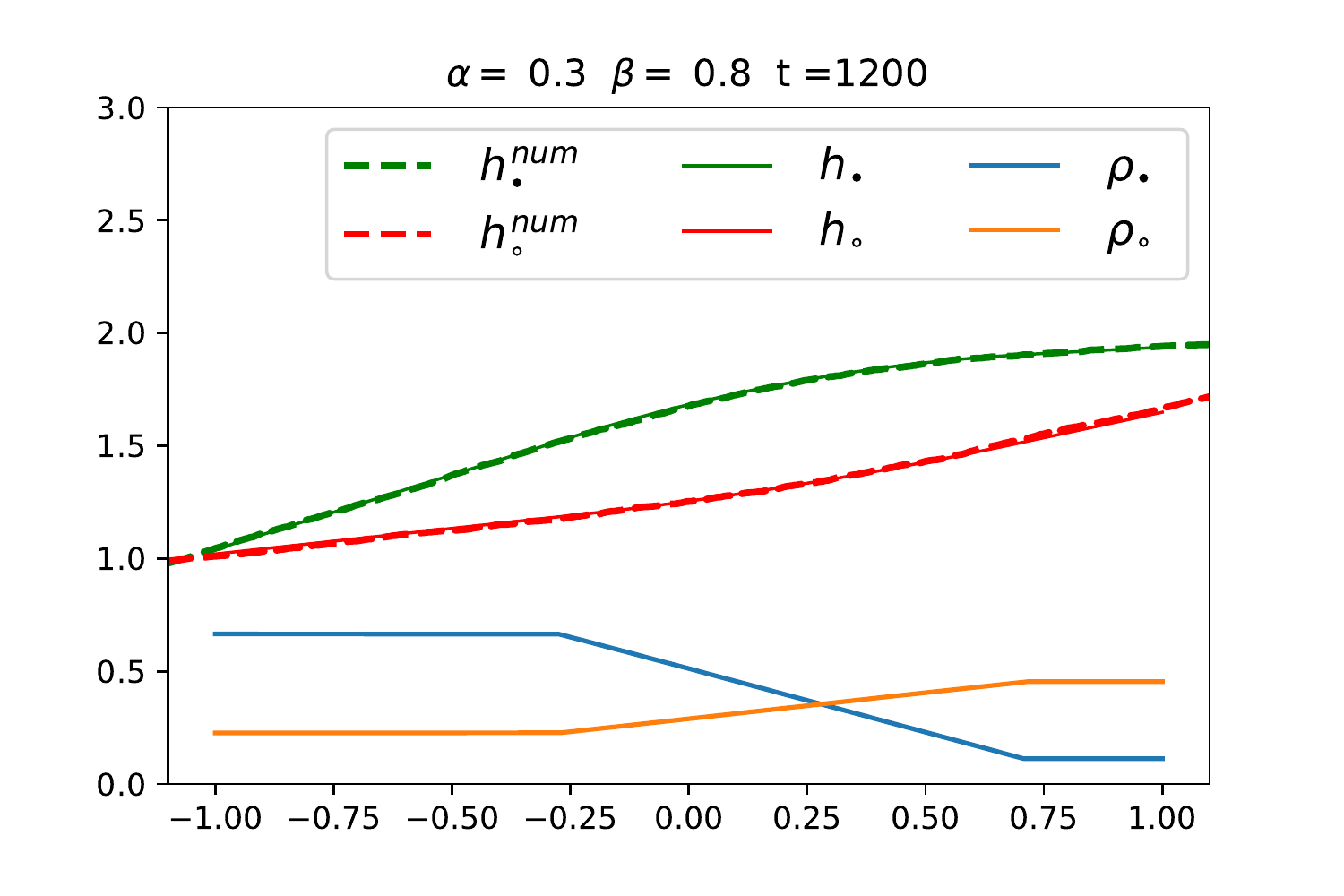}};
		\end{tikzpicture}
		\caption{\scriptsize $(\rho_\circ^L=0.23,\rho_\bullet^L=0.67)$, $(\rho_\circ^R=0.45,\rho_\bullet^R=0.11)$}
		\label{fig:sub2}
	\end{subfigure}
	
	\begin{subfigure}{.45\textwidth}
		\centering
		\begin{tikzpicture}
		\node (0,0) {\includegraphics[width=6cm]{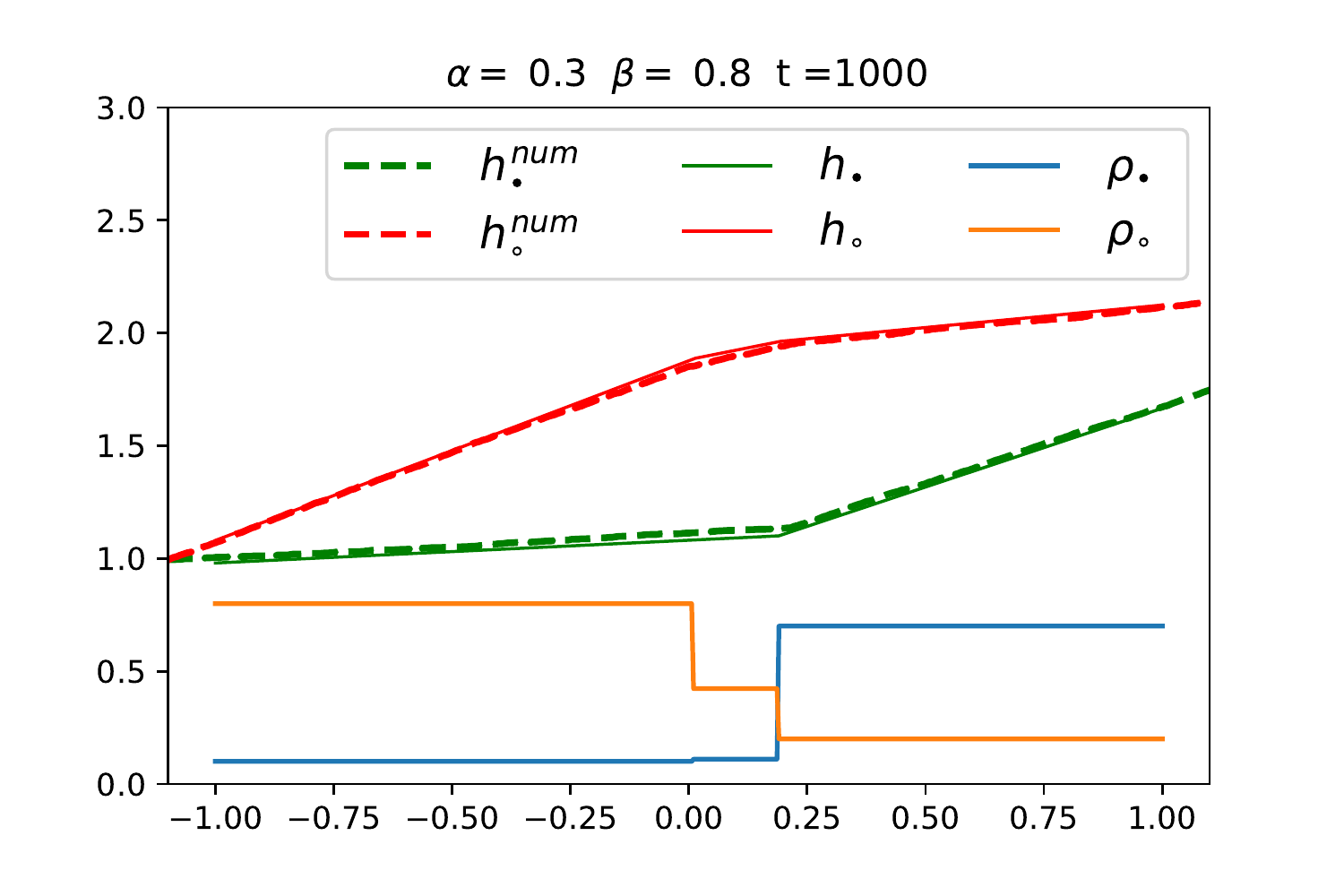}};
		\end{tikzpicture}
		\caption{\scriptsize$(\rho_\circ^L=0.8,\rho_\bullet^L=0.1)$, $(\rho_\circ^R=0.2,\rho_\bullet^R=0.7)$}
		\label{fig:sub3}
	\end{subfigure}
	\begin{subfigure}{.45\textwidth}
		\centering
		\begin{tikzpicture}
		\node (0,0) {\includegraphics[width=6cm]{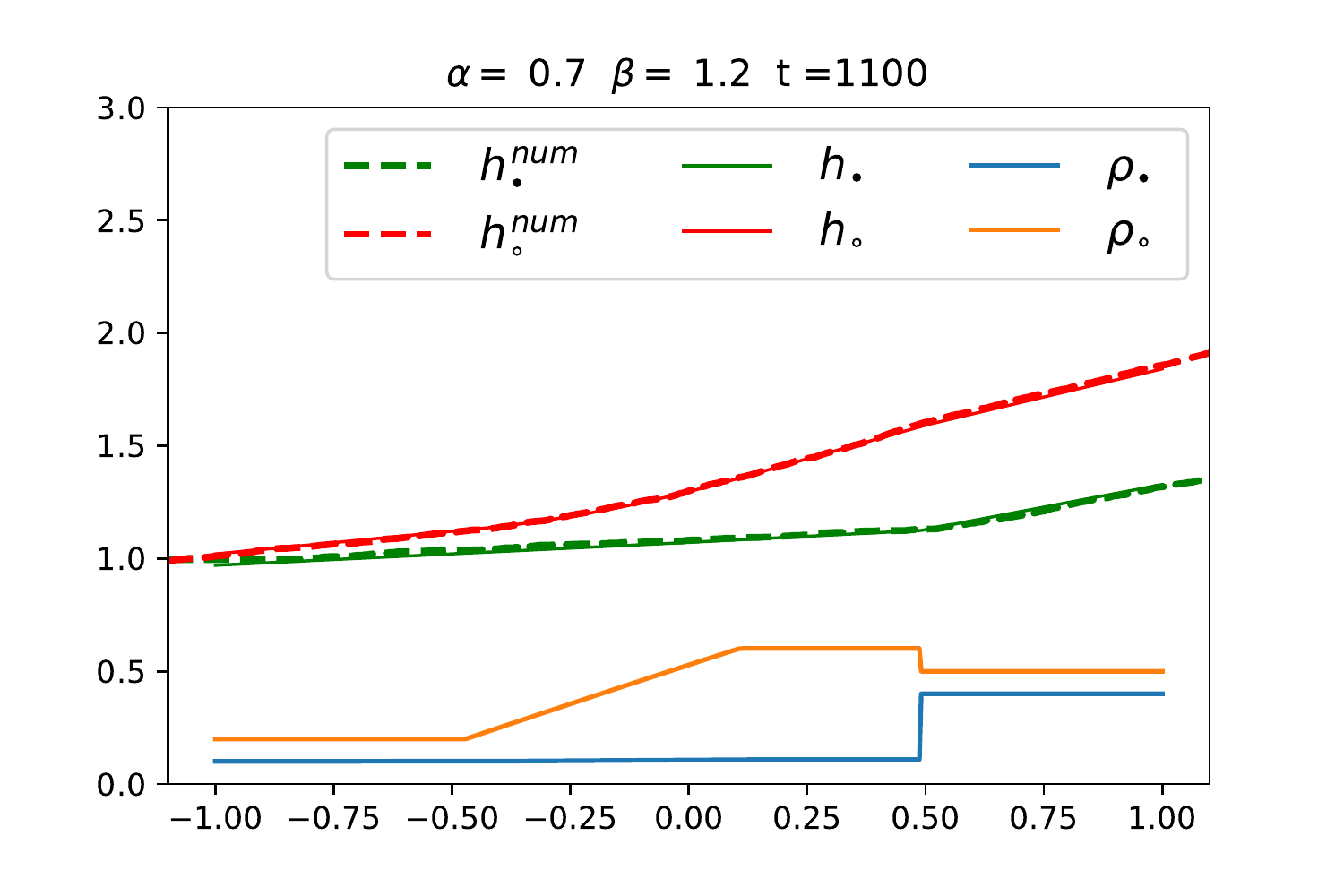}};
		\end{tikzpicture}
		\caption{\scriptsize$(\rho_\circ^L=0.2,\rho_\bullet^L=0.1)$, $(\rho_\circ^R=0.5,\rho_\bullet^R=0.4)$}
		\label{fig:sub4}
	\end{subfigure}
	\caption{Height function of the $\bullet$-particles (green lines) and of the $\circ$-particles (red lines). Dashed lines represent the numerical values, continuous lines represent the theoretical prediction. At the bottom of each plot we have reported the predicted densities.}\label{fig-sim-tutto}
\end{figure}

\begin{figure}[h!]
	
	\begin{subfigure}{.55\textwidth}
		\centering
		
		\includegraphics[width=7cm]{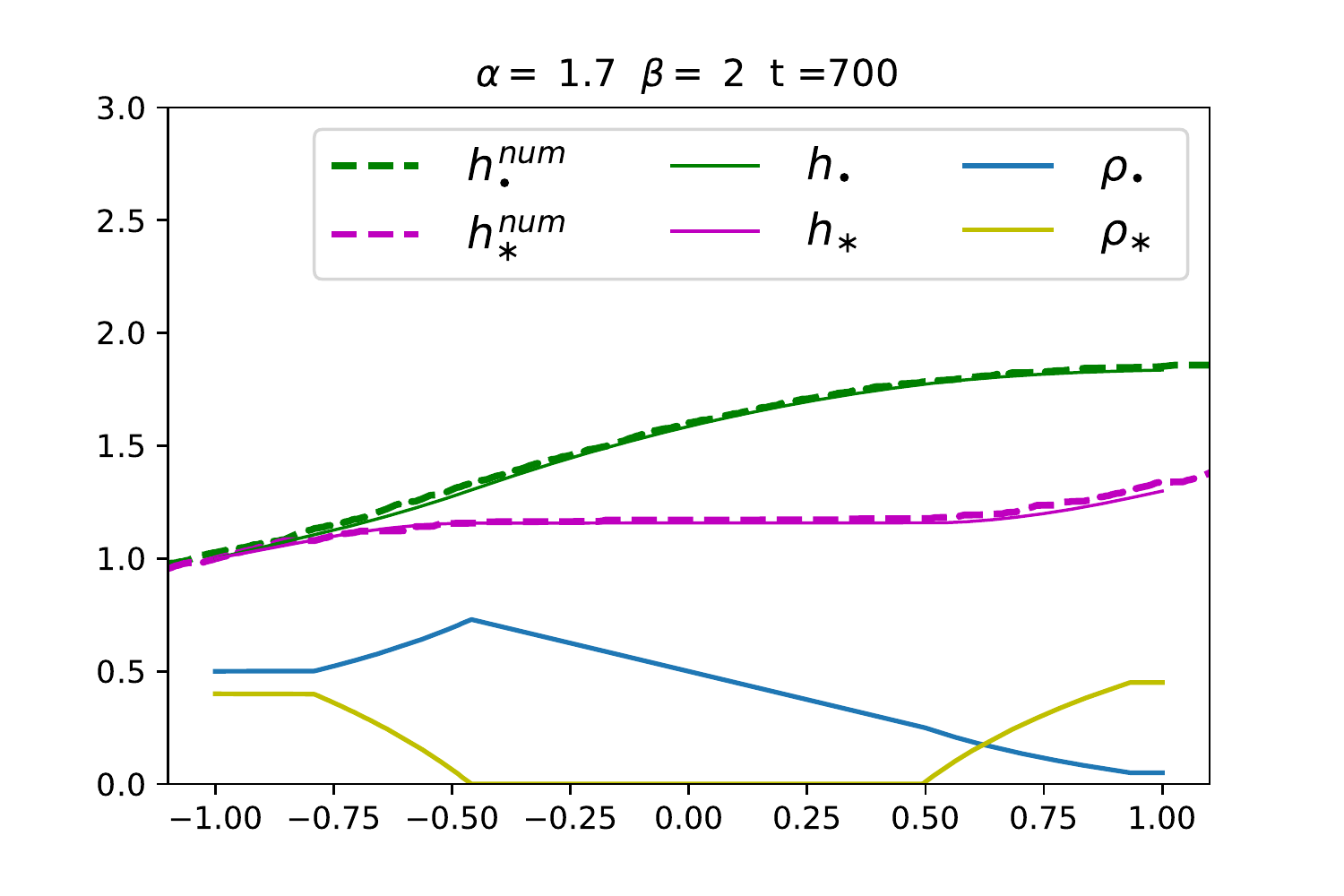}
		
	\end{subfigure}
	\begin{subfigure}{.44\textwidth}
		\centering
		
		\begin{tikzpicture}
		
		\begin{scope}[scale = .5]
		\draw[thick,->] (0,0) -- (7,0) node[below] {\scriptsize $\rho_\circ$};
		\draw[thick,->] (0,0) -- (0,7) node[left] {\scriptsize $\rho_\bullet$};

		\foreach \rzL in {0.1*6}
		\foreach \roL in {0.5*6}
		\foreach \rzR in {0.5*6}
		\foreach \roR in {0.05*6}
		\foreach \rof in {6*0.729}
		\foreach \rzf in {6*0.2701}
		\foreach \ros in {6*0.2532}
		\foreach \rzs in {6*0.7467}
		{
			
			\draw [line width=0.7 mm, blue, ->]  (\rzL,\roL) -- (\rzL*0.5 + \rzf*0.5, \rof*0.5 + \roL*0.5);
			
			\draw [line width=0.7 mm, blue]  (\rzL*0.5 + \rzf*0.5, \rof*0.5 + \roL*0.5)-- (\rzf,\rof);

			\draw [line width=0.7 mm, green, ->]  (\rzf,\rof) -- (\rzf*0.5 + \rzs*0.5,\rof*0.5 + \ros*0.5);
			\draw [line width=0.7 mm, green] (\rzf*0.5 + \rzs*0.5,\rof*0.5 + \ros*0.5)  -- (\rzs,\ros);

			\draw [line width=0.7 mm, red, ->]  (\rzs,\ros) -- (\rzR*0.5+\rzs*0.5,\roR*0.5+\ros*0.5);
			\draw [line width=0.7 mm, red]  (\rzR*0.5+\rzs*0.5,\roR*0.5+\ros*0.5) -- (\rzR,\roR);

			\node[circle,fill=black,inner sep=0pt,minimum size=5pt,label=south:{\scriptsize $L$} ]  at (\rzL,\roL)  {};
			
			\node[circle,fill=black,inner sep=0pt,minimum size=5pt,label=west:{\scriptsize $R$} ]  at (\rzR,\roR) {};
			
		}

		\draw (6,0) -- (0,6);
		
		\draw [thick](2pt,6 cm) node[left] {\scriptsize $1$};
		\draw [thick](6 cm,-2pt) node[below] {\scriptsize $1$};
		\end{scope}
		
		\end{tikzpicture}

	\end{subfigure}
	\caption{Situation diplaying an $\alpha$ and a $\beta$--fan with a depletion region in the middle, $(\rho_\circ^L=0.1,\rho_\bullet^L=0.5,\rho_\ast^L=0.4),(\rho_\circ^R=0.5\rho_\bullet^R=0.05,\rho_\ast^R=0.45)$. \emph{On the left:} height function of $\bullet$-particles (green lines) and of $\ast$-particles.  Continuous lines represent the numerical values, dashed lines represent the theoretical prediction. At the bottom we have reported the predicted densities. Notice the flat central region of $h_\ast$ corresponding to a region of vanishing $\rho_\ast$. \emph{On the right:} the projection of this configuration in the densities plane: in blue the $\beta$ fan, in green the TASEP-like fan with vanishing $\rho_\ast$ and in red the $\alpha$ fan.}\label{fig:fin}
	%
	%
\end{figure}
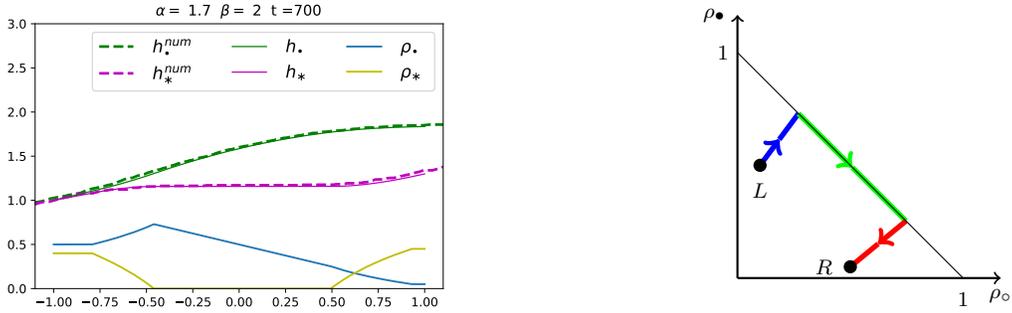

In Figs. \ref{fig-sim-tutto},\ref{fig:fin} we illustrate some representative results of our simulations. 
We have run each simulation up to a time  $t$ indicated at the top of each plot. At that time we find that the left boundary wave has not yet reached the site $i=-t$ and the right boundary wave has not reached the site $i=t$, so we can safely use eq.(\ref{integrated-density}) as a definition of the height function and plot it for $ -1\leq \xi \leq 1$ (for convenience we chose $h_{i}(-1)=1$).
First we illustrate the simulation of the system  
for $\alpha=0.3$, $\beta=0.8$  (Figs. \ref{fig:sub1}-\ref{fig:sub3}). In Fig. \ref{fig:sub1} the left  densities are $(\rho_\circ^L=0.45,\rho_\bullet^L=0.11)$ and the right densities are $(\rho_\circ^R=0.23,\rho_\bullet^R=0.67)$, which in terms of the $\bfz$ variables   correspond to a uniform  $z_\alpha^L=z_\alpha^R=0.15  $ and $z_1^L=0.1 < z_1^R=0.6$. This choice of boundary densities is expected to lead to the formation  of a single $\beta$-shock. Indeed the plot of  
the height functions for the $\bullet$-particles and $\circ$-particles shows two linear regions, corresponding to the two regions of constant densities, separated at the predicted shock speed, situated at $\xi=v_{s,\beta}=0.217$. 
In Fig. \ref{fig:sub2} the boundary densities are inverted w.r.t. Fig. \ref{fig:sub1}. In this case we don't expect any shock to persist, and indeed the result is compatible with a single fan.
In Fig. \ref{fig:sub3} we explore the separation of two shocks.
The boundary densities are $(\rho_\circ^L=0.8,\rho_\bullet^L=0.1)$, $(\rho_\circ^R=0.2,\rho_\bullet^R=0.7)$, which correspond to 
$z_\alpha^L=0.265> z_\alpha^R=0.139 $ and $z_\beta^L=0.097 < z_\beta^R=0.627$. In this case the solution of the Riemann problem predicts an $\alpha$--shock of speed $v_{s,\alpha}= 0.008$ followed by a $\beta$-shock of speed $v_{s,\beta}= 0.186$. The $\beta$ shock is clearly visible at the bend of $h_\bullet$ (green line), the $\alpha$ shock is less visible because of the small bends in both height functions.
In Fig. \ref{fig:sub4}, we have $\alpha=0.7, \beta=1.2$ and  we show the result of the simulation where the theoretical analysis predicts an
$\alpha$--fan and a $\beta$--shock, the shock being visible at the bend of $h_\bullet$ (green line), located at $\xi=v_{s,\beta}= 0.49$.

Finally in Fig.\ref{fig:fin} we explore the formation of a TASEP--like fan corresponding to a region where the density of $\ast$--particles vanishes. We chose $\alpha=1.7,\beta=2$, $(\rho_\circ^L=0.1,\rho_\bullet^L=0.5,\rho_\ast^L=0.4),(\rho_\circ^R=0.5,\rho_\bullet^R=0.05,\rho_\ast^R=0.45)$ so that initially the $\ast$-particles are present everywhere in the system. The plot of $h_\ast$ (purple line) shows a central flat region, which corresponds to a region of vanishing $\ast$-particles density.

\section{Conclusion}\label{section:conclusion}

In this article we have investigated the hydrodynamic behavior of an exactly solvable two species exclusion process, consisting of two kinds of particles moving in opposite directions on a one dimensional lattice  and swapping their position when adjacent. 
By making the assumption of local stationarity and exploiting the knowledge of the particle currents on periodic geometry, we have written down the hydrodynamic conservation  laws of the model and investigated their solutions: rarefaction fans, shocks and the solution of the Riemann problem. Then such predictions have been shown to be in agreement with numerical simulations of the microscopic model.
The macroscopic conservation laws are shown to belong to a class of conservation laws called Temple systems, which possess coinciding rarefaction and shock curves.

While this property has important implications for the mathematical analysis of the conservation equations, it is not clear to us what is  its physical underpinning. We believe it would be interesting to investigate how generic this property is among exclusion processes, and whether it is somehow related to the integrability of the underlying microscopic model. On this line of thoughts, natural candidates to investigate seem to be the multispecies integrable  generalizations (with more than $2$ species) introduced in 
\cite{cantini2016inhomogenous}.
%
%
%
%

Another interesting issue would be to understand the behavior of our model when restricted to a finite lattice in contact with boundary particle reservoirs.
%
On general grounds, novel feautures are expected to emerge like for example the phenomenon of boundary induced phase transitions \cite{krug1991boundary}.
For the case of a single conserved quantity the well-known max--min principle of Krug \cite{krug1991boundary}, later  generalized by Popkov and Sch\"utz \cite{popkov1999steady}  allows to determine the dependence of the current on the boundary couplings and hence the phase diagram of the model. However, it is not known how to generalize the max--min principle in the case of more than one conserved species. Preliminary numerical investigations based on the model presented here seems to indicate that  the Riemann problem may play a relevant role in order to tackle this problem  \cite{W-P}.

\chapter{Integrable tools for the exclusion process}
\label{integ}
Exactly solvable models are rare gems in theoretical physics. In the domain of far from equilibrium statistical physics, they are paving the way for its exploration. In that regard, the exclusion process within its different variants is playing this role in a way often compared to the role the Ising model played for the equilibrium counterpart. In this chapter we are considering two popular boundary conditions for the exclusion process: The infinite $\mathbb{Z}$ lattice, and the periodic boundary conditions. For the former, we are interested in the finite time probability distribution of the position of a finite number of particles conditioned on a given initial configuration. In the latter, the main objective is to find the expression of the currents in the steady state in a system with multiple species with arbitrary rates. The two problems seem quite distinct at the first glance, in particular, because the first does not possess a stationary state while the second does. Yet, it happens that a similar technique works for both as a starting point: writing the stationary measure as Bethe vector. This measure will not be a probability measure for the first and will not lead to Bethe equations, while for the second it will.
Exact probability calculations on the line gained importance after the seminal work of \cite{johansson2000shape} revealing a connection between properly scaled finite probabilities for TASEP and the largest eigenvalue distribution for the Gaussian Unitary Ensemble.

The novelty of this chapter is the section \ref{Multispecies exclusion} where we present a general framework for calculating the finite time conditional probability for an arbitrary number of species with arbitrary hopping rates. This leads to explicit determinantal formulas in particular cases. The rest of section \ref{Exclusion process on the line} serves a pedagogical purpose: we start by reviewing the simplest version of the problem, which is TASEP with a single species, this roughly follows \cite{schutz1997exact}, the complexity is then increased gradually by visiting briefly the problem with second class particles of unity rates, that has been discussed in \cite{chatterjee2010determinant} before reaching the general multi-species situation in section ~\ref{Multispecies exclusion}. This section will the core of a near-future independent publication.

Section \ref{Exclusion process on the ring} is devoted to reviewing how the Bethe Ansatz is applied for finding the currents in the exclusion process on the ring, most of the results obtained here are needed for the other chapters. The complexity is again increased gradually by first considering the problem of a single defect and treating it with Coordinate Bethe Ansatz following \cite{mallick1996shocks}, and then reviewing the arbitrary number of defects using the Algebraic Bethe Ansatz, following \cite{cantini2008algebraic}.

\section{Exclusion process on the line}
\label{Exclusion process on the line}
\subsection{Exact solution for TASEP on the line}
Consider a single component TASEP with a finite number of particles $N$, defined on the integers line $\mathbb{Z}$. Let $\{x_{1},...,x_{N}\} \subset \mathbb{Z}$ the initial positions of particles arranged in increasing order: $x_{i} < x_{i+1}$. Let $\{y_{1},...,y_{N}\} \subset \mathbb{Z}$ the final position of these particles at time $t$.
The state space will be: $\mathcal{S} = \{X \subset \mathbb{Z}, \#X=N \}$.

We would like to calculate the conditional probability $P(\{y_{1},...,y_{N}\};t|\{x_{1},...,x_{N}\}$. By abuse of notation, we omit the initial condition dependence. This probability evolves according to the Master equation: If there is no neighboring particles, $y_{i} < y_{i+1} -1$ for all $i$, then:

\begin{equation}\label{ME}
\frac{d}{dt} P(\{y_{1},...,y_{N}\};t) = \sum_{i=1}^{N}  P(\{y_{1},..,y_{i}-1,..,y_{N}\};t) - N P(\{y_{1},...,y_{N}\};t)
\end{equation}

In case two particles are neighbors, say for instance: $y_{j+1} = y_{j}+1$, and all the others are apart, then there will be two terms missing on right-hand side:

\begin{equation}\label{equ:ME2}
\frac{d}{dt} P(\{y_{1},..,y_{j},y_{j}+1,..,y_{N}\};t) = \sum_{i \neq j+1}  P(\{y_{1},..,y_{i}-1,..,y_{N}\};t) - (N-1) P(\{y_{1},...,y_{N}\};t) 
\end{equation}

It's possible to make equation \ref{ME} account for equation \ref{equ:ME2} if we allow the probability function to assign values for non-physical configurations with consecutive particles, these values have to be:
\begin{equation}
P(\{y_{1},..,y_{j},y_{j+1}=y_{j},..,y_{N}\};t) = P(\{y_{1},..,y_{j},y_{j}+1,..,y_{N}\};t)
\end{equation}
Now with this condition, it's easy to check that equation 1 becomes valid even with neighboring particles.
\subsubsection{Bethe Ansatz Solution}
The probability distribution can be seen as a vector in the infinite dimension space of functions $ P(\{.\}) \in \mathbb{R}^{\mathcal{S}}$. Solving the ordinary differential equation is equivalent to diagonalizing the Markov matrix. So we need to consider the spectral problem:
\begin{equation}\label{equ:ME}
\lambda P(\{.\}) = M  P(\{.\})
\end{equation}

\textbf{One Particle:} Let's first examine this equation for the almost trivial case of one particle $N=1$, No Ansatz is required here. We have a functional equation:
\begin{equation}
\lambda P(\{y\}) = P(\{y-1\}) - P(\{y\})
\end{equation}
Which admits plane waves as solutions:
\begin{equation}
P_{p}(\{y\}) = e^{i py}
\end{equation}
The parameter $p$ is analogous to the momentum and is associated with the eigenvalue:
\begin{equation}
\lambda_{p} =  e^{-ip} - 1
\end{equation}
And the associated time evolution will be:
\begin{equation}
P_{p}(\{y\};t) = e^{\lambda_{p} t} e^{i py}
\end{equation}
Of course, these solutions are not physical, since they are not normalizable, and not even real. However, we can decompose our initial condition in terms of these waves: P({x},t=0)= $ \delta_{x,y} = \frac{1}{2 \pi} \int_{0}^{2 \pi} e^{-ipx} e^{ipy} dp $. The bounded interval of the integral is due to the integer dependence of the left side, it's a Fourier series decomposition.  Now we can write the time evolution of this initial condition:
\begin{equation}
\begin{split}
P(\{y\};t) & =  \frac{1}{2 \pi i} \int_{0}^{2 \pi} e^{-ipx} P_{p}(\{y\};t) dp \\
& =   \frac{1}{2 \pi i} \int_{0}^{2 \pi}  e^{ (e^{-ip} - 1) t} e^{-i p(x-y)} dp \\
& =   \frac{1}{2 \pi i} \oint_{0}  e^{(z - 1)t} z^{(x-y-1)} dz \\
& =  e^{-t} \frac{t^{y-x}}{(y-x)!}
\end{split}
\end{equation}
Where the last step integral was performed by residues theorem.

\textbf{Two particles:} The situation is a bit more complicated for $N=2$. The eigenvalue problem for the Markov matrix is:
\begin{equation}
\lambda P(\{y_{1},y_{2}\}) = P(\{y_{1}-1,y_{2}\}) + P(\{y_{1},y_{2}-1\}) - 2P(\{y_{1},y_{2}\})
\end{equation}
We can again check that $e^{i(p_{1}y_{1}+p_{2}y_{2})}$ is indeed a family of solutions for the equation. However, this family is not compatible with the boundary condition:
\begin{equation}
P(\{y_{1},y_{2} = y_{1}\}) = P(\{y_{1},y_{2}=y_{1}+1\})
\end{equation}
At this point, the Bethe Ansatz is needed. The solutions can be expressed as a superposition of the possible permutations of the momentum of the plane waves:
\begin{equation}
P_{\{p_{1},p_{2}\}}(\{y_{1},y_{2}\}) = A_{1,2}e^{i(p_{1}y_{1}+p_{2}y_{2})} + A_{2,1}e^{i(p_{2}y_{1}+p_{1}y_{2})}
\end{equation}
With this combination, the boundary condition can be verified for a right choice of the coefficients, namely:

\begin{equation}
\frac{A_{1,2}}{A_{2,1}} = - \frac{1-e^{ip_{1}}}{1-e^{ip_{2}}}
\end{equation}
And the corresponding eigenvalue would be:
\begin{equation}
\lambda_{\{p_{1},p_{2}\}} = e^{-ip_{1}} + e^{-ip_{2}}  - 2 = \lambda_{p_{1}} + \lambda_{p_{2}}
\end{equation}
Now we need to write the initial condition as a superposition of this family:
\begin{equation}
\delta_{x_{1},y_{1}}\delta_{x_{2},y_{2}} = \frac{1}{(2 \pi)^{2}}\int_{0}^{2 \pi} \int_{0}^{2 \pi} f(p_{1},p_{2}) (e^{i(p_{1}y_{1}+p_{2}y_{2})} - \frac{1-e^{ip_{2}}}{1-e^{ip_{1}}}e^{i(p_{2}y_{1}+p_{1}y_{2})})dp_{1}dp_{2}
\end{equation}
Actually this integral is not well defined because of the presence of a singularity at $p_{1}=0$. Let's write it as an integral in the complex plane:
\begin{equation}
\delta_{x_{1},y_{1}}\delta_{x_{2},y_{2}} =
\oint_{0}
\oint_{0}
f(z_{1},z_{2}) (z_{1}^{y_{1}-1}z_{2}^{y_{2}-1} - \frac{1-z_{2}}{1-z_{1}}z_{1}^{y_{2}-1}z_{2}^{y_{1}-1})dz_{1}dz_{2}
\end{equation}
To define this integral, we need to tell whether the 1 is included in the unit circle. If we choose to exclude it, which means taking the contour integral only around zero, then the naive expression for $f$: $f(z_{1},z_{2}) = z_{1}^{-x_{1}} z_{2}^{-x_{2}} $ would yield a solution for the previous equation. Finally, we can write the time-evolved solution:

\begin{equation}
\begin{split}
P(\{y_{1},y_{2}\};t) = &
\oint \oint e^{(z_{1}^{-1} + z_{2}^{-1} - 2)t} (z_{1}^{y_{1}-x_{1}-1}z_{2}^{y_{2}-x_{2}-1} - \frac{1-z_{2}}{1-z_{1}}z_{1}^{y_{2}-x_{1}-1}z_{2}^{y_{1}-x_{2}-1})dz_{1}dz_{2} \\
=  &  (e^{-t} \oint e^{\frac{t}{z_{1}}} z_{1}^{y_{1}-x_{1}-1}dz_{1})(e^{-t} \oint e^{\frac{t}{z_{2}}} z_{2}^{y_{2}-x_{2}-1}dz_{2})\\
&-(e^{-t} \oint e^{\frac{t}{z_{1}}} \frac{z_{1}^{y_{2}-x_{1}-1}}{1-z_{1}}dz_{1})(e^{-t} \oint e^{\frac{t}{z_{2}}} (1-z_{2})z_{2}^{y_{1}-x_{2}-1}dz_{2})
\end{split}
\end{equation}
This suggests to define the function:

\begin{equation}
F_{p}(n;t) := e^{-t}  \oint e^{\frac{t}{z}} \frac{z^{n-1}}{(1-z)^{p}}dz
\end{equation}
So that we can write the previous expression as:

\begin{equation}
\begin{split}
P_{\{p_{1},p_{2}\}}(\{y_{1},y_{2}\};t) &=
F_{0}(y_{1}-x_{1};t)F_{0}(y_{2}-x_{2};t) - F_{1}(y_{2}-x_{1};t)F_{-1}(y_{1}-x_{2};t)\\
& = Det((F_{i-j}(y_{i}-x_{j};t))_{ij})
\end{split}
\end{equation}

\subsubsection*{Properties of the functions $ F_{p}(n,t) $}
First of all these functions can be expressed in series form, by developing the exponential in the integral, exchanging the integral and the sum, and then integrating each term by the residues theorem. Consider first that $p>0$

\begin{equation}
\begin{split}
F_{p}(n;t) &= e^{-t}\oint
\sum_{k=0}^{\infty}\frac{t^{k}}{z^{k}k!}
z^{n-1}
\sum_{m=0}^{\infty} C_{m+p-1}^{p-1} z^{m}dz\\
&=e^{-t}
\sum_{k=0}^{\infty}
\sum_{m=0}^{\infty}
\oint
\frac{t^{k}}{k!}
C_{m+p-1}^{p-1} z^{m+n-k-1}
dz\\
&= e^{-t}
\sum_{k=n}^{\infty}
\frac{t^{k}}{k!}
C_{k-n+p-1}^{p-1} = e^{-t}
\sum_{k=0}^{\infty}
C_{k+p-1}^{p-1}
\frac{t^{k+n}}{(k+n)!}
\end{split}
\end{equation}
For $p\leq 0$ the power series expansion above is still actually valid, but one has to extend the definition of $C_{a}^{b}$ to negative numbers. This is possible using the $\Gamma$ function:
$$C_{a}^{b} = \frac{\Gamma(a+1)}{\Gamma(b+1) \Gamma (a-b+1)}$$ 
This can be used to show that for $p\leq 0$:
\begin{equation}
C_{p+m-1}^{m}=\begin{cases}
(-1)^m C_{-p}^{m}  \quad &\text{if} \quad  m < p+1  \\
0 \quad &\text{if} \quad   m \geq p+1 \\
\end{cases}
\end{equation}
This can be used to show that the power series expansion of $\frac{1}{(1-z)^{p}}$ is reduced to the finite expected Binomial expansion for $p \leq 0$. We can as well rewrite the expression of $F_{p}(n,t)$:
\begin{equation}
F_{p}(n,t) =  e^{-t}
\sum_{k=0}^{-p}
(-1)^{k}C_{-p}^{k}
\frac{t^{k+n}}{(k+n)!}
\end{equation}
\textbf{Remark:} The function $F_{p}(n,t)$ is related to the confluent hypergeometric function ${}_{1}F_{1}$
\begin{equation}
F_{p}(n;t) = 	\frac{e^{-1} t^{n}}{n!}{}_{1}F_{1}(p,n+1;t)
\end{equation}
Where ${}_{1}F_{1}(a,b;t) = \sum_{k=0}^{\infty} \frac{(a)_{k}}{(b)_{k}} \frac{t^{k}}{k!}$
And $(a)_{k} = a(a+1)...(a+k-1)$.

Other properties which are easy to show:
\begin{itemize}
	\item
	\begin{equation}\label{key}
	\frac{d}{dt} F_{p}(n,t) = F_{p-1}(n-1;t) = F_{p}(n-1,t) - F_{p}(n,t)
	\end{equation}
	
	\item
	
	\begin{equation}\label{key}
	\int_{0}^{t} F_{p}(n,s) ds = F_{p+1}(n+1;t) - F_{p+1}(n+1;0) = F_{p+1}(n+1;t) - C_{-n+p-1}^{p-1} 
	\end{equation} 
	
	\item 
	
	\begin{equation}\label{key}
	\sum_{n=n_{1}}^{n_{2}} F_{p}(n;t) = F_{p+1}(n;t) - F_{p+1}(n+1;t)
	\end{equation}

\end{itemize}

\textbf{Arbitrary number of particles:}
An $N$ particle calculation follows the same logic as the two particles, but with some subtle details. The Bethe wave function would be composed of $N!$ term:

\begin{equation}
P_{\mathbf{p}}(\mathbf{y}) =
\sum_{\sigma \in \mathcal{S}_{N}}
A_{\sigma} \prod_{i=1}^{N} e^{ip_{\sigma(i)}y_{i}}
\end{equation}
Where $\mathbf{p} = {\{p_{1},...,p_{N}\}}$, $\mathbf{y} = {\{y_{1},...,y_{N}\}}$ and $S_{N}$ is the symmetric group. The restriction on the coefficients generalizes to:

\begin{equation}\label{exch}
\frac{A_{\sigma}}{A_{\sigma \tau_{i,i+1}}} = - \frac{1-e^{ip_{\sigma{(i)}}}}{1-e^{ip_{\sigma{(i+1)}}}}
\end{equation}

Where $\tau_{i,i+1}$ is the transposition applied on positions $i$ and $i+1$.

Since each permutation can be written as a product of the transposition of neighboring sites, then it's enough to fix one of the coefficients in order to fix all the others. Say for instance that $A_{Id} = 1$. We know that the decomposition of a permutation in terms of transpositions is not unique, so for this approach to be self-consistent, the value of the coefficient should be independent of the transposition path.
The most trivial check is the invariance under a double application of the same transposition, which is obviously verified here. It happens that we don't need to check for all the possible paths and it's enough to verify that the path is taken by next neighbor transpositions:

\begin{equation}
\tau_{1,3} = \tau_{1,2} \tau_{2,3} \tau_{1,2} = \tau_{2,3} \tau_{1,2} \tau_{2,3}
\end{equation} 
This is the Yang-Baxter equation. We can check that it is verified:
\begin{equation}
A_{\tau_{1,3}} = 
A_{\tau_{1,2} \tau_{2,3} \tau_{1,2}} = A_{\tau_{2,3} \tau_{1,2} \tau_{2,3}} = - \frac{1-e^{ip_{3}}}{1-e^{ip_{1}}}
\end{equation}
This can be generalized to any permutation $\tau_{i,j}$. One can notice that the map $\sigma \rightarrow A_{\sigma}$ is not a group homomorphism(if it were, there would have been no need to check for Yang-Baxter, it would be trivially verified), the homomorphism property applies only for permutations with independent support. Say $\sigma_{1}$ and  $\sigma_{2}$ are two such permutations, then:
\begin{equation}
A_{\sigma_{1} \sigma_{2}} = A_{\sigma_{1}} A_{\sigma_{2}}
\end{equation} 
By deduction on cyclic permutations, one can reach an explicit formula for $A_{\sigma}$ that is valid for any permutation:
\begin{equation}
A_{\sigma} = \epsilon(\sigma) \prod_{i=1}^{N}(1-z_{\sigma(i)})^{\sigma(i)-i}
\end{equation}
Where $\epsilon(\sigma)$ is the signature of $\sigma$. It's easy to show that this formula verifies indeed eq. \ref{exch}.
Now we need to check the validity of the (naive) composition of the initial condition in terms of the Bethe wave vectors. For this we need to show that:
\begin{equation}
\sum_{\sigma \neq Id}
\oint_{0}
...
\oint_{0}
A_{\sigma} \prod_{i=1}^{N}
z_{\sigma(i)}^{y_{i} - x_{\sigma(i)}-1}
dz_{1}...dz_{N}
=0
\end{equation} 
It's actually possible to show that each term in the sum is zero. Take for instance the term corresponding to some $\sigma \neq Id$. there must exists for it an index $ \tilde{i} $ such that $ \tilde{i}  > \sigma(\tilde{i} )$ so it follows:
\begin{equation}
y_{\tilde{i} } > y_{\sigma(\tilde{i} ) }\geq x_{\sigma(\tilde{i} )}
\end{equation}
So the factor $ z_{\sigma(\tilde{i})}^{y_{\tilde{i}} - x_{\sigma(\tilde{i})}-1} $ would have no poles, and since $A_{\sigma}$ is holomorphic too in a neighborhood of zero, the corresponding contour integral around zero  would vanish.
Now we need to examine the time evolution.
Let's write down the eigenvalue equation:

\begin{equation}
\begin{split}
\lambda_{\mathbf{p}} P_{\mathbf{p}}(\mathbf{y}) &=
\sum_{i=1}^{N}
\sum_{\sigma \in \mathcal{S}_{N}}
e^{-ip_{\sigma(i)}}
A_{\sigma} \prod_{i=1}^{N} e^{ip_{\sigma(i)}y_{i}}
-N P_{\mathbf{p}}\\
& = 
\sum_{i=1}^{N}
\sum_{j=1}^{N}
\sum_{\sigma: \sigma(i) = j}
e^{-ip_{j}}
A_{\sigma} \prod_{i=1}^{N} e^{ip_{\sigma(i)}y_{i}}
-N P_{\mathbf{p}} \\
& = 
\sum_{j=1}^{N}
e^{-ip_{j}}
\sum_{i=1}^{N}
\sum_{\sigma: \sigma(i) = j}
A_{\sigma} \prod_{i=1}^{N} e^{ip_{\sigma(i)}y_{i}}
-N P_{\mathbf{p}}
\end{split}
\end{equation}
So the second and the third sum can be united again into a sum over all the permutations and we find again that the eigenvalue for a Bethe wave has the same form as for two particles:
\begin{equation}
\lambda_{\mathbf{p}} =
\sum_{i=1}^{N}
(e^{-ip_{i}} - 1) =
\sum_{i=1}^{N}
\lambda_{p_{i}}
\end{equation}
All the ingredients are now ready for the time-evolved vector:
\begin{equation}\label{Evolved}
\begin{split}
P(\{\mathbf{y}\};t) &= 
\oint_{0}
...
\oint_{0}
(\prod_{j=1}^{N}e^{\frac{t}{z_{j}} - t})
\sum_{\sigma \in \mathcal{S}_{N}}
\epsilon(\sigma) 
\prod_{i=1}^{N}
(1-z_{\sigma(i)})^{\sigma(i)-i}
z_{\sigma(i)}^{y_{i} - x_{\sigma(i)}-1}
dz_{1}...dz_{N} \\
&=
\sum_{\sigma \in \mathcal{S}_{N}}
\epsilon(\sigma)
\prod_{i=1}^{N}
\oint_{0}
e^{\frac{t}{z_{\sigma(i)}} - t}
(1-z_{\sigma(i)})^{\sigma(i)-i}
z_{\sigma(i)}^{y_{i} - x_{\sigma(i)}-1}
dz_{\sigma(i)}\\
&=
\sum_{\sigma \in \mathcal{S}_{N}}
\epsilon(\sigma)
\prod_{i=1}^{N}
F_{i-\sigma(i)}(y_{i} - x_{\sigma(i)};t)\\
& = Det((F_{i-j}(y_{i}-x_{j};t))_{ij})
\end{split}
\end{equation}

\textbf{Remarks}
\begin{enumerate}
	\item It's possible to prove the final result of \ref{Evolved} is indeed the solution of the Master equation just by using the properties of the functions $F_{p}(n;t)$ and elementary operations on the determinant. This is done in \cite{schutz1997exact}
	\item It's obviously possible to extend this method to ASEP. The computations are a bit more tedious, partly because the plane waves are not the only eigenvectors of the Markov matrix. Families of bound states appear as eigenvectors as well. To see this, one has to get a bit into the details of the procedure: the extra term in the master equation will require an extra term in the boundary condition, so it becomes: $P(y,y) + q P(y+1,y+1)=(1+q)P(y,y+1)$ where $q$ is the backward hopping parameter. Applying this condition on the Bethe wave vector yields: $ A_{12}(1+q z_{1}z_{2} - z_{2}(1+q) )= - A_{21}(1+q z_{1}z_{2} - z_{1}(1+q) ) $, beside the usual solutions similar to the TASEP case, one can choose to satisfy the previous equation for instance by setting $A_{21} = 0$ and $ 1+q z_{1}z_{2} - z_{2}(1+q)  = 0 $ this leads to a constraint between $z_{1}$ and $z_{2}$ forcing them to leave the unit circle and producing a one-parameter family of solutions that would extend the base on which one needs to decomposing the initial condition.
	This has been done for two particles in\cite{schutz1997exact}, and generalized to an arbitrary number of particles in  \cite{tracy2008integral}.

\end{enumerate}

\subsection{Adding second class particles}
We consider TASEP on the line with a finite number ofirst-class particles, denoted by $A$ and second class particles, denoted by $B$, with unity hopping rates. The space state is now $\mathcal{S} \times \{A,B\}^{ N} $. Where $N$ is the total number of particles. For a given set of positions of the particles, the probabilities of the different permutations can be represented by a vector in $(\mathbb{C}^{2})^{\otimes N}$ (Notations don't assume particles type conservation). 
The master equation is the same as single species TASEP when neighboring particles are absent. Let's write the equation for 2 adjacent particles:

\begin{equation}
\frac{d}{dt} \mathbf{P}(y,y+1;t) = \mathbf{P}(y-1,y+1;t)- M \mathbf{P}(y,y+1;t)
\end{equation}

Where
$ \mathbf{P}(y_{1},y_{2};t) = 
\begin{pmatrix}
P^{AA}(y_{1},y_{2};t) \\
P^{AB}(y_{1},y_{2};t) \\
P^{BA}(y_{1},y_{2};t) \\
P^{BB}(y_{1},y_{2};t) 
\end{pmatrix} $ and $M = \begin{pmatrix}
1 & 0 & 0 & 0 \\
0 & 2 & 0 & 0 \\
0 & -1 & 1 & 0 \\
0 & 0 & 0 & 1
\end{pmatrix}$

This requires new boundary conditions for $P^{AB}$ and $P^{BA}$ so that the non-neighboring master equation gets reduced to the neighboring one on the boundaries. These conditions are:

\begin{equation}
P^{AB}(y,y) = 0
\end{equation}
\begin{equation}
P^{BA}(y,y) = P^{BA}(y,y+1) + P^{AB}(y,y+1)
\end{equation}

As usual we write the Bethe vector as:
\begin{equation}
\mathbf{P}_{\{p_{1},p_{2}\}}(y_{1},y_{2}) = \mathbf{A}_{1,2}e^{i(p_{1}y_{1}+p_{2}y_{2})} + \mathbf{A}_{2,1}e^{i(p_{2}y_{1}+p_{1}y_{2})}
\end{equation}

Where the coefficients are vector now. The matrix that allows transiting between them is found thanks to the boundary restrictions:

\begin{equation}
\mathbf{A}_{2,1} =
R(p_{1},p_{2})
\mathbf{A}_{1,2}
\end{equation}

Where:
\begin{equation}
R(p_{1},p_{2}) =
\begin{pmatrix}
-\frac{1-e^{ip_{2}}}{1-e^{ip_{1}}} & 0 & 0 & 0 \\
0 & -1 & 0 & 0 \\
0 & -\frac{e^{ip_{1}}-e^{ip_{2}}}{1-e^{ip_{1}}} & -\frac{1-e^{ip_{2}}}{1-e^{ip_{1}}} & 0 \\
0 & 0 & 0 & -\frac{1-e^{ip_{2}}}{1-e^{ip_{1}}}
\end{pmatrix}
\end{equation}
More generally, we can conclude how to apply a transposition on a coefficient:
\begin{equation}
\mathbf{A}_{\sigma. \tau_{i,i+1}}=R(p_{\sigma(i)},p_{\sigma(i+1)})\mathbf{A}_{\sigma}
\end{equation}

This is enough to provide all the coefficients in terms of $\mathbf{A}_{Id}$. To make sure that for each permutation $\sigma$, there is a well-defined $\mathbf{A}_{\sigma}$, one can show that the $R$ matrix verifies the Yang-Baxter Equation:
\begin{equation}
R_{23}(p_{1},p_{2})R_{12}(p_{1},p_{3})R_{23}(p_{2},p_{3}) = R_{12}(p_{2},p_{3})R_{23}(p_{1},p_{3})R_{12}(p_{1},p_{2}) 
\end{equation}
Where $  R_{23} = Id \otimes R $ and $  R_{12} = R\otimes Id$

It's not hard to verify that the initial condition can still be written as an integral of Bethe vectors in a similar fashion to single species. Let $\mathbf{P}(x_{1},x_{2};0)$ the initial probability vector, then we have: 
\begin{equation}
\delta_{x_{1},y_{1}}\delta_{x_{2},y_{2}} \mathbf{A}_{Id} =
\oint_{0}
\oint_{0}
(z_{1}^{x_{1}} z_{2}^{x_{2}} ) (z_{1}^{y_{1}-1}z_{2}^{y_{2}-1} - R(p_{1},p_{2})z_{1}^{y_{2}-1}z_{2}^{y_{1}-1}))\mathbf{A}_{Id} dz_{1}dz_{2}
\end{equation}

\begin{equation}
\delta_{\mathbf{x},\mathbf{y}} \mathbf{P}(\mathbf{x};0) =
\oint_{0}
\oint_{0}
(\prod_{i=1}^{N} z_{i}^{x_{i}} ) (z_{1}^{y_{1}-1}z_{2}^{y_{2}-1} - R(p_{1},p_{2})z_{1}^{y_{2}-1}z_{2}^{y_{1}-1})\mathbf{A}_{Id} dz_{2}
\end{equation}

\begin{equation}
\mathbf{P}_{\mathbf{p}}(\mathbf{y}) =
\sum_{\sigma \in \mathcal{S}_{N}}
\mathbf{A}_{\sigma} \prod_{i=1}^{N}
e^{i p_{\sigma(i)}y_{i}}
\end{equation}

So the time evolution:

\begin{equation} \mathbf{P}(\mathbf{y};t) =\oint...\oint
(\prod_{j=1}^{N}e^{\frac{t}{z_{j}} - t})
\sum_{\sigma \in \mathcal{S}_{N}}
\mathbf{A}_{\sigma} \prod_{i=1}^{N}
z_{\sigma(i)}^{y_{i}-x_{i}-1} dz_{i}
\end{equation}

It's possible to write an explicit formula for $\mathbf{P}$ in the case where the order of species of the final configuration is the same as the initial one, this has been done in \cite{chatterjee2010determinant}(No exchange theorem).
In the following section, we will be examining the case where the hopping rates are arbitrary per species, and the number of species is arbitrary too.

\subsection{Multispecies exclusion process with arbitrary hopping rates}
\label{Multispecies exclusion}
Let's treat the most general situation with $N$ particles of multiple species. The position of the particles shall be denoted by Latin letters, while their species by Greek ones, so Let $I = \{i_{1} <... < i_{N}\} $ be the set of positions corresponding to the species $\mathbf{\alpha} = \{\alpha_{1} <... < \alpha_{N}\}$. Let $r_{\alpha}$ the hopping rate of particles of species $\alpha$ and $r_{\alpha, \beta}$ the rate of the ordered swap $\alpha \leftrightarrow \beta $. of a particle of type $\alpha$ followed by one of type $\beta$.

Let $G(I,\mathbf{\alpha}|J,\mathbf{\beta};t)$ be the probability of having the state $(I,\mathbf{\alpha})$ at time $t$ starting from the initial  state $(J,\mathbf{\beta})$ The master equation for non neighboring particles:

\begin{equation}\label{equ:ME}
\frac{d}{dt} G(I,\mathbf{\alpha}|J,\mathbf{\beta};t) = \sum_{k=1}^{N} r_{\alpha_{k}}
G({i_{1}..,i_{k}-1,..i_{N}},\mathbf{\alpha}|J,\mathbf{\beta};t) - (\sum_{k=1}^{N}\alpha_{k}) G(I,\mathbf{\alpha}|J,\mathbf{\beta};t)
\end{equation}

For two neighboring particles:
\begin{equation}\label{equ:ME}
\frac{d}{dt} G(i,i+1,12) = r_{\alpha}
G(i-1,i+1,12) +
r_{\beta \alpha}
G(i,i+1,21) -
(r_{\alpha,\beta}+ r_{\beta}) G(i,i+1,12)
\end{equation}

\begin{equation}\label{equ:ME}
\frac{d}{dt} G(i,i+1,12) = r_{\alpha}
G(i-1,i+1,12) +
r_{\beta}
G(i,i,12) -(r_{\beta}+r_{\alpha}) G(i,i+1,12)
\end{equation}

So the boundary condition needs to be:

\begin{equation}\label{BC}
r_{\beta} G(i,i,12) = (r_{\alpha}-r_{\alpha,\beta}) G(i,i+1,12) + r_{\beta \alpha} G(i,i+1,21)
\end{equation}

For the Bethe vector, let's first notice that vector  $z_{1}^{i}z_{2}^{j}$ does not yield the same eigenvalue as $z_{2}^{i}z_{1}^{j}$ (for non-neighboring particles). To conserve this property, one needs to resale the Bethe vector in the following way:

\begin{equation}
\psi^{I}_{\bm{\alpha}} = r_{\bm{\alpha}}^{I}
\sum_{\sigma \in \mathcal{S}_{N}}
F_{\bm{\alpha}}^{\sigma} \sigma[\mathbf{z}^{I}]
\end{equation}
Where $r_{\bm{\alpha}}^{I} = \prod_{k=1}^{n} r_{\alpha_{k}}^{i_{k}}$, $\mathbf{z}^{I} = \prod_{k} z_{k}^{i_{k}}$ and $ \sigma[\mathbf{z}^{I}] = \prod_{k} z_{\sigma(k)}^{i_{k}} $. Applied on the Markov matrix, the corresponding eigenvalue is:

\begin{equation}
\lambda = \sum_{k} (z_{k}^{-1} - r_{k})
\end{equation}

As usual, we need to establish how the coefficients of the Bethe vector are related so that the boundary conditions are respected. Let's consider a 2 particle Bethe vector:

\begin{equation}
\psi_{\alpha, \beta}^{i,j} =
r_{\alpha}^{i} r_{\beta}^{j}
(F_{\alpha, \beta}^{()}z_{1}^{i}z_{2}^{j} + F_{\alpha, \beta}^{(12)}z_{2}^{i}z_{1}^{j})
\end{equation}
Inserting it into \ref{BC}, we get:

\begin{equation}
-
(F_{\alpha, \beta}^{()} +F_{\alpha,\beta}^{(12)})+
(r_{\alpha}-r_{\alpha,\beta})
(F_{\alpha, \beta}^{()}z_{2} + F_{\alpha,\beta}^{(12)}z_{1})+
\frac{r_{\alpha}}{r_{\beta}}
r_{\beta,\alpha}
(F_{\beta,\alpha}^{()}z_{2} + F_{\beta,\alpha}^{(12)}z_{1})
=0
\end{equation}

This can be written in a matrix form:

\begin{equation}
\begin{pmatrix}
F_{\alpha,\beta}^{(12)} & F_{\beta,\alpha}^{(12)} 
\end{pmatrix}
\begin{pmatrix}
(r_{\alpha}-r_{\alpha,\beta}) z_{1} - 1  \\
\frac{r_{\alpha}}{r_{\beta}}r_{\beta,\alpha} z_{1}
\end{pmatrix}
+
\begin{pmatrix}
F_{\alpha,\beta}^{()} & F_{\beta,\alpha}^{()}
\end{pmatrix}
\begin{pmatrix}
(r_{\alpha}-r_{\alpha,\beta}) z_{2} - 1  \\
\frac{r_{\alpha}}{r_{\beta}}r_{\beta,\alpha} z_{2}
\end{pmatrix} = 0
\end{equation}

And by the symmetry $\alpha \leftrightarrow \beta$ we get another equation:
\begin{equation}
\begin{pmatrix}
F_{\alpha,\beta}^{(12)}  & F_{\beta,\alpha}^{(12)}
\end{pmatrix}
\begin{pmatrix}
\frac{r_{\beta}}{r_{\alpha}}r_{\alpha,\beta} z_{1} \\
(r_{\beta}-r_{\beta,\alpha}) z_{1} - 1 
\end{pmatrix}
+
\begin{pmatrix}
F_{\alpha,\beta}^{()} & F_{\beta,\alpha}^{()}
\end{pmatrix}
\begin{pmatrix}
\frac{r_{\beta}}{r_{\alpha}}r_{\alpha,\beta} z_{2} \\
(r_{\beta}-r_{\beta,\alpha}) z_{2} - 1 
\end{pmatrix} = 0
\end{equation}
The two previous equations can be regrouped:

\begin{equation}
\begin{split}
\begin{pmatrix}
F_{\alpha,\beta}^{(12)}  & F_{\beta,\alpha}^{(12)}
\end{pmatrix}
\begin{pmatrix}
(r_{\alpha}-r_{\alpha,\beta}) z_{1} - 1 & \frac{r_{\beta}}{r_{\alpha}}r_{\alpha,\beta} z_{1} \\
\frac{r_{\alpha}}{r_{\beta}}r_{\beta,\alpha} z_{1} & (r_{\beta}-r_{\beta,\alpha}) z_{1} - 1 
\end{pmatrix}
= \\ -
\begin{pmatrix}
F_{\alpha,\beta}^{()} & F_{\beta,\alpha}^{()}
\end{pmatrix}
\begin{pmatrix}
(r_{\alpha}-r_{\alpha,\beta}) z_{2} - 1 & \frac{r_{\beta}}{r_{\alpha}}r_{\alpha,\beta} z_{2} \\
\frac{r_{\alpha}}{r_{\beta}}r_{\beta,\alpha} z_{2} & (r_{\beta}-r_{\beta,\alpha}) z_{2} - 1 
\end{pmatrix}
\end{split}
\end{equation}
In vector notation:
\begin{equation}
F^{(12)} M(z_{2}) + F^{()}M(z_{1}) = 0
\end{equation}

For two species It's convenient to define $F^{\sigma} := (F^{\sigma}_{\alpha,\alpha},F^{\sigma}_{\alpha,\beta},F^{\sigma}_{\beta,\alpha},F^{\sigma}_{\beta,\beta} ) \in \mathbb{C}^{2} \otimes \mathbb{C}^{2}  $ and $M$ will become:

\begin{equation}
M(z) = 
\begin{pmatrix}
r_{\alpha} z - 1 & 0&0&0 \\
0&(r_{\alpha}-r_{\alpha,\beta}) z - 1 & \frac{r_{\beta}}{r_{\alpha}}r_{\alpha,\beta} z & 0 \\
0& \frac{r_{\alpha}}{r_{\beta}}r_{\beta,\alpha} z & (r_{\beta}-r_{\beta,\alpha}) z - 1 & 0 \\
0&0&0 & r_{\beta} z - 1
\end{pmatrix}
\end{equation}

We need to generalize this to an arbitrary number of species $m$, but with only 2 particles, then it's convenient to define the line vector $ F^{\sigma} \in \mathbb{C}^{m} \otimes \mathbb{C}^{m}  $

So we can define the $\check{R}$ matrix:

\begin{equation}
F^{(12)} =  F^{()}\check{R}(z_{1},z_{2})
\end{equation}
Where:
\begin{equation}
\check{R}(z_{1},z_{2}) = - M(z_{1})M(z_{2})^{-1} 
\end{equation}
In a generalized tensor form:
\begin{equation}
M(z)_{\alpha \beta}^{\mu \nu} = ((r_{\alpha}-r_{\alpha,\beta}) z - 1 ) \delta_{\alpha}^{\mu} \delta_{\beta}^{\nu} + \frac{r_{\alpha}}{r_{\beta}}r_{\beta,\alpha} z \delta_{\alpha}^{\nu} \delta_{\beta}^{\mu}
\end{equation}
\begin{equation}
F^{\sigma \tau_{i, i+1}} =  F^{\sigma}\check{R}(z_{\sigma(i)},z_{\sigma(i+1)})
\end{equation}
\subsubsection{Integrability and restrictions on the rates}
The matrix $\check{R}$ needs to obey the Braided Yang-Baxter equation, namely:
\begin{equation}
\check{R}_{23}(z_{1},z_{2})\check{R}_{12}(z_{1},z_{3})\check{R}_{23}(z_{2},z_{3}) = \check{R}_{12}(z_{2},z_{3})\check{R}_{23}(z_{1},z_{3})\check{R}_{12}(z_{1},z_{2}) 
\end{equation}
Explicit expansion of this equation shows that we need to impose hierarchy over the species. A given species can hope only over lower ones in the hierarchy:
\begin{equation}
r_{\alpha, \beta} = 0 \quad \text{if} \quad \alpha > \beta 
\end{equation}
In addition to this restriction, an additional one is needed, for $\alpha > \beta > \gamma$

\begin{equation}
r_{\gamma, \alpha} - r_{\beta, \alpha} = r_{\gamma} - r_{\beta}
\end{equation}
Which means:
\begin{equation}
r_{\alpha, \beta } = (r_{\alpha} + \nu_{\beta}) \mathds{1}_{\beta > \alpha}
\end{equation}

with $\nu_{\beta}$ being a parameter depending only on $\beta$

With these restrictions, the $M$ matrix simplifies to:

\begin{equation}
M(z) = 
\begin{pmatrix}
r_{\alpha} z - 1 & 0 & 0 & 0 \\
0&-1 - \nu_{\beta } z & (r_{\beta} (r_{\alpha} + \nu_{\beta }) z)/ra & 0 \\
0& 0 & -1 + r_{\beta} z & 0 \\
0&0&0 & r_{\beta} z - 1
\end{pmatrix}
\end{equation}

And the $\check{R}$ matrix:

$$
\check{R}(z_{1},z_{2}) = 
\left(
\begin{array}{cccc}
-\frac{z_2 r_{\alpha }-1}{z_1 r_{\alpha }-1} & 0 & 0 & 0 \\
0 & -\frac{z_2 \nu_{\beta }+1}{z_1 \nu_{\beta }+1} & \frac{r_{\beta} (r_{\alpha} + \nu_{\beta }) (z_{1} - z_{2})}{r_{\alpha} (-1 + r_{\beta} z_{1}) (1 + \nu_{\beta } z_{1})} & 0 \\
0 & 0 & -\frac{z_2 r_{\beta }-1}{z_1 r_{\beta }-1} & 0 \\
0 & 0 & 0 & -\frac{z_2 r_{\beta }-1}{z_1 r_{\beta }-1} \\
\end{array}
\right)
$$

\subsubsection{Representation in terms of operators}
Let's define the action of a permutation over a variable:
\begin{equation}
\sigma x_{i} = x_{\sigma(i)} \sigma
\end{equation}

This allows us to write
\begin{equation}
F^{\sigma \tau_{i}}
\sigma \tau_{i} =  F^{\sigma}\check{R}_{i}(z_{\sigma(i)},z_{\sigma(i+1)}) \sigma \tau_{i} =
F^{\sigma} \sigma \check{R}_{i}(z_{i},z_{i+1}) \tau_{i} 
\end{equation}

We define the operator:
\begin{equation}
\pi_{i} :=
\check{R}_{i}(z_{i},z_{i+1}) \tau_{i}
\end{equation}

Which acts from the left on an $F^{\sigma} \sigma$:

\begin{equation}
F^{\sigma \tau_{i}} \sigma \tau_{i} =  F^{\sigma} \sigma \pi_{i}
\end{equation}

So this operator obviously allows to construct $F^{\sigma} \sigma$ starting from $ F^{Id} Id$. Let $\sigma = \prod_{i} \tau_{k_{i}}$ a decomposition of $\sigma$ in terms of transpositions, then, by defining:

\begin{equation}
\pi_{\sigma} := \prod_{i} \pi_{k_{i}}
\end{equation}

We can write:

\begin{equation}
F^{\sigma} \sigma =  F^{e} e \pi_{\sigma}
\end{equation}

Of course, the operators $\pi$ need to abide by the Braided equation:

\begin{equation}
\pi_{i} \pi_{i+1} \pi_{i} = \pi_{i+1} \pi_{i} \pi_{i+1}
\end{equation}

We can notice that $\sigma \rightarrow \pi_{\sigma} $ is a group morphism: $  \pi_{\sigma \acute{\sigma}} = \pi_{\sigma} \pi_{\acute{\sigma}} $

Now we can write the Bethe wave vector:

\begin{equation}
\begin{split}
\psi^{I}_{\bm{\alpha}} & =  r_{\bm{\alpha}}^{I}
\sum_{\sigma \in \mathcal{S}_{N}}
F_{\bm{\alpha}}^{\sigma} \sigma \mathbf{z}^{I} \\
& =  
r_{\bm{\alpha}}^{I}
F_{\bm{\alpha}}^{e}
(\sum_{\sigma \in \mathcal{S}_{N}}\pi_{\sigma})
\mathbf{z}^{I} \\
& = 
r_{\bm{\alpha}}^{I}
F_{\bm{\alpha}}^{e}
\Pi_{0}
\mathbf{z}^{I}
\end{split}
\end{equation}
Where $\Pi_{0} = \sum_{\sigma \in \mathcal{S}_{N}}\pi_{\sigma}$
One can show easily that $ \Pi_{0} $ satisfies the following property:
\begin{equation}
\Pi_{0} \pi_{\sigma} = \pi_{\sigma} \Pi_{0} = \Pi_{0}
\end{equation}

Now let's write the propagator:

\begin{equation}
\begin{split}
G(I,\mathbf{\alpha}|J,\mathbf{\beta};t) &=
r_{\bm{\alpha}}^{I} r_{\bm{\beta}}^{-J}
\oint...\oint
e^{\lambda(\mathbf{z})t} \mathbf{z}^{-J}
(\Pi_{0})_{\bm{\alpha}}^{\bm{\beta}}
\mathbf{z}^{I}
\prod_{i=1}^{N} \frac{dz_{i}}{2\pi i z_{i}} \\
& = 
r_{\bm{\alpha}}^{I} r_{\bm{\beta}}^{-J}
\oint...\oint
e^{\lambda(\mathbf{z})t}
\Phi(I,\mathbf{\bm{\alpha}}|J,\mathbf{\bm{\beta}})
\prod_{i=1}^{N} \frac{dz_{i}}{2\pi i z_{i}}
\end{split}
\end{equation}

Where
\begin{equation}
\Phi(I,\mathbf{\bm{\alpha}}|J,\mathbf{\bm{\beta}}) :=
\mathbf{z}^{-J}
(\sum_{\sigma \in \mathcal{S}_{N}}\pi_{\sigma})_{\bm{\alpha}}^{\bm{\beta}}
\mathbf{z}^{I}
\end{equation}

\subsubsection*{Graphical representation of the operator $\Pi_{0}$}

It's possible to see $ (\Pi_{0})_{\bm{\alpha}}^{\bm{\beta}} $ as a partition function of a vertex model constrained by an initial configuration $ \bm{\beta} $ and a final configuration $\bm{\alpha}$. To illustrate this idea, let's first consider two arbitrary species $\color{green} \bullet$  and $\color{red} \bullet$ , with $  \gp >  \rp $ we can identify 5 non zero elements of the $\check{R}$ matrix in the base$\{ \rp \rp , \rp \gp, \gp \rp,
\gp
\gp  \}$

\begin{tikzpicture}

\begin{scope}[shift=({0,0})]

\draw[thick,dashed] (0,0) -- (2,0) -- (2,2) -- (0,2) -- (0,0);
\draw[thick,very thick] [red] (0,0) -- (1,1);

\draw[thick,very thick] [green] (1,1) -- (2,2);
\draw[thick,very thick] [green] (2,0) -- (1,1);
\draw[thick,very thick] [red] (1,1) -- (0,2);
\draw (2,1) node[anchor=west] {$ = h(x,y)$};
\draw (0,2) node[anchor=south] {$  x$};
\draw (2,2) node[anchor=south] {$  y$};
\draw (0,0) node[anchor=north] {$  y$};
\draw (2,0) node[anchor=north] {$  x$};
\end{scope}

\begin{scope}[shift=({0,4})]

\draw[thick,dashed] (0,0) -- (2,0) -- (2,2) -- (0,2) -- (0,0);
\draw[thick,very thick] [green] (0,0) -- (1,1);
\draw[thick,very thick] [red] (1,1) -- (2,2);
\draw[thick,very thick] [red] (2,0) -- (1,1);
\draw[thick,very thick] [green] (1,1) -- (0,2);
\draw (2,1) node[anchor=west] {$ = g_{\color{green} \bullet}(x,y)$};
\draw (0,2) node[anchor=south] {$  x$};
\draw (2,2) node[anchor=south] {$  y$};
\draw (0,0) node[anchor=north] {$  y$};
\draw (2,0) node[anchor=north] {$  x$};
\end{scope}

\begin{scope}[shift=({-5,0})]
\draw[thick,dashed] (0,0) -- (2,0) -- (2,2) -- (0,2) -- (0,0);
\draw[thick,very thick] [red] (0,0) -- (1,1);
\draw[thick,very thick] [red] (1,1) -- (2,2);
\draw[thick,very thick] [red] (2,0) -- (1,1);
\draw[thick,very thick] [red] (1,1) -- (0,2);
\draw (2,1) node[anchor=west] {$ = g_{\color{red} \bullet}(x,y)$};
\draw (0,2) node[anchor=south] {$  x$};
\draw (2,2) node[anchor=south] {$  y$};
\draw (0,0) node[anchor=north] {$  y$};
\draw (2,0) node[anchor=north] {$  x$};

\end{scope}

\begin{scope}[shift=({-5,4})]
\draw[thick,dashed] (0,0) -- (2,0) -- (2,2) -- (0,2) -- (0,0);
\draw[thick,very thick] [green] (0,0) -- (1,1);
\draw[thick,very thick] [green] (1,1) -- (2,2);
\draw[thick,very thick] [green] (2,0) -- (1,1);
\draw[thick,very thick] [green] (1,1) -- (0,2);
\draw (2,1) node[anchor=west] {$ = g_{\color{green} \bullet}(x,y)$};
\draw (0,2) node[anchor=south] {$  x$};
\draw (2,2) node[anchor=south] {$  y$};
\draw (0,0) node[anchor=north] {$  y$};
\draw (2,0) node[anchor=north] {$  x$};

\end{scope}

\begin{scope}[shift=({-10,0})]
\draw[thick,dashed] (0,0) -- (2,0) -- (2,2) -- (0,2) -- (0,0);
\draw[thick,very thick] [red] (0,0) -- (1,1);
\draw[thick,very thick] [red] (1,1) -- (2,2);
\draw[thick,very thick] [green] (2,0) -- (1,1);
\draw[thick,very thick] [green] (1,1) -- (0,2);
\draw (2,1) node[anchor=west] {$ = 0$};
\draw (0,2) node[anchor=south] {$  x$};
\draw (2,2) node[anchor=south] {$  y$};
\draw (0,0) node[anchor=north] {$  y$};
\draw (2,0) node[anchor=north] {$  x$};

\end{scope}

\begin{scope}[shift=({-10,4})]

\draw[thick,dashed] (0,0) -- (2,0) -- (2,2) -- (0,2) -- (0,0);
\draw[thick,very thick] [green] (0,0) -- (1,1);
\draw[thick,very thick] [green] (1,1) -- (2,2);
\draw[thick,very thick] [red] (2,0) -- (1,1);
\draw[thick,very thick] [red] (1,1) -- (0,2);
\draw (2,1) node[anchor=west] {$ = f(x,y)$};

\draw (0,2) node[anchor=south] {$x$};
\draw (2,2) node[anchor=south] {$  y$};
\draw (0,0) node[anchor=north] {$  y$};
\draw (2,0) node[anchor=north] {$  x$};

\end{scope}


\end{tikzpicture}

Where the weights are:

\begin{equation}
f(x,y) := \check{R}^{ \rp \gp }_{\color{green} \bullet \rp} =
\frac{r_{\gp} (r_{\rp} + \nu_{\gp}) (x - y)}{r_{\rp} (-1 + r_{\gp} x) (1 + \nu_{\gp} x)}
\end{equation}

\begin{equation}
g_{\color{black} \bullet}(x,y) :=
\check{R}^{ \textcolor{black}{\bullet} \textcolor{black}{\bullet} }_{\color{black} \bullet \textcolor{black}{\bullet}} =   -\frac{y r_{\color{black} \bullet }-1}{x r_{\color{black} \bullet }-1}
\end{equation}

\begin{equation}
h(x,y) =: \check{R}^{ \rp \gp }_{\color{red} \bullet \gp} = 
-\frac{y \nu_{\gp }+1}{x \nu_{\gp }+1}
\end{equation}

So if we have $m$ species, we would have $3 \tbinom{m}{2} + 2 m$ non zero vertices.

The quantity that we can want to compute is $\Phi(I,\mathbf{\bm{\alpha}}|J,\mathbf{\bm{\beta}})$ which is a sum over permutations. Each non-zero term of this sum corresponds to one or more than one possible cconnection between the initial and final order of species using the above building blocks such that there is a path of one color connecting particles of the same color. This can be best understood through examples. So we will consider two examples on three particles.

\subsubsection*{Examples on 3 particles}

Consider two red particles and one green with $  \gp >  \rp $. We would like to calculate: $\Phi(I, \gp \rp \rp |J, \rp \rp \gp )$
So we need to build diagrams that connect the initial configuration $ \rp \rp \gp $ drawn on the top of the diagram to the final configuration $ \gp \rp \rp $ drawn at the bottom of the diagram. The diagrams should be so that there is a path of green color connecting green particles and red paths connecting red particles. Besides the building blocks illustrated above, we can use vertical lines. In this example, the only permutations that can achieve this are the ones for which $\sigma(3)=1$, so there are two permutations $\tau_{12}\tau_{23}$ and $\tau_{13}$

$$\Phi(I, \gp \rp \rp |J, \rp \rp \gp) =  \mathbf{z}^{-J}(\confa + \confb )\mathbf{z}^{I} $$

Now we can compute each term:

\begin{equation}
\mathbf{z}^{-J} \confa \mathbf{z}^{I}= z_{1}^{-j_{1}} z_{2}^{-j_{2}} z_{3}^{-j_{3}}
f(z_{2},z_{3})f(z_{1},z_{3})z_{3}^{i_{1}} z_{1}^{i_{2}} z_{2}^{i_{3}}
\end{equation}

\begin{equation}
\mathbf{z}^{-J} \confb \mathbf{z}^{I}= z_{1}^{-j_{1}} z_{2}^{-j_{2}} z_{3}^{-j_{3}}
f(z_{2},z_{3})
f(z_{1},z_{3})
g_{\color{red} \bullet}(z_{1},z_{2})
z_{3}^{i_{1}} z_{2}^{i_{2}} z_{1}^{i_{3}}
\end{equation}

This example was relatively simple due to two reasons, first, the only possible permutations were the ones that conserve the colors. Second, there was a single diagram at most per permutation. However, this is not always the case. Let's consider for instance this final configuration: $ \rp \gp \rp $, then in order to calculate $ \Phi(I, \rp \gp \rp  |J, \rp \rp \gp) $, we notice that we can construct non zero diagrams not only by using permutations that conserve the colors (ie. verifying $ \sigma(3)=2 $).
Where we have two of such permutations $\tau_{23}$ and $\tau_{23}\tau_{12} = (123)$, and two others that don't conserve the colors: $\tau_{12}\tau_{23} = (321)$ and $\tau_{13}$. In addition to that, we can construct two diagrams that are associated with $\tau_{13}$:

\begin{equation}
\begin{split}
&\Phi(I, \rp \gp \rp  |J, \rp \rp \gp) = \\ &\mathbf{z}^{-J}(\underbrace{\confc}_{{(\pi_{(23)})}^{\rp \rp \gp}
	_{\rp \gp \rp}} + \underbrace{\confd}_{{(\pi_{(123)})}^{\rp \rp \gp}
	_{\rp \gp \rp}} + \underbrace{\confe}_{{(\pi_{(321)})}^{\rp \rp \gp}
	_{\rp \gp \rp}} + \underbrace{\conff+ \confg}_{{(\pi_{(13)})}^{\rp \rp \gp}
	_{\rp \gp \rp} }
\mathbf{z}^{I} 
\end{split}
\end{equation}

Each term acts on $\mathbf{z}^{I}$ according to the associated permutation, which is given by the final ordering of the $z's$

\begin{equation}
\mathbf{z}^{-J} \confc \mathbf{z}^{I}= z_{1}^{-j_{1}} z_{2}^{-j_{2}} z_{3}^{-j_{3}}
f(z_{2},z_{3})
z_{1}^{i_{1}} z_{3}^{i_{2}} z_{2}^{i_{3}}
\end{equation}

\begin{equation}
\mathbf{z}^{-J} \confd \mathbf{z}^{I}= z_{1}^{-j_{1}} z_{2}^{-j_{2}} z_{3}^{-j_{3}}
f(z_{2},z_{3})
h(z_{1},z_{3})
z_{3}^{i_{1}} z_{1}^{i_{2}} z_{2}^{i_{3}}
\end{equation}

\begin{equation}
\mathbf{z}^{-J} \confe \mathbf{z}^{I}= z_{1}^{-j_{1}} z_{2}^{-j_{2}} z_{3}^{-j_{3}}
g_{\color{red} \bullet}(z_{1},z_{2})
f(z_{1},z_{3})
z_{2}^{i_{1}} z_{3}^{i_{2}} z_{1}^{i_{3}}
\end{equation}

\begin{equation}
\mathbf{z}^{-J} \conff \mathbf{z}^{I}= z_{1}^{-j_{1}} z_{2}^{-j_{2}} z_{3}^{-j_{3}}
h(z_{2},z_{3})
g_{\color{red} \bullet}(z_{1},z_{3})
f(z_{1},z_{2})
z_{3}^{i_{1}} z_{2}^{i_{2}} z_{1}^{i_{3}}
\end{equation}

\begin{equation}
\mathbf{z}^{-J} \confg \mathbf{z}^{I}= z_{1}^{-j_{1}} z_{2}^{-j_{2}} z_{3}^{-j_{3}}
f(z_{2},z_{3})
h(z_{1},z_{3})
g_{\color{green} \bullet}(z_{1},z_{2})
z_{3}^{i_{1}} z_{2}^{i_{2}} z_{1}^{i_{3}}
\end{equation}

\subsubsection*{Exchange equations}
Note that if we define the quantity:

\begin{equation}\label{key}
M^{J}(z_{1},...,z_{n}) := \mathbf{z}^{J} \Pi_{0}
\end{equation}
Where the permutation acts on its left by its inverse.
Now we can write this exchange equation:

\begin{equation}
M^{J}(z_{1},...,z_{k},z_{k+1},...,z_{n}) \check{R}_{k}(z_{k},z_{k+1}) = 
M^{J}(z_{1},...,z_{k+1},z_{k},...,z_{n})
\end{equation}

The exchange equations in component:

\begin{equation}
M^{J}_{... ,\alpha_{k},\alpha_{k},...}(z_{1},...,z_{k},z_{k+1},...,z_{n}) \frac{1-r_{\alpha_{k}} z_{k+1}}{1-r_{\alpha_{k}} z_{k}} =
M^{J}(z_{1},...,z_{k+1},z_{k},...,z_{n})
\end{equation}

\subsubsection{Proof of the initial condition}
One needs to check the initial condition:
\begin{equation}
\delta_{\bm{\alpha}, \bm{\beta}}
\delta_{J, J}
=
r_{\bm{\alpha}}^{I} r_{\bm{\beta}}^{-J}
\oint...\oint
\mathbf{z}^{-J}
(\sum_{\sigma \in \mathcal{S}_{N}}\pi_{\sigma})_{\bm{\alpha}}^{\bm{\beta}}
\mathbf{z}^{I}
\prod_{i=1}^{N} \frac{dz_{i}}{2\pi i z_{i}}
\end{equation}
We will prove that the terms cancel one by one:
\begin{equation}
\sigma \neq e  \implies \oint_{0} \prod_{k=1}^{n}\frac{dx_{k}}{2\pi i x_{k}} \mathbf{x}^{-J} \pi_{\sigma} \mathbf{x}^{I} = 0
\end{equation}
Let $S$ be the support of $\sigma$,  $S := \{l : \sigma(l) \neq l \} $, and let's define:
$$k = \max(S)$$
$$m = \sigma^{-1}(k)$$
Obviously, since $\sigma \ne e$, $S$ is not empty, so the previous elements exist.

\begin{figure}
	\centering
	
	\begin{tikzpicture}[scale = 0.8]
	
	\draw[black,fill=black] (2,9) circle (.5ex) node[anchor=south] {};
	\draw[black,fill=black] (0,9) circle (.5ex) node[anchor=south] {$z_{2}$};
	\draw[black,fill=black] (-2,9) circle (.5ex) node[anchor=south] {$z_{1}$};
	\draw[black,fill=black] (4,9) circle (.5ex) node[anchor=south] {$z_{m}$};
	\draw[black,fill=black] (6,9) circle (.5ex) node[anchor=south] {};
	\draw[black,fill=black] (8,9) circle (.5ex) node[anchor=south] {$z_{k}$};
	
	\draw[black,fill=black] (2,5) circle (.5ex) node[anchor=north] {$
		z_{k}$};
	\draw[black,fill=black] (0,5) circle (.5ex) node[anchor=north] {};
	
	\draw[black,fill=black] (-2,5) circle (.5ex) node[anchor=north] {};
	\draw[black,fill=black] (4,5) circle (.5ex) node[anchor=north] {$z_{1}$};
	\draw[black,fill=black] (6,5) circle (.5ex) node[anchor=north] {};
	\draw[black,fill=black] (8,5) circle (.5ex) node[anchor=north] {$z_{m}$};

	\draw[very thick] [black] (6,9) -- (-2,5) ;
	\draw[very thick] [black] (0,9) -- (0,5) ;
	\draw[very thick] [black] (2,9) -- (6,5) ;
	\draw[very thick] [blue] (8,9) -- (2,5)  ;
	\draw[very thick] [black] (-2,9) -- (4,5) ;
	\draw[very thick] [orange] (4,9) -- (8,5)  ;

\end{tikzpicture}
	
\caption{A permutation in a reduced form. The blue line corresponds to a variable $z_{k}$ contributing as a polynomial to the matrix elements of $\pi_{\sigma}$. The orange line corresponds to $z_{m}$}
\label{fig:reduced}
\end{figure}
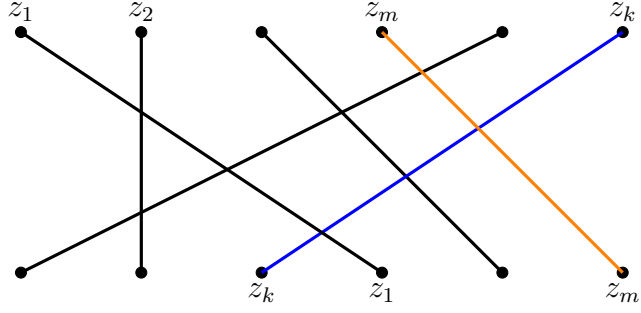

It's possible to write $\sigma$ in terms of transpositions in a reduced form so that all the non-constant elements propagate in a monotonous manner, figure \ref{fig:reduced}.
Let's on the other hand notice that all the weight functions $f, g_{\bullet}$ and $h$ are affine in their second variable and  $z_{k}$ will always be the second variable for all the weight factors of $\pi_{\sigma}$. This implies that $\pi_{\sigma}$ will be a polynomial of order $k-\sigma(k)$ in $z_{k}$. If we integrate with respect to $z_{k}$, then the integral will be necessarily zero except for $i_{\sigma(k)} - j_{k} \in \{-(k-\sigma(k)),...,-1,0 \} $. Since $m<k$, we have:
\begin{equation}
i_{\sigma(m)}- j_{m} > i_{\sigma(m)-1}- j_{m} > ... > i_{\sigma(k)}- j_{m} > i_{\sigma(k)}- j_{k}
\end{equation}
The strictness of the previous inequalities implies:
\begin{equation}
i_{\sigma(m)}- j_{m} - (i_{\sigma(k)}- j_{k}) > k-\sigma(k)
\end{equation}
So, for the situations where the integral with respect to $z_{k}$ is not zero, we can integrate with respect to $z_{m}$ and we have:

$$i_{\sigma(m)}- j_{m} > 0$$ which makes the integral vanishes for $z_{m}$ thanks to analyticity on the neighborhood of zero. 

\subsubsection{Two species}

A situation where it is relatively easy to bring the calculations to the end is for an initial condition of a single second-class particle in front of a finite number of first-class particles, and a final configuration where all the first-class particles have jumped over the second-class particle. This is a generalization of the first example given in the previous section. Let $J =\{j_{1}< ...< j_{n}<j\}$ the initial positions of particles, with $j$ being the one of the second class. Similarly, $I = \{i < i_{1} <... <i_{n}\}$ are the final positions. We need first to calculate:

\begin{equation}
\Phi(I, \gp \rp... \rp |J, \rp... \rp \gp) =
z^{i-j}
\prod_{i=1}^{n} f(z_{i},z) 
\sum_{\sigma \in \mathcal{S}_{n}} \epsilon(\sigma)
\prod_{k=1}^{n}
(1-z_{\sigma(k)} r_{\color{red} \bullet })^{\sigma(k)-k} z_{\sigma(k)}^{i_{k}-j_{\sigma(k)}}
\end{equation}

$$ =
\Big(\frac{r_{\gp} (r_{\rp} + \nu_{\gp})}{r_{\rp}}\Big)^{n}
z^{i-j}
\sum_{\sigma \in \mathcal{S}_{n}}
\epsilon(\sigma)
\prod_{k=1}^{n}
\frac{(1-z_{\sigma(k)} r_{\color{red} \bullet })^{\sigma(k)-k} z_{\sigma(k)}^{i_{k}-j_{\sigma(k)}}
	(z_{k} - z)}
{(-1 + r_{\gp} z_{k}) (1 + \nu_{\gp} z_{k})}
$$
$$ =
\Big(\frac{r_{\gp} (r_{\rp} + \nu_{\gp})}{r_{\rp}}\Big)^{n}
z^{i-j}
det\Big[
\frac{(1-z_{k} r_{\color{red} \bullet })^{h-k} z_{k}^{i_{k}-j_{h}}
	(z_{k} - z)}
{(-1 + r_{\gp} z_{k}) (1 + \nu_{\gp} z_{k})}\Big]_{hk}
$$

Now we can calculate the conditional probability: 

\begin{equation}
G(I, \gp \rp... \rp |J, \rp... \rp \gp) =
r_{\bm{\alpha}}^{I} r_{\bm{\beta}}^{-J}
\oint...\oint
e^{\lambda(\mathbf{z})t}
\Phi(I, \gp \rp... \rp |J, \rp... \rp \gp)
\frac{dz}{2\pi i z}\prod_{i=1}^{n} \frac{dz_{i}}{2\pi i z_{i}}
\end{equation}
Where:
\begin{equation}
\lambda(\mathbf{z}) = \sum_{k=1}^{n} z_{k}^{-1} - n r_{\rp} + z^{-1} - r_{\gp}
\end{equation}

\begin{equation}
G(I, \gp \rp... \rp |J, \rp... \rp \gp) = 
\Big(\frac{r_{\gp} (r_{\rp} + \nu_{\gp})}{r_{\rp}}\Big)
\oint e^{\frac{t}{z} - r_{\gp}t}
(r_{\gp}z)^{i-j}
det[
m_{k,h}(z,t)
]_{hk}
\frac{dz}{2\pi i z}
\end{equation}

Where:

\begin{equation}
m_{k,h}(z,t) := e^{-r_{\rp}t}	\oint \frac{e^{\frac{t}{x}} (1-x r_{\rp})^{h-k} (r_{\rp}x)^{i_{k}-j_{h}}
	(x - z)}
{(-1 + r_{\gp} x) (1 + \nu_{\gp} x)}  \frac{dx}{2\pi i x}
\end{equation}

\subsubsection{}
We would like to find probability that the second particle at time $t$ has been over jumped by the other particles regardless of their positions:
\begin{equation}
G(i, \gp \rp... \rp |J, \rp... \rp \gp) = \sum_{i_{n} = i+n}^{\infty}...\sum_{i_{2} = i+2}^{i_{3}-1}\sum_{i_{1} = i+1}^{i_{2}-1} G(I, \gp \rp... \rp |J, \rp... \rp \gp)
\end{equation}
Performing elementary operations on the determinant:

\begin{equation}
G(i, \gp \rp... \rp |J, \rp... \rp \gp) =
\oint e^{\frac{t}{z} - r_{\gp}t}
(r_{\gp}z)^{i-j}
det[
h_{k,h}(z,t)
]_{hk}
\frac{dz}{2\pi i z}
\end{equation}

With

$$
h_{k,h}(z,t) :=
\Big(\frac{r_{\gp} (r_{\rp} + \nu_{\gp})}{r_{\rp}}\Big)
e^{-r_{\rp}t}	\oint \frac
{e^{\frac{t}{x}} (1-x r_{\rp})^{h-k-1}
	(r_{\rp}x)^{i+k-j_{h}}
	(x - z)}
{(-1 + r_{\gp} x) (1 + \nu_{\gp} x)}  \frac{dx}{2\pi i x}
$$
\subsubsection{Escaping probability}
In the case $n=1$

\begin{equation}
G(i, \gp \rp |J, \rp \gp; t) =
\Big(\frac{r_{\gp} (r_{\rp} + \nu_{\gp})}{r_{\rp}}\Big)
e^{-(r_{\rp}+r_{\gp})t}
\oint
\oint
e^{(\frac{1}{x}+ \frac{1}{z})t}
\frac
{
	(r_{\gp}z)^{i-j}
	(r_{\rp}x)^{i+1-j_{1}}(x - z)}
{(1-x r_{\rp})(-1 + r_{\gp} x) (1 + \nu_{\gp} x)}  \frac{dx}{2\pi i x}
\frac{dz}{2\pi i z}
\end{equation}

One can show that if $i < j_{1}$ then the integral will be zero with respect to the $z$ variable. We want to sum over $i \geq j_{1}$:

\begin{equation}
\begin{split}
& G( \gp \rp |J, \rp \gp; t) = \\
&\Big(\frac{r_{\gp} (r_{\rp} + \nu_{\gp})}{r_{\rp}}\Big)
e^{-(r_{\rp}+r_{\gp})t}
\oint_{0}
\oint_{0}
e^{(\frac{1}{x}+ \frac{1}{z})t}
\frac
{
	(r_{\gp}z)^{j_{1}-j}
	(r_{\rp}x)(x - z)}
{(1-r_{\gp}r_{\rp}x z)
	(1-x r_{\rp})(-1 + r_{\gp} x) (1 + \nu_{\gp} x)}  \frac{dx}{2\pi i x}
\frac{dz}{2\pi i z}
\end{split}
\end{equation}
We can integrate over $z$. Since zero is an essential pole, and since $j_{1}-j < 0$, there is no pole at infinity (imagine the function defined on the Riemann sphere), so we can take into account the residue of the pole outside the integral path, i.e. the pole at $z = (r_{\gp}r_{\rp}x)^{-1}  $

\begin{equation}
G( \gp \rp |J, \rp \gp; t) =
\Big(\frac{r_{\gp} (r_{\rp} + \nu_{\gp})}{r_{\rp}}\Big)
\oint
e^{\frac{(1-r_{\gp} x)(1-r_{\rp} x)}{x} t}
\frac
{
	(r_{\rp}x)^{j-j_{1}+1} ( x - (r_{\gp}r_{\rp}x)^{-1} )}
{(1-x r_{\rp})(-1 + r_{\gp} x) (1 + \nu_{\gp} x)}  \frac{dx}{2\pi i x}
\end{equation}

For large $t$ we can perform a saddle point analysis, by writing the previous integral as 
\begin{equation}
\oint_{0} f(z) e^{g(z)t} dz
\end{equation}
with:
\begin{equation}
f(z) =
\frac
{r_{\gp} (r_{\rp} + \nu_{\gp})
(r_{\rp}x)^{j-j_{1}+1} ( x - (r_{\gp}r_{\rp}x)^{-1} )}
{
	r_{\rp}(1-x r_{\rp})(-1 + r_{\gp} x) (1 + \nu_{\gp} x)
} 
\end{equation}
and
\begin{equation}
g(z) = \frac{(1-r_{\gp} x)(1-r_{\rp} x)}{x}
\end{equation}
The saddle point $z_{c}$ is a verifying $g'(z_{c}) = 0$. We have two of them:
\begin{equation}
	z_{c}^{\pm} = \pm \frac{1}{\sqrt{r_{\rp}r_{\gp}}}
\end{equation}
The saddle point method is based on deforming the integral path so that it passes by the saddle point and such that the new path satisfies $Im(g(z))$ is constant. This constant is obviously the imaginary part of the saddle point.
In our case we can notice that circle centered at zero and with radios $ \frac{1}{\sqrt{r_{\rp}r_{\gp}}}$ has a zero imaginary part for $g(z)$. We cannot deform the path integral into that circle without getting one of the poles: $\frac{1}{r_{\gp}}$ or $ \frac{1}{r_{\rp}}$ depending on the relative values of $r_{\gp}$ and $r_{\rp}$. Since $f(z_{c}^{\pm}) = 0$, the contribution of the saddle point will be zero and the contribution of the pole inside will be dominant:

\begin{itemize}
	\item if $r_{\gp} > r_{\rp}$, then the pole $\frac{1}{r_{\gp}}$ is inside the circle, and the we have:

	\begin{equation}
G( \gp \rp |J, \rp \gp; t) =
\frac{ r_{\rp} + \nu_{\gp}}{r_{\gp} + \nu_{\gp}}
\big(\frac{r_{\rp}}{r_{\gp}}\big)^{j-j_{1}+1}
	\end{equation}

	\item if $r_{\rp} > r_{\gp} $, then the pole $\frac{1}{r_{\rp}}$ is inside the circle, and we get
		\begin{equation}
	G( \gp \rp |J, \rp \gp; t) = 1
	\end{equation}
\end{itemize}

This gives the same results for the same asymptotic escaping probability as the one obtained by an elementary method in chapter 5.

\section{Exclusion process on the ring}
\label{Exclusion process on the ring}
1D lattice models with periodic boundary conditions have a long tradition in mathematical physics. Such boundaries provide a mathematical simplicity besides their physical relevance. In the thermodynamic limit, they share properties with systems defined on the line. In our case, we are using results obtained on the ring in the steady state for a dynamic system on the line. In particular, in chapter 5, we will need the expression of the speed of a defect as a function of the density field, this was obtained via a Matrix Product Ansatz in \cite{mallick1996shocks} and independently using the coordinate Bethe Ansatz in \cite{derrida1999bethe}. We will be reviewing the latter in section 2.1. On the other hand, in chapter 2 we will be using the expression of the currents of different species as a function of the densities, these expressions were again obtained in the thermodynamic limit for a system on a ring using the Nested algebraic Bethe Ansatz \cite{cantini2008algebraic}. Section 2.2 will be mainly devoured for reviewing this. The choice of reviewing these two works in this order serves as well a pedagogical purpose. They provide together a simple setting for explaining the Bethe Ansatz on the Ring.

\subsection{Coordinate Bethe Ansatz for a defect in the ring}

Consider a lattice of $L$ sites with periodic boundary condition (the site $L+1$ is identified with the site $1$), with $M-1$ first class particles and a single defect, i.e.  a second-class particle with arbitrary rates:
\begin{equation}
\begin{split}
\alpha \quad 20 \rightarrow 02
\\
\beta \quad 12 \rightarrow 21
\end{split}
\end{equation}
Let $Y_{t}$ be the distance traveled by the defect up to time $t$, i.e. the number of forward jumps minus the number of backward jumps. Our objective is to determine the statistical properties of this random variable in the steady state (for large $t$).
\subsubsection{Derrida-Lebowitz trick.} 
An idea that is useful whenever we have a Markov process and we get intersected in the statistical properties of a sub process is to introduce a counter for this sub-process, i.e. a random variable that represents the number of times this sub process occur to to a time $t$. It's $Y_{t}$ in our case, and then to try to write down a time evolution equation for its generating function. This was first used in \cite{derrida1998exact}, \cite{derrida1999universal}. Let's see how does this work in our case. Let $P_{t}(C,Y)$ be the probability for the system to be at the configuration $C$ and having $Y_{t} = Y$. We can classify the transitions among the different configurations into three types:
\begin{itemize}
	\item A transition $C^{'} \rightarrow C $ that increases $Y_{t}$ by one. Denote its rate by $M_{1}(C,C^{'})$
	\item A transition $C^{'} \rightarrow C $ that decreases $Y_{t}$ by one. Denote its rate by $M_{-1}(C,C^{'})$
	\item A transition $C^{'} \rightarrow C $ that leaves $Y_{t}$ unchanged. Denote its rate by $M_{0}(C,C^{'})$
\end{itemize}

Now we can write the evolution equation for $P_{t}(C,Y)$, which is a generalized master equation: 

\begin{equation}\label{gen master}
\frac{dP_{t}(C,Y)}{dt} = \sum_{C^{'}} M_{0}(C,C^{'})P_{t}(C^{'},Y) + M_{1}(C,C^{'})P_{t}(C^{'},Y-1) + M_{-1}(C,C^{'})P_{t}(C^{'},Y+1)
\end{equation}
Note that $M_{1}(C^{'},C^{'}) = M_{-1}(C,C) = 0$ and $ M_{0}(C,C) = \sum_{C}( M_{1}(C,C^{'}) + M_{-1}(C,C^{'})) $.

The conditional generating function for $Y_{t}$ is:

\begin{equation}
F_{t}^{\nu}(C) := E(e^{\nu Y_{t}} | C ) = \sum_{Y} e^{\nu Y} P_{t}(C,Y)
\end{equation}

Deriving with respect to time and using eq. \ref{gen master} ,then exchanging the sums allows to find out that it obeys the evolution equation:

\begin{equation}
\frac{dF^{\nu}_{t}(C)}{dt} := \sum_{C^{'}} (M_{0}(C,C^{'})+ e^{\nu} M_{1}(C,C^{'}) + e^{-\nu} M_{-1}(C,C^{'})) F^{\nu}_{t}(C^{'})
\end{equation}\label{evo gen fun}

This suggest to define the Matrix:
\begin{equation}
M^{\nu} = M_{0} + e^{\nu}M_{1} + e^{-\nu}M_{-1}
\end{equation}

So that we can write the previous equation in a compact form:
\begin{equation}
\frac{d F^{\nu}_{t}}{dt} = M^{\nu} F^{\nu}_{t}
\end{equation}

The full generating function is the sum over the components of $F^{\nu}_{t}$, and can be written as a linear combination of exponential functions:

\begin{equation}
\mathcal{F}_{t}^{\nu} = \sum_{C} F_{t}^{\nu}(C) = \sum_{i} \alpha_{i} e^{\lambda_{i}t}
\end{equation}
Where the $\lambda_{i}$ are eigenvalues of the matrix $M^{\nu}$.

The Matrix $M^{\nu} + Id$ has positive entries so it has a non-degenerate real eigenvalue greater than the module of all other eigenvalues according to Perron-Frobinus theorem. Same holds for $M^{\nu}$. Let $\lambda(\nu)$ be this largest eigenvalue for $M^{\nu}$. The generating function behaves for large time as:

\begin{equation}
F_{t} \sim e^{\lambda(\nu) t}
\end{equation}

The objective is to find the speed of the defect for large times:

\begin{equation}
\frac{1}{t} E(Y_{t}) = \frac{1}{t} \frac{d F_{t}}{d \nu}|_{\nu = 0} \sim e^{\lambda(0) t} \frac{d \lambda}{d \nu} 
\end{equation}
Of course the largest eigenvalue for $M$ is zero thanks to its stochasticity, so we have $\lambda(0)=0$
\begin{equation}
\frac{1}{t} E(Y_{t}) = \frac{d \lambda}{d \nu}|_{\nu = 0} = \lim_{\nu \rightarrow 0} \frac{\lambda(\nu)}{\nu}
\end{equation}
To diagonalise $M^{\nu}$ and find its largest eigenvalue, Once can use the Bethe Ansatz

\subsubsection{Bethe Ansatz}
Let's label the configurations by the positions of the particles with the convention: $ x_{1} < x_{2} < ..<x_{N}+L$ Where $1 \leq x_{1} \leq L$ is the position of the defect. Let $\psi(x_{1}, ... x_{M})$ be an eigenvector of $M^{\nu}$ with an eigenvalue $\lambda(\nu)$ (for the moment, it can be any eigenvalue),
The master equation gives rise to a functional equation verified by $\psi$
This equation is simple for a configuration where the particles are not neighbors, i.e: $x_{i+1}-x_{i}\geq2$:
\begin{equation}\label{MasterE}
\lambda \psi(x_{1}, ... x_{M}) = \sum_{i=2}^{N} \psi(x_{1},...,x_{i}-1,..., x_{M}) + e^{\nu} \alpha \psi(x_{1}-1, ... x_{M}) - (N-1 + \alpha) \psi(x_{1}, ... x_{M})
\end{equation}
If we want this equation to be valid for all possible positions, we can do that by tolerating that the function $\psi$ assigns values to non-physical configurations, precisely configurations where two neighboring particles have the same positions. These assigned values need to be chosen so that eq. \ref{MasterE} reduces to the correct form for configurations with neighboring particles. One can find this way the following boundary conditions:
\begin{equation}
\psi(x_{1}, ...,x_{i},x_{i}..., x_{M}) = \psi(x_{1}, ...,x_{i},x_{i}+1..., x_{M}) \quad for \quad 1 < i < M
\end{equation}
\begin{equation}
\psi(x_{1},x_{1},..., x_{M}) =  e^{-\nu} \beta \psi(x_{1}+1,x_{3},...,x_{M}, x_{1}+L) + \alpha \psi(x_{1},x_{1}+1,..., x_{M})
\end{equation}
\begin{equation}
e^{\nu} \alpha \psi(x_{1}-1,..., x_{1}+L-1) = (1 - \beta) \psi(x_{1},..., x_{1}+L-1) 
\end{equation}
If we plug a plane wave of the form $(z_{1} e^{\nu} \alpha)^{x_{1}} \prod_{i=2}^{M} z_{i}^{x_{i}}$ in \ref{MasterE} we get a solution whenever:
\begin{equation}
\lambda = \sum_{i=1}^{M} \frac{1}{z_{i}} - (N-1+\alpha)
\end{equation}
However this solution will not verify the boundary condition. We use the Bethe Ansatz for a more general form of the solutions:
\begin{equation}
\psi(x_{1},..., x_{M}) =
(e^{\nu} \alpha)^{x_{1}}
\sum_{\sigma \in S_{M}} A_{\sigma}
\prod_{i=1}^{M}
z_{\sigma(i)}^{x_{i}}
\end{equation}
The three boundary conditions can be satisfied by imposing respectively the following constraints on the coefficients:
\begin{equation}
A_{\sigma}(1-z_{\sigma(j+1)}) = - A_{\sigma \tau_{j, j+1}}(1-z_{\sigma(j)}) \quad for  \quad 1<j<M
\end{equation}
\begin{equation}
A_{\sigma}(1- \alpha z_{\sigma(2)})
+ A_{\sigma \tau_{2,3}...\tau_{M-1,M}} (\alpha \beta z_{\sigma(1)} z_{\sigma(2)}^{L})
=
- A_{\sigma \tau_{1,2}}(1-\alpha z_{\sigma(1)}) 
- A_{\sigma \tau_{2,3}...\tau_{M-1,M} \tau_{1,M}} (\alpha \beta z_{\sigma(2)} z _{\sigma(1)}^{L})
\end{equation}

\begin{equation}
A_{\sigma} z_{\sigma(M)}^{L} [(1-\beta) z_{\sigma(1)} - 1]
=
- A_{\sigma \tau_{1,M}} z_{\sigma(1)}^{L} [(1-\beta) z_{\sigma(M)} - 1]
\end{equation}
Applying the third constraint on the second:
\begin{equation}
A_{\sigma}(1-\alpha z_{\sigma(2)})
=
- A_{\sigma \tau_{1,2}}(1- \alpha z_{\sigma(1)}) 
-
\alpha \beta
A_{\sigma \tau_{2,3}...\tau_{M-1,M}}
( z_{\sigma(1)} z_{\sigma(2)}^{L}
-  z_{\sigma(2)} z_{\sigma(2)}^{L}\frac{b z_{\sigma_{1}} - 1}{b z_{\sigma_{2}} - 1} )
\end{equation}
With $b = 1-\beta$.
Applying successively the first constraint to the third, one can write a relation between  $A_{\sigma }$ and $A_{\sigma \tau(2,3)... \tau(M-1,M)}$:
\begin{equation}
A_{\sigma \tau_{2,3}... \tau_{M-1,M}}
=A_{\sigma }
\prod_{k=1}^{M-2}
\frac{z_{\sigma(M-k)} - 1}{z_{\sigma(2)}-1}
=A_{\sigma }
\frac{\prod_{k=1}^{M}(z_{k} - 1)}{(z_{\sigma(2)}-1)^{M-1} (z_{\sigma(1)}-1)}
\end{equation}
Applying this to the second constraint:
\begin{equation}
A_{\sigma \tau_{1,2}}(1-\alpha z_{\sigma(1)}) 
=
-A_{\sigma}(1- \alpha z_{\sigma(2)})
-\alpha \beta
A_{\sigma \tau_{2,3}...\tau_{M-1,M}}
z_{\sigma(2)}^{L} ( z_{\sigma(1)}
-  z_{\sigma(2)}\frac{b z_{\sigma_{1}} - 1}{b z_{\sigma_{2}} - 1} )
\end{equation}
\begin{equation}
A_{\sigma \tau_{1,2}}(1-\alpha z_{\sigma(1)}) 
=
-A_{\sigma}((1- \alpha z_{\sigma(2)})
+\alpha \beta
\frac{z_{\sigma(2)}^{L} \prod_{k=1}^{M}(z_{k} - 1)}{(z_{\sigma(2)}-1)^{M-1} (z_{\sigma(1)}-1)}
( z_{\sigma(1)}
-  z_{\sigma(2)}\frac{b z_{\sigma_{1}} - 1}{b z_{\sigma_{2}} - 1} ))
\end{equation}
\begin{equation}
A_{\sigma \tau_{1,2}}(1-\alpha z_{\sigma(1)}) 
=
-A_{\sigma}\bigg[(1- \alpha z_{\sigma(2)})
+\alpha \beta
\frac{
	(z_{\sigma(2)} - z_{\sigma(1)})
	z_{\sigma(2)}^{L} \prod_{k=1}^{M}(z_{k} - 1)}{(z_{\sigma(2)}-1)^{M-1} (z_{\sigma(1)}-1)
	(b z_{\sigma(2)} - 1)
}\bigg]
\end{equation}
\subsubsection*{Bethe Equations}
Applying the last equation twice leads to the following constraints on the moments $z's$:

\begin{equation}
\bigg [ \frac{(\alpha z_{i} - 1) (b z_{i} - 1) (z_{i} - 1)^{M-1}}{
	- \alpha \beta \prod_{k=1}^{M}(1-z_{k})
	z_{i}^{L}} +1 \bigg]\frac{1}{1-z_{i}} 
=
\bigg [ \frac{(\alpha z_{j} - 1) (b z_{j} - 1) (z_{j} - 1)^{M-1}}{  - \alpha \beta \prod_{k=1}^{M}(1-z_{k}) z_{j}^{L}
} + 1 \bigg]\frac{1}{1-z_{j}} 
\end{equation}
In words, the left side of the equality does not depend on $i$, so it is constant:

\begin{equation}\label{BE1}
E =
\bigg [ \frac{(\alpha z_{i} - 1) (b z_{i} - 1) (z_{i} - 1)^{M-1}}
{ C
	z_{i}^{L}}
+1
\bigg]\frac{1}{1-z_{i}} 
\end{equation}
Where $C$ is:
\begin{equation}\label{BE2}
C = - \alpha \beta \prod_{k=1}^{M}(1-z_{k})
\end{equation}
Finally the wave function $\psi(x_{1},...,x_{M})$ has to be invariant under translation, which leads to:
\begin{equation}\label{BE3}
e^{\nu} \alpha \prod_{k=1}^{M} z_{k} = 1
\end{equation}
\subsubsection*{Analysis of Bethe equations}
We will go sketchy in this paragraph as the more technical details can be found in \cite{derrida1999bethe}.The objective is to compute $\lambda(\nu)$.
The last Bethe equation eq. \ref{BE3} can be written as:
\begin{equation}
\nu = -\ln(\alpha) - \sum_{k=1}^{M}
\ln(z_{k})
\end{equation}
Both $\lambda$ and $\nu$ depend on the $z$ variables through quantities of the form:
\begin{equation}
\sum_{k=1}^{M} h(z_{k})
\end{equation}
Where $h(z) = \frac{1}{z}$ for $\lambda$ and $h(z) = \ln(z)$ for $\nu$.
It's possible to find a general formula for this quantity:
\begin{equation}
\sum_{k=1}^{M} h(z_{k}) = (M-1)h(1) + h(\frac{1}{\alpha} ) + \sum_{n=1}^{\infty} \frac{C^{n}}{n}
\bigg[ \oint_{1} + \oint_{\frac{1}{\alpha}} \bigg]
\frac{dz}{2 \pi i} h^{'}(z) [Q(z)]^{n}
\end{equation}

Where:

\begin{equation}
Q(z) = \frac{-z^{L} (1+(z-1)E)}{(bz-1)(\alpha z - 1)(z-1)^{M-1}}
\end{equation}

This leads to the following expressions for $\lambda$ and $\nu$

\begin{equation}
\lambda(\nu) = -
\sum_{n=1}^{\infty} \frac{C^{n}}{n}
\bigg[ \oint_{1} + \oint_{\frac{1}{\alpha}} \bigg]
\frac{dz}{2 \pi i} \frac{1}{z^{2}} [Q(z)]^{n}
\end{equation}

\begin{equation}
\nu = -
\sum_{n=1}^{\infty} \frac{C^{n}}{n}
\bigg[ \oint_{1} + \oint_{\frac{1}{\alpha}} \bigg]
\frac{dz}{2 \pi i} \frac{1}{z} [Q(z)]^{n}
\end{equation}

Beside this, it is possible to show that:

\begin{equation}
0=
\sum_{n=1}^{\infty} \frac{C^{n}}{n}
\bigg[ \oint_{1} + \oint_{\frac{1}{\alpha}} \bigg]
\frac{dz}{2 \pi i} \frac{1}{z-1} [Q(z)]^{n}
\end{equation}

\subsubsection*{Asymptotics for the speed}

We first notice that the limit $\nu \rightarrow 0$ corresponds to $z_{i} \rightarrow 0$ for $2 \leq i \leq M$ 	and $z_{1} \rightarrow \frac{1}{\alpha}$. In this limit, we have as well $C \rightarrow 0$. Let $Q^{(0)} = \lim_{\nu \rightarrow 0} Q$. The speed can be written using only the first terms in the expressions of $\nu$ and $\lambda$:

\begin{equation}
v = \lim_{\nu \rightarrow 0} \frac{\lambda}{\nu} = 
\frac{\big[ \oint_{1} + \oint_{\frac{1}{\alpha}} \big]
	\frac{dz}{2 \pi i} \frac{1}{z^{2}} Q^{(0)}(z)}
{\big[ \oint_{1} + \oint_{\frac{1}{\alpha}} \big]
	\frac{dz}{2 \pi i} \frac{1}{z}Q^{(0)}(z)}
\end{equation}

\begin{equation}
0=
\bigg[ \oint_{1} + \oint_{\frac{1}{\alpha}} \bigg]
\frac{dz}{2 \pi i} \frac{1}{z-1}
Q^{(0)}(z)
\end{equation}

\begin{equation}
Q^{(0)}(z) = \frac{-z^{L} (1+(z-1)E^{(0)})}{(bz-1)(\alpha z - 1)(z-1)^{M-1}}
\end{equation}

From the last two equations:

\begin{equation}
0=
\bigg[ \oint_{1} + \oint_{\frac{1}{\alpha}} \bigg]
\frac{dz}{2 \pi i} \frac{1}{z-1}
\frac{-z^{L} (1+(z-1)E^{(0)})}{(bz-1)(\alpha z - 1)(z-1)^{M-1}}
\end{equation}

So if we define:
\begin{equation}\label{XLM}
X_{L,M} := \bigg[ \oint_{1} + \oint_{\frac{1}{\alpha}} \bigg]
\frac{dz}{2 \pi i}
\frac{z^{L}}{(bz-1)(\alpha z - 1)(z-1)^{M}}
\end{equation}
We get:
\begin{equation}
X_{L,M} + X_{L,M-1} E^{(0)} = 0
\end{equation}
And the speed will be:
\begin{equation}
v = \frac{X_{L,M-1} X_{L-2,M-1} - X_{L-2,M-2} X_{L,M} }
{X_{L,M-1} X_{L-1,M-1} - X_{L-1,M-2} X_{L,M} }
\end{equation}

\subsubsection*{Hydrodynamic limit}
Assume $\alpha \neq 1$. We need to find the limit of $X_{L,M}$ as  $L,M \rightarrow \infty $ with the ratio $\frac{M}{L} = \rho$ fixed. To find the asymptotic of the integral, one needs to use the method of saddle point. Substituting $M= \rho L$ We can write the integral of the form $\oint_{\gamma} f(z)(g(z))^{L}dz$, and we are interested in the limit $L \rightarrow \infty$. The method is based on deforming $\gamma$, if possible, so that the phase of $g(z)$ is fixed and that it passes by the saddle point $z_{c}$, which is a point that verifies $g^{'}(z_{c}) = 0 $, assume there is only one for simplicity. The phase being constant, the saddle point method applies in a similar fashion as in the real case. At the saddle point the gradient of $Re(g)$ is perpendicular to the gradient of $Im(g)$, hence the name of the saddle point.

Note first that $\frac{1}{\alpha}$ is a simple pole, so the integral around it is:

\begin{equation}
\oint_{\frac{1}{\alpha}}
\frac{dz}{2 \pi i}
\frac{z^{L}}{(bz-1)(\alpha z - 1)(z-1)^{M}}
=
\frac{1}{(b-\alpha)}
(\frac{1}{(1-\alpha)^{\rho} \alpha^{(1-\rho)}})^{L}
\end{equation}
And same for the pole $\frac{1}{b}$
\begin{equation}
\oint_{\frac{1}{b}}
\frac{dz}{2 \pi i}
\frac{z^{L}}{(bz-1)(\alpha z - 1)(z-1)^{M}}
=
\frac{1}{(\alpha-b)}
(\frac{1}{(1-b)^{\rho} b^{(1-\rho)}})^{L}
\end{equation}

The saddle point is $z_{s} = \frac{1}{1-\rho}$

The contribution from the saddle point is:

\begin{equation}
\frac{(1-\rho)^{2}}{(\rho-1+b)(\rho-1+\alpha)} (\frac{1}{\rho^{\rho}(1-\rho)^{1-\rho}})^{L}
\end{equation}

Now the evaluation of the integral \ref{XLM} will depend on the relative position of the saddle point and the poles. A comparison between the different configurations yields the known results, let's discuss one of them:

 $\frac{1}{\alpha} < z_{c} < \frac{1}{b}$ is equivalent to $1-\alpha < \rho < \beta$
then the contour around the pole $\frac{1}{\alpha} $ and the saddle point can be merged into one. A small computation shows that the contribution of the saddle point will dominate the one from the pole. And the speed will be can be computed

\begin{equation} v=1-2\rho
\end{equation}

Analyzing the rest of cases lead to the different regimes for the speed:

\begin{equation}
For \; \; \beta < \rho \; and \; 1-\alpha < \rho \quad \quad v = 1-\rho -\beta 
\end{equation}
\begin{equation}
For \; \; \beta > \rho \; and \; 1-\alpha > \rho \quad \quad v = \alpha -\rho 
\end{equation}
\begin{equation}
For \; \; \beta < \rho < 1- \alpha \quad \quad v= \alpha - \beta
\end{equation}

\subsection{Algebraic Bethe Ansatz for The Exclusion Process on the ring}

The basic idea for solving a quantum system is to find operators that commute. Hopefully, one has sufficiently many so that the common eigenspaces are all uni-dimensional. For systems defined on a 1D lattice with local interactions, a formalism known as the Algebraic Bethe Ansatz (ABA) provides a procedure for generating such commuting operators besides finding the common eigenvectors and the corresponding eigenvalues. However, for this mechanism to function, some implicit conditions on the local interactions have to be met, if so, we speak of an integrable system in the sense of Yang-Baxter. Although this was originally developed for quantum systems, it can be used for systems sharing similar mathematical structures. In our case, our modal is not Hamiltonian but stochastic, so we have a Markov matrix that replaces the Hamiltonian, and a master equation that replaces Schrodinger's equation.
This part is a review of some selected known literature treating the exclusion process on the ring. The objective is to reach the expression of the currents used in chapter 2. In section 2.2.1 we describe how ABA work for ASEP with one single species to obtain the statistical properties of the current. The objective of this is two-fold: first to provide a rather simple setting for explaining the procedure of the ABA. Secondly, to use the results obtained here for the next section.  Originally the problem was treated with the coordinate Bethe Ansatz in \cite{derrida1998exact}, where the deviation function of the current was obtained. For our needs, we will stop at the Bethe equations. This presentation can be seen as a detailed version of appendix A of \cite{cantini2008algebraic}

In section 2.2.2, we treat the case of an arbitrary number of defects (multi-species TASEP). Although the coordinate Bethe Ansatz dealt successfully with a system with a single defect, using it for an arbitrary number of defects would be quite cumbersome. ABA is a more elegant and, in a sense, efficient technique for this case. this has been done in \cite{cantini2008algebraic}, where the nested ABA was used to diagonalize the deformed Markov matrix so to provide for analytical expressions for the currents. we mainly here review this with more details.

Numerous introductory monologues for ABA exist. For a very short, very elementary one \cite{nepomechie1998spin}. For a detailed course \cite{slavnov2018algebraic}. We will be using here the nested ABA. This version is needed whenever the local Hilbert space has a dimension higher than 2.  For an introduction to the nested ABA \cite{babelon2007short, babelon2003introduction, slavnov2020introduction}. Finally, \cite{golinelli2006asymmetric} provides a compact elegant review for ABA applied to the exclusion process.

\subsubsection{ABA for ASEP with one species}
Consider ASEP with $m$ particles on a ring with $N$ sites. Each particle can hop forward with a rate $p$ and backward with a rate $q$. The state of the system is described by a vector in the space $\mathcal{H} = (\mathbb{C}^{2})^{\otimes N}$, where each base element corresponds to a configuration of the system and is composed of an $N$ tensor product of elements from the local base $\{ \bra{0}, \bra{1} \}$. Of course, the particles conservation will make only a subspace of $\mathcal{H}$ accessible.
The markov matrix of the system can be written as a sum of local operators, each is acting non trivially only on two neighboring sites:
\begin{equation}
M = \sum_{i=1}^{N} 
M_{i,i+1}
\end{equation}

Where the site $N+1$ identified with the site $1$, and $M_{i,i+1}$ is given by:

\begin{equation}
M_{i,i+1} =
\mathbf{1}_{1} \otimes...\otimes\mathbf{1}_{i-1}
\otimes
\begin{pmatrix}
0 & 0 & 0 & 0 \\
0 & -p & q & 0 \\
0 & p & -q & 0 \\
0 & 0 & 0 & 0
\end{pmatrix}
\otimes
\mathbf{1}_{i+2} \otimes...\otimes\mathbf{1}_{N}
\end{equation}
It's quite known that this local operator can be written in terms of Pauli matrices so that the Markov matrix can be seen as a non-Hermitian spin chain:

\begin{equation}
M = \sum_{i=1}^{N} \big(p S_{i}^{-}S_{i+1}^{+} + qS_{i}^{+}S_{i+1}^{-} + \frac{1}{4} S_{i}^{z}S_{i+1}^{z}  \big) - \frac{N}{4}
\end{equation}
This can be mapped to an XXZ quantum spin chain with twisted boundary conditions \cite{essler1996representations}. and explains the relevance of Bethe ansatz for diagonalizing the Markov operator, which was famously used to find the spectral gap of the model \cite{gwa1992bethe}. However, if we are interested in the statistical properties of the current then a generalized master equation with a deformed Markov matrix is required similarly to the previous section. Let $Y_{t}$ be the random variable counting the number of forward jumps of all particles minus the number of their backward jumps up to time $t$. The probability $P_{t}(C,Y)$ of the system being at configuration $C$ and having $Y_{t}=Y$ verifies an evolution equation that has the same form as eq. \ref{gen master} in the previous section, and it results in a generating function for $Y_{t}$ that has the form of eq. \ref{evo gen fun}. So its behavior is determined by the knowledge of the largest eigenvalue of the deformed Matrix:

\begin{equation}
M^{\nu} = M_{0}+ e^{\nu} M_{1} + e^{-\nu} M_{-1}
\end{equation} 
This matrix can be written as a sum of local operators:
\begin{equation}
M^{\nu} = \sum_{i=1}^{N} 
M^{\nu}_{i,i+1}
\end{equation}

Where:

\begin{equation}
M^{\nu}_{i,i+1} =
\mathbf{1}_{1} \otimes...\otimes\mathbf{1}_{i-1}
\otimes
\begin{pmatrix}
0 & 0 & 0 & 0 \\
0 & -p & q e^{-\nu} & 0 \\
0 & p e^{\nu} & -q & 0 \\
0 & 0 & 0 & 0
\end{pmatrix}
\otimes
\mathbf{1}_{i+2} \otimes...\otimes\mathbf{1}_{N}
\end{equation}
Obviously the limit $\nu \rightarrow 0$ is usual markov matrix: $M^{0}=M$.

The integrability of the operator $M^{\nu}$ is equivalent to the existence of a matrix $\check{R}(x,y) \in End(\mathbb{C}^{2} \otimes \mathbb{C}^{2})$ with the following properties:
\begin{enumerate}
	\item It's a solution to the braided Yang-Baxter Equation:
	\begin{equation}
	\check{R}_{23}(x,y)\check{R}_{12}(x,z)\check{R}_{23}(y,z) = \check{R}_{12}(y,z)\check{R}_{23}(x,z)\check{R}_{12}(x,y) 
	\end{equation}
	
	\item Its derivative is the local operator: $ M^{\nu}_{12} = \partial_{x}\check{R}_{12}(x,y)|_{x=y=0} $. The relevance of this requirement will become clear in what follows, precisely eq. \ref{M der}.
	
	\item It satisfied the the inversion relation: $\check{R}_{12}(x,y)\check{R}_{12}(y,x) = \mathbf{1} $. The interpretation of this will be again clear latter.
	
\end{enumerate}

A natural candidate is the baxterized form:

\begin{equation}
\check{R}_{12}(x,y) = 1 + \lambda(x,y) M^{\nu}_{12}
\end{equation} 

The constraints determine $\lambda$ up to a parameterization, we choose:

\begin{equation}
\lambda(x,y) = \frac{e^{\frac{x-y}{2}} - e^{\frac{y-x}{2}}}{ q e^{\frac{x-y}{2}} - p e^{\frac{y-x}{2}}}
\end{equation}

So the $\check{R}$ matrix acts on a two neighboring local spaces as:

\begin{equation}
\check{R}(x,y) =
\begin{pmatrix}
1 & 0 & 0 & 0 \\
0 & 1-p\lambda(x,y) & q e^{-\nu}\lambda(x,y) & 0 \\
0 & p e^{\nu}\lambda(x,y) & 1-q\lambda(x,y) & 0 \\
0 & 0 & 0 & 1
\end{pmatrix}
\end{equation}

Let $R_{ab}:= P_{ab}\check{R}_{ab}$. Where $ P_{ab}$ is the permutation operator applied on the local spaces $a$ and $b$, it permutes the corresponding components of product states. So $R$ acts on two neighboring sites as:

\begin{equation}
R(x,y) =
\begin{pmatrix}
1 & 0 & 0 & 0 \\
0 & p e^{\nu}\lambda(x,y) & 1-q\lambda(x,y) & 0 \\
0 & 1-p\lambda(x,y) & q e^{-\nu}\lambda(x,y) & 0  \\
0 & 0 & 0 & 1
\end{pmatrix}
\end{equation}

This matrix verifies a slightly different version of YBE:
\begin{equation}
R_{12}(x,y)R_{13}(x,z)R_{23}(y,z) = R_{23}(y,z)R_{13}(x,z)R_{12}(x,y)
\end{equation}

The next usual step is to define the monodromy matrix that acts on the space $a \otimes \mathcal{H}$, where $ a= \mathbb{C}^{2} $ is an auxiliary space.

\begin{equation}
T_{a,\mathcal{H}}(x, \boldsymbol{\eta}) = R_{a,N}(x,\eta_{N})...R_{a,1}(x,\eta_{1})
\end{equation}

Where $R_{a,i}(x,\eta_{N})$ acts non trivially only on the space $a$ and the site $i$

The matrix $T$ satisfies the fundamental commutation relation.

\begin{equation}\label{RTT C2}
R_{a,b}(x,y)T_{a,\mathcal{H}}(x, \boldsymbol{\eta})T_{b,\mathcal{H}}(y, \boldsymbol{\eta})
=
T_{b,\mathcal{H}}(y, \boldsymbol{\eta})T_{a,\mathcal{H}}(x, \boldsymbol{\eta})R_{a,b}(x,y)
\end{equation}
which is again equivalent to YBE. It's possible to show that simply by using the commutation $[R_{a,i}(x,\eta_{i}), R_{b,j}(y,\eta_{j})] = 0$ for $i \ne j$.

We can write the monodromy matrix in the base of the auxiliary space:

\begin{equation}
T_{a,\mathcal{H}}(x, \boldsymbol{\eta}) =
\begin{pmatrix}
A(x, \boldsymbol{\eta}) & B(x, \boldsymbol{\eta})  \\
C(x, \boldsymbol{\eta}) & D(x, \boldsymbol{\eta})  \\
\end{pmatrix}
\end{equation}
Where $A,B,C,D$ are operators acting on the space $\mathcal{H}$. Expanding the fundamental commutation relation eq. \ref{RTT C2} will tell us how these operators commute. What will be relevant to our needs are: 

\begin{equation}
[A(x, \boldsymbol{\eta}),A(y, \boldsymbol{\eta})]=0
\end{equation}

\begin{equation}
[B(x, \boldsymbol{\eta}),B(y, \boldsymbol{\eta})]=0
\end{equation}

\begin{equation}\label{com AB}
A(x, \boldsymbol{\eta}) B(y, \boldsymbol{\eta}) 
=
\frac{e^{\nu}}{q \lambda(y,x)}
B(y, \boldsymbol{\eta}) A(x, \boldsymbol{\eta})
- \frac{e^{\nu} (1-q \lambda(y,x)) }{q \lambda(y,x)}
B(x, \boldsymbol{\eta})A(y, \boldsymbol{\eta})
\end{equation}

\begin{equation}\label{com DB}
D(x, \boldsymbol{\eta}) B(y, \boldsymbol{\eta})
=
\frac{e^{\nu}}{q \lambda(x,y)}
B(y, \boldsymbol{\eta}) D(x, \boldsymbol{\eta})  - \frac{e^{\nu} (1-p \lambda(x,y)) }{q \lambda(x,y)}
B(x, \boldsymbol{\eta}) D(y, \boldsymbol{\eta})
\end{equation}

Now the transfer matrix is obtained by tracing out the auxiliary space of the monodromy matrix :
\begin{equation}
t(x, \boldsymbol{\eta})= tr_{a}(T_{a,\mathcal{H}}(x, \boldsymbol{\eta}))
\end{equation}
Which simply means:
\begin{equation}
t(x, \boldsymbol{\eta})= A(x, \boldsymbol{\eta})
+ D(x, \boldsymbol{\eta})
\end{equation}
This matrix has the advantage of commuting with itself for different parameters:

\begin{equation}
[	t(x, \boldsymbol{\eta}), 	t(y, \boldsymbol{\eta})] = 0
\end{equation}
This is again a result of the fundamental commutation relation eq. \ref{RTT C2}. multiplying both of its sides from the right by $R^{-1}_{a,b}(x,y)$ and tracing out the two auxiliary spaces, we get:
\begin{equation}\label{RTT C2}
tr_{a,b}
(T_{a,\mathcal{H}}(x, \boldsymbol{\eta})T_{b,\mathcal{H}}(y, \boldsymbol{\eta}))
=
tr_{a,b}
(T_{b,\mathcal{H}}(y, \boldsymbol{\eta})T_{a,\mathcal{H}}(x, \boldsymbol{\eta}))
\end{equation}
Where the $R$ matrices disappeared thanks to the cyclicity of the trace.
Now we need to uncover one of the sides let's say the left one by writing its coordinates:
\begin{equation}\label{RTT C2}
T^{i_{1}j_{1},i_{3}j_{3}}_{a,\mathcal{H}}(x, \boldsymbol{\eta})
\delta^{i_{2}j_{2}}
T^{j_{2}i_{2},j_{3}k_{3}}_{b,\mathcal{H}}(x, \boldsymbol{\eta})
\delta^{j_{1}i_{1}}
=
T^{j_{1}j_{1},i_{3}j_{3}}_{a,\mathcal{H}}(x, \boldsymbol{\eta})
T^{j_{2}j_{2},j_{3}k_{3}}_{b,\mathcal{H}}(x, \boldsymbol{\eta})
\end{equation}
The right side is the the coordinate version of $tr_{a}
(T_{a,\mathcal{H}}(x, \boldsymbol{\eta}))
tr_{b}
(T_{b,\mathcal{H}}(y, \boldsymbol{\eta}))$, which leads to the desired commutation.

So the family of operators $\{t(x, \boldsymbol{\eta}), x \in \mathbb{C} \}$ can be simultaneously diagonalized. The operator  $M^{\nu}$ can be derived:
\begin{equation}\label{M der}
M^{\nu} = t^{-1}(0,\boldsymbol{0}) \frac{d}{dx}t(x,\boldsymbol{0})|_{x=0}
\end{equation}
To show this let $x_{0}$ be value for the spectral parameter $x$ for which  $R_{12}(x=x_{0},0) = P_{12}$. (in our case $x_{0}=0$). At these values, the transfer matrix is:

\begin{equation}
t(x_{0},\boldsymbol{0}) = tr_{a}(P_{a,1}...P_{a,N}) = P_{1,2,...,N}
\end{equation}
\begin{equation}
t^{-1}(x_{0},\boldsymbol{0}) = tr_{a}(P_{a,N}...P_{a,1})= P_{N,N-1,...,1}
\end{equation}
\begin{equation}
\begin{split}
\frac{d}{dx}
t(x,\boldsymbol{0})|_{x=x_{0}}
=& \sum_{i=1}^{N}
tr_{a}(P_{a,1}...P_{a,i-1} M_{a,i}(x_{0},0)P_{a,i+1}...P_{a,N}) \\
=&
\sum_{i=1}^{N} M_{i-1,i}(x_{0},0)
\end{split}
\end{equation}

\subsubsection*{The reference state}
The Bethe vector is constructed starting from a reference state and using a creation operator. The natural choice for this reference state is an empty system with no particles:

\begin{equation}
\ket{0} = \ket{0}_{1} \otimes ... \otimes \ket{0}_{N} \quad \text{with} \quad \ket{0}_{i} =
\begin{pmatrix}
1  \\
0  
\end{pmatrix}
\end{equation}

Let's examine the action of the Monodromy operators on $\ket{0}$

\begin{equation}
A (x, \boldsymbol{\eta}) \ket{0} = \ket{0}
\end{equation}
\begin{equation}
C (x, \boldsymbol{\eta}) \ket{0} = 0
\end{equation}
\begin{equation}
D (x, \boldsymbol{\eta}) \ket{0} = (q e^{-\nu})^{N}
\prod_{i=1}^{N}
\lambda(x,\eta_{i})
\ket{0}
\end{equation}

To understand the previous relations, it's enough to write the $R$ matrix in the base of the auxiliary space:

\begin{equation}
R_{a,i}(x, \eta_{i}) =
\begin{pmatrix}
A^{(i)}(x, \eta_{i}) & B^{(i)}(x, \eta_{i})  \\
C^{(i)}(x, \eta_{i}) & D^{(i)}(x, \eta_{i})  \\
\end{pmatrix}
\end{equation}
Where the entry operators act non trivially only on the $i$ space. The monodromy matrix is then:
\begin{equation}
T_{a,\mathcal{H}}(x, \boldsymbol{\eta}) =
\begin{pmatrix}
A^{(1)}(x, \eta_{1}) & B^{(1)}(x, \eta_{1})  \\
C^{(1)}(x, \eta_{1}) & D^{(1)}(x, \eta_{1})  \\
\end{pmatrix}
...
\begin{pmatrix}
A^{(N)}(x, \eta_{N}) & B^{(N)}(x, \eta_{N})  \\
C^{(N)}(x, \eta_{N}) & D^{(N)}(x, \eta_{N})  \\
\end{pmatrix}
\end{equation}
The action of the entry operators on $\ket{0}$ is simple, for instance, in particular:
\begin{equation}
A^{(N)}(x, \eta_{N}) \ket{0} = \ket{0}, \quad
C^{(N)}(x, \eta_{N}) \ket{0} = 0, \quad
D^{(N)}(x, \eta_{N}) \ket{0} = q e^{-\nu} \lambda(x,\eta_{N})  \ket{0}
\end{equation}
Now applying it consecutively on the Monodromy matrix will give triangular matrices whose product is the desired result. 
Note that $ \ket{0} $ is not an eigenvector of the operator $B$. However, following the same logic as previously, it's possible by recurrence to conclude the action of the operator $B$ on $ \ket{0} $.

\begin{equation}
B (x, \boldsymbol{\eta}) \ket{0} = \sum_{i=1}^{N}
(-q)^{i}
(qe^{-\nu})^{i-1}
\prod_{j=1}^{i} \lambda(x,\eta_{j})
\ket{0}_{1} \otimes ...\otimes\ket{1}_{i}\otimes... \otimes \ket{0}_{N}
\end{equation}

So applying $B$ will create a linear combination of single particle states. It will be our creation operator.

\subsubsection*{Bethe vector}
The objective is to have an eigenvector of the transfer matrix. Let's search for an $m$ particles vector of the form:
\begin{equation}
\ket{\Psi^{m}(y_{1},...,y_{m})} = B(y_{1})...B(y_{m}) \ket{0}
\end{equation}
By applying the operator $A$ and $D$ on the Bethe vector, we get in general a term that is proportional to it and other terms that are not. The proportional term is called the wanted term, and the other terms are not wanted. Bethe equations are obtained such that the unwanted terms cancel out. Let's first examine the application of $A$ on $\Psi^{m}$. The idea is to use the fundamental commutation relation \ref{com AB} consecutively to bring the operator $A$ to the end of the $B(y_{1})...B(y_{m})$ chain so that it will be converted to a scalar when applied on $\ket{0}$. This will generate $2^{m}$ terms that can be classified into $m+1$ categories according to the spectral parameter of the operator $A$ after having reached the end of the $B's$ chain. The wanted term is straightforward, it's enough to retain the first term of the commutation relation at each time:

\begin{equation}
A(x) \ket{\Psi^{m}(y_{1},...,y_{m})} |_{\text{wanted}} = 
\prod_{i=1}^{m}
\frac{e^{\nu}}{q \lambda(y_{i},x)}
\ket{\Psi^{m}(y_{1},...,y_{m})}
\end{equation}

To find the $j$ unwanted term, we use the commutation of the $B's$ operators, since this term will not depend on the order of the $B's$, we bring $B(y_{j})$ to the left of the $B$ chain before applying $A$. now there is a unique way for $y_{j}$ it to reach the last position, which is by following the second term of the fundamental commutation at the beginning, and then by sticking to the first term for the rest, so we get:

\begin{equation}
\begin{split}
&A(x) \ket{\Psi^{m}(y_{1},...,y_{m})} |_{\text{unwanted}} = \\
&\sum_{j=1}^{m}
(- \frac{e^{\nu} (1-q \lambda(y_{j},x)) }{q \lambda(y_{j},x)})
\prod_{i \neq j}^{m}
\frac{e^{\nu}}{q \lambda(y_{i},y_{j})}
\ket{\Psi^{m}(x,y_{1},...,y_{j-1},y_{j+1},...,y_{m})}
\end{split}
\end{equation}
And in a similar fashion, we have the action of the operator $D$
\begin{equation}
D(x) \ket{\Psi^{m}(y_{1},...,y_{m})} |_{\text{wanted}} = 
(q e^{-\nu})^{N}
\prod_{i=1}^{N}
\lambda(x,\eta_{i})
\prod_{i=1}^{m}
\frac{e^{\nu}}{q \lambda(x,y_{i})}
\ket{\Psi^{m}(y_{1},...,y_{m})}
\end{equation}

\begin{equation}
\begin{split}
&D(x) \ket{\Psi^{m}(y_{1},...,y_{m})} |_{\text{unwanted}} = \\
&\sum_{j=1}^{m}
(- \frac{e^{\nu} (1-p \lambda(x,y_{j})) }{q \lambda(x,y_{j})})
\prod_{i \neq j}^{m}
\frac{e^{\nu}}{q \lambda(y_{j},y_{i})}
(q e^{-\nu})^{N}
\prod_{i=1}^{N}
\lambda(y_{j},\eta_{i})
\ket{\Psi^{m}(x,y_{1},...,y_{j-1},y_{j+1},...,y_{m})}
\end{split}
\end{equation}
Now we can write the eigenvalue for transfer matrix:
\begin{equation}
\Lambda(x) = \prod_{i=1}^{m}
\frac{e^{\nu}}{q \lambda(y_{i},x)}
+
(q e^{-\nu})^{N}
\prod_{i=1}^{N}
\lambda(x,\eta_{i})
\prod_{i=1}^{m}
\frac{e^{\nu}}{q \lambda(x,y_{i})}
\end{equation}

\subsubsection*{Bethe equations}

\begin{equation}
(- \frac{e^{\nu} (1-q \lambda(y_{j},x)) }{q \lambda(y_{j},x)})
\prod_{i \neq j}^{m}
\frac{e^{\nu}}{q \lambda(y_{i},y_{j})}
+
(- \frac{e^{\nu} (1-p \lambda(x,y_{j})) }{q \lambda(x,y_{j})})
\prod_{i \neq j}^{m}
\frac{e^{\nu}}{q \lambda(y_{j},y_{i})}
(q e^{-\nu})^{N}
\prod_{i=1}^{N}
\lambda(y_{j},\eta_{i})
=0
\end{equation}
Which gives after simplifications the $m$ equations

\begin{equation}
e^{\nu N}
=
\prod_{i \neq j}^{m}
\frac{\lambda(y_{i},y_{j})}{ \lambda(y_{j},y_{i})}
\prod_{i=1}^{N}
q
\lambda(y_{j},\eta_{i})
\quad
1 \leq j \leq m
\end{equation}

These equations are simpler to analyze in the TASEP limit $(p,q) \rightarrow (1,0)$ This is however beyond the objective of this section and was done in ... to extract ...

\subsubsection*{Twisted Monodromy Matrix}
Let $w$ be a matrix acting on $\mathbb{C}^{2}$ such that $w_{1}w_{2} = w \otimes w$ commutes with the matrix $R$:
\begin{equation}\label{com twist}
[w_{1}w_{2} , R_{1,2}(x,y) ] = 0
\end{equation}
Then it is straightforward to show that the matrix $w_{a}T_{a,\mathcal{H}}$ satisfies the fundamental commutation relation:
\begin{equation}
R_{a,b}(x,y)
(w_{a}T_{a,\mathcal{H}}(x, \boldsymbol{\eta}))
(w_{b}T_{b,\mathcal{H}}(y, \boldsymbol{\eta}))
=
w_{b}T_{b,\mathcal{H}}(y,\boldsymbol{\eta})
w_{a}T_{a,\mathcal{H}}(x, \boldsymbol{\eta})
R_{a,b}(x,y)
\end{equation}
Where $w_{a}$ acts non trivially on the auxiliary space $a$. $w_{a}T_{a,\mathcal{H}}$ is called the twisted monodromy matrix. It naturally appears in systems with twisted periodic boundary condition, which is the same as the periodic one except that the coupling between the first site and the last site differs by a phase factor. One asks: how does a twisted monodromy matrix impact the ABA procedure?
First note that it changes the properties of the reference state, this is easy to understand if we write it in the auxiliary space:
\begin{equation}
w_{a}T_{a,\mathcal{H}}=
\begin{pmatrix}
w_{11}A + w_{12}C & w_{11}B + w_{12}D \\
w_{21}A + w_{22}C & w_{21}B + w_{22}D \\
\end{pmatrix}
\end{equation}
So $(w_{a}T_{a,\mathcal{H}})_{21}$ is not an annihilation operator for the vacuum  and $(w_{a}T_{a,\mathcal{H}})_{21}$ doesn't admit it as an eigenvector. One in principle has to search for another reference state than the vacuum. However, if we choose $w$ to be diagonal, (i.e. $w_{21}=w_{12}=0$), then these properties are conserved, except that the eigenvalues get a factor for the diagonal operators. We note as well that our $R$ matrix is invariant under the commutation \ref{com twist} for an arbitrary diagonal $w$ operator. For this case, the new Bethe equations become:

\begin{equation}
e^{\nu N}
=
\frac{w_{22}}{w_{11}}
\prod_{i \neq j}^{m}
\frac{\lambda(y_{i},y_{j})}{ \lambda(y_{j},y_{i})}
\prod_{i=1}^{N}
q
\lambda(y_{j},\eta_{i})
\quad
1 \leq j \leq m
\end{equation}
And the corresponding twisted eigenvalue:

\begin{equation}
\Lambda(x) = w_{11}\prod_{i=1}^{m}
\frac{e^{\nu}}{q \lambda(y_{i},x)}
+
w_{22}
(q e^{-\nu})^{N}
\prod_{i=1}^{N}
\lambda(x,\eta_{i})
\prod_{i=1}^{m}
\frac{e^{\nu}}{q \lambda(x,y_{i})}
\end{equation}
\textbf{TASEP limit:} in the limit $(p,q) \rightarrow (1,0)$ the function $\lambda$ becomes: $ \lambda(x,y) = 1 - e^{x-y}$. However, we get a singularity with the Bethe equations where some spectral parameters need to be infinite. Since the spectral parameters are just intermediate parameters, we can parameterize them to avoid the singularity: $Z_{i} =  e^{-y_{i}}/q$. So the Bethe equations become:

\begin{equation}
e^{\nu N}
=
\frac{w_{22}}{w_{11}}
\prod_{i \neq j}^{m}
\frac{-Z_{j}}{Z_{i}}
\prod_{k=1}^{N}
\frac{e^{-\eta_{k}}}{e^{-\eta_{k}} - Z_{j}}
\quad
1 \leq j \leq m
\end{equation}
And the eigenvalue:

\begin{equation}
\Lambda(x) = w_{11}\prod_{i=1}^{m}
(e^{\nu} (1-e^{x}Z_{i}))
\end{equation}
This will be useful for the following section

\subsubsection{Algebraic Bethe Ansatz for arbitrary number of defects:}

Consider a lattice of $N$ sites with periodic boundary conditions with $M_{1}$ first class particles, and $M_{2}$ second class particles of arbitrary rates using the same notation as the previous section.
\begin{equation}
\begin{split}
\alpha \quad 20 \rightarrow 02
\\
\beta \quad 12 \rightarrow 21
\end{split}
\end{equation}
A probability wave vector of the system is an element of the space: $\mathcal{H} = (\mathbb{C}^{3})^{\otimes N}$ Where each base element is an $N$ tensor product of elements from the set $\{ \bra{0}, \bra{1}, \bra{2} \}$ and corresponds to a configuration of the system. Of course, the particles conservation will make only a subspace of $\mathcal{H}$ accessible. 
The markov matrix of the system can be written as a sum of local operators, each is acting non trivially only on two neighboring sites:
\begin{equation}
M = \sum_{i=1}^{N} 
M_{i,i+1}
\end{equation}
the site $N+1$ is identified with the site $1$ and $M_{i,i+1}$ is given in the base:

$\bra{00},\bra{01},\bra{02},
\bra{10},\bra{11},\bra{12},
\bra{20},\bra{21},\bra{22} $ 
\begin{equation}
M_{i,i+1} =
\left(
\begin{array}{ccc:ccc:ccc}
0 & 0 & 0 & 0 & 0 & 0 & 0 & 0 & 0 \\
0 & 0 & 0 & 1 & 0 & 0 & 0 & 0 & 0  \\
0 & 0 & 0 & 0 & 0 & 0 & \alpha & 0 & 0  \\
\hdashline
0 & 0 & 0 & -1 & 0 & 0 & 0 & 0 & 0 \\
0 & 0 & 0 & 0 & 0 & 0 & 0 & 0 & 0 \\
0 & 0 & 0 & 0 & 0 & -\beta & 0 & 0 & 0 \\
\hdashline
0 & 0 & 0 & 0 & 0 & 0 & -\alpha & 0 & 0 \\
0 & 0 & 0 & 0 & 0 & \beta & 0 & 0 & 0 \\
0 & 0 & 0 & 0 & 0 & 0 & 0 & 0 & 0 \\
\end{array}
\right)
\end{equation}
The tensor product by identity operator is implied on the sites where the local operator doesn't act.
Being interested in the currents of the first and second class particles, we need to deform the Markov matrix in a similar fashion as the previous section. We introduce the random variables:
\begin{itemize}
	\item $Y^{10}_{t}$ the number of times a first class particle jumped over a void particle up to time $t$.
	\item $Y^{12}_{t}$  the number of times a first class particle jumped over a second class particle up to time $t$. 
	\item $Y^{20}_{t}$  the number of times a second class particle jumped over a void up to time $t$.
\end{itemize}
hence the modified Markov matrix needs three parameters:
\begin{equation}
M^{\nu_{10},\nu_{12},\nu_{20}}_{i,i+1} =  E^{10}_{i,i+1}  + \beta E^{12}_{i,i+1}  + \alpha E^{20}_{i,i+1} 
\end{equation}

Where:

\begin{equation}
E^{10}_{i,i+1} = e^{\nu_{10}} \ket{01} \bra{10} - \ket{10} \bra{10} 
\end{equation}
\begin{equation}
E^{20}_{i,i+1} = e^{\nu_{20}} \ket{02} \bra{20} - \ket{20} \bra{20} 
\end{equation}
\begin{equation}
E^{12}_{i,i+1} = e^{\nu_{12}} \ket{21} \bra{12} - \ket{12} \bra{12} 
\end{equation}

So the deformed Markov matrix is:

\begin{equation}
M^{\nu_{10},\nu_{20},\nu_{12}} = 
\sum_{i=1}^{L}
M^{\nu_{10},\nu_{20},\nu_{12}}_{i,i+1}
\end{equation}
Obviously the limit ${(\nu_{10},\nu_{20},\nu_{12})} \rightarrow (0,0,0)$ gives rise to the usual non deformed Markov matrix:
\begin{equation}
M = M^{0,0,0}
\end{equation}

In a similar fashion to the single defect case, the conditioned generating function for the random vector $(Y^{10},Y^{20},Y^{12})$ has to obey the evolution equation:

\begin{equation}
\frac{d}{dt}F_{t}^{\nu_{10},\nu_{20},\nu_{12}}(C) := \sum_{C^{'}}
M^{\nu_{10},\nu_{20},\nu_{12}}(C,C^{'})
F_{t}^{\nu_{10},\nu_{20},\nu_{12}}(C^{'})
\end{equation}

The full generating function is given by a sum over the configurations: $F_{t}^{\nu_{10},\nu_{20},\nu_{12}} = \sum_{C} F_{t}^{\nu_{10},\nu_{20},\nu_{12}}(C)$ and is estimated at large time by an exponential function with a parameter $\lambda(\nu_{10},\nu_{20},\nu_{12})$ that is the largest eigenvalue of the matrix $M^{\nu_{10},\nu_{20},\nu_{12}}$:
\begin{equation}
F_{t}^{\nu_{10},\nu_{20},\nu_{12}} \sim e^{\lambda(\nu_{10},\nu_{20},\nu_{12})t}
\end{equation}
In the limit $(\nu_{10},\nu_{20},\nu_{12}) \rightarrow (0,0,0) $ this eigenvalue is the one of the matrix $M$ which is zero, and it is non degenerate for all the values of $(\nu_{10},\nu_{20},\nu_{12})$, a result that stems from Perron-Frobenius theorem. In the next section, we will see how to diagonalize the operator $M^{\nu_{10},\nu_{20},\nu_{12}} $ in order to find this eigenvalue.

\subsubsection{The Nested Algebraic Bethe Ansatz}

The model can be thought of as an $SU(3)$ spin chain. The integrability of the operator $M^{\nu_{10},\nu_{20},\nu_{12}}$ is equivalent to the existence of a matrix $\check{R}(x,y) \in End(\mathbb{C}^{3} \otimes \mathbb{C}^{3})$ with the following properties:
\begin{enumerate}
	\item It's a solution to the braided Yang-Baxter Equation:
	\begin{equation}
	\check{R}_{23}(x,y)\check{R}_{12}(x,z)\check{R}_{23}(y,z) = \check{R}_{12}(y,z)\check{R}_{23}(x,z)\check{R}_{12}(x,y) 
	\end{equation}
	
	\item Its derivative is the local operator: $ M^{\nu_{10},\nu_{20},\nu_{12}}_{12} = \partial_{y}\check{R}_{12}(x,y)|_{x \rightarrow y} $
	
	\item It satisfied the the inversion relation.
\end{enumerate}

Since the local operator $M^{\nu^{10},\nu^{12},\nu^{20}}_{i,i+1}$ is a sum of more elementary operators, one can search for an $ \check{R}$ matrix of the Baxterized form:

\begin{equation}
\check{R}_{i,i+1}(x,y) =
1 + 
g_{10}(x,y) E^{10}_{i,i+1}  + g_{12}(x,y) E^{12}_{i,i+1}  + g_{20}(x,y) E^{20}_{i,i+1} 
\end{equation}

A solution:

\begin{equation}
g_{10}(x,y) = 1-e^{x-y}
\end{equation} 
\begin{equation}
g_{12}(x,y) = 1-\frac{1+\beta (e^{-y}-1)}{1+\beta (e^{-x}-1)}
\end{equation} 
\begin{equation}
g_{20}(x,y) = 1-\frac{1+\alpha (e^{x}-1)}{1+\beta (e^{y}-1)}
\end{equation} 

Let $R_{ab}:= P_{ab}\check{R}_{ab}$. This matrix verifies a slightly different version of YBE:
\begin{equation}
R_{12}(x,y)R_{13}(x,z)R_{23}(y,z) = R_{23}(y,z)R_{13}(x,z)R_{12}(x,y)
\end{equation}

The next usual step is to define the monodromy matrix that acts on the space $a \otimes \mathcal{H}$, where $ a= \mathbb{C}^{3} $ is an auxiliary space.

\begin{equation}
T_{a,\mathcal{H}}(x, \boldsymbol{\eta}) = R_{a,N}(x,\eta_{N})...R_{a,1}(x,\eta_{1})
\end{equation}

This matrix satisfies the fundamental commutation relation:

\begin{equation}
R_{a,b}(x,y)T_{a,\mathcal{H}}(x, \boldsymbol{\eta})T_{b,\mathcal{H}}(y, \boldsymbol{\eta})
=
T_{a,\mathcal{H}}(y, \boldsymbol{\eta})T_{b,\mathcal{H}}(x, \boldsymbol{\eta})R_{a,b}(x,y)
\end{equation}
Now the transfer matrix is obtained by tracing out the auxiliary space of the monodromy matrix :
\begin{equation}
t(x, \boldsymbol{\eta})= tr_{a}(T_{a,\mathcal{H}}(x, \boldsymbol{\eta}))
\end{equation}
This matrix has the advantage of commuting with itself for different parameters:

\begin{equation}
[	t(x, \boldsymbol{\eta}), 	t(y, \boldsymbol{\eta})] = 0
\end{equation}
So the family of operators $\{t(x, \boldsymbol{\eta}), x \in \mathbb{C} \}$ can be simultaneously diagonalized. The operator  $M^{\nu_{10},\nu_{20},\nu_{12}}$ can be derived:
\begin{equation}
M^{\nu_{10},\nu_{20},\nu_{12}} = -t^{-1}(0,\boldsymbol{0}) \frac{d}{dx}t(x,\boldsymbol{0})|_{x=0}
\end{equation}

\subsubsection*{Commutation relations}
Let's contemplate the $\check{R}$ matrix:
\begin{equation}
 \check{R} =
\left(
\begin{array}{ccc:ccc:ccc}
1 & 0 & 0 & 0 & 0 & 0 & 0 & 0 & 0 \\
0 & 1 & 0 & e^{\nu_{10}} g_{10} & 0 & 0 & 0 & 0 & 0  \\
0 & 0 & 1 & 0 & 0 & 0 & e^{\nu_{20}} g_{20} & 0 & 0  \\ 
\hdashline
0 & 0 & 0 & 1-g_{10} & 0 & 0 & 0 & 0 & 0 \\
0 & 0 & 0 & 0 & \textcolor{blue}{1} & \textcolor{blue}{0} & 0 & \textcolor{blue}{0} & \textcolor{blue}{0} \\
0 & 0 & 0 & 0 & \textcolor{blue}{0} & \textcolor{blue}{1-g_{12}} & 0 & \textcolor{blue}{0} & \textcolor{blue}{0} \\ \hdashline
0 & 0 & 0 & 0 & 0 & 0 & 1-g_{20} & 0 & 0 \\
0 & 0 & 0 & 0 & \textcolor{blue}{0} & \textcolor{blue}{e^{\nu_{12 }}g_{12}} & 0 & \textcolor{blue}{1} & \textcolor{blue}{0} \\
0 & 0 & 0 & 0 & \textcolor{blue}{0} & \textcolor{blue}{0} & 0 & \textcolor{blue}{0} & \textcolor{blue}{1} \\

\end{array}
\right)
\end{equation}

Where the elements colored in blue form a lower dimension $\check{R}$ matrix acting on the space $\mathbb{C}^{2}\otimes\mathbb{C}^{2}$:

\begin{equation}
\check{R}^{(1)} = 
\begin{pmatrix}
1 & 0 & 0 & 0 \\
0 & 1-g_{12} & 0 & 0 \\
0 & e^{\nu_{12}}g_{12} & 1 & 0 \\
0 & 0 & 0 & 1
\end{pmatrix}
\end{equation}
This matrix can be shown to solve a model with only first and second-class particles, so it can be seen as one species ASEP with particles hopping forward at rate $\alpha$ and backwards at rate $\beta$.

Let's write the Monodromy matrix in the base of the auxiliary space:

\begin{equation}
T_{a,\mathcal{H}}(x, \boldsymbol{0}) =
\begin{pmatrix}
A(x) & B_{1}(x) & B_{2}(x)  \\
C_{1}(x) & D_{11}(x) & D_{12}(x) \\
C_{2}(x) & D_{21}(x) & D_{22}(x) 
\end{pmatrix}
\end{equation}

So the transfer matrix can be written as:

\begin{equation}
t(x) = A(x) + D_{11}(x) + D_{22}(x)
\end{equation}

Using the fundamental commutation relation, we can find how these operators commute.

We are looking for eigenvectors for the transfer matrix. As in the case of $\mathbb{C}^{2}$ local space, one has to start with a reference state and find adequate creation operators.

\subsubsection*{Reference state}
A natural possible reference state is an empty system with no particles, so it's a tensor product of empty sites:

\begin{equation}
\ket{0} = \ket{0}_{1} \otimes ... \otimes \ket{0}_{N} \quad with \quad \ket{0}_{i} =
\begin{pmatrix}
1  \\
0  \\
0
\end{pmatrix}
\end{equation}

The Monodromy matrix can be written as:

\begin{equation}
T_{a,\mathcal{H}}(x) =
\begin{pmatrix}
A^{(1)}(x) & B^{(1)}_{1}(x) & B^{(1)}_{2}(x)  \\
C^{(1)}_{1}(x) & D^{(1)}_{11}(x) & D^{(1)}_{12}(x) \\
C^{(1)}_{2}(x) & D^{(1)}_{21}(x) & D^{(1)}_{22}(x)
\end{pmatrix}...
\begin{pmatrix}
A^{(N)}(x) & B^{(N)}_{1}(x) & B^{(N)}_{2}(x)  \\
C^{(N)}_{1}(x) & D^{(N)}_{11}(x) & D^{(N)}_{12}(x) \\
C^{(N)}_{2}(x) & D^{(N)}_{21}(x) & D^{(N)}_{22}(x)
\end{pmatrix}
\end{equation}
Where upper index $(i)$ refers to an operator acting non trivially only on the site $(i)$ of the lattice, and its expression is given by the corresponding block of the $\check{R}$ matrix.
Examining the structure of these matrices, its easy to verify that $\ket{0}$ is an eigenvector of the three operators constituting the transfer matrix:


\begin{equation}
A \ket{0} = \ket{0}
\end{equation}
\begin{equation}
D_{11} \ket{0} = (e^{\nu_{10}}(1-e^{x}))^{N} \ket{0}
\end{equation}
\begin{equation}
D_{22} \ket{0} = (\alpha e^{\nu_{20}}(1-e^{x}))^{N} \ket{0}
\end{equation}

\subsubsection*{The Bethe vector}
For a two dimension local space system, there is a single creation operator among the Monodromy operators, usually called $B$ that can create a one particle Bethe state with a momentum $\mu$ by applying it to the reference empty state $\ket{\mu}=B(\mu)\ket{0}$. To get an $n$ particle state, it's enough to apply the creation operator $n$ times with the corresponding moments, requiring it to be eigenvector to the transfer matrix generates the Bethe equations. In our case, we need a creation operator for the first-class particles and another for the second class particles. Examining the structure of the $\check{R}$ matrix, we can understand that $B_{1}$ is the first and $B_{2}$ is the second. These two operators don't commute, so the Bethe vector has to be written as a linear combination of all the possible ordering of operators:

\begin{equation}
\ket{\Psi^{M_{1},M_{2}} (y_{1},...,y_{r})} = \sum_{i_{1},...i_{r}\in \{1,2\}} \Psi^{M_{1},M_{2}}_{i_{1},...i_{r}} B_{i_{1}}(y_{1})...B_{i_{r}}(y_{r}) \ket{0}
\end{equation}
Where $M_{1},M_{2}$ is the number of first and second class particles respectively. Of course the coefficients that $\Psi^{M_{1},M_{2}}_{i_{1},...i_{r}}$ with lower indices that are not composed of $M_{1}$ ones and $M_{2}$ twos have to be zero.
This vector has to be an eigenvector of the transfer matrix, it is not in general an eigenvector of the operators $A$,$D_{11}$,$D_{22}$. by applying each of these operators on the Bethe vector, we get a wanted term that is proportional to it, and unwanted term that is not. The three unwanted terms should cancel out. The conditions for this will constitute the Bethe equations.

The cancellation of the unwanted terms will require the denationalization of a matrix of the form $w_{1}(\mathbf{y}) T_{11}^{(1)} + w_{2}(\mathbf{y}) T_{22}^{(1)}$ which is a twisted monodromy matrix for TASEP with one species (the second class particles) in a lattice composed of the first and second class particles. This leads to Bethe equations with two sets of spectral parameters $M_{1}+M_{2}$ $y's$ and another set of $M_{2}$ $Z's$ coming from the lower order transfer matrix. The details can be found in \cite{cantini2008algebraic} as well as the derivation of the currents in the hydrodynamic limit which is similar to the one defect case.

\chapter{Boundary-induced phase transitions in multi-species driven diffusive systems}
\label{open}
	
\section{Introduction}

	Driven diffusive systems are archetypes for non-equilibrium statistical mechanics. They appear in various areas of physics, chemistry and theoretical biology \cite{chou2011non} \cite{blythe2007nonequilibrium}. To have a general idea, one can imagine a gas of identical particles in a 1D lattice that is coupled to reservoirs from both sides. The driven aspect of the system is obtained by breaking the space symmetry through an external field so that there is a current of particles even if the two reservoirs on the boundaries are identical. Such systems are known to exhibit shock solutions, in contrast to their purely diffusive counterparts. Once the current as a function of the coarse-grained density in a homogeneous state is known, the phase diagram for the steady state of the open-boundary system can be determined by a simple general principle known as the extremal current principle. Its first version, dealing with the maximum current phase, was proposed by Krug \cite{krug1991boundary} \cite{krug1991steady}. A more general version taking into account the minimum phase was elaborated by Schüz et al \cite{popkov1999steady}, \cite{hager2001minimal}.
Despite the success of this principle in treating open boundary problems of numerous models, its validity is restricted to systems with a single species of particles. A generalization to interacting multi-species systems is far from being obvious. One needs to define multiple coarse-grained densities corresponding to the different species. The expressions of the corresponding currents as a function of the densities are derived from the local dynamic for a given model and are assumed to be known in a homogeneous system.

	In some particular cases of multi-species systems, it's still possible to do an exact analysis. For instance, in \cite{evans1995asymmetric} \cite{evans1995spontaneous}, 2-species TASEP is considered with a restriction on the boundary rates so that the hole-particle symmetry is preserved. The steady state is exactly solved using the Matrix Product Ansatz (MPA) for special values of the parameters, the mean-field approximation is needed to continue the analysis and sketch a phase transition that identifies a phase with a power law decay and another with exponential decay. More generally: Under unity hopping rates in the bulk and boundary rates preserving the hole-particle symmetry, it is possible to decompose the 2-species TASEP into two 1-species systems by viewing the second-class particles as void for one system and first-class particles for the other system, this sometimes called the coloring argument \cite{ayyer2010some}.
	A Colorable model is in general not integrable (in the sense that no matrix representation for the steady state exists) except for special values of the boundary rates. \cite{crampe2015open}
	
	Another example is given in \cite{khorrami2000exact} where a multi-species generalization of TASEP with open boundaries was treated exactly with MPA with hopping rates of particles drawn from a distribution with hierarchical priority. No hole-particles symmetry is conserved here, however, the applicability of the generalization of the MBA required that only one parameter expression is allowed for injecting particles and another for extracting particles, which results in a two parameters phase transition similar to one species TASEP for a class of distributions, while the HD phase is missing for the rest of distributions.

	More generally, a quadratic algebra MPA description of the steady state of a multi-species stochastic system will always lead to a constraint on the rates of the boundaries \cite{alcaraz1998n}
	
	A simplifying special situation is when the current of a species (or a combination of species) is null. This is the case in \cite{arita2006phase}, \cite{arita2006exact} where 2-species TASEP is considered with confined second-class particles with equal boundary hopping rates. Stationary state and phase transition is obtained, using MPA, and is shown to be composed of three regions similar to one-species TASEP. In \cite{uchiyama2008two}  2-species ASEP with confined first class particles was analyzed and phase transition was obtained with exact methods. This model was generalized in \cite{cantini2016koornwinder} to multi-species ASEP with again semi-permeable boundaries. Phase transition was analyzed in \cite{ayyer2017exact}.
	A full classification of 2-species ASEP with integrable boundaries can be found in \cite{crampe2015open}.

A class of models which admit product invariant measures in the homogeneous state was studies in the literature. This restriction provides the advantage of making the mean field currents exact, and most importantly, it allows through a restriction on the hopping rates on the boundaries to find a simple relation between the effective densities for a boundary and the corresponding hopping rates at that boundary. Such relations are unknown for a general non-product measure. Examples for such systems are treated in \cite{rakos2004exact} 
\cite{popkov2004spontaneous}
\cite{popkov2004infinite} \cite{popkov2011hierarchy}
\cite{popkov2004hydrodynamic}.
Although in principle, it is possible to take into account the correlations by defining a projection measure as argued in \cite{popkov2004spontaneous}, however we are not aware of any model where this possibility has been tested.

Our final example for the special cases is a recent one: in \cite{bonnin2021two} 2-speed TASEP is treated (the two species can't swap) with entry and exit rates for each species. Mean field analysis is used to give an approximate prediction of the bulk behavior that works well only when the boundary rates are close to each other.
	
The objective of this chapter is to investigate a method allowing to find the steady state bulk and boundaries for systems with multi-species for arbitrary non zero boundary rates without a prior knowledge with relation between boundary rates and the reservoirs densities. We show through an example that such a relation doesn't exist in general: the effective density of a reservoir does not depend only on the corresponding boundary rates, but as well on what happens in the whole systems, in other words, the reservoirs should be seen as coupled rather than independent. In addiction, we show the relevance of the normal modes(the Riemann variable) in providing for a physical interpretation of the behavior of the system. In the arguments for our approach, we assumed a diagonal relation between the diffusive currents and the density gradient. Despite the fact that this has been disputed in \cite{popkov2004infinite}, \cite{popkov2004hydrodynamic}, \footnote{I thank the reviewer that made me aware of these papers}, the method still seems to give reasonably good results for the models we tested on. A further investigation is needed to understand whether this is du to a particularity of our our models. The model we will be testing on presents a weakly hyperbolic point (un umbilic point), which provides for another example where the analysis in \cite{popkov2013boundary} can be applied.

	 We apply our method to 2-species TASEP with arbitrary non-zero hopping rates both on the bulk and on the boundaries. This requires taking into account the subtle interplay between bulk dynamics and the density effect of coupling constants on the boundaries.A Rather good agreement with Monte Carlo simulations is obtained. Preliminary investigations, not included here, show the validity of the method for a higher number of species.  
	
	\subsection{Outlines and main result:}
	
		Although our main interest is the hydrodynamics of short-range interaction particles system with multiple conserved species. We can place ourselves in a more general framework and consider $n$ abstract quantities with local densities that are functions of space and time:
	
	$$ \boldsymbol{\rho}  (x,t) = (\rho_{1}(x,t),...,\rho_{n}(x,t))  \quad x \in \mathbb{R} \quad t \geq 0  $$
And associated currents:
	$$\boldsymbol{J}(\boldsymbol{\rho}) = (J_{1}(\boldsymbol{\rho}),.., J_{n}(\boldsymbol{\rho}))$$
	Where $J_{i}$ is the current of $\rho_{i}$. These densities evolve according to a set of coupled partial differential conservation laws:

	\begin{equation}
	\partial_{t}\boldsymbol{\rho} + \partial_{x}\boldsymbol{J} = 0
	\label{eqn:con}
	\end{equation}

		The aim of this chapter is to point out and make use of a connection between the following, a priori independent, two problems:
	
	\begin{itemize}
		\item[$\blacktriangleright$] \textbf{The Riemann Problem:} which is defined on an infinite line with initial uniform densities except for a discontinuity at 0:
		\begin{equation}\label{eqn:RP}
		\boldsymbol{\rho}  (x,0) =  \boldsymbol{\rho}^{L} \mathds{1}_{x<0}(x) + \boldsymbol{\rho}^{R} \mathds{1}_{x>0}(x) \quad x \in \mathbb{R}
		\end{equation}
		The equations with this initial condition will be invariant under the transformation $(x,t) \rightarrow (\lambda x, \lambda t)$ and therefore the solution can be expressed as a function of one variable $ \frac{x}{t}$. In particular, we have:
		\begin{equation}\label{key}
		\boldsymbol{\rho}  (0,t) = \boldsymbol{\rho}  (0,1)  := \boldsymbol{\rho}|_{0} := \boldsymbol{R}(0,\boldsymbol{\rho}^{L},\boldsymbol{\rho}^{R}) \quad \text{for} \quad t>0
		\end{equation}
		We will call this constant value: the solution to the Riemann problem at zero.

		\item[$\blacktriangleright$] \textbf{The open boundaries problem:} Now let's consider a finite system of size $ L $ coupled to two reservoirs so that the densities on the extremities are given by the same corresponding densities of the Riemann problem, namely:
		
		\begin{equation}
		\begin{aligned}
		& \boldsymbol{\rho}  (0,t) =\boldsymbol{\rho}^{L} \\
		& \boldsymbol{\rho}  (L,t) =  \boldsymbol{\rho}^{R} \\
		\end{aligned}
		\end{equation}

			These boundary conditions are a-priori ill-posed, they cannot in general be fulfilled point-wise. A weak sense formulation is needed and was first introduced in \cite{bardos1979first}. It’s based on the vanishing viscosity method. A second order diffusion term  $\epsilon \partial_{x x}\boldsymbol{\rho}$ when added to the conservation equation makes it parabolic, and thus the boundary conditions well defined, the limit $\epsilon \rightarrow 0$ is taken in $L^{1}_{loc}$. An equivalent entropic formulation exists \cite{bardos1979first} \cite{le1988explicit} \cite{ancona1999scalar}. For fixed boundary conditions, the system is expected to reach a steady state in the limit $t \rightarrow \infty$, in this state, the densities profile becomes only functions of the space variable. In the bulk, the conservation equation is reduced to $\partial_{x} \mathbf{J} = 0$ which leads to constant bulk densities for non-degenerate currents. We shall call $\boldsymbol{\rho}^{B}$ the bulk density in the steady state. For what follows, unless stated otherwise, we assume the presence of a residual viscosity that plays a role only near the boundaries and allows to consider that the boundary conditions are verified point-wise.
	\end{itemize}
	

	\subsubsection*{Failure of the independent boundaries approach}
	In systems out of equilibrium, the concept of a reservoir of fixed densities is not always straightforward. The density on one boundary is not necessarily a function of the dynamics on that boundary but can be a function of the behavior of all of the system, i.e. it depends on the dynamics on the other boundary too.
	For the model we considered, our first attempts were to to produce a reservoir density as a function of boundary rates for each boundary independently. We tried to model a reservoirs as an external site where particles from different species are created and and annihilated at a much higher frequencies than the hopping rates of the model, so that the life-time of a particle  belonging to a each species is proportional to the desired density of that species on the boundary. This approached combined with the principle failed to give a satisfactory results. Another approach is to choose the boundary rates that are the same as the average effective hopping rates of particles in a uniform system with the desired densities. This again has failed. Finally, we tried to take into account the correlations in a similar fashion as proposed by \cite{popkov2004spontaneous},although this has improved the results, it was still not sufficiently satisfactory. In the applications,  we introduce a method that allows us to determine the boundaries, as well as the bulk, as one function of all the coupling parameters on both sides. For the time being we simply assume that we measure the boundaries by the measuring the first and the last site. Conceptually, one can consider these sites as part of the reservoirs.

		
	The analysis for a class of multi-densities hyperbolic systems is significantly simplified in terms of quantities known as the normal modes or the Riemann variables. These are transformation functions of the densities that allow to write the conservation laws in a simpler form.
	
	The principle that we suggest allows obtaining the bulk densities of the open boundaries system once the corresponding Riemann problem is solved.

		\subsubsection*{The principle}
		The steady-state phase diagram of the bulk for the open boundary problem is governed by the solution of the associated Riemann problem at zero. We have:
		\begin{enumerate} [label=(\alph*)]
			\item 
			$ 	\boldsymbol{\rho}^{B} = \boldsymbol{\rho}|_{0} $
			\item The sings of the eigenvalues $ v_{k} $  of the Jacobian matrix $[\frac{\partial J_{i}}{\partial \rho_{j}}(\boldsymbol{\rho}^{B})]_{ij}$ at the bulk governs the behavior of the normal modes at the boundaries in the following way:
			\begin{itemize}
				\item $ v_{k} > 0 $, the corresponding normal mode is induced from the left and exhibits an exponential convergence on the right
				\item $v_{k} < 0$ the other way around
				\item $v_{k} = 0$ represents a transition point where the convergence is polynomial on both sides, and induced by neither of the boundaries.
			\end{itemize}
		\end{enumerate}

	Note that the idea of the signs of eigenvalues governing the phase transition is already discusses in \cite{popkov2011hierarchy}  where the phase transitions were classified among continuous and non-continuous, leading to a diagram in the order parameter of the bulk densities. 

	
	We start by showing that this principle is an equivalent reformulation of the extremal current principle for the case of a single conserved quantity, then we will be treating the general case of multi-component density.

	


	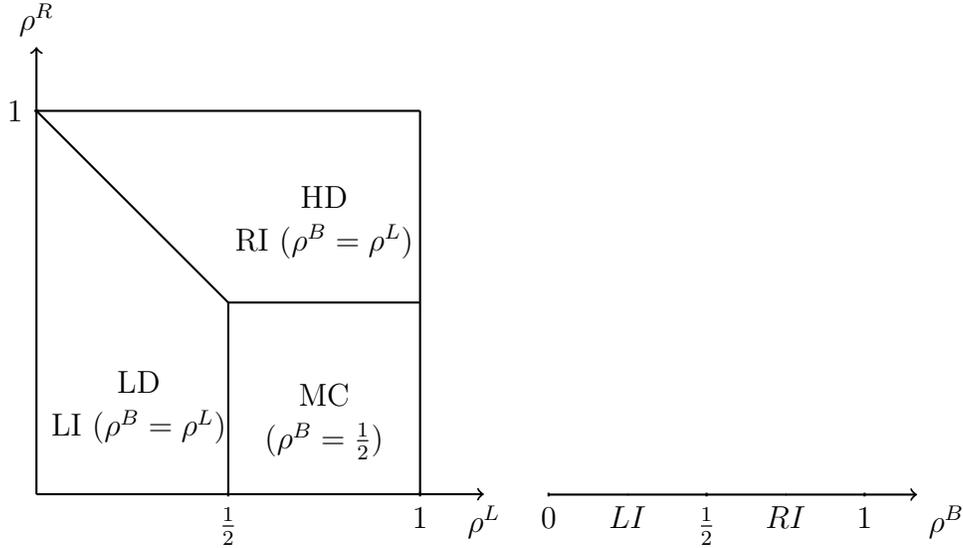
\begin{figure}[h!]
		\centering
		\begin{subfigure}[b]{0.4\linewidth}
			\begin{center}
				\begin{tikzpicture}[thick, scale=0.85]
				\draw[thick,->] (0,0) -- (7,0) node[anchor=north] {$\rho^{L}$};
				\draw[thick,->] (0,0) -- (0,7) node[anchor=south ]{$\rho^{R}$};	
				
				\draw (3 cm,1pt) -- (3 cm,-1pt) node[anchor=north] {$\frac{1}{2}$};
				
				\draw (6 cm,1pt) -- (6 cm,-1pt) node[anchor=north] {$1$};

				\draw (1pt,6 cm) -- (-1pt,6 cm) node[anchor=east] {$1$};
				
				\draw (3,3) -- (6,3);
				
				\draw (0,6) -- (6,6);
				\draw (0,6) -- (3,3);
				\draw (3,0) -- (3,3);
				\draw (6,0) -- (6,6);
				
				\draw (5.7,1) node[anchor=east] {};
				
				\draw (4.5,5) node[anchor=north] {HD};
				\draw (4.5,4.4) node[anchor=north] {RI $ (\rho^{B}=\rho^{L}) $};
				
				\draw (1.6,2.1) node[anchor=north] {LD};
				\draw (1.6,1.5) node[anchor=north] {LI ($\rho^{B} = \rho^{L}$)};

				\draw (4.5,1.9) node[anchor=north] {MC};
				\draw (4.5,1.3) node[anchor=north] {($\rho^{B} = \frac{1}{2}$)};
				
				\end{tikzpicture}
			\end{center}
		\end{subfigure}
		\begin{subfigure}[b]{0.4\linewidth}
			\begin{center}
				\begin{tikzpicture}[thick, scale=0.7]
				\draw[thick,->] (0,0) -- (7,0) node[anchor=north west] {$\rho^{B}$};
				
				\draw (3 cm,1pt) -- (3 cm,-1pt) node[anchor=north] {$\frac{1}{2}$};
				
				\draw (6 cm,1pt) -- (6 cm,-1pt) node[anchor=north] {$1$};
				\draw (0 cm,1pt) -- (0 cm,-1pt) node[anchor=north] {$0$};
				
				\draw (1.5 cm,0pt) -- (1.5 cm,0pt) node[anchor=north] {$LI$};	
				\draw (4.5 cm,0pt) -- (4.5 cm,0pt) node[anchor=north] {$RI$};
				
				\end{tikzpicture}
			\end{center}
		\end{subfigure}
		\label{fig:phaseTASEP}
		\caption{Phase diagram of one species TASEP in terms of controlling parameters on the left, and in terms of the bulk density on the right. The low density (LD) phase can be identified as the Left induced(LI).high density(HD) phase can be identified as (RI). The maximal current phase is identified with a bulk induced phase, with $\rho^{B}=\frac{1}{2}$ that represents a transition point between the two previous phases as illustrated on the right.}
		\label{TASEPopen}
	\end{figure}

	\section{Extremal current principle revisited}
	Driven diffusive systems coupled to reservoirs with a single conserved driven quantity are considered in the literature. Krug \cite{krug1991boundary} first studied a system with concave current expression $J(\rho)$ and coupled to a vanishing density reservoir on the right, and postulated that independently of the microscopic dynamics, the system tries to maximize its current over the interval $[0, \rho^{L}]$. He described a phase transition occurring when $\rho^{R}$ passes through the density for which the current is maximal. Based on insights obtained from the exact solution of TASEP, and KLS \cite{schutz1993phase}, this maximal current principle was generalized by Schütz and others \cite{popkov1999steady}, \cite{hager2001minimal}, \cite{katz_nonequilibrium_1984}  to the extremal current principle where the current expression needs not be concave and the reservoir on the right can be arbitrary, and it was finally proved rigorously in \cite{bahadoran2012hydrodynamics}.

	According to it the current in the system is given by:

	\begin{equation}
	\label{eqn:ecp}
	j =
	\begin{cases}
	\max_{\rho \in [\rho^{R}, \rho^{L}]}(J(\rho))  & \text{if $\rho^{L}>\rho^{R}$} \\
	\min_{\rho \in [\rho^{L}, \rho^{R}]}(J(\rho))  & \text{if $\rho^{L}<\rho^{R}$} \\
	\end{cases}
	\end{equation}
	
	This principle allows sketching a phase diagram that exhibits both first order and second order nonequilibrium phase transitions. The phases are typically named Low-density, high-density, and maximal/minimal current phases. This terminology is more conveniently replaced by Left induced, Right Induced, and bulk-induced phases. Check figure \ref{TASEPopen} for TASEP as an example. The bulk density can be used to uniquely identify the selected steady state of the system. That allows sketching a lower dimensional diagram in terms of the bulk density which will play a more important role for systems with multiple conserved quantities.
	\subsection{The Riemann problem perspective}
	
	Let's show that the extremal principle is compatible with the solution at zero of the corresponding Riemann problem starting with an initial data of densities $\rho^{L}$ on the left and $\rho^{R}$ on the right. We need to consider the two cases:
	\begin{itemize}[leftmargin=0cm]
		\item [] \fbox{$\rho^{L} < \rho^{R}$} This is to be compared with the minimum current branch of the extremal principle. We first assume that $J$ is strictly convex over the interval  $[\rho^{L}, \rho^{R}]$. Its derivative $v(\rho) := \frac{dJ}{d\rho}  $ will then be increasing on that interval so it is reversible on it. The solution to the Riemann problem can be expressed as a function of $\xi = \frac{x}{t}$:

		\begin{equation}\label{key}
		\rho  (\xi) = \rho^{L} \mathds{1}_{\xi< v(\rho^{L})  } + \rho^{R} \mathds{1}_{\xi> v(\rho^{R})} +   v^{-1}(\xi)   \mathds{1}_{ v(\rho^{L}) < \xi < v(\rho^{R})}
		\end{equation}
		
		To compare the solution at zero with density predicted by the minimum current phase, we can identify three situations:
		
		\begin{itemize}
			\item if $ v(\rho^{L}) > 0 $ The solution of the Riemann problem at zero has the value $\rho^{L}$. On the other hand, since $v$ is increasing, it will keep being positive over the interval $ [\rho^{L}, \rho^{R}]$, this ensures $J$ is increasing and the minimum current over the same interval to be attained at $\rho^{L}$. This implies that the bulk density will have the same value: $\rho^{B} = \rho^{L}$. We say that we are in the  \textit{left induced phase}.
			\item if $ v(\rho^{R}) < 0 $. With similar reasoning, we find that $\rho^{R}$ will be simultaneously the density of the bulk and the solution of the Riemann problem at zero. We say that we are in the \textit{right induced phase}.
			\item If neither of the two previous statements is true, then thanks to the monotonicity of $v$, there exists a unique value $ \rho^{*} \in [\rho^{L}, \rho^{R}] $ for which $v(\rho^{*}) = 0$. This value is both the Riemann solution at zero and the value at which the minimum of $J(\rho) $ is attained. This makes it as well the density of the bulk: $\rho^{B} = \rho^{L}$. We say that the system is in the \textit{bulk induced phase} .
		\end{itemize}

		Notice that if $J$ is only broadly convex, which means that it has a linear part, then the inverse of the derivative will have a discontinuity that corresponds to a discontinuity of the Riemann solution, however, the above reasoning will still be valid except if this linear part is a constant, then this discontinuity will be located at zero, and both the extremal current principle and our principle will fail to predict what happens. For the case of TASEP, an analysis based on domain wall dynamics \cite{santen2002asymmetric} predicts the absence of the bulk and a linear profile joining the two boundaries resulting from a symmetric random walk performed by the shock over the lattice.
		
		So far, we have only considered a convex current. The Riemann problem for an arbitrary smooth $J$ was discussed in \cite{osher1983riemann}, the solution at $\xi$ in the regime  $\rho^{L} < \rho^{R} $ should satisfy:
		
		\begin{equation}
		\xi \rho(\xi) - J(\rho(\xi)) = \max_{v \in [\rho^{L}, \rho^{R}]} \{ \xi v - J(v) \}
		\end{equation}
		
		With the help of some elementary geometrical operations, one can get convinced that this solution could be obtained by replacing on the interval $ [\rho^{L}, \rho^{R}]$ the current $J$ by its convex hull defined as:

		\begin{equation}\label{key}
		\breve{J}(\rho) =  \max \{I \; convex, and \; I \leq J \}
		\end{equation}
		
		Since the minimum of the current is the same as its convex hull over the interval, this allows repeating all the analysis mentioned above using $\breve{J}$ instead of $J$.

		\item [] \fbox{$\rho^{L} > \rho^{R}$} We need here to require $J$ to be concave, and if it's not then, it should be replaced by its concave hull on the interval $ [\rho^{R}, \rho^{L}]$  and we can again make the same arguments comparing the solution of the Riemann problem at zero with the bulk density predicted by the maximum current regime of the extremal principle.
		
	\end{itemize}

	For the sake of simplicity, we required $J$ (or its hull) to be smooth. If it is not derivative at some density, then the Riemann solution will have a constant part equal to this density, on an interval determined by the left and right derivatives. If the zero happens to belong to this interval, we get a hybrid phase that is not boundary induced but yet shares its properties in terms of the exponential convergence on the boundaries.
	
	\subsection{Vanishing viscosity approach:}

	We will now review a simple proof of the extremal current principle (\ref{eqn:ecp}) that will as well serve a pedagogical purpose for the multi-species case. 
	
	It is known that solutions for conservation systems are not unique and that physical solutions are obtained in the vanishing viscosity limit, where a diffusive component is the current expression. At the steady state, the total current is the same all over the system and can be written as:
	
	\begin{equation}\label{key}
	J^{total} = J(\rho) - D \frac{\partial \rho}{\partial x} 
	\end{equation}
	
	In general $D$ is a function of $\rho$, but since we are only interested in the limit $D \rightarrow 0$, this dependence will be irrelevant. The behavior of the system can be obtained by analyzing the ODE:
	
	\begin{equation}\label{key}
	\frac{\partial \rho}{\partial x}  = (J(\rho) - J^{total}  )/ D
	\end{equation}
	
	If we are looking for a solution with a trajectory joining $\rho^{L}$ and $\rho^{R}$ then such a trajectory in 1D exists only if the flow of the ODE is oriented from  $\rho^{L}$ to $\rho^{R}$ all over the segment joining them. This means that the solution should always be monotonous.  
	
	If $\rho^{L} < \rho^{R}$ then this means that:
	
	\begin{equation}\label{key}
	J(\rho) - J^{total} \geq 0
	\end{equation}
	Since the zero will be asymptotically attained at the bulk, we recover the minimum current phase of \ref{eqn:ecp}, and in a similar manner, we can find the other phase.

	This ODE has at least one stationary point in the interval between $\rho^{L}$ and $\rho^{R}$. Let's assume it is the only one (this is equivalent to assuming that the hull of the current doesn't have a constant part). The type of this stationary point falls into one of these three categories:

	\begin{itemize}
		
		\item \textbf{Sink}: then the system is in the right induced phase, and we have an exponential decay on the left boundary. One can simply see why it is an exponential decay by linearizing the ODE in the neighborhood of $\rho^{R}$:
		$ \partial_{x} \rho = v(\rho^{R}) (\rho- \rho^{R})/D $ whose solution is:

		\begin{equation}\label{key}
		\rho(x) = \rho^{R} + (\rho^{L} - \rho^{R}) e^{\frac{v(\rho^{R})}{D}(x-x_{L})}
		\end{equation}
		
		Of course $v(\rho^{R})  < 0$, so $\lim_{D \rightarrow 0}(\rho(x_{R})) = \rho^{R}$ and  $ v(\rho^{B}) <0 $
		
		\item \textbf{Source}: this corresponds to a left induced phase, and we have again an exponential decay as previously, but $ v(\rho^{B}) > 0$
		
		\item \textbf{Second order singularity}: The derivative of the current is zero at the stationary point. We have a power law decay to the bulk, it's easy to show that: let $n>1$ be the first order for which the derivative of the current at the bulk is non-zero. This order has to be even. We expand the flow to this order : $\rho^{'} =J^{(n)}(\rho^{B}) (\rho- \rho^{B})^{n}/D n! $ that admits a decay as:
		\begin{equation}\label{key}
		\rho(x) \sim \rho^{B} + \sqrt[n-1]{\frac{D n!}{(1-n) J^{(n)}(\rho^{B})} \frac{1}{x}}
		\end{equation}

		Note that strictly speaking, The ODE does not admit a trajectory that goes from one side to the other of a second-order stationary point. A proper description of the system in this phase requires adding a stochastic noise term to the equation. We can say that the noise allows the trajectory to jump over the stationary point.

		\subsection*{Remark}
		The behavior of $\rho^{B}$ is associated with what class $v(\rho^{B})$ belongs to within the set: $\{-,0,+\}$. This association is consistently identical to the behavior of $\rho|_{0}$ with $v(\rho|_{0})$. We will see how this idea will be generalized in the multi-species case by applying it to the Riemann variables instead.

	\end{itemize}

	\section{The case of a multi-species driven diffusive system}

	The non trivial case is when we have $n>1$ coupled densities. We can rewrite the conservation laws \ref{eqn:con}:
	
	\begin{equation}\label{eqn:c}
	\partial_{t}\boldsymbol{\rho} +  V \partial_{x}\boldsymbol{\rho} = 0
	\end{equation}
	Where $V_{ij} = \frac{\partial J_{i}}{\partial \rho_{j}}$. Let's assume that the Riemann variables exist (this is always the case for $n=2$), These variables are defined as the transformation of the densities:
	
	\begin{equation}
	\boldsymbol{\rho} \in \mathcal{D}_{\rho} \rightarrow \boldsymbol{z} \in \mathcal{D}_{z} \subset \mathbb{R}^{n}
	\end{equation}
	
	Such that if the conservation laws are written in terms of these variables, the matrix $V$ will become diagonal: 
	\begin{equation}\label{key}
	\partial_{t}z_{i} +  v_{i}(\boldsymbol{z}) \partial_{x}z_{i} = 0 \quad \quad  1 \leq i \leq n
	\end{equation}

	Where $ v_{i}(\boldsymbol{z}) = \frac{\partial J_{k}}{\partial z_{i}} / \frac{\partial \rho_{k}}{\partial z_{i}} $. For more details, look at the annex.
	Let's first consider the Riemann problem with the initial condition $ z(x) = \boldsymbol{z}^{L} \mathds{1}_{x<0} + z^{R} \mathds{1}_{x>0}$. Let $\boldsymbol{z}|_{0}$ be the solution at zero.
	Let's sketch a phase diagram in the  $\boldsymbol{z}|_{0}$ space. Consider first in this space the hyper-surfaces defined by $v_{i}(\boldsymbol{z}) = 0$, it will partition $ \mathcal{D}_{z} $ into three regions:  $ \mathcal{D}_{z} $ = $\{ \boldsymbol{z}: v_{i}(\boldsymbol{z}) > 0\} \cup  \{ \boldsymbol{z}: v_{i}(\boldsymbol{z}) < 0\} \cup  \{ \boldsymbol{z}: v_{i}(\boldsymbol{z}) = 0\} $.
	
	Each of these three regions defines the behavior of $z_{i}|_{0}$

	\begin{itemize}
		\item $v_{i}(\boldsymbol{z}|_{0}) > 0$ that leads to $z_{i}|_{0} = z_{i}^{L}$, so $z_{i}|_{0}$ is left induced.
		\item $v_{i}(\boldsymbol{z}|_{0}) < 0$ that leads to $z_{i}|_{0} = z_{i}^{R}$, so $z_{i}|_{0}$ is right induced.
		\item$v_{i}(\boldsymbol{z}|_{0}) = 0$ and we say that $z_{i}|_{0}$ is bulk induced.
	\end{itemize}

	If we keep partitioning the $ \boldsymbol{z} $-space for each of the $z_{i}$, we end up in general with $ 3^{n} $ regions, each is defined by a choice for each of the $v_{i} \in \{+,0,-\}$. However, in a strictly hyperbolic system, the eigenvalues are strictly ordered, which adds a restriction on these choices, forbidding some of the phases.

	Let's now move to the open boundary problem.
	The total current of the particles of type $i$ can be written as:
	\begin{equation}\label{key}
	J^{total}_{i} = J_{i}(\boldsymbol{z}) - D_{i} \frac{\partial \rho_{i}}{\partial x}
	\end{equation}
	
	Where $D_{i}>0$. We assumed here that the diffusive component resulting from the gradient of the other types of particles is negligible compared to the one from the same type.
	
	Let's rewrite the previous equation as:
	\begin{equation}\label{key}
	\frac{\partial \boldsymbol{z}}{\partial x}  = M^{-1}D^{-1}(J(\boldsymbol{z}) - J^{total}) := F(\boldsymbol{z})
	\end{equation}

	Where $ M_{ij} = \frac{\partial \rho_{i}}{\partial z_{j}} $, $D$ is a diagonal matrix $D_{ii} = D_{i}$.
	
	To obtain the phase diagram, one has to analyze this ODE. Once again, we will use the bulk $\boldmath$ variables as order parameters for the phase diagram. First let's notice that: $F(\boldsymbol{z}^{B}) = 0$. 
	
	The bulk is a stationary point for the ODE. The phase will be determined by the type of this stationary point, i.e. the signs of the eigenvalues of the Jacobin $ \frac{\partial F_{i}}{\partial z_{j}} (\boldsymbol{z}^{B}) $ 
	From the properties of the Riemann variables one can show that this Jacobin is a diagonal matrix in the bulk and is simply: 
	
	\begin{equation}\label{key}
	\frac{\partial F_{i}}{\partial z_{j}} (\boldsymbol{z}^{B}) = D_{i}^{-1} v_{i} \delta_{ij}
	\end{equation}

	So the phase diagram is again governed by the set $ {v_{i}} $
	
	We can have:
	
	\begin{itemize}
		\item \textbf{A Sink} if all the $v_{i}$ are negative, this means that the bulk is driven from right
		\item \textbf{A source} if all the $v_{i}$ are positive, this means that the bulk is driven from left.
		\item \textbf{A Saddle point} if some $v_{i}$ are negative and some are positive, this means that the bulk is mixed-driven, each $z_{i}$ will be driven according to the sign of the corresponding $v_{i}$
		
		\item \textbf{Second order singularity} if some $v_{i}$ are zero. The bulk will belong to the intersection of the manifolds $v_{i} = 0$
	\end{itemize}

	We conclude that  $ \boldsymbol{z}|_{0} $ and $ \boldsymbol{z}^{B}$  This doesn't constitute proof of our principle but rather a self-consistency test.
	
	\subsection{Proof of the principle}
	Consider the conservation laws defined on a half-space $\mathbb{R^{+}}$ with a single boundary condition located at the origin:
	\begin{equation}
	\boldsymbol{\rho}  (0,t) =\boldsymbol{\rho}^{L} \quad t>0
	\end{equation}
	The set of admissible limit values at this boundary in the zero-viscosity limit are the one that verify a boundary entropy inequality. \cite{bardos1979first} \cite{dubois1988boundary} \cite{mazet1986analyse}. Namely this set is:
	
	\begin{equation}
	\begin{split}
	E^{+}(\boldsymbol{\rho}^{L}) = \{ \boldsymbol{\rho} \in \mathbb{R}^{n}, q(\boldsymbol{\rho}) - q(\boldsymbol{\rho}^{L}) - D\eta(\boldsymbol{\rho}^{L}) (f(\boldsymbol{\rho})-f(\boldsymbol{\rho}^{L}) ) \leq 0, \quad \\  \forall  (\eta,q) \; \text{pair of entropy-flux}\}
	\end{split}
	\end{equation}

	For our open boundary problem, this set is as well the set of admissible bulk densities for a given left boundary condition.
	
	On the other hand, let's consider the set of all possible values of the solution Riemann problem at zero with a fixed left density:
	\begin{equation}
	V^{+}(\boldsymbol{\rho}^{L}) = \{ R(0 ; \boldsymbol{\rho}^{L}, \boldsymbol{\rho}^{R} ) , \; \boldsymbol{\rho}^{R} \; \text{Varing in} \; \mathbb{R}^{n}  \}
	\end{equation}

	It has been shown in \cite{dubois1988boundary} that for strictly hyperbolic systems, the two previous sets are equal:
	\begin{equation}
	E^{+}(\boldsymbol{\rho}^{L}) = V^{+}(\boldsymbol{\rho}^{L}) \quad \forall \boldsymbol{\rho}^{L} \in \mathbb{R}^{n}
	\end{equation}

	We can now obviously formulate this property for a system with a system of right boundary conditions:
	
	\begin{equation}
	E^{-}(\boldsymbol{\rho}^{R}) = V^{-}(\boldsymbol{\rho}^{R}) \quad \forall \boldsymbol{\rho}^{R} \in \mathbb{R}^{n}
	\end{equation}
	
	In our system with two boundary conditions, in the steady state, the bulk belongs to the intersection:
	
	\begin{equation}
	\boldsymbol{\rho}^{B} \in V^{+}(\boldsymbol{\rho}^{L})  \cap V^{-}(\boldsymbol{\rho}^{R}) 
	\end{equation}

	All what is left is to show that this intersection has only this unique element. This can be shown by a simple argumentum ad absurdum.

	\section{Applications}

	\subsection{Open boundaries 2-TASEP with arbitrary hopping rates}
	
	This model is a multi-species generalization of TASEP. it consists of two types of particles in addition to the void. The hopping rates in the bulk and on the boundaries are:

	\begin{center}
		
		\setlength{\tabcolsep}{10pt} 
		\renewcommand{\arraystretch}{1.5} 
		\begin{tabular}{l|c|r|c} 
			\textbf{Hopping} & \textbf{Left} & \textbf{Bulk} & \textbf{Right}\\
			\hline
			\hline
			$ \bullet\,\ast \rightarrow  \ast\,\bullet $ & $ \nu_{\bullet \ast}^{L} $ & $ \beta $ &  $ \nu_{\bullet \ast}^{R} $\\
			$ \ast\,\circ \rightarrow  \circ\,\ast $ & $ \nu_{\ast \circ}^{L} $ & $ \alpha $ &  $ \nu_{\ast \circ}^{R} $\\
			$ \bullet\,\circ \rightarrow  \circ \,\bullet $ & $ \nu_{\bullet \circ}^{L} $ & $ 1 $ &  $ \nu_{\bullet \circ}^{R} $\\
		\end{tabular}
		
	\end{center}

	The currents for this model with periodic boundary conditions were calculated in \cite{cantini2008algebraic}. These currents were used in \cite{cantini2022hydrodynamic} to study its hydrodynamic behavior and in particular to solve the corresponding Riemann problem.
	
	Let's restate the expression of the currents:
	
	\begin{gather}
	J_\circ= z_\alpha(z_\beta-1)+\rho_\circ(z_\alpha-z_\beta)\\
	\label{1Jrz}
	J_\bullet= z_\beta(1-z_\alpha)+\rho_\bullet(z_\alpha-z_\beta)
	\end{gather}

	where with $z_\alpha \in [0,\min(1,\alpha)]$ and $z_\beta \in [0,\min(1,\beta)]$ are solution of the saddle point equations
	\begin{gather}
	\frac{\rho_\circ}{z_\alpha}+\frac{\rho_\bullet}{z_\alpha-1}+\frac{1-\rho_\circ-\rho_\bullet}{z_\alpha-\alpha}=0\\ \label{chang-var2}
	\frac{\rho_\bullet}{z_\beta}+\frac{\rho_\circ}{z_\beta-1}+\frac{1-\rho_\circ-\rho_\bullet}{z_\beta-\beta}=0.
	\end{gather}

	The variables $z_{\alpha}, z_{\beta}$ happen to be the as well the Riemann variables. \cite{cantini2022hydrodynamic} for full details.

	In the one species TASEP, the density on each boundary is uniquely determined by the hopping rate at that boundary. In that sense the boundaries are independent. This is still true for a colorable two-species TASEP with equal hopping rates in the bulk. However, strong numerical evidence suggests that this stop being true for the general case, like our case. The boundaries become coupled and one has to solve simultaneously the boundaries and the bulk:
	
	\begin{equation}\label{key}
	(\boldsymbol{\nu}^{L}, \boldsymbol{\nu}^{R}) \longrightarrow  (\boldsymbol{\rho}^{L}, \boldsymbol{\rho}^{B}, \boldsymbol{\rho}^{R} )
	\end{equation}

	We provide two approaches to perform that, an iterative one and a direct one.
	
	\subsubsection{The iterative approach}

	%

	\begin{figure}[h!]
		\centering
		\begin{subfigure}[b]{0.4\linewidth}
			\includegraphics[width=\linewidth]{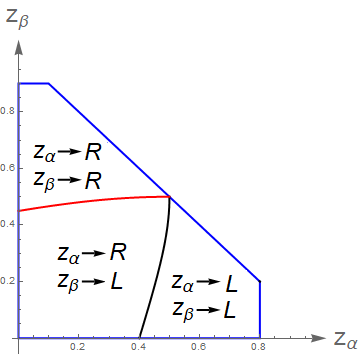}
			
		\end{subfigure}
		\begin{subfigure}[b]{0.4\linewidth}
			\includegraphics[width=\linewidth]{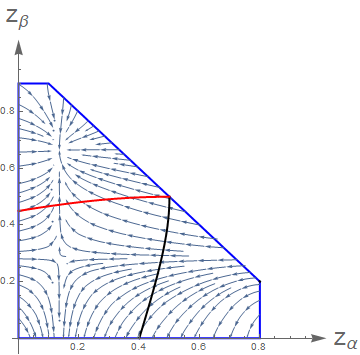}
			
		\end{subfigure}
		\caption{Phase diagram of 2-species TASEP with open boundaries on the left. an example of the ODE flow with a sink singularity in the left induced phase, and a saddle point in a mixed induced one. $v_{\alpha} = 0$ in Red,  $v_{\beta} = 0$ in black.}
		\label{fig:coffee3}
	\end{figure}

	Let's write the the currents on the left and on the right boundaries:
	
	\begin{equation}
	\begin{split}
	& J_{\bullet}^{L} = \nu_{\bullet \circ}^{L} \rho_{\circ}^{L} + \nu_{\bullet \ast}^{L} (1- \rho_{\circ}^{L} - \rho_{\bullet}^{L} ) \\
	& J_{\circ}^{L} = -(\nu_{\bullet \circ}^{L} + \nu_{\ast \circ}^{L}) \rho_{\circ}^{L} \\
	& J_{\circ}^{R} = -\nu_{\bullet \circ}^{R} \rho_{\bullet}^{R} - \nu_{\ast \circ}^{R} (1- \rho_{\circ}^{R} - \rho_{\bullet}^{R} ) \\
	& J_{\bullet}^{R} = (\nu_{\bullet \circ}^{R} + \nu_{\bullet \ast}^{R}) \rho_{\bullet }^{R} \\
	\end{split}
	\end{equation}
	
	Of course in the steady state, the currents are uniform all over the system, so we can write:
	\begin{equation}
	\boldsymbol{J}^{L}(\boldsymbol{\rho}^{L}) = \boldsymbol{J}(\boldsymbol{\rho}^{B})= \boldsymbol{J}^{R}(\boldsymbol{\rho}^{R})
	\end{equation}
	Where $\boldsymbol{J}(\boldsymbol{\rho}^{B})$ is the current in the bulk, a known function of the bulk densities.
	
	We have here 4 equations with 6 variables(the densities on the left, right and bulk).
	We can add to them two consistency equations resulting from principle:
	\begin{equation}\label{key}
	(\boldsymbol{\rho}^{L},  \boldsymbol{\rho}^{R} ) \xrightarrow[]{\text{Riemann}} \boldsymbol{\rho}^{B}
	\end{equation}

\begin{figure}[h!]
		\centering
		\begin{subfigure}[b]{0.49\linewidth}
			\includegraphics[width=\linewidth]{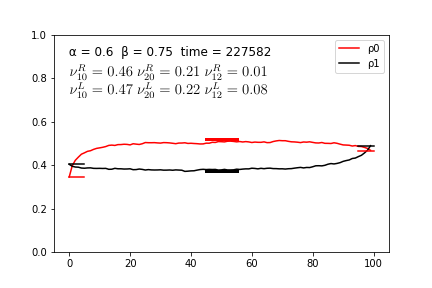}
			
		\end{subfigure}
		\begin{subfigure}[b]{0.49\linewidth}
			\includegraphics[width=\linewidth]{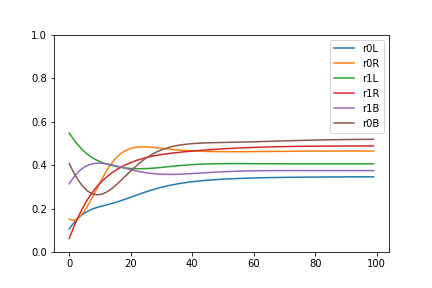}
			
		\end{subfigure}
		\begin{subfigure}[b]{0.49\linewidth}
			\includegraphics[width=\linewidth]{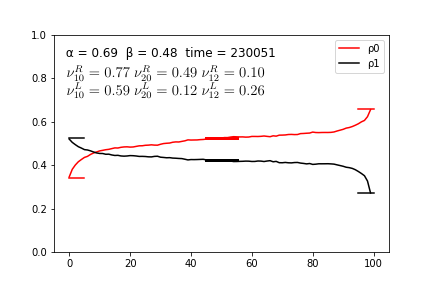}
			\caption{Densities profile. The horizontal segments represent the predicted values }
		\end{subfigure}
		\begin{subfigure}[b]{0.49\linewidth}
			\includegraphics[width=\linewidth]{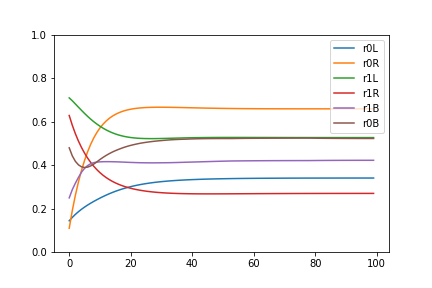}
			\caption{Iterative evolution of the different densities, with damping $\gamma = 0.1$}
		\end{subfigure}
		\caption{Two examples of the application of the algorithm for finding the densities on the boundaries and in the bulk of the 2-species TASEP}
		\label{fig:densitiesopen}
	\end{figure}

	This provides in principle a closed system of 6 equations, that one can solve them using an iterative method: we choose random initial densities for the boundaries, then we find the bulk density and then calculate the boundary densities, and we continue the iteration between the boundaries and the bulk till convergence. This algorithm applied in such a way can get stuck in cyclic trajectories, but this can simply be avoided by introducing some damping. i.e. a sufficiently small parameter $\gamma$ such that:
	$\boldsymbol{x}^{n+1} = \gamma \boldsymbol{f}(\boldsymbol{x}^{n}) + (1 - \gamma) \boldsymbol{x}^{n}  $ where  $ \boldsymbol{x} $ is the nth iteration of the variables and  $ \boldsymbol{f} $ is the set of functions governing the iterations. This method for this model gives results in good agreement with simulations. Examples figure \ref{fig:densitiesopen}
	
	\subsubsection*{Remark}
	One has to make sure that the variables don't leave their physical domain. 
	
	If the initial condition lies outside the basin of attraction of the fixed point, then the algorithm will not converge to the expected values. This has been observed in a marginal fraction of tests and required simply repetition with different initial values.

	\subsubsection{An equivalent approach}
	We have seen that in there is a finite number of possible scenarios for the bulk in terms of the Riemann variables. One can find the variables in the bulk that are compatible with each of these scenarios.

	\begin{itemize}
		\item \textbf{Left driven solution}: it can be obtained by solving two equations with two variables:
		\begin{equation}\label{key}
		\boldsymbol{J}^{L}(\boldsymbol{z}) = \boldsymbol{J}(\boldsymbol{z})
		\end{equation}
		Where $\boldsymbol{z}$ are the Riemann variable both in the bulk and on the left.
		
		\item A right driven solution: similarly, by solving the same two equations but using $\boldsymbol{J}^{R}$ instead of $\boldsymbol{J}^{L}$
		
		\item Mixed driven: $z_{\alpha}$ is driven from right, and $z_{\beta}$ is driven from the left. We need to solve four equations with four variables:
		\begin{equation}\label{key}
		\boldsymbol{J}^{L}(z_{\alpha}^{L},z_{\beta}) = \boldsymbol{J}(z_{\alpha},z_{\beta}) = \boldsymbol{J}^{R}(z_{\alpha},z_{\beta}^{R})
		\end{equation}

		\item Bulk driven for $z_{\alpha}$ and left driven for $z_{\beta}$: We have to solve three equations with three variables:
		\begin{equation}
		\begin{split}
		& v_{0}(z_{\alpha}, z_{\beta}) = 0 \\
		& \boldsymbol{J}^{L}(z_{\alpha}^{L}, z_{\beta}) = \boldsymbol{J}(z_{\alpha}, z_{\beta}) \\
		\end{split}
		\end{equation}
		
		\item Bulk driven for $z_{\beta}$ and right driven for $ z_{\alpha} $: similar to the previous case.
		\item Bulk driven for both. One point is possible here for the bulk:
		\begin{equation}
		(z_{\alpha}, z_{\beta}) = (\frac{1}{2}, \frac{1}{2})
		\end{equation}

	\end{itemize}

\begin{figure}[h!]
		\centering
		\begin{subfigure}[b]{0.49\linewidth}
			\includegraphics[width=\linewidth]{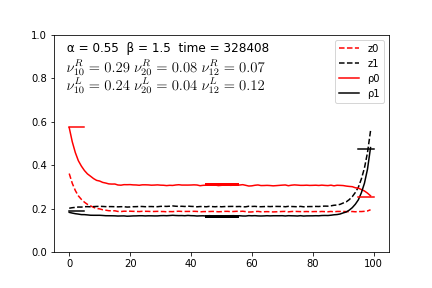}
			
		\end{subfigure}
		\begin{subfigure}[b]{0.49\linewidth}
			\includegraphics[width=\linewidth]{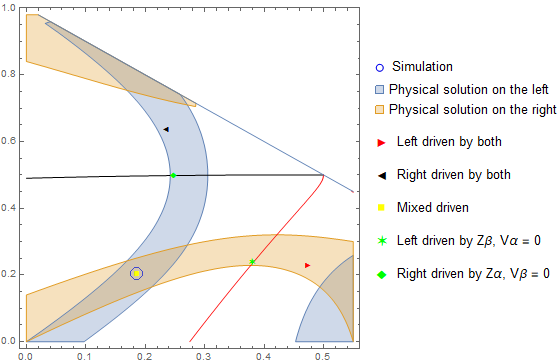}
			
		\end{subfigure}
		\begin{subfigure}[b]{0.49\linewidth}
			\includegraphics[width=\linewidth]{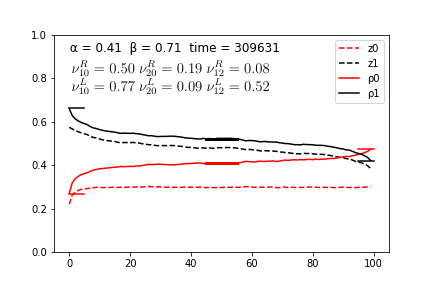}
			\caption{Simulation of the density profile. The horizontal segments represent predicted values.}
		\end{subfigure}
		\begin{subfigure}[b]{0.49\linewidth}
			\includegraphics[width=\linewidth]{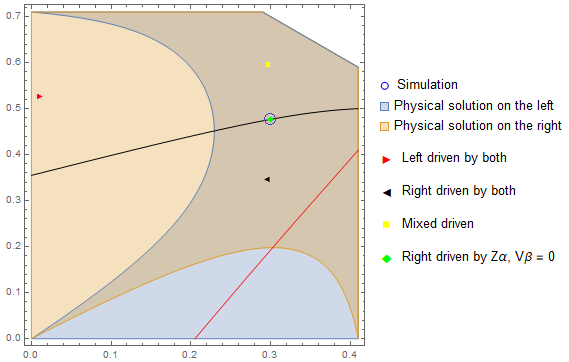}
			\caption{Possible solutions in the bulk. The blue (red) region represents the bulk z variables compatible with physical densities on the left (right)}
		\end{subfigure}
		\caption{Two examples for the application of the equivalent approach for solving the 2-species Tasep with open boundaries }
		\label{fig:densitiesequivalent}
	\end{figure}

Figure \ref{fig:densitiesequivalent} illustrate how this approach is applied to 2-TASEP.

\subsection{Limits of the method and open questions}

When repeated for thousands of realizations for different random boundary rates, the method gives accurate results for approximately 95\% of realizations. For the rest, we observe some miss-match with the simulations. We are still investigating into this. A few hypothesis are possible: It might be an effect of a non-diagonal elements of diffusion matrix similarly to the one observed for the model considered in \cite{popkov2004hydrodynamic} \cite{popkov2004infinite} However, it's a bit strange that these non-diagonal terms affect only a small minority of realizations.
another potential hypothesis is a failure of the hydrodynamic when getting too close from the singular points of the model. In deed, we observed that this miss-match often happen when the bulk is too close of a singularity. We count to check if by any change the break of integrability has any role to play in addition to the break of the product measure. It's as well worth investigating whether this miss-match is induced by a spontaneous symmetry breaking as the one observed in \cite{evans1995spontaneous}. Finally, We think that models with Temple class hydrodynamics have a particular status. The models we considered are indeed Temple class and same is true for the model defined in \cite{popkov2003shocks} and considered with open boundaries in \cite{popkov2004hydrodynamic} \cite{popkov2004infinite}. It would be really interesting to test these methods on examples that are not Temple class.

\chapter{Effect of a single second class particle}
\label{effect}
\section{Introduction}

Second-class particles with unity hopping rates have played a significant role in understanding the exclusion process. There was no shortage of motivation behind introducing them. They can be perceived as tracer particles, if one is added to a position $x$, it follows the characteristics of Burgers equation emanating from $x$ \cite{ferrari1992shocks, rezakhanlou1995microscopic}. In other words, it follows a trajectory such that its surrounding density field is constant. In a shock, where different characteristics join, the second-class particle gets stuck which provides a possible microscopic definition for the position of the shock. \cite{andjel1988shocks, derrida1993exact}. Studying the behavior of second-class particles gives insights into the propagation of an excess of mass perturbation for burgers equation \cite{van1991fluctuations, ferrari1991microscopic}. A rather fascinating property of a second-class particle is that if it is added to a point from where characteristics emerge (i.e. a decreasing discontinuity), then it will choose one of the available characteristics at random with a uniform distribution, in other words, it will pick up uniformly an asymptotic speed within the possible ones, and stick to it \cite{ferrari1995second}. second-class particles provide as well a microscopic way of describing density fluctuations for the Burgers equation \cite{van1991fluctuations}.

Since second-class particles with unity rates are seen as void by the first-class particles, they have no impact on the surrounding density field. The picture is drastically different for second-class particles with arbitrary rates: an interplay between the behavior of the particle and the surrounding density is observed and studied. These particles, besides their theoretical importance in extending the ones of unity rates, can have a direct interpretation as defects in transport models. We will be referring sometimes to a second-class particle with arbitrary rates as a defect. They represent the main interest of this chapter.

We start by reviewing some basic properties for TASEP and unity second-class particles. These are mainly the ones that we will need for the following sections. 
In section \ref{A defect in a step initial profile}, we treat rather heuristically the interaction between the second-class particle with the density field on the line arising from a step initial condition, where new phenomenology is reported and analyzed.
In section \ref{Speed process of a defect in a step initial configuration}, we prove rather rigorously some properties of the asymptotic speed distribution of a second-class particle in a fan.

\subsection{Notations}
Consider TASEP on $ \mathbb{Z} $. The space of configurations is $\{0,1\}^{Z}$ with elements $\eta := (\eta(k))_{k\in \mathbb{Z} }$.
The generator of the process formally reads:

\begin{equation}
A g(\eta) = \sum_{i \in \mathbb{Z}} (g(\tau_{i,i+1} \eta) - g(\eta)) \eta(i)(1-\eta(i+1))
\end{equation}

where $g$ is a bounded real function on the configuration space, $\tau_{i,i+1} \eta$ is the configuration obtained from $\eta$, by swapping the values of sites $i$ and $i+1$.
If $\eta_{0}$ is the initial configuration, we call $\eta_{t}$ the time evolved configuration according to this dynamics.

\subsection{Invariant measure for TASEP}
The first reflex when dealing with a Markov process is to question the existence of stationary states. This is a probability measure over the configuration space that will remain constant under time evolution. Visually, if one imagines $N$ copies of the system populated with configurations chosen according to this measure, then the proportions of populations stay the same in the infinite $N$ limit.

\begin{lem}[Spitzer]
	For $0 \leq \rho \leq 1$, let $\nu_{\rho}$ be the product measure on the configuration space with uniform Bernoulli marginals of density $\rho$, ie. $\nu_{\rho}(\eta(x) = 1) = \rho$. then $\nu_{\rho}$ is an invariant measure for TASEP.
\end{lem}

\textit{Sketch Proof}: it is possible to understand this property by making use of some elementary results from queuing theory. Each particle can be thought of as a server, and each void as a customer. When a particle jumps, a customer is served and sent to the next queue.
At the initial time, the length of queues follows a geometrical distribution with parameter $\rho$. The probability that a given queue has $i$ customers is $ (1-\rho)^{i}\rho$. This is identical to the stationary state of a queue with arrivals following a Poisson process of intensity $1-\rho$. Since the server is serving as well with Poisson times at intensity 1, according to Burke’s theory, the serving will be at rate $1-\rho$, which is the rate of arrivals for the next server. Thus, this steady state will be maintained.
\qed
\begin{coro}
	A tagged particle in TASEP on the line with $\rho$ uniform density jumps as a free particle with waiting times determined by a Poisson process of the rate of $1-\rho$.
\end{coro}

\textbf{Remark:}
while the invariant measure for TASEP is a product measure, the measure is not any more a product once more species are introduced, however Angel \cite{angel2006stationary} showed that the stationary measure for 2-TASEP can understood as a collapse process of two independent processes with a uniform product measure. This picture was interpreted by Ferrari and Martin \cite{ferrari2007stationary} in terms of a queuing process with discrete times besides being generated to arbitrary number of species. Based on these ideas, Evans, Ferrari and Mallick constructed the stationary measure for N-species TASEP in a matrix product formulation. These ideas stand for the unity rates of the additional species. It would be interesting to investigate the potential generalization to arbitrary rates.

\subsection{Convergence of density field:}
Let $u(x,t)$ be the entropy solution of Burgers equation with initial data $u_{0}(x)$ and for $\epsilon > 0$, consider TASEP with the initial product measure of marginals: $ \nu_{0}^{\epsilon}( \eta_{0}^{\epsilon} (x)=1 ) = u_{0}(\epsilon x) $

Then we have:

\begin{equation}
\lim_{\epsilon \rightarrow 0} E( \eta_{t/\epsilon}^{\epsilon}([x/ \epsilon]) ) 
= u(x,t)  \quad a.e
\end{equation}

One way to visualize this is to imagine a lattice with constant $\epsilon$, initiated according to $u_{0}$ and evolving with a scaled accelerated time $t/\epsilon$.

For a step initial data $u_{0}(\epsilon x)$ does not depend on $\epsilon$, and so it is possible to formulate the previous statement as a long time limit shape:
\begin{equation}
\lim_{t \rightarrow \infty} E( \eta^{t}([v t]) ) 
= u(v,1)  \quad a.e
\end{equation}

with an  initial measure having constant $\rho$ density on the left and $\lambda$ on the right, $\nu_{\rho, \lambda}(\eta(x) = 1) =  \rho \mathbf{1}_{x<0} + \lambda \mathbf{1}_{x\geq0} $.

This limit was first proven for a particular case of a decreasing step initial condition by Rost \cite{rost1981non}, using subadditive ergodic theory, and then generalized to arbitrary initial condition by many others: Seppäläinen \cite{seppalainen1996hydrodynamic,seppalainen1999existence,seppalainen2008translation}, Benassi and Fouque \cite{benassi1987hydrodynamical}, Fouque, Saada, and Vares \cite{benassi1991asymmetric} Andjel and Vares \cite{andjel1987hydrodynamic}. It can be insightful to state the original statement expressed by Rost: Let $N(v_{1},v_{2},t)$ be the number of particles between $v_{1}t$ and $v_{2}t$ at time t, then

\begin{equation}
\lim_{t \rightarrow \infty}	\frac{N(v_{1},v_{2},t)}{t} = \int_{v_{1}}^{v_{2}} u(v,1) dv \quad a.s
\end{equation}

\textbf{Local equilibrium}
We can notice that the only possible invariant measures are the uniform product measure and measures producing a frozen system, i.e. all sites are full starting from some site. However, it is possible to formulate a local invariant measure, in the following sense: Let $A$ be a finite set in $\mathbb{Z}$, then:

\begin{equation}
\lim_{\epsilon \rightarrow 0} 
E(
\prod_{x \in A}
\eta_{t/\epsilon}^{\epsilon}([x   +  r/\epsilon  ]) ) 
= u(r,t)^{|A|}  \quad a.s
\end{equation}

where $|A|$ is the cardinal of $A$. So locally around the position $r/\epsilon$ at time $t/\epsilon$ ,the measure is approximately uniform product measure with density $u(r,t)$.

For a Riemann initial condition, it is possible to express the local equilibrium in an even more intuitive manner. If one places itself in a moving frame of reference of velocity $v$, then the observer sees the system converge to a uniform measure of density $u(v,1)$, after a long time.

In a more complex system of particles, the invariant measure corresponding to a constant density might not be a product measure, but it is still possible to express the local equilibrium in the sense that the system converges locally around a particular speed to that invariant measure.

\subsection{Harris graphical representation}
One useful way to visualize the time evolution of TASEP, which was introduced by Harris \cite{harris1978additive}, is to imagine a Poisson clock attached to each site of the lattice. The set of clocks define a Poisson point process $\omega$ on $\mathbb{Z} \times \mathbb{R}^{+}$ with rate 1. Each point $(n,t) \in w$ is represented by an arrow $(n,t) \rightarrow (n+1,t)$ figure.

\begin{figure}[h!]
	\centering
	\begin{subfigure}[b]{0.49\textwidth}
		\centering
		
		\begin{tikzpicture}[scale = 0.8]
		\draw[very thick] (0.5,5) -- (9.5,5) node[anchor=south] {$t=0$};
		
		\foreach \i in {1,...,9}
		{
			\draw (\i,0) -- (\i,5);
		}
		
		\draw[very thick] (0.5,0) -- (9.5,0) node[anchor=north] {$t=T$}; ;
		
		\draw [->] (1,2) -- (2,2);
		\draw [->] (1,4) -- (2,4);
		
		\draw [->] (2,3.2) -- (3,3.2);
		\draw [->] (2,0.9) -- (3,0.9);
		
		\draw [->] (3,4.4) -- (4,4.4);
		\draw [->] (3,1.2) -- (4,1.2);
		\draw [->] (3,0.3) -- (4,0.3);
		
		\draw [->] (4,2.2) -- (5,2.2);
		
		\draw [->] (5,3.5) -- (6,3.5);
		\draw [->] (5,1.9) -- (6,1.9);
		
		\draw [->] (6,4.5) -- (7,4.5);
		\draw [->] (6,2.7) -- (7,2.7);
		\draw [->] (6,0.4) -- (7,0.4);
		
		\draw [->] (7,4.1) -- (8,4.1);
		\draw [->] (7,1.4) -- (8,1.4);
		
		\draw [->] (8,2.4) -- (9,2.4);
		
		\draw[black,fill=black] (1,5) circle (.5ex);
		\draw[black,fill=black] (2,5) circle (.5ex);
		
		\draw[black,fill=black] (5,5) circle (.5ex);
		\draw[black,fill=black] (6,5) circle (.5ex);
		\draw[black,fill=black] (7,5) circle (.5ex);
		\end{tikzpicture}

		\caption{}
		\label{fig:y equals x}
	\end{subfigure}
	\hfill
	\begin{subfigure}[b]{0.49\textwidth}
		\centering
		
		\begin{tikzpicture}[scale = 0.8]
		\draw[very thick] (0.5,5) -- (9.5,5) node[anchor=south] {$t=0$};
		\draw[very thick] (0.5,5) -- (9.5,5) node[anchor=south] {};
		
		\foreach \i in {1,...,9}
		{
			\draw (\i,0) -- (\i,5);
		}
		
		\draw[very thick] (0.5,0) -- (9.5,0) node[anchor=north] {$t=T$}; ;
		\draw [->] (1,2) -- (2,2);
		\draw [->] (1,4) -- (2,4);
		
		\draw [->] (2,3.2) -- (3,3.2);
		\draw [->] (2,0.9) -- (3,0.9);
		
		\draw [->] (3,4.4) -- (4,4.4);
		\draw [->] (3,1.2) -- (4,1.2);
		\draw [->] (3,0.3) -- (4,0.3);
		
		\draw [->] (4,2.2) -- (5,2.2);
		
		\draw [->] (5,3.5) -- (6,3.5);
		\draw [->] (5,1.9) -- (6,1.9);
		
		\draw [->] (6,4.5) -- (7,4.5);
		\draw [->] (6,2.7) -- (7,2.7);
		\draw [->] (6,0.4) -- (7,0.4);
		
		\draw [->] (7,4.1) -- (8,4.1);
		\draw [->] (7,1.4) -- (8,1.4);
		
		\draw [->] (8,2.4) -- (9,2.4);
		
		\draw[black,fill=black] (1,5) circle (.5ex);
		\draw[black,fill=black] (2,5) circle (.5ex);
		
		\draw[black,fill=black] (5,5) circle (.5ex);
		\draw[black,fill=black] (6,5) circle (.5ex);
		\draw[black,fill=black] (7,5) circle (.5ex);
		
		\draw [line width=0.8mm] (7,5) -- (7,4.1);
		\draw [line width=0.8mm] (8,4.1) -- (8,2.4);
		\draw [line width=0.8mm] (9,2.4) -- (9,0);
		\draw [line width=0.8mm] (8,2.4) -- (9,2.4);
		\draw [line width=0.8mm] (7,4.1) -- (8,4.1);
		
		\draw [line width=0.8mm] (6,5) -- (6,2.7) -- (7,2.7) -- (7,1.4) -- (8,1.4) -- (8,0) ;
		
		\draw [line width=0.8mm] (5,5) -- (5,1.9) -- (6,1.9) -- (6,0) ;

		\draw [line width=0.8mm]  (2,5) -- (2,3.2) -- (3,3.2) --  (3,1.2) -- (4,1.2) -- (4,0);
		
		\draw [line width=0.8mm] (1,5) -- (1,2) -- (2,2) -- (2,0.9) -- (3,0.9) -- (3,0);
		\end{tikzpicture}
		
		\caption{}
		\label{fig:five over x}
	\end{subfigure}
	\caption{Harris graphical construction}
	\label{fig:three graphs}
\end{figure}
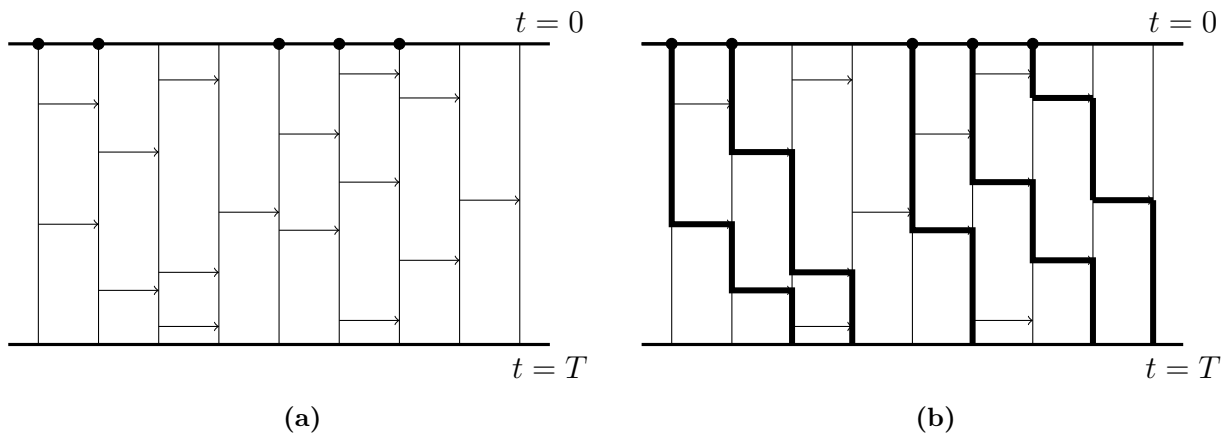

Particles follow a vertical path and try to pass through arrows whenever the next path is empty, figure \ref{fig:three graphs} 

A problem might appear in this definition since the concept of the next particle that will jump can be ill-defined whenever there is an infinite sequence of arrows converging to a point of time, and this will happen with a  probability 1  for each moment. However, for any finite time interval, with probability 1, $\mathbb{Z}$ will be partitioned into finite intervals with no arrows connecting the corresponding blocks, which makes the construction well defined.

\subsection{Basic tool: Coupling}
Coupling is a basic proof tool in probability theory in general and a very frequently used technique for systems of interacting particles, for which it was introduced by Liggett \cite{liggett1976coupling} \cite{liggett1985interacting}. The basic idea consists of conceiving a common realization of two or more random processes in such a way that each process independently does not "feel" any difference from it natural time evolution. Formally, this amounts to the construction of a joint process $(X_{t},Y_{t})$ with marginals measures identical to those of the two processes. For TASEP, this technique was used to prove many of its macroscopic properties. We will give an example in the next paragraph regarding the speed distribution of a second-class particle in a rarefaction fan. Contently, consider two initial conditions for TASEP, $\eta_{0}^{1}$ and $\eta_{0}^{2}$, one common way to couple them is to use the same clocks on the sites, i.e. to let the two processes follow the same Harris flow. Another common way would be to label the particles on each of the configurations and to attach the same Poisson clock to particles with the same label. We will be introducing in section \ref{Speed process of a defect in a step initial configuration} a new coupling scheme for systems involving a second-class particle of arbitrary rates.

\subsection{Second class particle with unity rates in a rarefaction fan}
As it was previously mentioned, the first class particles follow the characteristics of Burgers equation. However, if we consider a decreasing step initial condition $\nu_{\rho, \lambda}$ with $\rho > \lambda$ with a second class particle at the origin, then right after the initial moment, the discontinuity collapses to a linear profile, and the second class particle can find itself a priori at any position of this rarefaction fan. This situation has been studied by Ferrari and Kipnis \cite{ferrari1995second}.

\begin{thm}(Ferrari-Kipnis)
	
	Let $X_{t}$ be the position of the second class particle at time $t$ and let $X^{\epsilon}_{t} = \epsilon X_{t/\epsilon}$, then:
	
	\begin{equation}\label{F-K}
	\lim_{\epsilon \rightarrow 0} X^{\epsilon}_{t} \overset{\text{}}{=} U_{t} \quad \text{in distribution},
	\end{equation}
	where $U_{t}$ is a random variable with uniform distribution over the interval: $[(1-2\rho)t, (1-2 \nu)t ]$
	
\end{thm}

\begin{proof}
	The general proof is provided in \cite{ferrari1995second}. For a pedagogical purpose, we will detail a particular case of the $1-0$ step initial condition which already contains the core ideas. Let $\eta_{0}$ be this step initial configuration with particles on all the negative sites including the origin, and let $\tilde{\eta}_{0}$ be the same initial configuration except having a void at the origin.
	If we consider a coupling between the process $\eta_{t}$ and $\tilde{\eta}_{t}$, where the clocks are attached to the sites, then the unique discrepancy initially present between $\eta_{0}$ and $\tilde{\eta}_{0}$ will stay unique and it is rather easy to understand that it will behave like a second class particle, figure \ref{fig:coupled1}.
	
	Now let $N(x,t)$, $\tilde{N}(x,t)$ be the number of particles whose positions are strictly greater than $x$ for the configurations $\eta_{t}$, $\tilde{\eta}_{t}$ , respectively. Finally, let $N^{\epsilon}(x,t) = N(x/ \epsilon,t/ \epsilon)$, $\tilde{N}^{\epsilon}(x,t) = \tilde{N}(x/ \epsilon,t/ \epsilon)$. We have obviously $\tilde{N}(x,t) \leq N(x,t) \leq \tilde{N}(x,t) + 1 $, and similarly for $N^{\epsilon}$ and $\tilde{N}^{\epsilon}$.
	Our objective is to prove eq. \ref{F-K}, which is equivalent to:
	
	\begin{equation}
	\lim_{\epsilon \rightarrow 0} P( X_{t}^{\epsilon} > x)  = \frac{t-x}{2t}
	\end{equation}

	We first notice that the event $X_{t}^{\epsilon} > x$ is equivalent to $ N^{\epsilon}(x,t) = \tilde{N}^{\epsilon}(x,t) + 1 $, this means:

	\begin{equation}\label{PE}
	P( X_{t}^{\epsilon} > x)  =
	E(
	N^{\epsilon}(x,t)) - \tilde{E}(\tilde{N}^{\epsilon}(x,t))
	\end{equation}

	In order to calculate the right side of the previous equation, we can perform another coupling with the same initial conditions $\eta_{0}$ and $\tilde{\eta}_{0}$, but instead: attaching the clocks to the particles rather than to the sites. This requires prior labeling that we choose according to their order from the left, figure \ref{fig:coupled2}.
	Within this coupling, the number of discrepancies will not be conserved under time evolution and hence the loss of second-class particles interpretation. On the other hand, what we have is merely a translation by one step,  $\eta_{t}(k+1) = \tilde{\eta}_{t}(k) $. The advantage of this is that the event $ N^{\epsilon}(x,t) = \tilde{N}^{\epsilon}(x,t) + 1 $ becomes equivalent to $ \eta_{t/\epsilon}((x+1)/ \epsilon) = 1 $, so the right side of eq. \ref{PE} will be equal to:
	
	\begin{equation}
	E(
	N^{\epsilon}(x,t)) - \tilde{E}(\tilde{N}^{\epsilon}(x,t))=
	E(\eta_{t/\epsilon}((x+1)/ \epsilon)
	\end{equation}
	Here we can use the convergence of the density field to have the limit $\epsilon \rightarrow 0$
	
	\begin{equation}
	\lim_{\epsilon \rightarrow 0} P( X_{t}^{\epsilon} > x) = u(x,t) = \frac{t-x}{2t}
	\end{equation}

	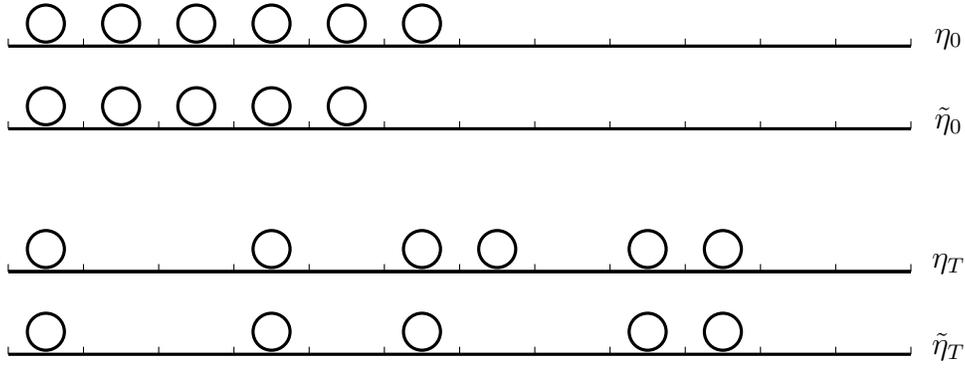
\begin{figure}[h!]
		\centering
		
		\begin{tikzpicture}
		\draw [very thick] (0,0) -- (12,0);
		\draw (12.5,0.1) node {$ \eta_{0} $} [anchor=north] ;

		\foreach \i in {0,...,12}
		{
			\draw (\i,0) -- (\i,0.1) ;
		}
		
		\foreach \i in {0,...,5}
		{
			\node at (\i+0.5,0.3){};
			\draw [very thick] (\i+0.5,0.3) circle (7pt);
		}

		\begin{scope}[shift=({0,0.4})]
		
		\foreach \i in {0,...,4}
		{
			\node at (\i+0.5,0.3-1.5){};
			\draw [very thick] (\i+0.5,0.3-1.5) circle (7pt);
		}
		
		\draw [very thick] (0,-1.5) -- (12,-1.5);
		\draw (12.5,0.1-1.5) node {$ \tilde{\eta}_{0} $} [anchor=north] ;
		
		\foreach \i in {0,...,12}
		{
			\draw (\i,-1.5) -- (\i,0.1-1.5) ;
		}

		\end{scope}
		
		\begin{scope}[shift=({0,-3})]
		
		\draw [very thick] (0,0) -- (12,0);
		
		\foreach \i in {0,...,12}
		{
			\draw (\i,0) -- (\i,0.1) ;
		}
		
		\foreach \i in {0,3,5,6,8,9}
		{
			\node at (\i+0.5,0.5){};
			\draw [very thick] (\i+0.5,0.3) circle (7pt);
		}
		
		\draw [very thick] (0,0) -- (12,0);
		\draw (12.5,0.1) node {$ \eta_{T} $} [anchor=north] ;
		
		\begin{scope}[shift=({0,0.4})]
		
		\foreach \i in {0,3,5,8,9}
		{
			\node at (\i+0.5,0.3-1.5){};
			\draw [very thick] (\i+0.5,0.3-1.5) circle (7pt);
		}
		\draw [very thick] (0,-1.5) -- (12,-1.5);
		\draw (12.5,0.1-1.5) node {$ \tilde{\eta}_{T} $} [anchor=north] ;
		
		\foreach \i in {0,...,12}
		{
			\draw (\i,-1.5) -- (\i,0.1-1.5) ;
		}
		
		\end{scope}
		
		\end{scope}
		
		\end{tikzpicture}

		\caption{Time evolution of two coupled systems with Poisson clocks attached to the sites. The discrepancy behaves as a second-class particle.}

		\label{fig:coupled1}
		
	\end{figure}

	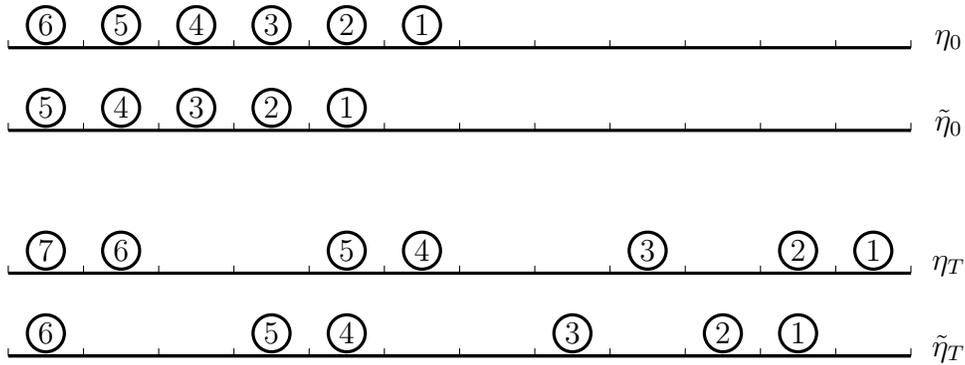
\begin{figure}[h!]
		\centering
		
		\begin{tikzpicture}
		\draw [very thick] (0,0) -- (12,0);
		\draw (12.5,0.1) node {$ \eta_{0} $} [anchor=north] ;

		\foreach \i in {0,...,12}
		{
			\draw (\i,0) -- (\i,0.1) ;
		}
		
		\foreach \i in {1,...,6}
		{
			\node at (5-\i+1.5,0.3){\i};
			\draw [very thick] (5-\i+1.5,0.3) circle (7pt);
		}

		\begin{scope}[shift=({0,0.4})]
		
		\foreach \i in {0,...,12}
		{
			\draw (\i,-1.7+0.2) -- (\i,0.1-1.7+0.2) ;
		}

		\foreach \i in {1,...,5}
		{
			\node at (5-\i+0.5,0.5-1.7){\i};
			\draw [very thick] (5-\i+0.5,0.5-1.7) circle (7pt);
		}
		\draw [very thick] (0,-1.5) -- (12,-1.5);
		\draw (12.5,0.1-1.5) node {$ \tilde{\eta}_{0} $} [anchor=north] ;
		
		\end{scope}
		
		\begin{scope}[shift=({0,-3})]

		\draw [very thick] (0,0) -- (12,0);
		\draw (12.5,0.1) node {$ \eta_{T} $} [anchor=north] ;
		
		\foreach \i in {0,...,12}
		{
			\draw (\i,0) -- (\i,0.1) ;
		}
		
		\node at (0.5 + 11, 0.5- 0.2){1};
		\draw [very thick] (0.5 + 11,0.5- 0.2) circle (7pt);
		\node at (0.5 + 10, 0.5- 0.2){2};
		\draw [very thick] (0.5 + 10,0.5- 0.2) circle (7pt);
		\node at (0.5 + 8,0.5- 0.2){3};
		\draw [very thick] (0.5 + 8,0.5- 0.2) circle (7pt);
		\node at (0.5 + 5,0.5- 0.2){4};
		\draw [very thick] (0.5 + 5,0.5- 0.2) circle (7pt);
		\node at (0.5 + 4,0.5- 0.2){5};
		\draw [very thick] (0.5 + 4,0.5- 0.2) circle (7pt);
		\node at (0.5 + 1,0.5- 0.2){6};
		\draw [very thick] (0.5 + 1,0.5- 0.2) circle (7pt);
		\node at (0.5 ,0.5- 0.2){7};
		\draw [very thick] (0.5,0.5- 0.2) circle (7pt);

		\begin{scope}[shift=({0,0.4})]
		
		\draw [very thick] (0,-1.5) -- (12,-1.5);
		\draw (12.5,0.1-1.5) node {$ \tilde{\eta}_{T} $} [anchor=north] ;
		\foreach \i in {0,...,12}
		{
			\draw (\i,-1.5) -- (\i,0.1-1.5) ;
		}

		\node at (-0.5 + 11,0.5-1.5 - 0.2){1};
		\draw [very thick] (-0.5 + 11,0.5-1.5- 0.2) circle (7pt);
		
		\node at (-0.5 + 10,0.5-1.5- 0.2){2};
		\draw [very thick] (-0.5 + 10,0.5-1.5- 0.2) circle (7pt);
		
		\node at (-0.5 + 8,0.5-1.5- 0.2){3};
		\draw [very thick] (-0.5 + 8,0.5-1.5- 0.2) circle (7pt);

		\node at (-0.5 + 5,0.5-1.5- 0.2){4};
		\draw [very thick] (-0.5 + 5,0.5-1.5- 0.2) circle (7pt);
		
		\node at (-0.5 + 4,0.5-1.5- 0.2){5};
		\draw [very thick] (-0.5 + 4,0.5-1.5- 0.2) circle (7pt);

		\node at (-0.5 + 1,0.5-1.5- 0.2){6};
		\draw [very thick] (-0.5 + 1,0.5-1.5- 0.2) circle (7pt);

		\draw [very thick] (0,-1.5) -- (12,-1.5);

		\end{scope}

		\end{scope}
		
		\end{tikzpicture}

		\caption{Time evolution of two coupled systems with Poisson clocks attached to the particles, particles with the same label attempt to make a jump at the same time.}

		\label{fig:coupled2}
		
	\end{figure}

\end{proof}

Note that it's possible to prove the previous convergence in a stronger sense, it was proven almost surely in \cite{mountford2005motion}, using Seppäläinen’s variational formula \cite{seppalainen1999existence}.
This uniform distribution will not in general stay uniform is the initial configuration is perturbed. However it's still possible to find the limit distribution for arbitrary initial condition using a formalism developed in \cite{cator2013busemann} \cite{cator2012busemann}. This method is based on a mapping between TASEP and the Last Passage Percolation (LPP) model, and uses an interpretation of the second class particle as an interface between two competing surfaces in the LPP picture, however it doesn't seem to be generalization once the rates of the second class particle are not unity.

\subsection{Matrix Product Ansatz for second class particle on the ring}

The statistical behavior of a second-class particle with arbitrary rates as well as the density profile was obtained in a system with periodic boundary condition, using exact methods, one of such is the Matrix Product Ansatz, (MPA), \cite{mallick1996shocks} Consider a ring with $L+1$ sites, $N$ first class particles and a single second-class particle with rates:
\begin{equation}
\begin{split}
\alpha \quad 20 \rightarrow 02
\\
\beta \quad 12 \rightarrow 21
\end{split}
\end{equation}
Let's place ourselves in the reference of the second class particle that will have a position zero, and all other particles have positions $\{1,2,...,L\}$. Then the main idea of the MBA is the Ansatz that the stationary measure can be written as a matrix element of the product of $L$ matrices belonging to two types: $D$ representing the particles, and $E$ representing the void:
\begin{equation}
p(\eta) =  \frac{1}{Z_{L,N}}
\bra{V}
\prod_{k=1}^{L}(\eta(k)D + (1-\eta(k))E )
\ket{W}
\end{equation}

In words, the weight of a configuration is obtained by a choice of a corresponding product through replacing $\bullet \leftrightarrow D$ and $\circ \leftrightarrow E$, for instance:
$ p(\bullet \circ \circ ... \bullet \bullet ) \propto \bra{V}DEE...DD\ket{W} $
, where $D$, $E$,$ \bra{V} $,$ \ket{W} $ being non commuting operators verifying the simple algebra:
\begin{equation}
\begin{split}
DE = D+E \\
D \ket{W} = \frac{1}{\beta} \ket{W} \\
\bra{V}E = \frac{1}{\alpha} \bra{V}
\end{split}
\end{equation}

It is not very hard to understand why this algebra allows for the weights to be stationary. The simplest way to get a flavor of it is through an example: Let us for instance check that the weight of the configuration $(\circ \circ \bullet \bullet \bullet \circ \circ$) is stationary. This weight is controlled by the following transitions:
\begin{center}
	\begin{tikzpicture}
	\draw (-1.2,0.3) node {$\circ \bullet \circ \bullet \bullet \circ \circ$} [anchor=east] ;

	\draw (-1.2,-0.3) node {$\circ \circ \circ \bullet \bullet \bullet  \circ $} [anchor=north] ;

	\draw [->] (0,0.3) -- (1,0.1);
	\draw (0.5,0.7) node [below] {$1$} ;
	\draw [->] (0,-0.3) -- (1,-0.1);
	\draw (0.5,-0.2) node [below] {$\alpha$} ;
	
	\draw (2.4,0) node {$\circ \circ \bullet \bullet \bullet \circ \circ$} [anchor=north] ;
	
	\draw [->] (3.7,0.1) -- (4.7,0.3);
	\draw (0.5+ 3.7,0.7) node [below] {$1$} ;
	\draw [->] (3.7,-0.1) -- (4.7,-0.3);
	\draw (0.5+ 3.7,-0.2) node [below] {$\alpha$} ;
	
	\draw (6,0.3) node {$\circ \circ \bullet \bullet \circ \bullet \circ$} [anchor=east] ;
	
	\draw (6,-0.3) node {$\circ \bullet \bullet \bullet \circ \circ \circ$} [anchor=north] ;
	\end{tikzpicture}
\end{center}
It is straightforward to check the stationarity after doing the appropriate decomposition:
\begin{equation}
p(\circ \bullet \circ \bullet \bullet \circ \circ) = \frac{Z_{5,3}}{Z_{6,3}} p(\circ \bullet  \bullet \bullet \circ \circ) + \frac{Z_{5,2}}{Z_{6,3}} p(\circ  \circ \bullet \bullet \circ \circ)
\end{equation}

\begin{equation}
p(\circ \circ \circ \bullet \bullet \bullet  \circ) = \frac{1}{\alpha}\frac{Z_{5,3}}{Z_{6,3}} p( \circ \circ \bullet \bullet \bullet  \circ) 
\end{equation}

\begin{equation}
p(\circ \circ \bullet \bullet \bullet \circ \circ) = \frac{Z_{5,2}}{Z_{6,3}} p(\circ \circ \bullet \bullet  \circ \circ) + \frac{Z_{5,3}}{Z_{6,3}} p(\circ \circ \bullet \bullet \bullet  \circ) = \frac{1}{\alpha}\frac{Z_{5,3}}{Z_{6,3}} p( \circ \bullet \bullet \bullet \circ \circ)
\end{equation}

The partition function $Z_{L,N}$ can be written as:
\begin{equation}
Z_{L,N} = \bra{V} \sum_{\eta \in \mathcal{C}}
\prod_{k=1}^{L}(\eta(k)D + (1-\eta(k))E )
\ket{W} := \bra{V} G_{L,N}
\ket{W}
\end{equation}
where $ \mathcal{C} $ is the space of configurations for the system: $ \{\mathcal{C} = \eta : \sum_{k=1}^{L} \eta(k)=N \} $.
There are known infinite matrix representations for $D$, $E$, $\bra{V}$ and $\ket{W}$, that allows finding the expression of $Z_{L,N}$. Once this is obtained, full calculations can be found in \cite{mallick1996shocks}, and any relevant observers can be expressed in terms of it. What is important for us is the speed of the second-class particle:

\begin{equation}
\begin{split}
v &= \alpha E(\tau_{1} = 0) - \beta E(\tau_{L} = 1) \\
& = \frac{1}{Z_{L,N}} \alpha \bra{V} E G_{L-1,N} \ket{W} - \beta \bra{V} G_{L-1,N-1} D \ket{W}
\\ 
& = \frac{Z_{L-1,N} - Z_{L-1,N-1}}{Z_{L,N}}
\end{split}
\end{equation}
Taking the asymptotic limit of large $N$ with $\frac{N}{L} = \rho$ fixed, it is possible to find simple expressions of this speed as a function of the density. Since these expressions will be useful to the rest of the chapter, let's state them here:
\begin{equation}\label{a} \tag{a}
For \; \; \beta > \rho > 1- \alpha \quad \quad v=1-2\rho
\end{equation}
\begin{equation}\label{b} \tag{b}
For \; \; \beta < \rho \; and \; 1-\alpha < \rho \quad \quad v = 1-\rho -\beta 
\end{equation}
\begin{equation}\label{c} \tag{c}
For \; \; \beta > \rho \; and \; 1-\alpha > \rho \quad \quad v = \alpha -\rho 
\end{equation}
\begin{equation}\label{d} \tag{d}
For \; \; \beta < \rho < 1- \alpha \quad \quad v= \alpha - \beta
\end{equation}

It is possible to find the expression of the density profile using this technique. The result that is relevant to our next section is that the second-class particle disrupts the density field only when $\beta < \rho < 1-\alpha $, creating a macroscopic density of $\beta$ on its right and $1-\alpha$ on its left.

Note that MBA is not the only method used to find this expression. in \cite{derrida1999bethe}, Bethe Ansatz was used to derive the expression of the previous speeds, with the advantage of providing expressions for the diffusion constant and in principle higher cumulants.

Let us finish here by noticing that it is possible to write the speed expressions in this compact form:

\begin{equation}\label{key}
v = v_{+} - v_{-}
\end{equation}
with:
$$ v_{+} = min(\alpha, 1- \rho ) $$
$$ v_{-} = min(\beta, \rho ) $$

This suggests an intuitive understanding in terms of queues. This idea will be exploited and elaborated further in section \ref{Speed process of a defect in a step initial configuration}.

\section{A defect in a step initial profile}
\label{A defect in a step initial profile}
We consider a second-class particle with arbitrary rates initially located at the origin of the $\mathbb{Z}$ lattice with a uniform Bernoulli product $\nu$ density for the negative sites and, likewise, $\mu$ for the positive ones. We are interested in both the dynamics of the second-class particle and the evolution of the density field. The two are coupled for the case of $\alpha + \beta < 1$ since we know from \cite{mallick1996shocks} that for this case the defect might macroscopically disturb the density profile in a ring, we will assume this to still be valid for our case, an assumption that will be supported by a mean-field analysis and numerical simulations. The asymptotic speed will the second-class particle will be deterministic in this case. We will show using self-consistency analysis in which situations the density profile is disturbed and we will find its limit shape by solving the Burgers equation with a moving interior boundary condition induced by the second-class particle. Rather rich zoology is observed according to the different values of the parameters. 

In the case of $\alpha + \beta > 1$, the second-class particle behavior is decoupled from the density profile, however, the interest here will be on the asymptotic behavior of the second-class particle that will not be deterministic.

We start with the case of a $1-0$ initial condition, which is simple but yet exhibits a particularity of an escaping particle phenomenon. 

\subsection{1-0 initial condition}
We choose the densities $\nu = 1$, and $\mu = 0$, and we start by considering the situation where one of $\alpha$ or $\beta$ (or both) is greater than 1.

\subsubsection{An escaping second class particle}

\begin{figure}[h!]
	\centering
	\begin{subfigure}[b]{0.45\linewidth}
		\includegraphics[width=\linewidth]{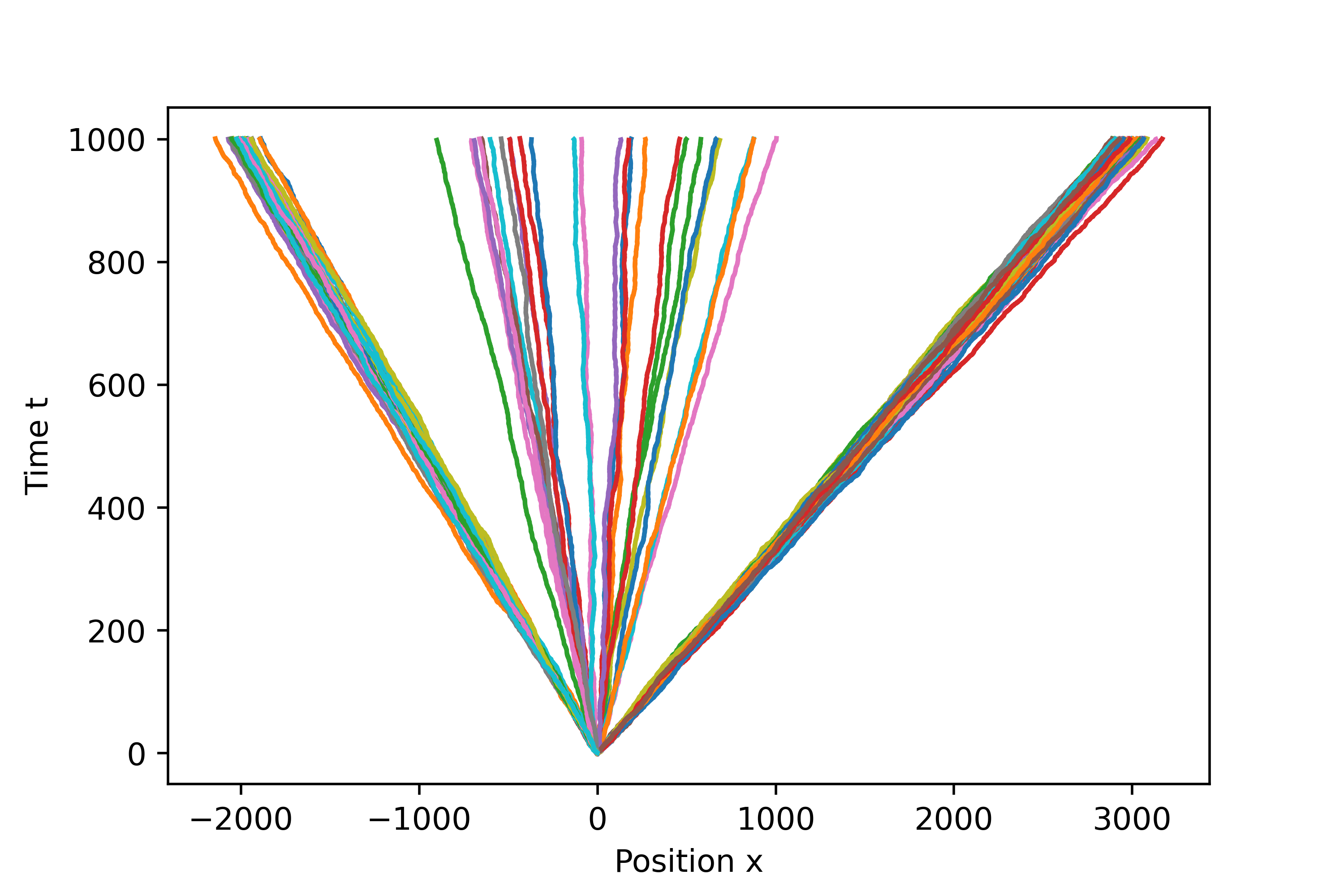}
		\caption{}
	\end{subfigure}
	\begin{subfigure}[b]{0.45\linewidth}
		\includegraphics[width=\linewidth]{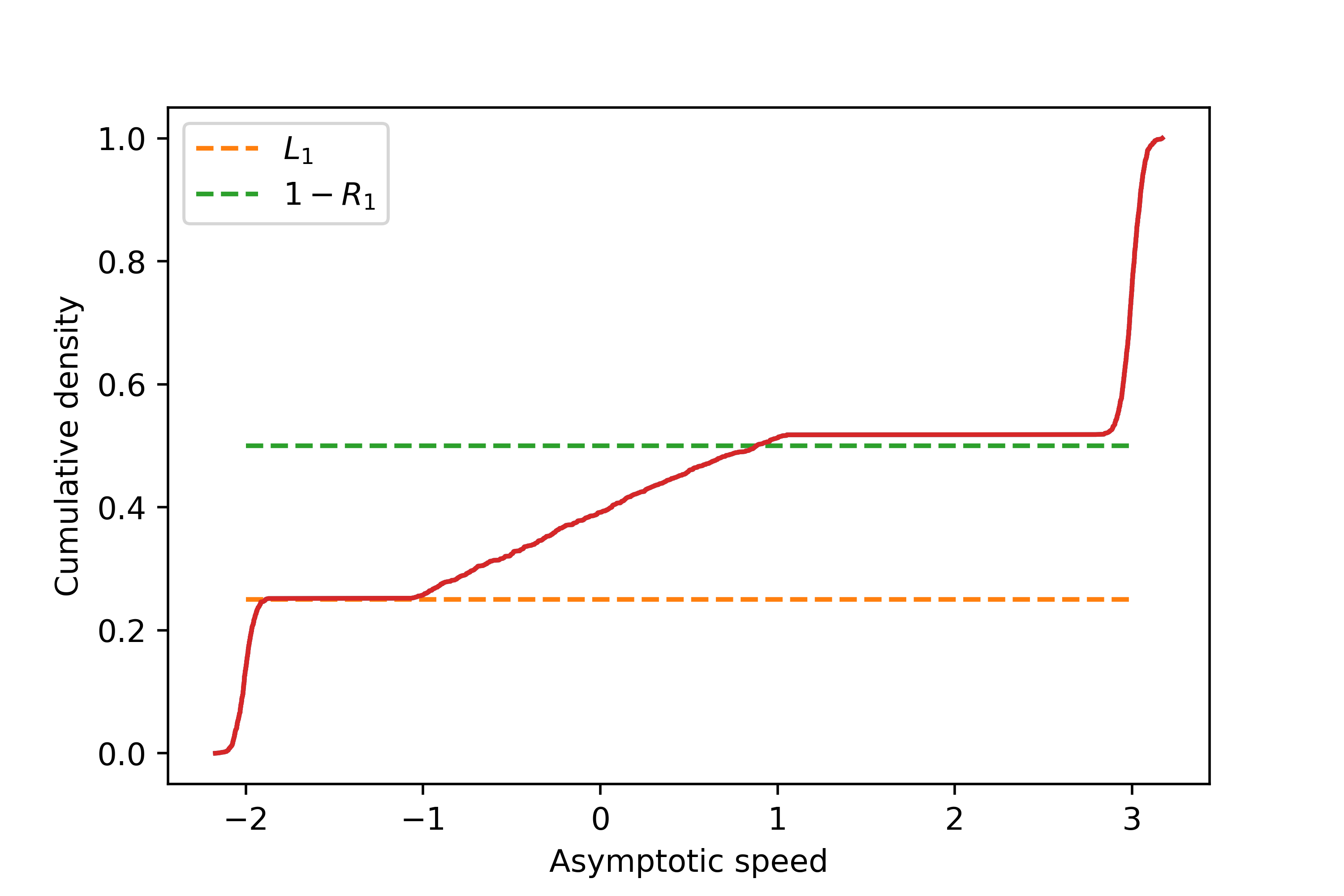}
		\caption{}
	\end{subfigure}
	\begin{subfigure}[b]{0.45\linewidth}
		\includegraphics[width=\linewidth]{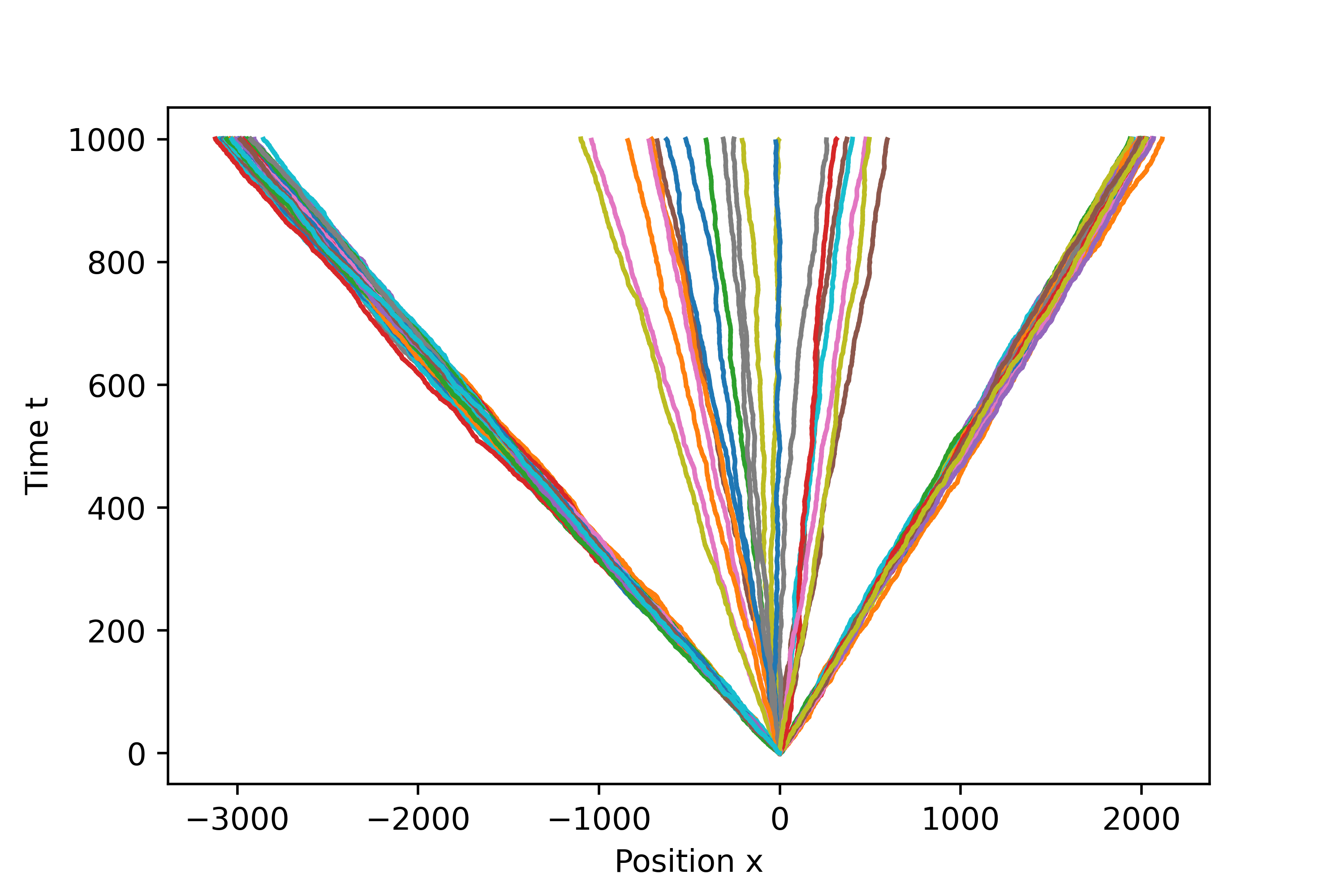}
		\caption{}
	\end{subfigure}
	\begin{subfigure}[b]{0.45\linewidth}
		\includegraphics[width=\linewidth]{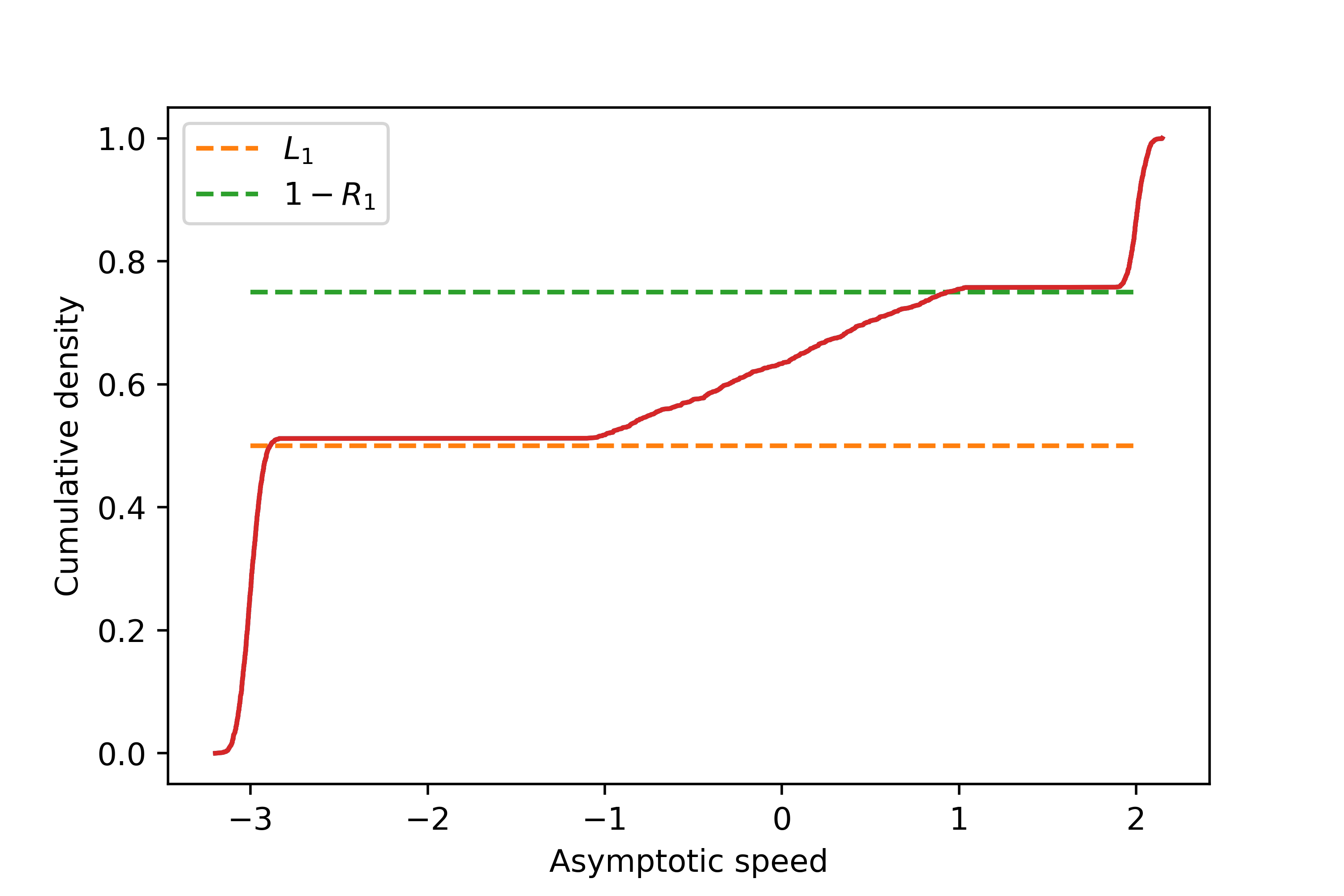}
		\caption{}
	\end{subfigure}
	\caption{Trajectories and asymptotic speeds cumulative distribution of the second class particle for $\alpha = 3$ and $\beta = 2$ in $(a)$ and $(b)$, and for $\alpha = 2$ and $\beta = 3$ in $(c)$ and $(d)$. Cumulative density is plotted over 2000 realization for time $t=1000$. Dashed lines represent theoretical values of the escaping probabilities. 100 trajectories are plotted on the left. The symmetries between the top figures and the bottom ones are due to the choice of $\alpha$ and $\beta$ figuring the hole-particle symmetry of the system.}
	\label{fig1}
\end{figure}

Numerical simulations, figure \ref{fig1}, suggest that in a fraction of realizations, the particle 2 moves at a speed $\alpha$ ($\beta$) in case $\alpha>1$($ \beta > 1$). It is easy to understand what happens: the second-class particle sometimes manages to get drawn in an environment with only holes, (with only first-class particles) and thus moves as a free particle.
In the case where a second class particle doesn't escape, the simulations suggest that the fixed limit speed will be chosen in the interval $[-1,+1]$. This is similar to the behavior of a second class particle with unit rates with the important difference that no information is known about the probability distribution from which this asymptotic speed is drawn. Numerical evidence suggests that it is not a uniform one in general. We will see later in which situations it will stay uniform.
We will be interested in the next paragraph in finding the escaping probability from the left and from the right.

\subsubsection{The right escaping probability}
The right scenario happens when the second-class particle finds itself eventually in a void environment with no first-class particle in front of it, and will hence be moving as a free particle with a speed $\alpha>1$. If one of the first-class particles manages to jump over it, then it will obviously limit it is speed to $1$.

Note that in case $ \alpha \leq 1 $, the probability of the right scenario is zero, because the distance between particle 2 and the first of the first-class particles will follow at best (for $\alpha =1$) a symmetric random walk, so it will hit zero with certainty.

Let $ X_{k} $ be the random variable that represents the distance between particle 2 and the first particle of the first-class particles after $k$ changes of this distance, taken positively when the second-class particle is in front of the first-class particle 1. $X$ follows an asymmetric random walk with:

\begin{align*}
&P(X_{0} = 1) = 1 \\
&P(X_{k+1} = 2|X_{k}= 1) = \frac{\alpha}{\alpha + \beta} := w \\
&P(X_{k+1} = -1|X_{k}= 1) = \frac{\beta}{\alpha + \beta} \\
&P(X_{k+1} = n+1|X_{k}=n>1) = \frac{\alpha}{\alpha + 1} := p > \frac{1}{2} \\
&P(X_{k+1} = n-1|X_{k}=n>1) = \frac{1}{\alpha + 1} = 1-p 
\end{align*}

Our problem is a problem of a random walker with absorbing boundaries \cite{kac1945random}.

Let $$R_{n} = P(X_{l} \geq 1 for \; all \; l >k | X_{k} = n \geq 1)$$

Our objective is to calculate $R_{1}$. We can establish a recursive relation that is verified by $R_{n}$

\begin{equation}
R_{n} = pR_{n+1} + (1-p)R_{n-1}  \;\; n \geq 2
\end{equation}

This recursive sequence can be solved by considering the linear system:

\begin{equation}
\begin{pmatrix} 
R_{n+1}  \\
R_{n}
\end{pmatrix}
=
\begin{pmatrix} 
\frac{1}{p} & \frac{p-1}{p} \\
1 & 0 
\end{pmatrix}
\begin{pmatrix} 
R_{n}  \\
R_{n-1} 
\end{pmatrix}
=\begin{pmatrix} 
\frac{1}{p} & \frac{p-1}{p} \\
1 & 0 
\end{pmatrix} ^{n-1}
\begin{pmatrix} 
R_{2}  \\
R_{1} 
\end{pmatrix}
\end{equation}

This matrix has 1 and $ \frac{1-p}{p} $ as eigenvalues, thus the general term of the sequence is:
\begin{equation}
R_{n} = A + B (\frac{1-p}{p})^{n} \; n \geq 2
\end{equation}

where A and B are constants to be determined from initial conditions. We can get A easily: $\lim_{n\to\infty} R_{n} = 1$, which yields $ A = 1 $. Let us express B in terms of $R_{2}$:

\begin{equation}
R_{n} = 1 + (R_{2}-1) (\frac{1-p}{p})^{n-2} \; n \geq 2 
\end{equation}

In particular: 

\begin{equation}
R_{3} = 1 + (R_{2}-1) (\frac{1-p}{p}) 
\end{equation}

On the other hand, we have:

\begin{equation}
R_{2} = pR_{3} + (1-p)R_{1}
\end{equation}
and 
\begin{equation}
R_{1} = wR_{2}
\end{equation}

From the last three relations, we can finally get $R_{1}$

\begin{equation}
R_{1} = \frac{w(1-2p)}{(1-p)(1+w) -1} = \dfrac{\alpha - 1}{\alpha + \beta - 1}
\end{equation}

We notice that this result matches the one obtained from the particular case of the formalism developed in chapter 3 using the Bethe Ansatz.

The general term for $ n \geq 2 $ reads:
\begin{equation}
\begin{split}
R_{n} & = 1 + \frac{(1-p)(1-w)}{(1-p)(1+w)-1}(\frac{1-p}{p})^{n-2} \\ 
& = 1- \dfrac{\beta}{(\alpha + \beta -1)}\dfrac{1}{\alpha^{n-1}}
\end{split}
\end{equation}

Note that it is almost straightforward to generalize the escaping probability for the case of an initial condition $\nu - 0$, with $\nu < 1$. The initial distance between the first class particle and the second is not anymore necessarily one, but it can take any value $n$ with a probability $\nu (1-\nu)^{n-1}$. So the escaping probability will become:

\begin{equation}
\begin{split}
R^{\nu-0} &= \sum_{n = 1}^{\infty} \nu (1-\nu)^{n-1} R_{n} \\
& = \frac{\nu}{(\alpha + \beta -1)}[ \dfrac{\beta  \alpha}
{(\alpha - 1 + \nu)} + \alpha-\beta - 1  ]
\end{split}
\end{equation}

\subsubsection{The left escaping probability}

The left scenario happens if and only if there is no void before the second-class particle. If $\beta \leq 1 $ then this probability is zero. If $\beta > 1$ we can calculate this probability using the hole-particle symmetry: viewing the void as a first-class particle moving backward and hopping over the second-class particle at rate $\alpha$ and viewing the first-class particles as a void in which the second-class particle can hop backward at rate $\beta$. That would come down to exchanging $\alpha$ and $\beta$ in the previous calculations. We call this probability $L_{1}$

\begin{equation}
L_{1} = \frac{\beta-1}{\alpha+ \beta-1} 
\end{equation}

And it can be generalized for an initial condition $1-\mu$ by replacing $\nu$ by $1-\mu$ beside $\alpha$ and $\beta$ in the expression of $R^{\nu-0}$

\begin{equation}
L^{1-\mu} = \frac{1-\mu}{(\alpha + \beta -1)}[ \dfrac{\beta  \alpha}
{(\beta - \mu)}  -\alpha +  \beta - 1  ]
\end{equation}

\subsection{Non escaping particles}
\begin{figure}[h!]
	\centering
	\includegraphics[width=0.5\linewidth]{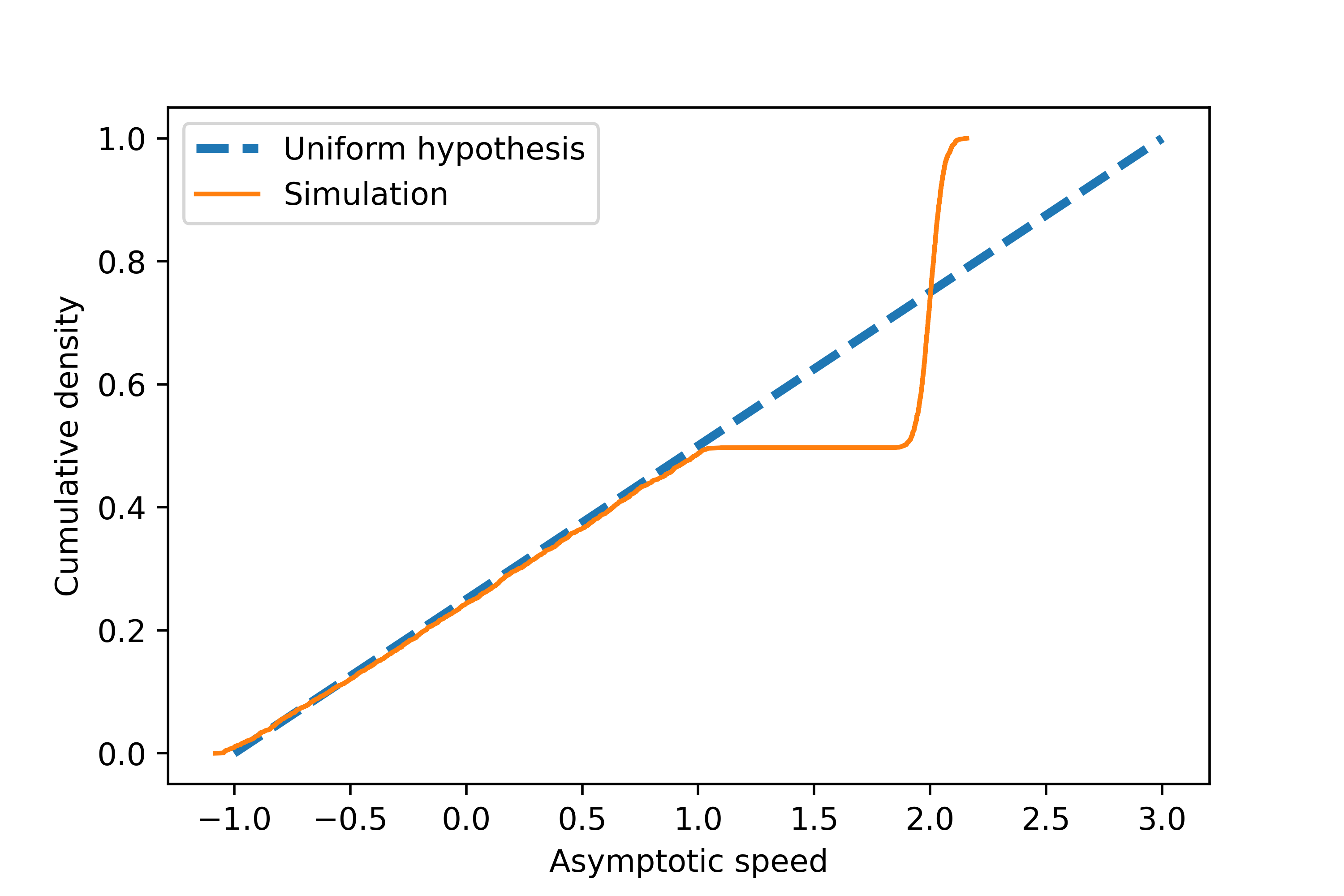}
	\caption{Comparison between a simulated cumulative density and a uniform hypothesis in the case of $\beta= 1, \alpha=2$. The simulated graph is plotted for 4000 realizations, each is evolved up to $t=1000$. The theoretical uniform distribution is plotted over the interval $[1-2 \beta, 2 \alpha - 1]$. This interval is as well valid for $\alpha, \beta < 1$. By giving the right slope, it suggests a consistency between the conjecture and the escaping probability formula.}
	\label{conjecture}
\end{figure}

Preliminary investigations lead us to the following conjuncture: 

\begin{conj}
	The probability distribution of the asymptotic speed of a non-escaping second-class particle is uniform in the interval  $[-1,1]$ for $\alpha,\beta > 1$.
\end{conj}
This is firstly supported by numerical evidence, figure \ref{conjecture}. It is as well understood in the limits $\alpha, \beta \in \{1,\infty\} $. To understand the infinity limit, take for instance $\beta=1$ and $\alpha= \infty$. There will be only an infinitesimal chance for a non-escaping scenario that will start with an initial  condition: $...11121000...$. The couple $21$ behaves as a second-class particle of unity rates, and will thus be uniformly distributed. This conjuncture as well will prolong the same result but for the case of $\alpha, \beta <1$ that we will prove in section \ref{Speed process of a defect in a step initial configuration}.

\subsection{Density field profile and the second class particle}
If we assume local equilibrium, the behavior of the second-class particle is dependent on the surrounding density field. This behavior can impact in its turn the density field creating an interplay between the two. In this section, we will show heuristically that the local equilibrium assumption and the asymptotic speed formulas for the different parameters are enough to predict the macroscopic evolution of the system. We call a density region: an interval for the density where the speed is given by one formula. Let's recall them:

\textbf{The Density Regions:}
\begin{equation}\label{a} \tag{a}
\text{For} \; \; \beta > \rho > 1- \alpha \quad \quad v=1-2\rho
\end{equation}
\begin{equation}\label{b} \tag{b}
\text{For} \; \; \beta < \rho \; \text{and} \; 1-\alpha < \rho \quad \quad v = 1-\rho -\beta 
\end{equation}
\begin{equation}\label{c} \tag{c}
\text{For} \; \; \beta > \rho \; \text{and}  \; 1-\alpha > \rho \quad \quad v = \alpha -\rho 
\end{equation}
\begin{equation}\label{d} \tag{d}
\text{For} \; \; \beta < \rho < 1- \alpha \quad \quad v= \alpha - \beta
\end{equation}
The general logic of our analysis relies on the following simple procedure:
\begin{enumerate}
	\item Assume the second class particle belongs to one of these regions.
	\item Solve the density field according to this assumption.
	\item Check the consistency between the solution and the assumption. If there is no consistency, try again with a different region.
\end{enumerate}
Applying this, we find a unique self-consistent situation for each set of parameters. We confirm it by simulation.

We will be discussing two cases: when $ \alpha + \beta <1 $
and when $ \alpha + \beta > 1 $

\subsubsection{The case of $\alpha + \beta < 1$}
Let's assume that the density profile is a decreasing function of space. Let's assume as well that this profile is initially continuous, even if this is clearly not true for exactly $ t=0 $ .

Since $ \beta < 1-\alpha $, only the regions (b), (c) and (d) of the density are meaningful.

If the particle finds itself in the region (b), then it moves at a speed $ 1- \rho -\beta > 1 - 2\rho $ so the particle will move into lower density till it reaches the region (d) and will cross its upper boundary because the inequality above still holds on this boundary.

If the particle finds itself in the region (c), it will move at a speed $\alpha - \rho < 1 - 2\rho$, so it will move into higher density and will continue till it reaches again the region (d) crossing also its lower boundary.
After being in region (d), the particle will move at a constant speed $ v= \alpha - \beta $
The condition $ \alpha + \beta < 1 $ implies:
$1 - 2(1-\alpha) < \alpha - \beta < 1- 2\beta $.

It means that there is a point of density $\rho_{0}$ inside the region (d) verifying $1 - 2\rho_{0} = \alpha - \beta$, and the particle will be attracted to this point. Actually $\rho_{0}$, is the average of the density of the boundaries of the region (d): $ \rho_{0} = (1-\alpha + \beta)/2$.

\subsubsection*{Formation of discontinuity}

So far, we have ignored the dynamic interaction between the particle and the density profile. If we choose the parameters $\alpha = \beta = 0$, the particle will become a wall and discontinuity will be created at that point. We can suspect a discontinuity for other values of $\alpha$ and $\beta$. To check this possibility heuristically, we will consider two different densities, on the left of the particle $\rho_{-}$ and on the right of the particle $\rho_{+}$. Let us try to evaluate these two values. If our hypothesis about the discontinuity happens to be wrong we should find  $\rho_{-} = \rho_{+}$.

The particle will be trapped in this discontinuity, so the discontinuity will move at the same speed as the particle. We can find $\rho_{-}$ and $\rho_{+}$ using two elementary equations:

The first one:
\begin{equation}\label{key}
\rho_{-}(1-\rho_{-}) - \rho_{+}(1-\rho_{+}) = (\alpha - \beta)(\rho_{-}-\rho_{+})
\end{equation}

This is a conservation equation that relates the current on the left and on the right of the discontinuity to the speed in a hydrodynamic manner.
It can be simplified:
\begin{equation}\label{key}
1 - \rho_{-} - \rho_{+} = \alpha - \beta
\end{equation}
The second equation relates the rate of first-class particles jumping over the particle and the rate of holes jumping backwards on the particle with its speed:

\begin{equation}\label{key}
\alpha - \beta = (1-\rho_{+})\alpha -  \beta\rho_{-}
\end{equation}

The two previous equations make it obvious that the only solution is:
\begin{equation}\label{key}
\begin{split}
& \rho_{-} = 1-\alpha \\
& \rho_{+} = \beta
\end{split}
\end{equation}

This is of course not surprising since it's already known on the ring using MPA \cite{mallick1996shocks}. 

\subsubsection{Dynamic density profile}
The natural next question is concerned with the influence of the presence of the particle on the rest of the density profile.

For simplicity, we place ourselves in the frame of the particle. In this frame, the density verifies the hydrodynamic conservation equation:

\begin{equation}\label{key}
\frac{\partial \rho}{\partial t} + (1 - 2\rho - \alpha + \beta) \frac{\partial \rho}{\partial x} = 0
\end{equation}

Let's bring it into an even more familiar form:

\begin{equation}\label{key}
\frac{\partial \tilde{\rho}}{\partial t} + (1 - 2\tilde{\rho}) \frac{\partial \tilde{\rho}}{\partial x} = 0
\end{equation}

with:
\begin{equation}\label{key}
\tilde{\rho} = \rho +\frac{\alpha - \beta}{2}
\end{equation}

\subsubsection*{The right part:}

The right part of the density profile has this boundary condition:

\begin{equation}\label{key}
\tilde{\rho}(0,t) = \beta +\frac{\alpha - \beta}{2} = \frac{\alpha + \beta}{2}
\end{equation}

Now we can guess the solution by comparing it to a model with an open left boundary and defined on a half-space with a density on the boundary  $ \dfrac{\alpha + \beta}{2} $ \cite{blythe2007nonequilibrium}.

Since we have $ \dfrac{\alpha + \beta}{2} < \dfrac{1}{2} $, this corresponds exactly to the phase where a kinetic wave of a constant density: $ \dfrac{\alpha + \beta}{2} $ propagates inside the system with a speed $1 -2(\frac{\alpha+\beta}{2}) = 1 - \alpha - \beta$

For the front of the kinetic wave, we expect it to be linear with a speed going from $ 1 - \alpha - \beta$ at the upper front to $ 1 - \alpha + \beta$ at the bottom.

These arguments lead us to check this solution for the right part of the density with respect to the particle:

\begin{equation}\label{key}
\rho_{R}(x,t) = 
\left\{
\begin{array}{ll}
\beta  & \mbox{if } 0 < x \leq (1-\beta -\alpha)t  \\
\dfrac{-x}{2t} + \frac{1+\beta - \alpha}{2} & \mbox{if }  (1-\beta -\alpha)t \leq x  \leq (1+\beta -\alpha)t \\
0  & \mbox{if }  x > (1+\beta -\alpha)t
\end{array}
\right.
\end{equation}
One can verify immediately that this is indeed a solution.

\subsubsection*{The left part}
Similar arguments as above made on the holes this time can lead us to this solution:
\begin{equation}\label{key}
\rho_{L}(x,t) = 
\left\{
\begin{array}{ll}
1-\alpha  & \mbox{if } (\beta +\alpha - 1)t \leq x < 0   \\
\dfrac{-x}{2t} + \frac{1-\alpha + \beta}{2} & \mbox{if }  (\beta -\alpha - 1)t \leq x  \leq (\beta +\alpha - 1)t \\
1  & \mbox{if }  x < (\beta -\alpha - 1)t
\end{array}
\right.
\end{equation}

Let's finally write the density profile with respect to the static reference:
\begin{equation}
\rho_{0}(x,t) = 
\left\{
\begin{array}{ll}
1  & \mbox{if }  x < -t \\

\frac{-x}{2t} + \frac{1}{2} & \mbox{if } -t \leq x  \leq ( 2\alpha - 1)t \\

1-\alpha  & \mbox{if } (2\alpha - 1)t \leq x < \alpha  - \beta    \\

\beta  & \mbox{if } 0 < x \leq (1-2\beta )t  \\

\frac{-x}{2t} + \frac{1}{2} & \mbox{if }  (1-2\beta)t \leq x  \leq t \\
0  & \mbox{if }    t < x

\end{array}
\right.
\end{equation}

So the second class particle can be seen as moving interior boundary condition.

\subsubsection{The case of $ \alpha + \beta > 1 $}
In this case, $ \beta > 1-\alpha $, so only the regions (a), (b) and (c) are meaningful.
The region (b) is accessible only if $\beta < 1$. The region (c) is accessible only if $\alpha < 1$

If the particle finds itself at the region (a) it will move at the speed $1 - 2\rho$ which is the same speed as the characteristics (the speed of perturbations), so it will not modify the usual density profile. With this speed, the particle will always be experiencing the same density, so this speed will not change.

If $\beta < 1$ then the particle might find itself at region (b), it will move at a speed $1-\beta -\rho > 1-2\rho$ so the particle will move towards the lower density till it reaches the density $ \rho = \beta $ where it can stabilize at a speed $1 - 2\beta \in [-1,1] $
If $\alpha < 1 $ the particle might find itself in the region (d), it will move at the speed $ \alpha - \rho < 1- 2\rho $ so it will move towards the higher density till it reaches $ \rho = 1- \alpha  $, where it stabilizes at the speed $1 - 2\rho = 2\alpha - 1 \in [-1, 1]$.

\subsubsection{Conclusion}
\begin{figure}[h!]
	\centering
	\begin{subfigure}[b]{0.45\linewidth}
		\includegraphics[width=\linewidth]{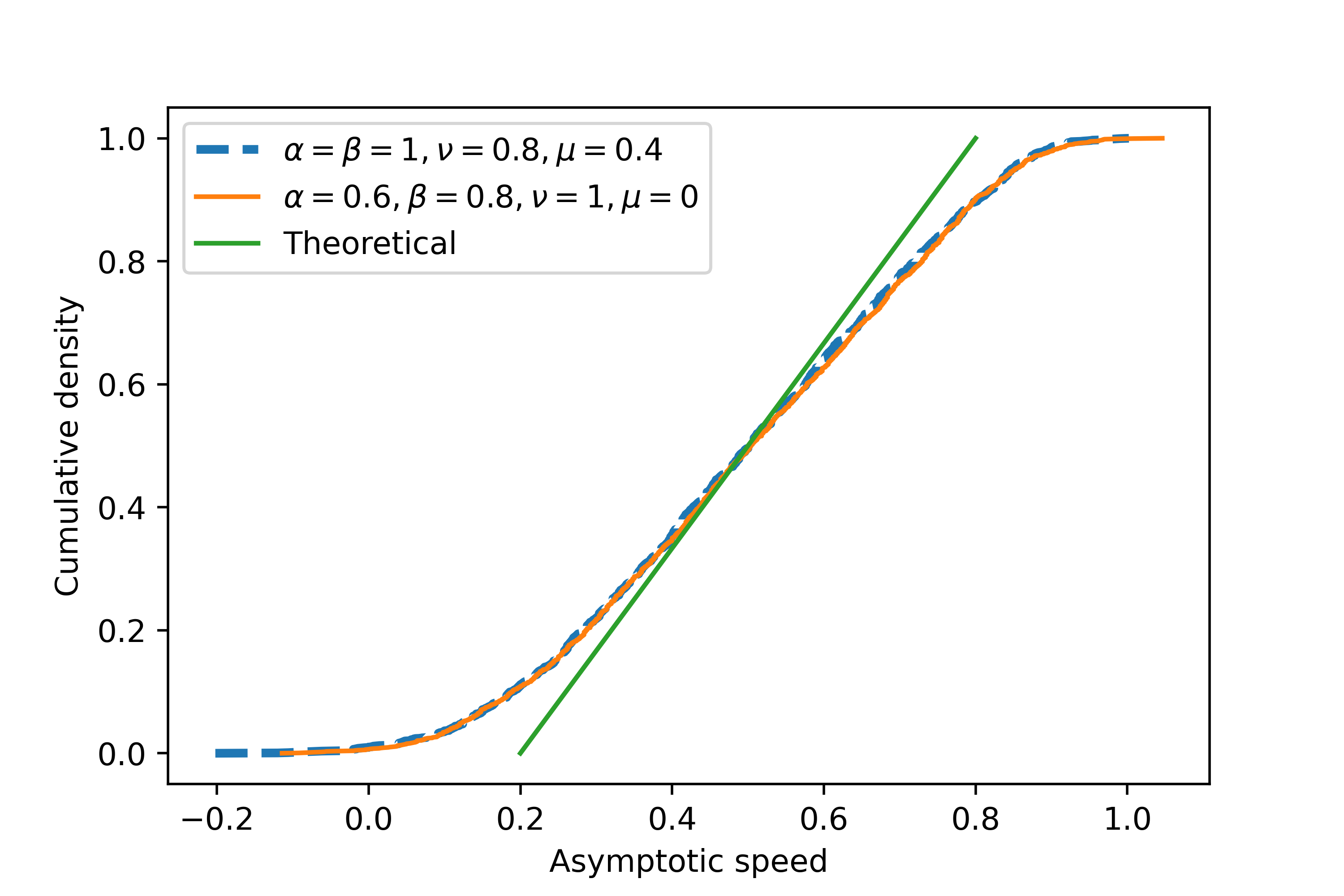}
	\end{subfigure}
	\begin{subfigure}[b]{0.45\linewidth}
		\includegraphics[width=\linewidth]{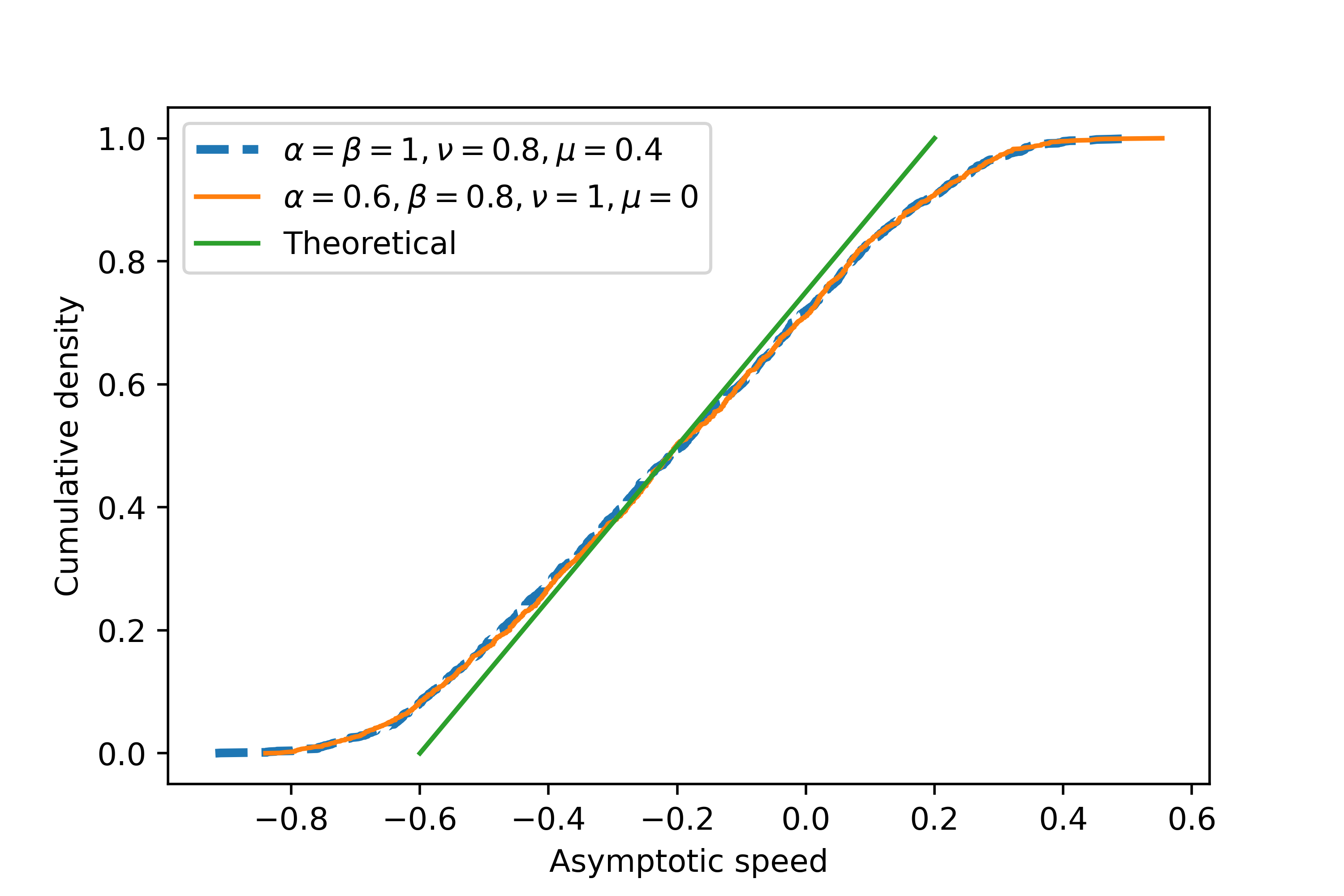}
	\end{subfigure}
	\caption{Two examples of the cumulative distribution of asymptotic speed of a second class particle in a 1-0 step initial profile, $\alpha, \beta <1$,$\alpha+\beta>1$ (continuous orange line). This distribution is identical to the one of unity rates but with initial profile $\nu-\mu$ with $ \nu = \beta $ and $\mu = 1- \alpha$ (dashed line), which is uniform. Green straight segment represents the theoretical uniform distribution. Smoothness of distributions and the divergence from the theoretical on the boundaries are due to finite time effect. Distributions are plotted for 2000 realizations, each is for $t=500$.}
	\label{xxyy}
\end{figure}

In the case where $\alpha + \beta > 1$, the particle will choose a speed that belongs to the interval $[max(-1,1-2\beta), min(1,2\alpha -1) ]$ and will stick to this speed. We will prove in section \ref{Speed process of a defect in a step initial configuration} that this speed is chosen according to a uniform distribution, as if the particle had unit rates but was put in a $\nu-\mu$ step initial profile, with $\nu = \beta$, $\mu = 1-\alpha$, figure \ref{xxyy}.

Note that the size of the previous interval will go to zero in the limit $\alpha + \beta \rightarrow 1 $. The particle will be at a speed $ 1 - 2\beta = 2\alpha - 1 = \alpha - \beta $, so the speed is continuous when passing between the two regimes.

\subsection{$\nu-\mu$ step initial configuration}
We come back again to the general situation of a $\nu-\mu$ initial condition and a second-class particle at the origin. We choose $ 0 < \mu$  and  $\nu < 1$ so we do not encounter the already escaping particles phenomena.

The qualitative behavior of the system will depend on the relative position of the four parameters: $\mu, \nu, 1-\alpha, \beta $. In all generality, we may encounter $4! = 24$ different regimes, however, the symmetry reduces this quantity by half.

\subsubsection*{Symmetries}
Each macroscopic configuration has a symmetric one with respect to zero, obtained by the transformation:
$$(\mu,\nu,\alpha,\beta) \longrightarrow (1-\nu,1-\mu,\beta,\alpha) $$
This is nothing but an extension of the hole-particle symmetry.

We will be dividing our discussion into two regimes: $ \nu > \mu$ and $\nu < \mu$

\subsection{The case of $ \nu > \mu$}
\label{The case of nu greater than mu}
This would be the rarefaction fan regime in the absence of the second-class particle.
We distinguish again the two cases: $\alpha + \beta <1$ and $\alpha + \beta > 1$.

\subsubsection{$\alpha + \beta < 1$}

In the case of $(\nu, \mu) = (1,0) $, we saw that the second-class particle will create a decreasing discontinuity in the rarefaction fan. This of course can still happen for generic $\mu, \nu$, however, we encounter new observations: under some conditions on parameters, the second-class particle might stay outside the rarefaction region without creating any discontinuity, for other set of parameters, it will create a shock wave to his left, or to his right or to both in addition to the decreasing discontinuity at its position. To distinguish all of these different cases, we proceed in a similar fashion as in 1.2: we assume the particle belongs to some density region, we compare the velocity of the particle with the velocity of the boundaries of the assumed region which will inform us about the stability of the assumption, and confirm by numerical simulations.

The diagram below recapitulates the different scenarios here according to the relative values of the parameters:

\begin{center}
	\begin{tikzpicture}
	\draw[thick,->] (0,0) -- (7,0) node[anchor=north west] {$1-\alpha$};
	\draw[thick,->] (0,0) -- (0,7) node[anchor=south east] {$\beta$};
	
	\draw (2 cm,1pt) -- (2 cm,-1pt) node[anchor=north] {$\mu$};
	
	\draw (4 cm,1pt) -- (4 cm,-1pt) node[anchor=north] {$\nu$};
	
	\draw (6 cm,1pt) -- (6 cm,-1pt) node[anchor=north] {$1$};
	
	\draw (1pt,2 cm) -- (-1pt,2 cm) node[anchor=east] {$\mu$};
	
	\draw (1pt,4 cm) -- (-1pt,4 cm) node[anchor=east] {$\nu$};
	
	\draw (1pt,6 cm) -- (-1pt,6 cm) node[anchor=east] {$1$};
	
	\draw (0,2) -- (6,2);
	\draw [dashed] (0,4) -- (4,4);
	\draw (4,4) -- (6,4);
	\draw (0,6) -- (6,6);
	\draw (2,0) -- (2,2);
	\draw [dashed] (2,2) -- (2,6);
	\draw (4,0) -- (4,6);
	\draw (6,0) -- (6,6);
	\draw [dashed] (0,0) -- (6,6);
	\draw (2,2) -- (4,4);
	
	\draw (2,0.5) node[anchor=east] {$C,R$};
	\draw (3.7,1) node[anchor=east] {$D,SL$};
	\draw (5.7,1) node[anchor=east] {$D,SR$};
	\draw (5.7,0.5) node[anchor=east] {$SL$};
	\draw (5.7,3) node[anchor=east] {$D,SR$};
	\draw (3.2,2.5) node {$D$};
	\draw (6,4.5) node[anchor=east] {$C,L$};
	
	\draw (0.5,2) node[anchor=north] {};

	\draw (0.6,1.7) node[anchor=north] {Det-R};
	\draw (1,3.3) node[anchor=north] {W-I};
	\draw (1,5.3) node[anchor=north] {W-II};
	\draw (3,5.3) node[anchor=north] {W-III};
	\draw (4.7,5.7) node[anchor=north] {Det-L};
	\draw (2.7,3.6) node[anchor=north] {W-IV};
	
	\end{tikzpicture}
\end{center}

\begin{itemize}
	\item C: Continuous profile, the presence of the particle does not affect the density profile. The particle either escapes to the left ($C,L$) of the rarefaction fan at a speed $\alpha - \nu$ or to the right ($C,R$) of the fan at a speed of $1-\beta -\mu$, figure \ref{fig:munu1a}.
	
	\item D: Discontinuity, the density profile presents a decreasing discontinuity located at the position of the second class particle, and both move at the speed of $\alpha - \beta$. This discontinuity can be located within a rarefaction fan splitting it into two, figure \ref{fig:munu1b}, which is similar to the $1-0$ initial condition case. Or, it can be accompanied by one shock or two.
	
	\item SR: A shock on the right of the discontinuity: The presence of the particle will generate, in addition to the discontinuity, a shock that is located on its right and moves at a higher speed $1-\beta-\mu$. This shock will replace the fan on the right, figure \ref{fig:munu1c}.
	
	\item SL: The shock is located on the left this time and has a speed $\alpha-\nu$, figure \ref{fig:munu1d}.
	
\end{itemize}

\begin{figure}[h!]
	\centering
	\begin{subfigure}[b]{0.4\linewidth}
		\includegraphics[width=\linewidth]{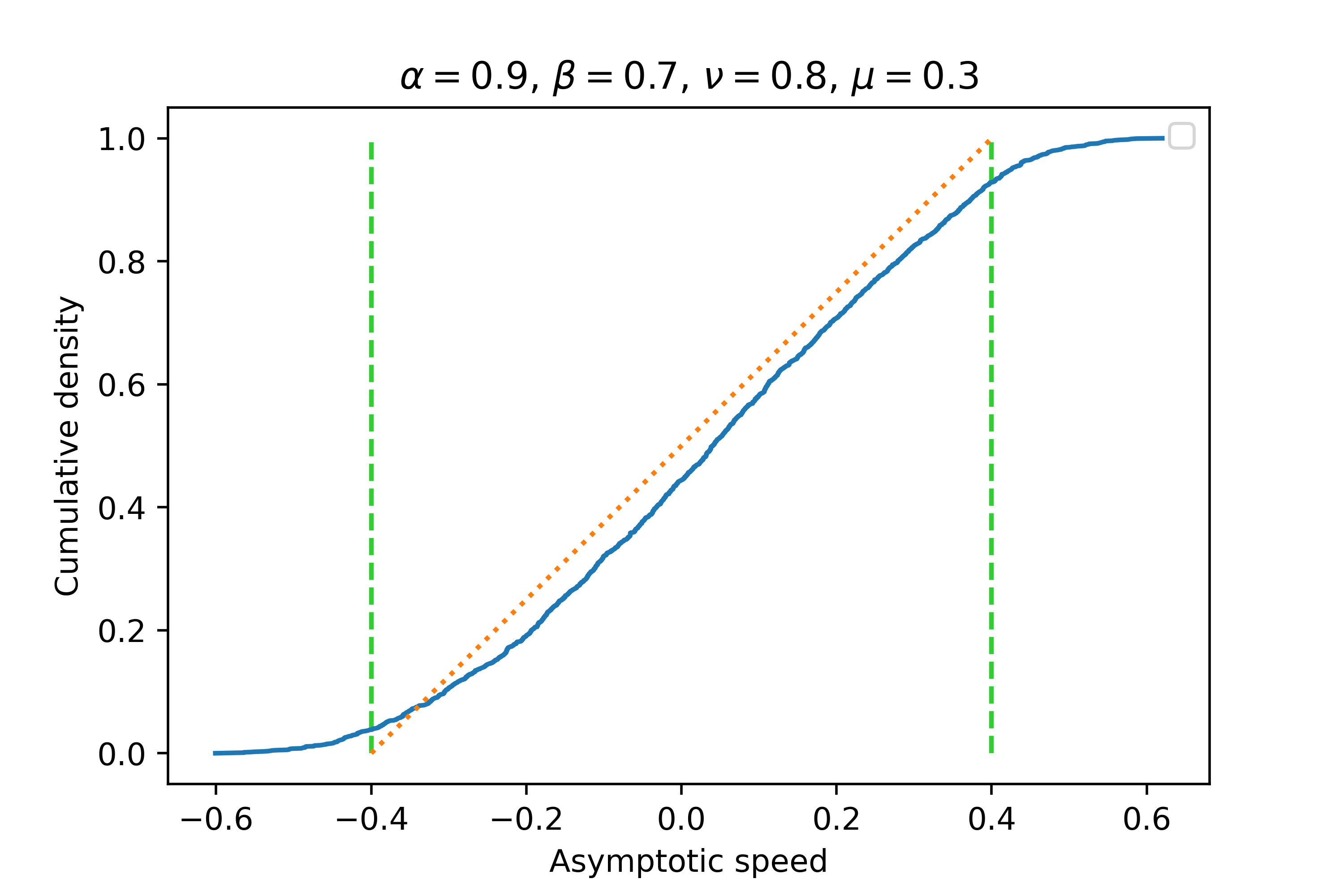}
		\caption{}
		\label{}
	\end{subfigure}
	\begin{subfigure}[b]{0.4\linewidth}
		\includegraphics[width=\linewidth]{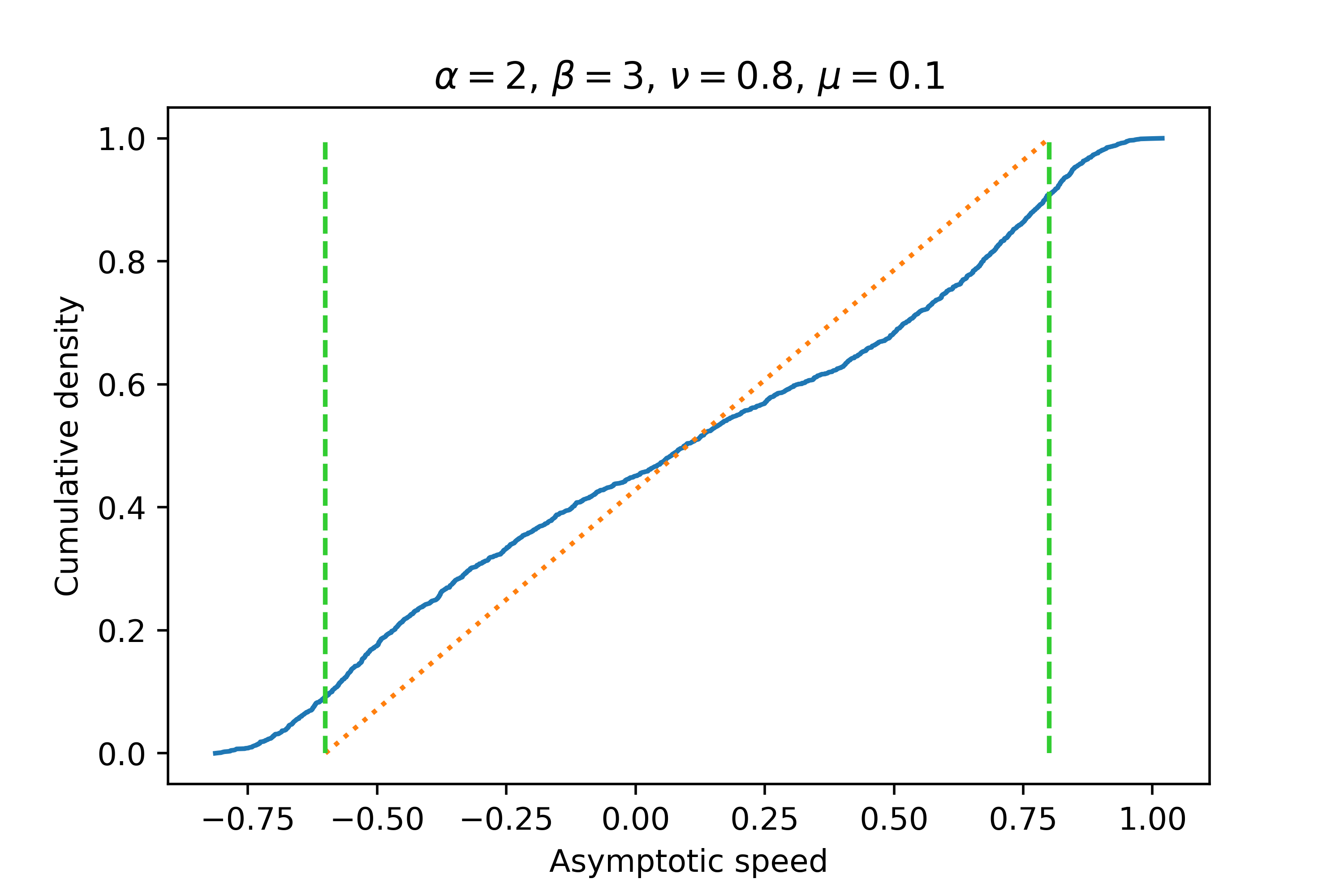}
		\caption{}
		\label{}
	\end{subfigure}
	\begin{subfigure}[b]{0.4\linewidth}
		\includegraphics[width=\linewidth]{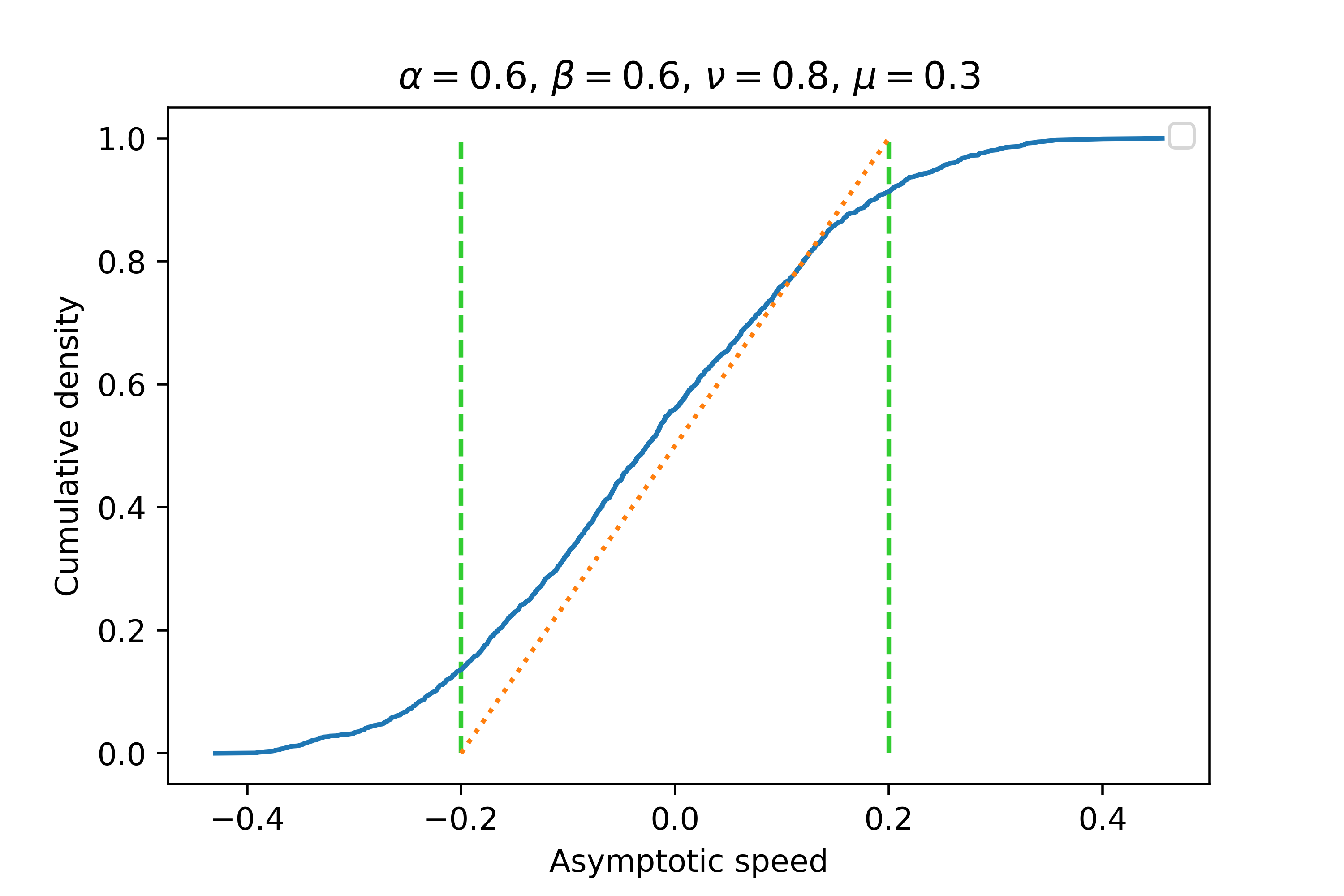}
		\caption{}
		\label{}
	\end{subfigure}
	\begin{subfigure}[b]{0.4\linewidth}
		\includegraphics[width=\linewidth]{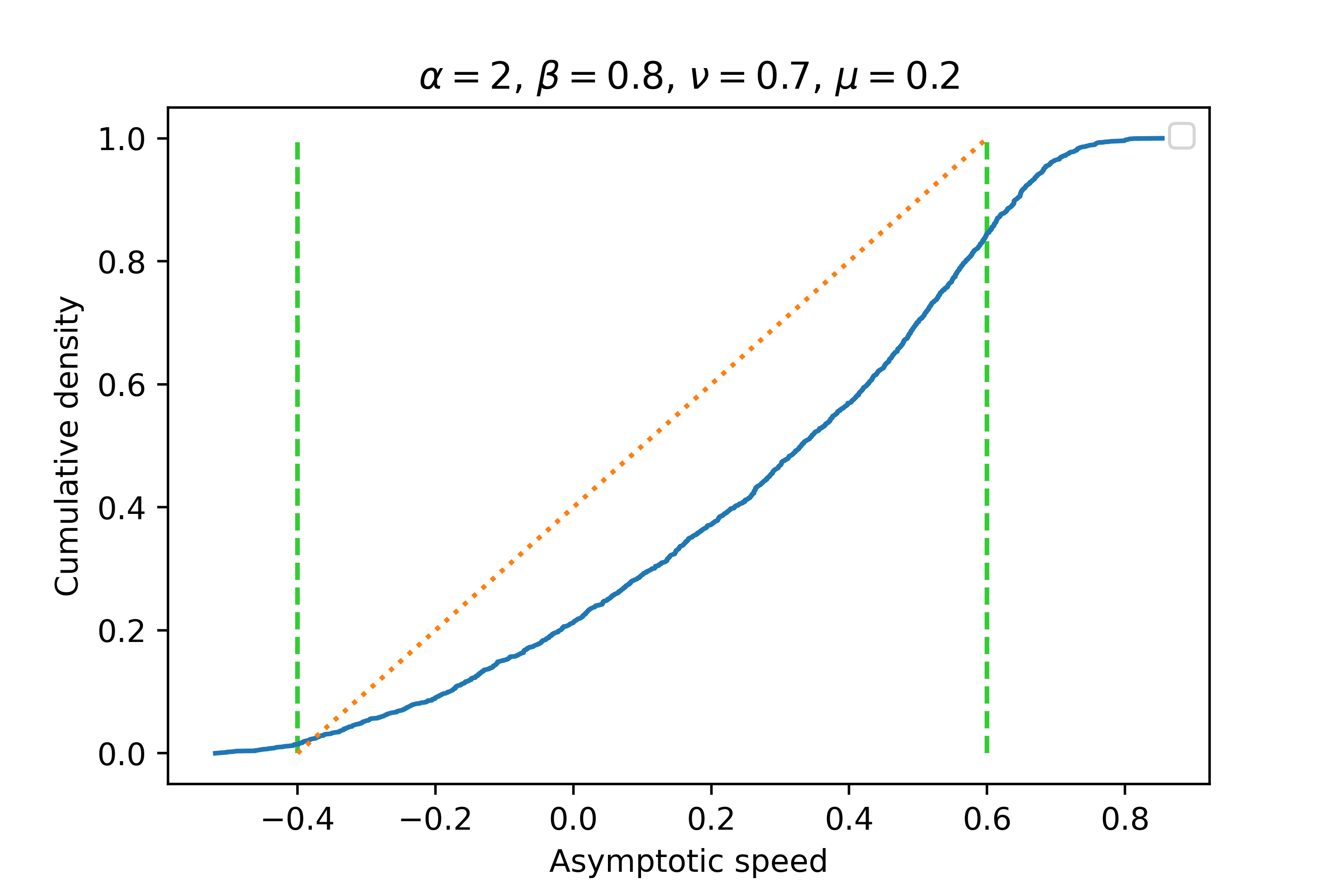}
		\caption{}
		\label{}
	\end{subfigure}
	\caption{Examples featuring non uniform distributions for the asymptotic speed of the second class particle, where $(\nu,\mu) \neq (1,0)$ and $(\alpha,\beta) \neq (1,1)$. Vertical dashed lines represent the predicted windows for the distribution. Dotted orange lines correspond to a hypothetical uniform distribution. (a) corresponds to W-I, and (c) to W-IV. Graphs are plotted for 1000 realizations.}
	\label{nonuniform}
\end{figure}

\begin{figure}[h!]
	\centering
	\begin{subfigure}[b]{0.4\linewidth}
		\includegraphics[width=\linewidth]{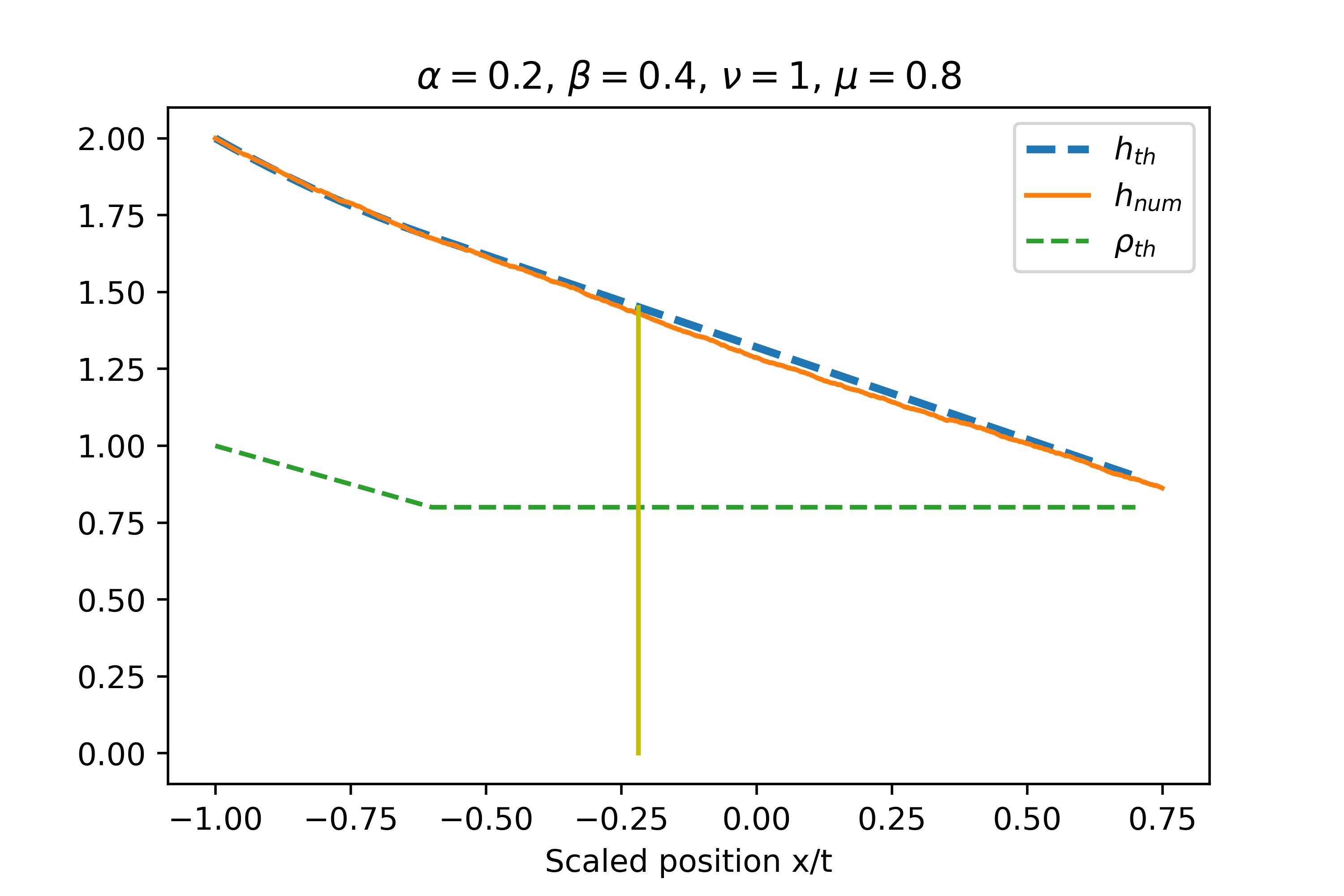}
		\caption{C,R}
		\label{fig:munu1a}
	\end{subfigure}
	\begin{subfigure}[b]{0.4\linewidth}
		\includegraphics[width=\linewidth]{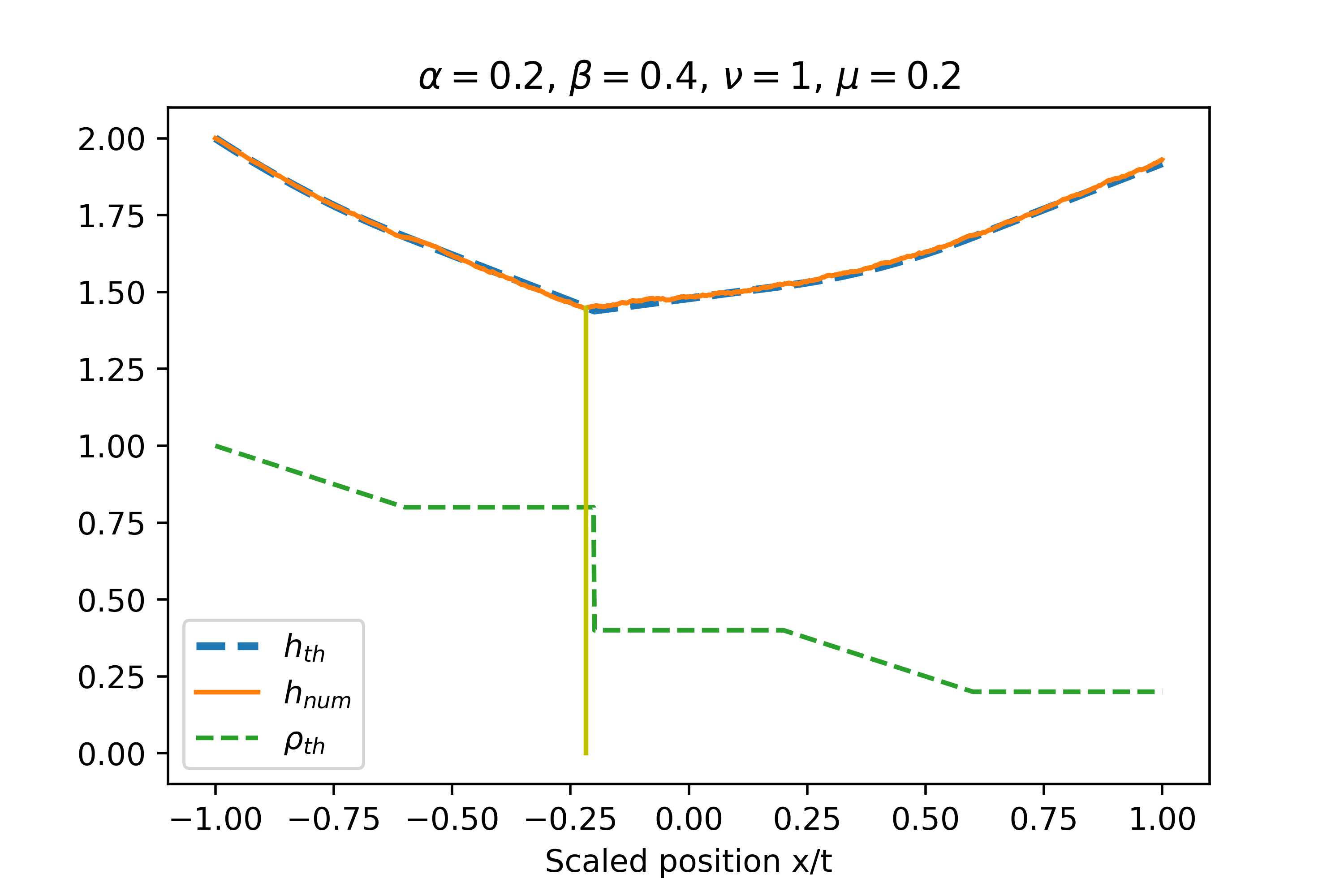}
		\caption{D}
		\label{fig:munu1b}
	\end{subfigure}
	\begin{subfigure}[b]{0.4\linewidth}
		\includegraphics[width=\linewidth]{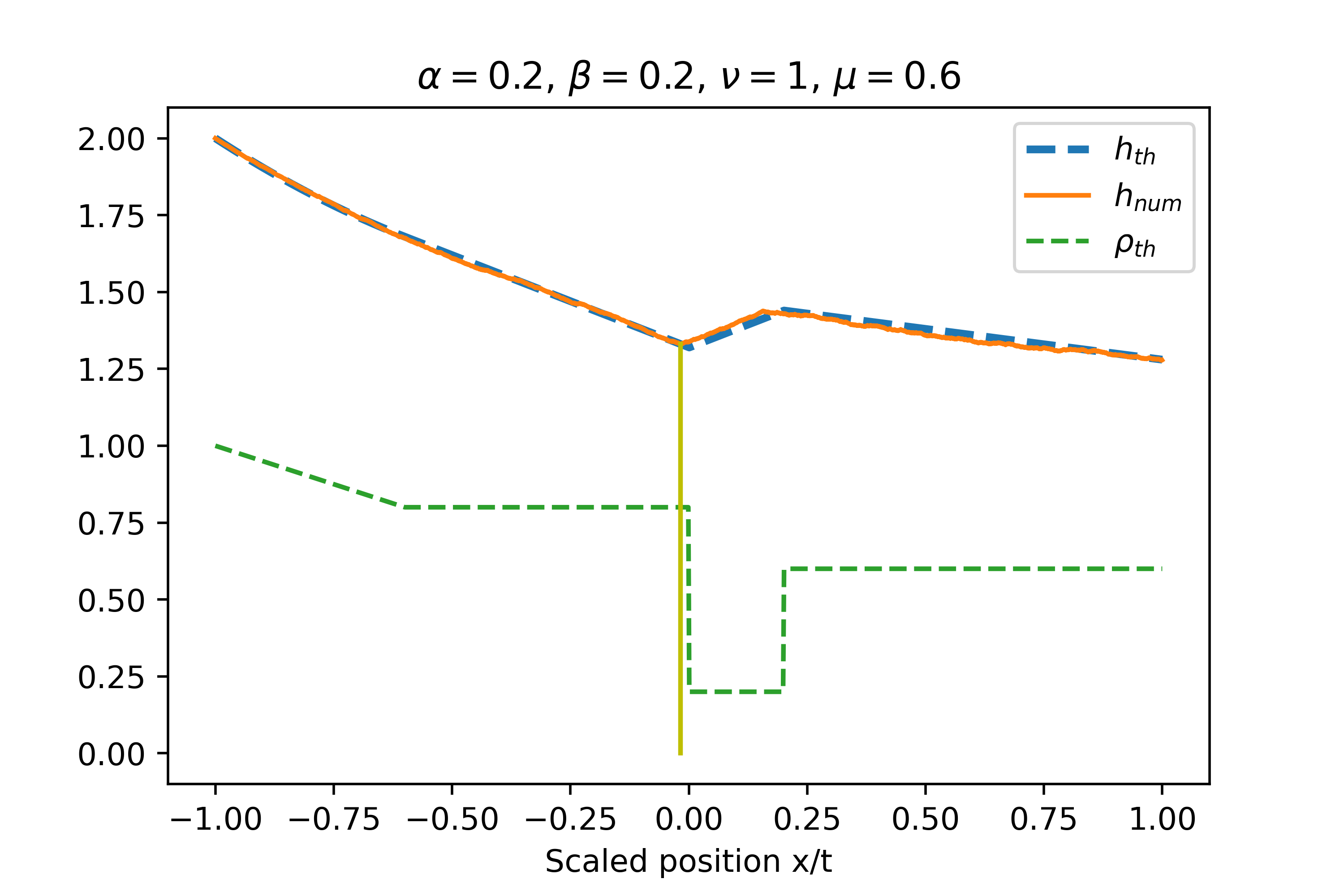}
		\caption{SL}
		\label{fig:munu1c}
	\end{subfigure}
	\begin{subfigure}[b]{0.4\linewidth}
		\includegraphics[width=\linewidth]{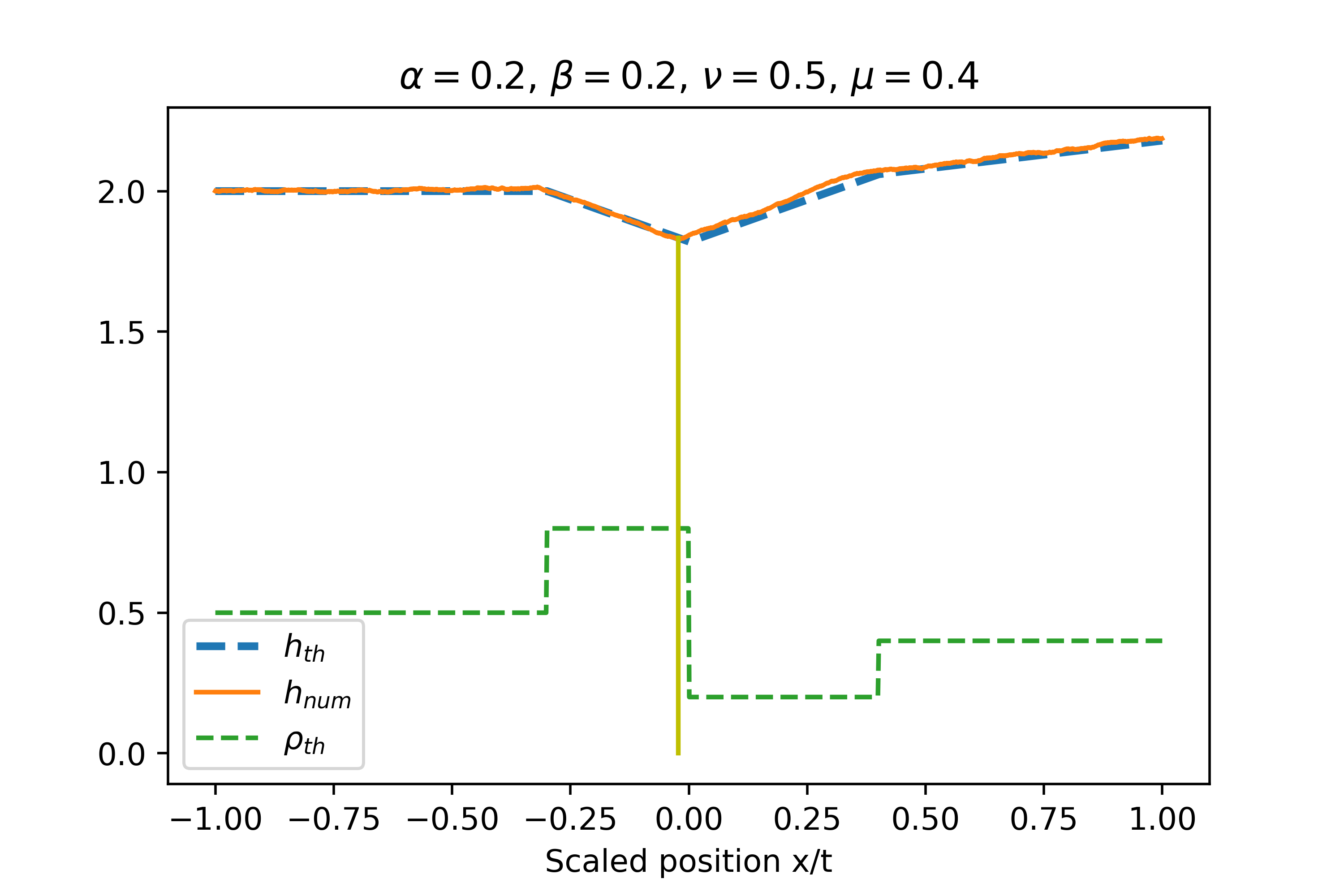}
		\caption{SL, SR}
		\label{fig:munu1d}
	\end{subfigure}
	\caption{Examples of the phenomenology of the density field in the case of $\mu < \nu$ and $\alpha+\beta < 1$. Dashed green line represent the theoretical density. Dashed blue line is the theoretical height defined as $h(x) = \int_{x_{0}}^{x}(1-2\rho(s))ds+ h(x_{0})$. Continuous orange line represents the numerically simulated height function. The vertical line is the simulated position of the second-class particle. The field is run from the step initial condition to $t=2000$. The acronyms of the sub-figures are explained in the section \ref{The case of nu greater than mu}}.
	\label{fig:munu1}
\end{figure}

\subsubsection{The case of $\alpha + \beta>1$}
We know that, in this case, the particle will not disturb the density profile. Its asymptotic speed can be either deterministic or random belonging to an interval. Again, this will depend on the parameters like shown in the diagram:

\begin{itemize}
	\item Det-L: the particle will have a deterministic speed lower than the lowest of the boundaries of the rarefaction fan, so it will be located on its left. This speed is: $\alpha - \nu$
	\item Det-R: the symmetric case of the previous one, the speed will be: $1-\beta - \mu$
	\item W-I: The limit speed is a random distribution within the window $ [1-2\nu , 1-2\mu] $
	\item W-II: The limit speed is a random distribution within the window $ [1-2\beta , 1-2\mu] $
	\item W-III: The limit speed is a random distribution within the window $ [1-2\beta ,  2\alpha-1] $
	\item W-IV The limit speed is a random distribution within the window	$ [1-2\nu , 2\alpha -1] $

\end{itemize}
It is sometimes useful to compactify the four windows with one expression:
$$[max(1-2\nu, 1-2\beta) , min(2\alpha -1, 1-2\mu)] $$

Numerical evidence suggest that these distributions are in general not uniform. figure \ref{nonuniform}. Further investigations are required to determine their forms.

One can check that the boundaries between the different regions are continuous.

\subsection{The case of $\nu < \mu$}
\label{The case of nu less than mu}

This is the case of a shock in the absence of a second-class particle. The shock profile is not affected by the presence of the particle except in the region $1-\alpha > \mu$ and $\beta < \nu $, where the shock bifurcates into two as a result of a decreasing discontinuity created by the particle. The diagram below illustrates all regions here.  

\begin{figure}[h!]
	\centering
	\begin{subfigure}[b]{0.4\linewidth}
		\includegraphics[width=\linewidth]{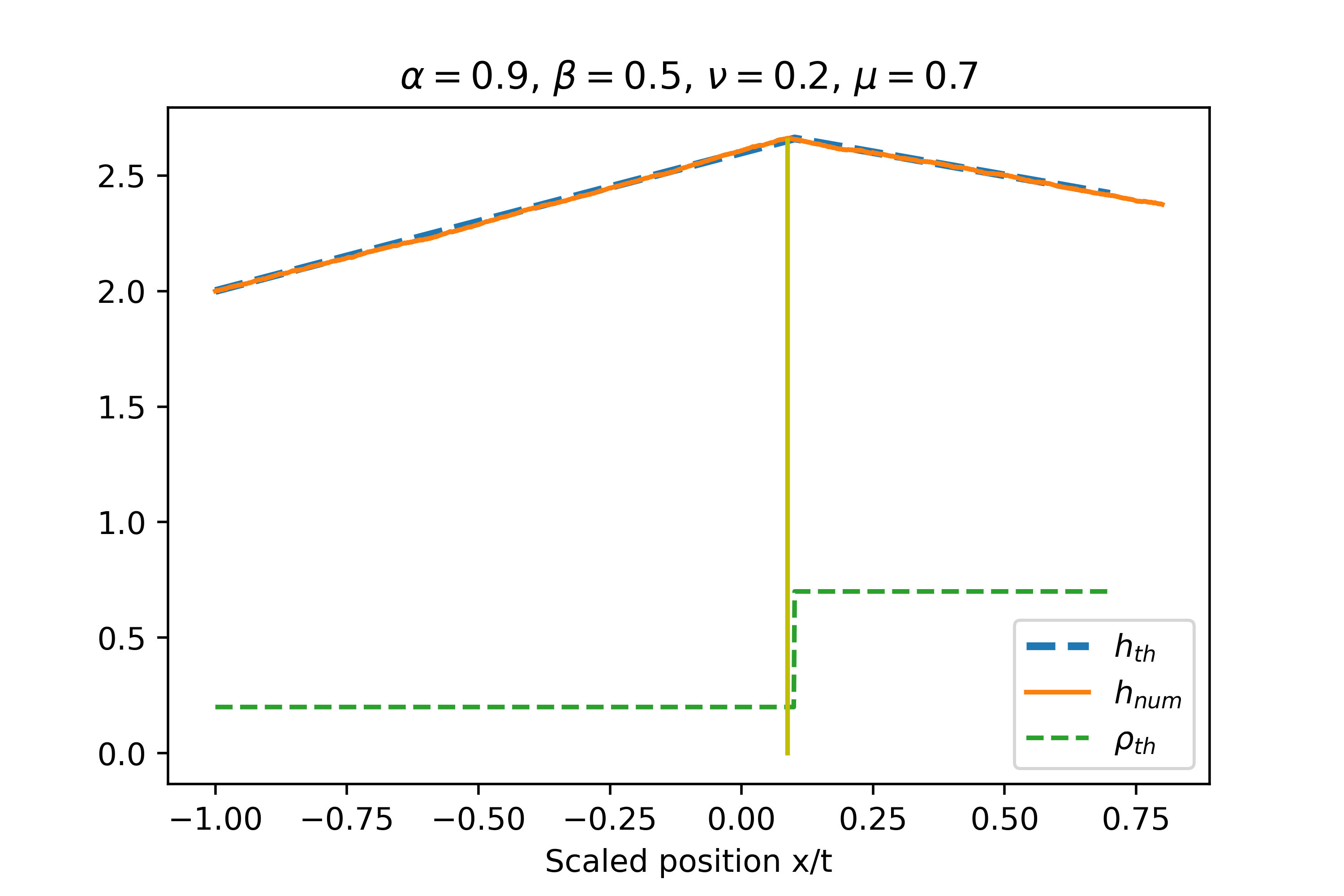}
		\caption{St}
		\label{shocksa}
	\end{subfigure}
	\begin{subfigure}[b]{0.4\linewidth}
		\includegraphics[width=\linewidth]{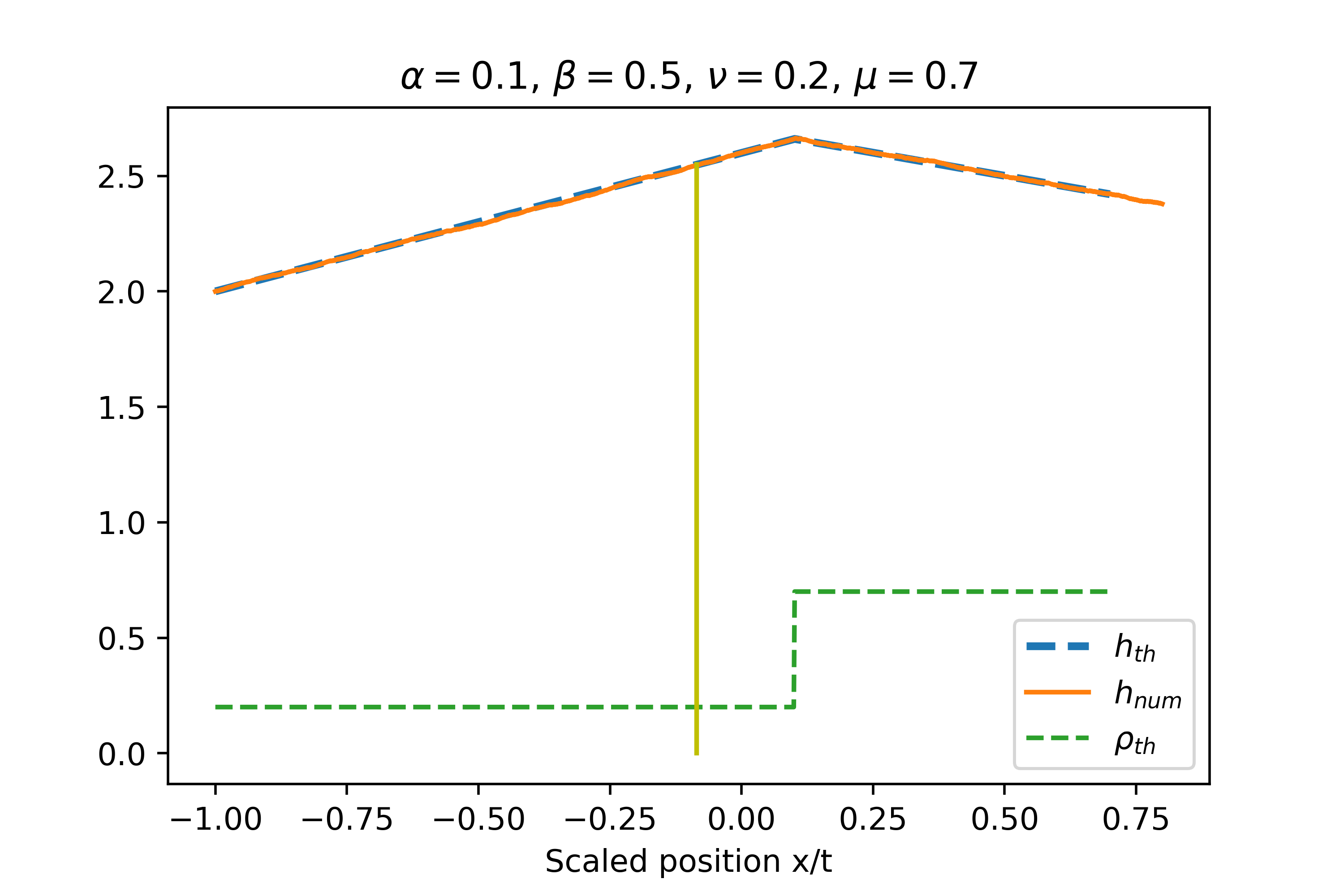}
		\caption{L}
		\label{shocksb}
	\end{subfigure}
	\begin{subfigure}[b]{0.4\linewidth}
		\includegraphics[width=\linewidth]{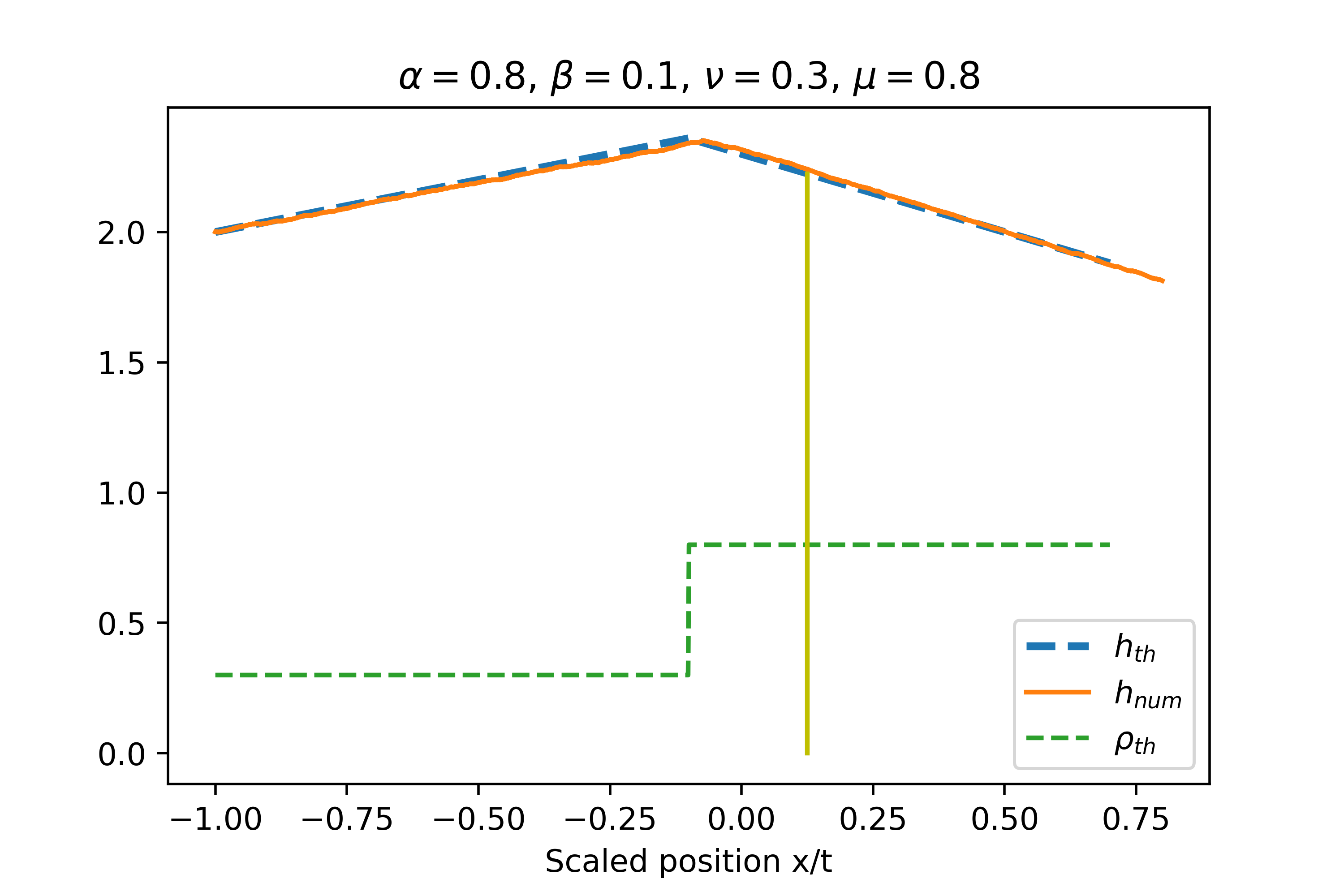}
		\caption{R}
		\label{shocksc}
	\end{subfigure}
	\begin{subfigure}[b]{0.4\linewidth}
		\includegraphics[width=\linewidth]{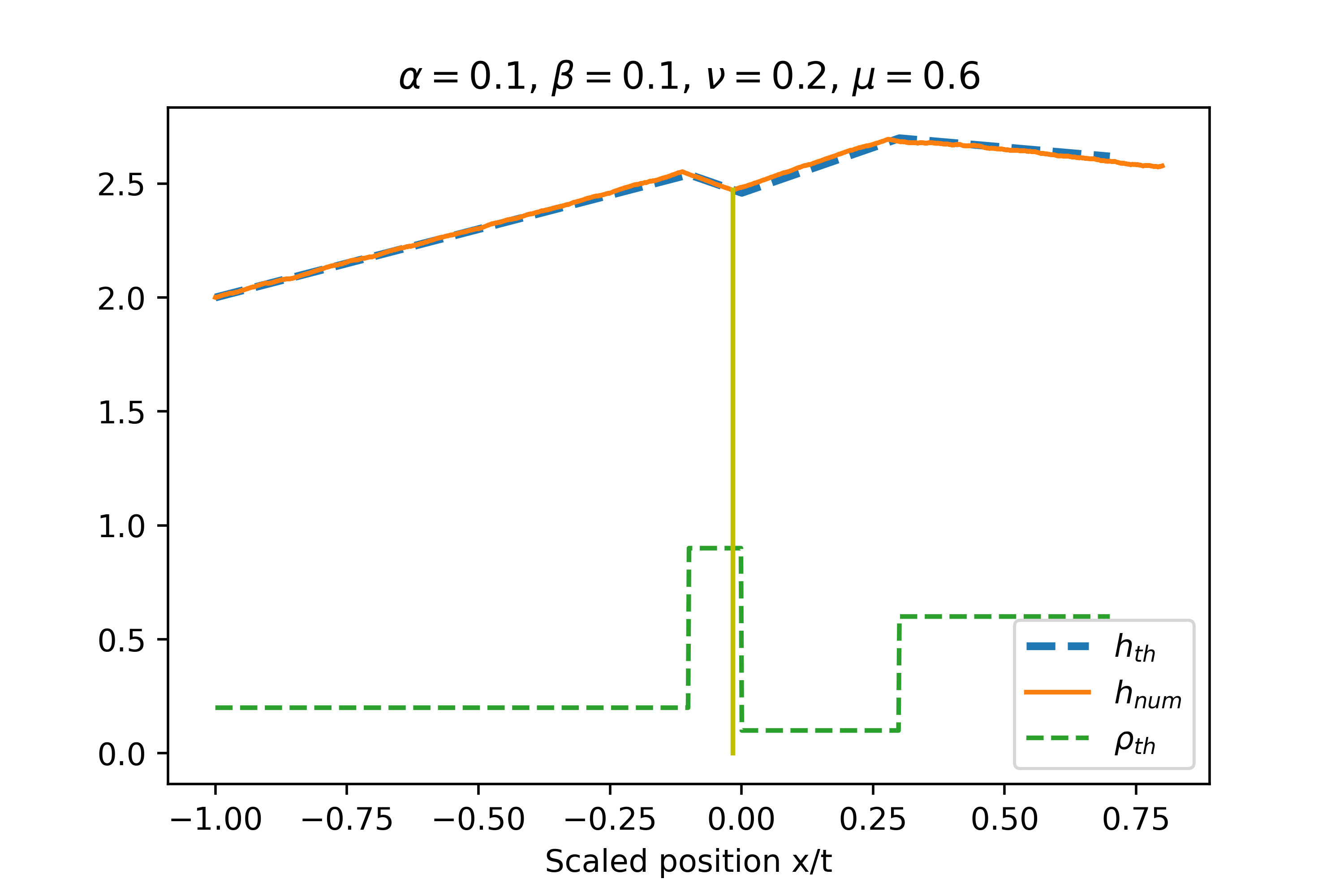}
		\caption{Sp}
		\label{shocksd}
	\end{subfigure}
	\caption{Examples of the phenomenology in the case of $\mu > \nu$ and $\alpha+\beta < 1$. The dashed green line represents the theoretical density. Dashed blue line is the theoretical height defined as $h(x) = \int_{x_{0}}^{x}(1-2\rho(s))ds+ h(x_{0})$. The continuous orange line represents the numerically simulated height function. The vertical line is the simulated position of the second-class particle. The field is run from the step initial condition to $t=2500$. The acronyms of the sub-figures are explained in section \ref{The case of nu less than mu}}
	\label{shocks}
\end{figure}

\begin{center}
	\begin{tikzpicture}[scale = 0.8]
	\draw[thick,->] (0,0) -- (7,0) node[anchor=north west] {$1-\alpha$};
	\draw[thick,->] (0,0) -- (0,7) node[anchor=south east] {$\beta$};

	\draw (2 cm,1pt) -- (2 cm,-1pt) node[anchor=north] {$\nu$};
	
	\draw (4 cm,1pt) -- (4 cm,-1pt) node[anchor=north] {$\mu$};
	
	\draw (6 cm,1pt) -- (6 cm,-1pt) node[anchor=north] {$1$};
	
	\draw (1pt,2 cm) -- (-1pt,2 cm) node[anchor=east] {$\nu$};
	
	\draw (1pt,4 cm) -- (-1pt,4 cm) node[anchor=east] {$\mu$};
	
	\draw (1pt,6 cm) -- (-1pt,6 cm) node[anchor=east] {$1$};
	
	\draw (0,2) -- (6,2);
	
	\draw (0,6) -- (6,6);
	
	\draw (4,0) -- (4,6);
	\draw (6,0) -- (6,6);
	
	\draw (5.7,1) node[anchor=east] {};
	
	\draw (5,4.3) node[anchor=north] {L};
	
	\draw (2,1.3) node[anchor=north] {R};
	
	\draw (2,4.3) node[anchor=north] {St};
	
	\draw (5,1.3) node[anchor=north] {Sp};
	
	\end{tikzpicture}
	
\end{center}

\begin{itemize}
	\item \textbf{St} The particle is stuck in the shock and at its speed of $1-\nu -\mu$, figure \ref{shocksa}
	\item \textbf{R} The particle has a higher speed than the shock. Its speed is $1-\mu-\beta$ figure \ref{shocksc}
	\item \textbf{L} The particle has a lower speed than the shock. Its speed is given by $\alpha - \nu$, figure \ref{shocksb}
	\item \textbf{Sp} The shock splits into two shocks separated by a discontinuity located at the particle position and moves at a speed $\alpha - \beta$. The shock on its right moves at a speed $1-\mu-\beta$. The shock on the left has a speed $\alpha - \nu$, figure \ref{shocksd}. This region is a continuation of its counterpart when $\nu > \mu$.

\end{itemize}

One can verify that the limits between the different regions are continuous.

\subsection{A uniform vanishing density of second class particles:}

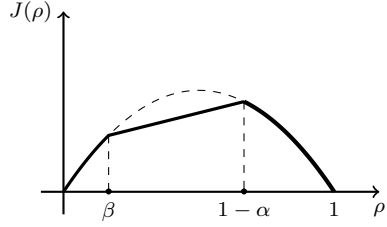
\begin{figure}
	\centering
	\begin{tikzpicture}[scale=1.2]
	
	\draw [thick,->] (-.25,0)--(3.5,0)node [below] {\scriptsize $\rho$};
	\draw [thick,->] (0,-.25)--(0,2)node [left] {\scriptsize $J(\rho)$};
	\def \r {1/2}
	\def \n {.5}
	\def \f {2}
	\draw[dashed,domain=0:3,smooth,variable=\t]plot (\t,{\t*(3-\t)*\r})  node[below] {\scriptsize $1$};
	\draw[very thick,domain=0:{\n},smooth,variable=\t]plot (\t,{\t*(3-\t)*\r})-- (\f,{\f*(3-\f)*\r});
	\draw[ultra thick,domain={\f}:3,smooth,variable=\t]plot (\t,{\t*(3-\t)*\r});
	\draw[dashed,fill=black] (\n,{\n*(3-\n)*\r})--(\n,0)circle  (.03) node[below] {\scriptsize $\beta$};
	\draw[dashed,fill=black] (\f,{\f*(3-\f)*\r})--(\f,0)circle  (.03) node[below] {\scriptsize $1-\alpha$};
	\end{tikzpicture}
	\caption{A vanishing density of second class particles makes the usual burgers current(dashed line), linear in the interval $[\beta, 1-\alpha]$ (thick line)}
	\label{current}
\end{figure}

\begin{figure}[h!]
	\centering
	\begin{subfigure}[b]{0.4\linewidth}
		\includegraphics[width=\linewidth]{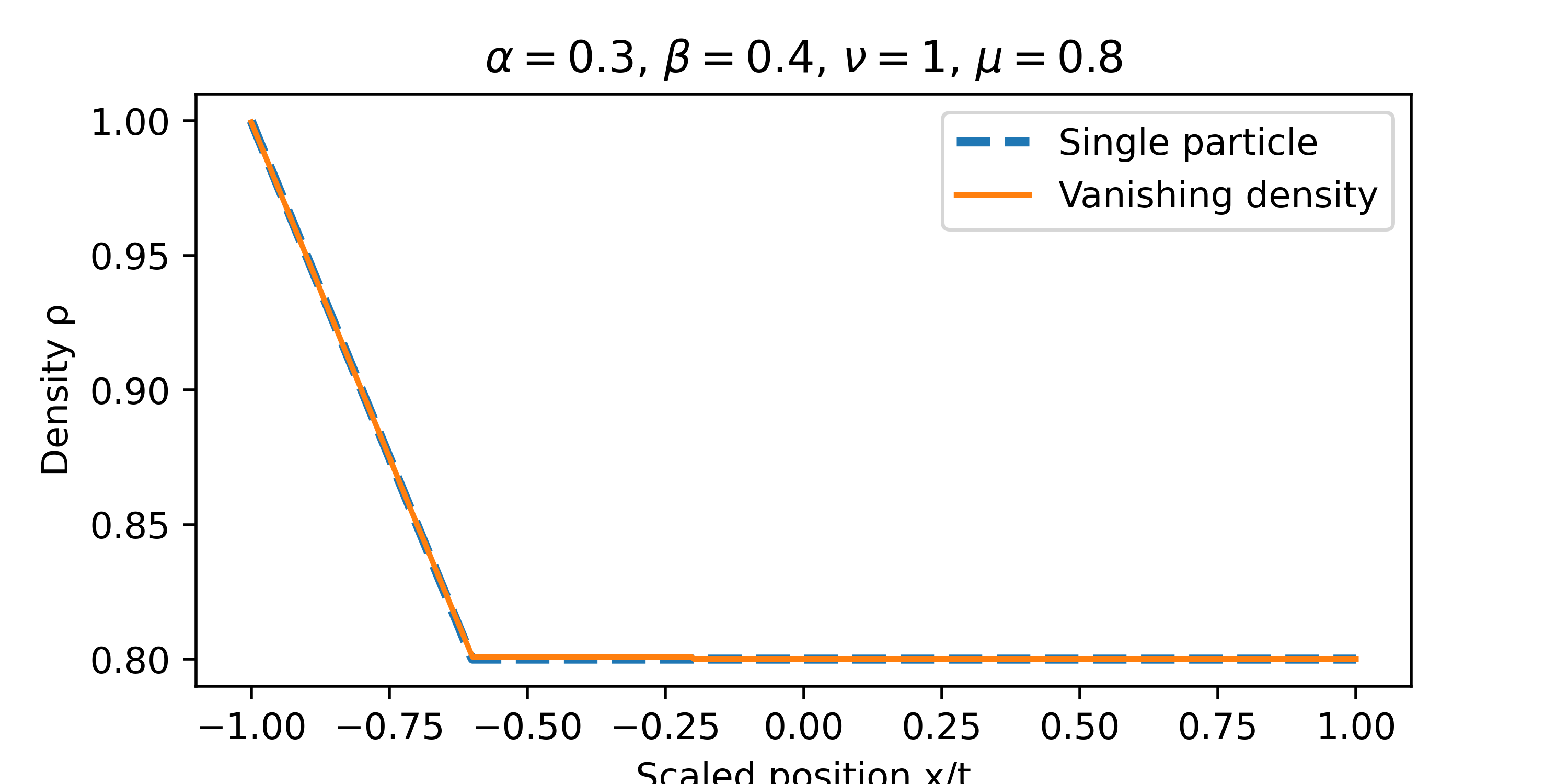}
		\caption{}
		\label{compare1}
	\end{subfigure}
	\begin{subfigure}[b]{0.4\linewidth}
		\includegraphics[width=\linewidth]{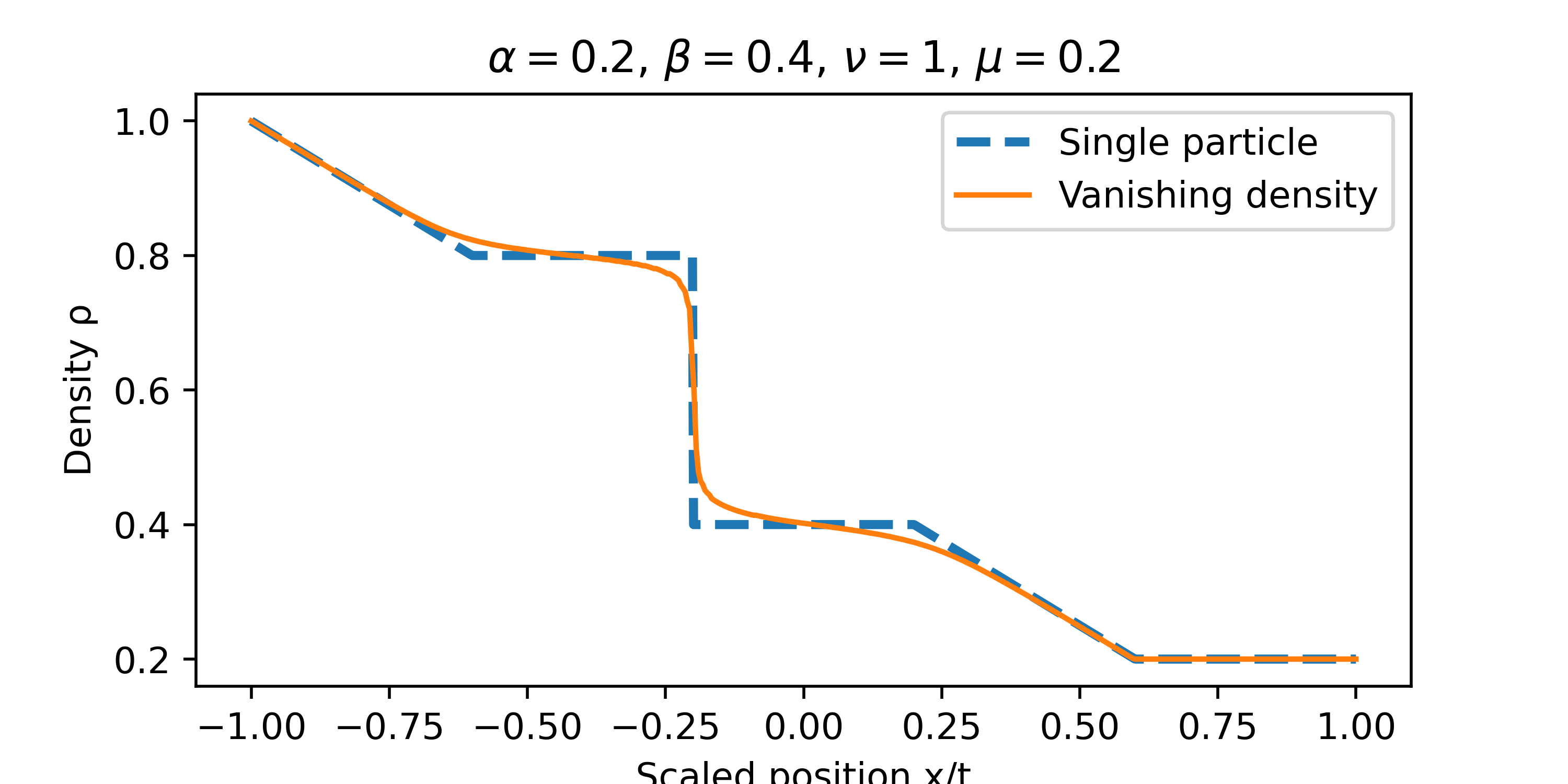}
		\caption{}
		\label{compare2}
	\end{subfigure}
	\begin{subfigure}[b]{0.4\linewidth}
		\includegraphics[width=\linewidth]{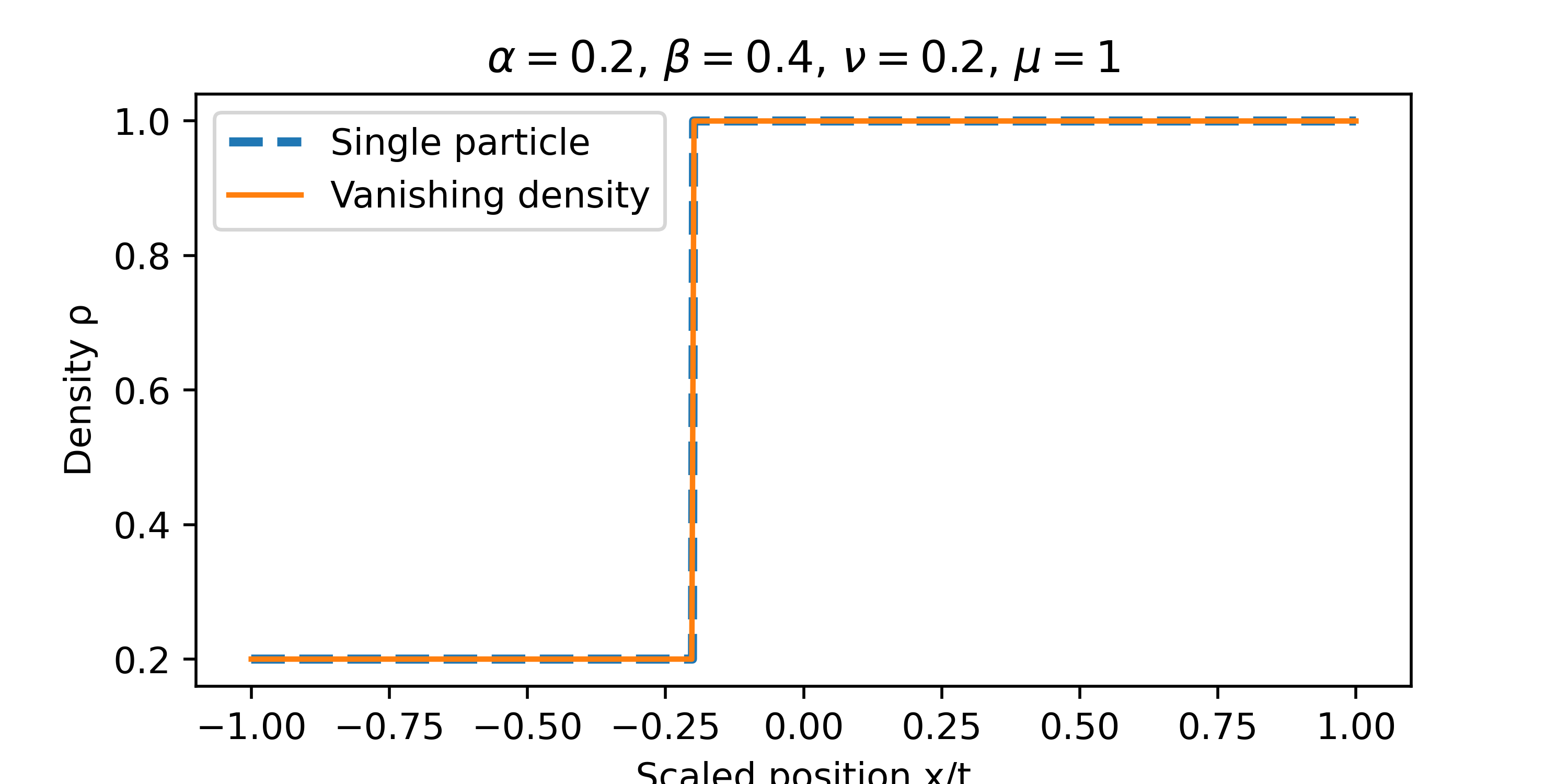}
		\caption{}
		\label{compare22}
	\end{subfigure}
	\begin{subfigure}[b]{0.4\linewidth}
		\includegraphics[width=\linewidth]{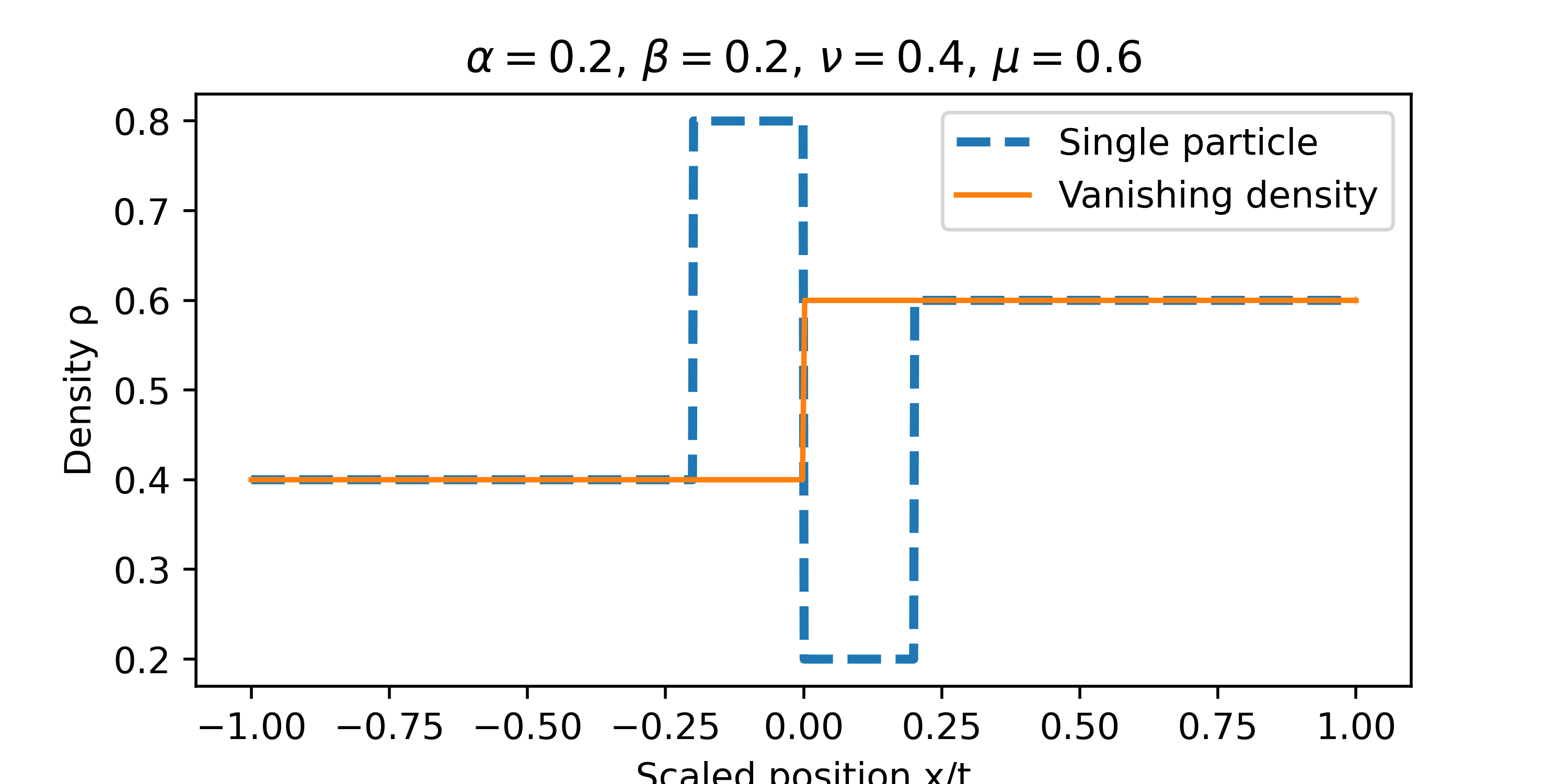}
		\caption{}
		\label{compare4}
	\end{subfigure}
	\begin{subfigure}[b]{0.4\linewidth}
		\includegraphics[width=\linewidth]{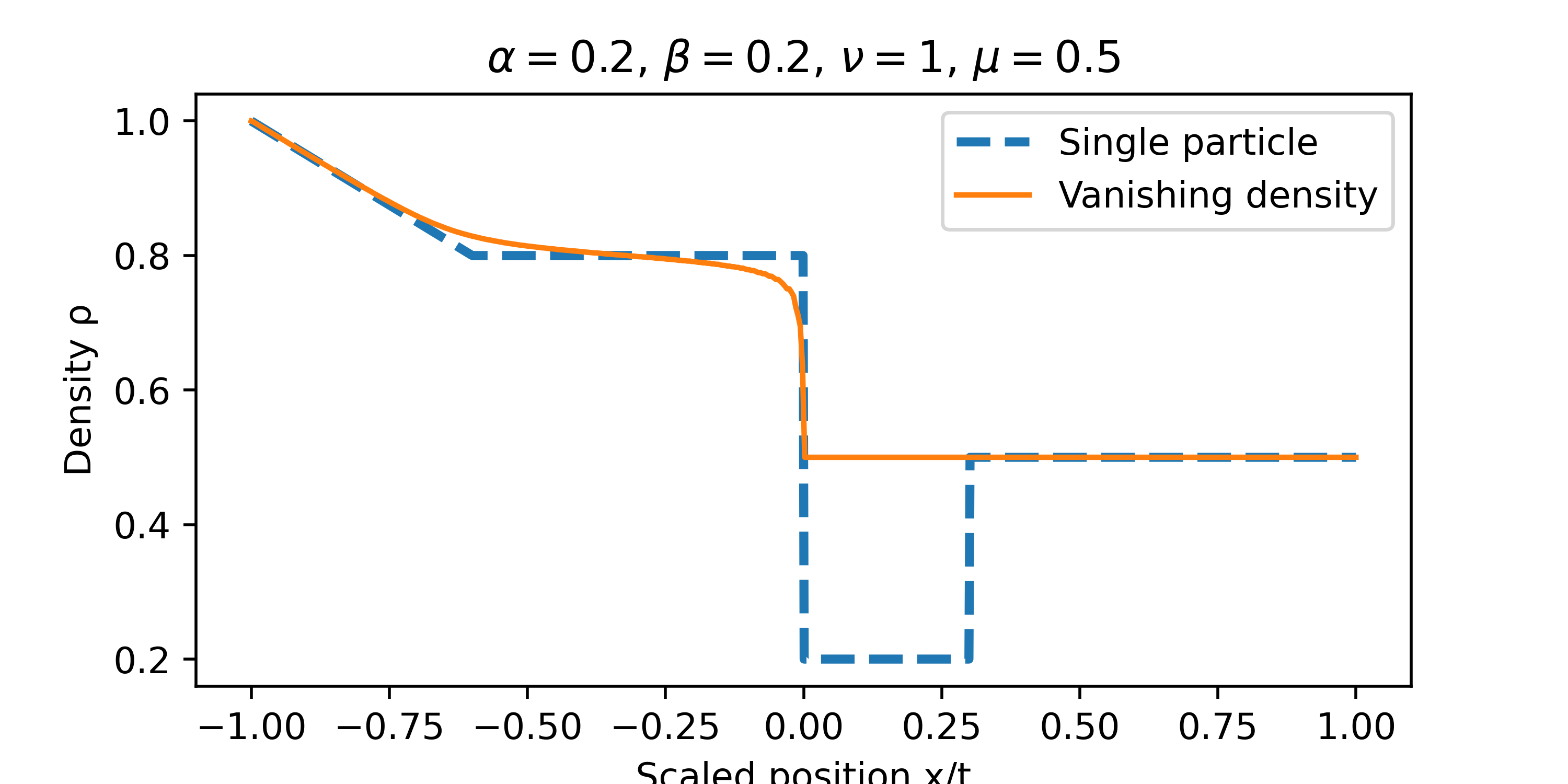}
		\caption{}
		\label{compare3}
	\end{subfigure}
	\begin{subfigure}[b]{0.4\linewidth}
		\includegraphics[width=\linewidth]{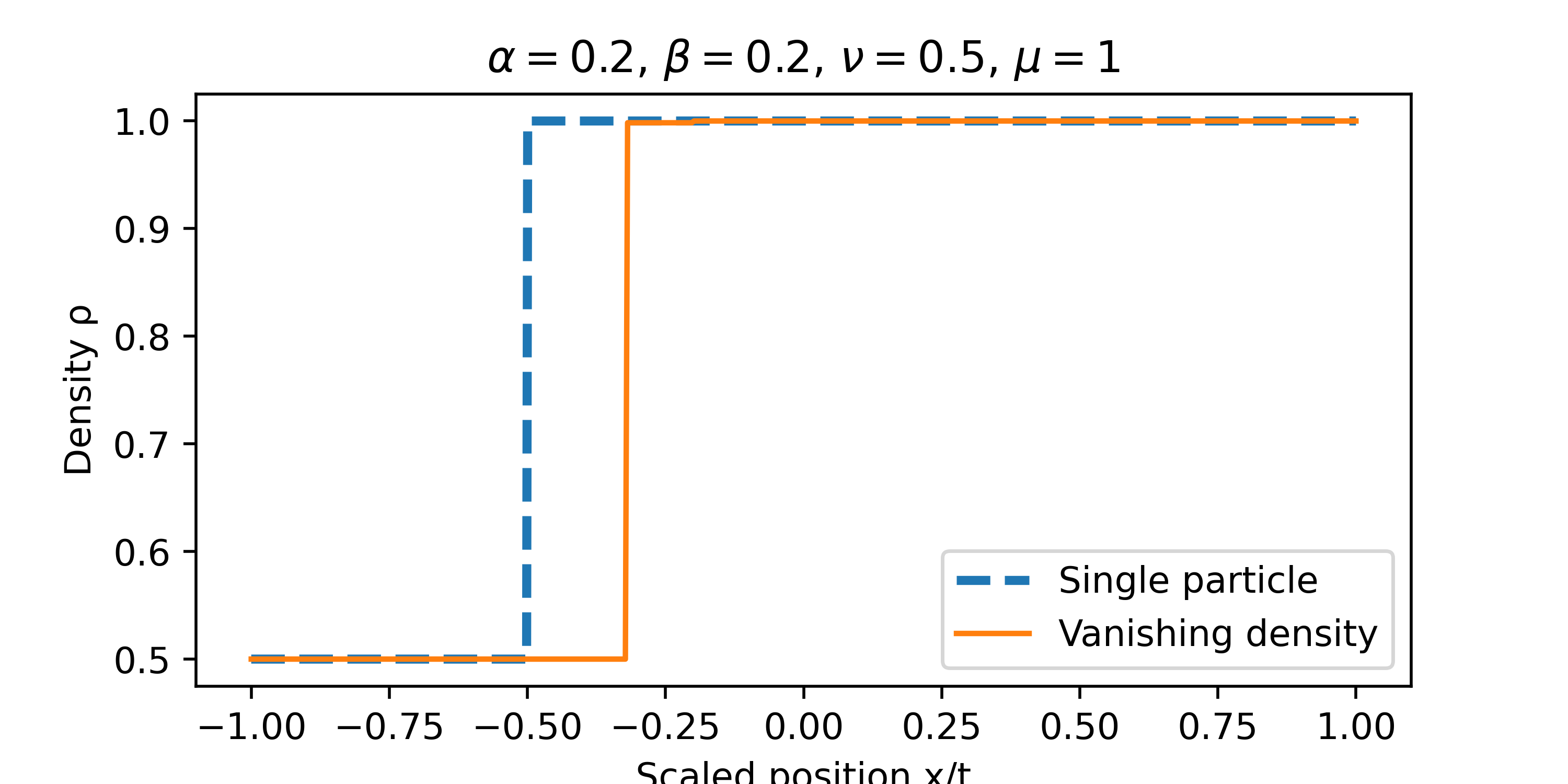}
		\caption{}
		\label{compare33}
	\end{subfigure}
	\caption{Comparison between the effect of a single particle and the effect of vanishing density of second class particles over the density field. Plots with the vanishing density are made for a uniform density of $10^{-3}$}.
	\label{compare}
\end{figure}

In this section, we would like to investigate the following question: under which circumstances one single second class particle is macroscopically equivalent to a vanishing uniform density of second-class particles on the line? The question is of course non trivial only in the case where $\alpha + \beta < 1$. This vanishing density can be obtained as a limit of the model introduced in \cite{cantini2022hydrodynamic}. This limit modified the usual current of the Burgers equation in a way illustrated in figure \ref{current}. One can identify the following situations:
\begin{itemize}
	\item This simplest case is when the intervals $[\beta, 1-\alpha]$ and $ [min(\mu,\nu), max(\mu,\nu) ] $ do not overlap. In this case, neither the vanishing density nor a single particle has an effect on the density field that behaves simply as a solution of Burgers equation, figure \ref{compare1}
	
	\item The interval $[\beta, 1-\alpha]$ is included in $ [min(\mu,\nu), max(\mu,\nu) ] $.The density profile here is the same for both a single particle and a vanishing density: in the case of $\mu<\nu$ the vanishing density has an impact over the density field which is the same as the perturbation created by a single particle. figure \ref{compare2} In the case of $\mu>\nu$ the shock does not feel neither a single particle nor a vanishing density, figure \ref{compare22}
	
	\item The interval$ [min(\mu,\nu), max(\mu,\nu) ]$ is included in  $[\beta, 1-\alpha]$. Here a single class particle will create two shocks while a vanishing density will create only a discontinuity regardless of the order of $\nu$ and $\mu$ figure \ref{compare4}

	\item The two intervals overlap without one of them being included in the other. The two situations do not give rise to the same density profile. In the case of the fan, the particle produces a shock, figure \ref{compare3}. In the case of the shock, its speed is affected by the vanishing density, since the linear part will affect Hugoniot condition, figure \ref{compare33}

\end{itemize}

\section{Speed process of a defect in a step initial configuration}
\label{Speed process of a defect in a step initial configuration}
As mentioned previously, a second-class particle of rates $\alpha+\beta>1$ does not have a deterministic asymptotic speed. The purpose of this section is to provide a rigorous proof that for the case of $\alpha < 1$ and $\beta < 1$ and a 1-0 initial step configuration, the distribution of the asymptotic speed is still uniform within the allowed window. The proof relies on some results from the queuing theory besides extending the usage of the coupling tool to systems with defect particles. In the next lemma, we will consider a system of first-class particles with two special tagged particles that have arbitrary hopping rates.

\begin{lem}\label{lemma1}
	Let $(\eta_{t},x_{1}^{\alpha_{1}}(t),x_{2}^{\alpha_{2}}(t)) \in \{0,1\}^{\mathbb{Z}} \times \mathbb{Z} \times \mathbb{Z} $ be a configuration of the system at time $ t $ with two tagged first class particles of forward hopping rates $\alpha_{1}$ and $\alpha_{2}$ located at positions $ x_{1}^{\alpha_{1}} (t) $ and $ x_{2}^{\alpha_{2}} (t)$, respectively.
	Assume that $ x_{2}^{\alpha_{2}} (0) = x_{1}^{\alpha_{1}} (0) + 1 $, then the process $x_{1}^{\alpha_{1}} (t)$ is symmetric with respect to the parameters $\alpha_{1}$ and $ \alpha_{2} $. 
\end{lem}

In other words: if $ (\eta_{0},x_{1}^{\alpha_{1}}(0),x_{2}^{\alpha_{2}}(0)) = (\tilde{\eta}_{0},\tilde{x}_{1}^{\alpha_{2}}(0),\tilde{x}_{2}^{\alpha_{1}}(0)) $, then for any set of fixed times $ (t_{i})_{1 \leq i \leq n} $, the joint probability distribution of

$$ (x_{1}^{\alpha_{1}}(t_{1}),..., x_{1}^{\alpha_{1}}(t_{n})) \stackrel{d}{=}  (\tilde{x}_{1}^{\alpha_{2}}(t_{1}),..., \tilde{x}_{1}^{\alpha_{2}}(t_{n})) $$

For later convenience and with no loss of generality we set: $\alpha_{1} = 1$ and $\alpha_{2} = \alpha$.

\textit{Sketch Proof}:
this lemma can be shown from results in the queuing theory literature. For that purpose, imagine the particles as servers and the void as customers. In \cite{weber1979interchangeability, lehtonen1986ordering}, it has been proven that if we have a series of consecutive exponential servers initially empty, and an arbitrary arrival statistical process, then the departure process (the movement of the most left particle) is independent of the ordering of the servers. We obviously need this property in our case for only two servers (the two servers’ case is anyway equivalent to any finite number of servers). Since the arrival process can be arbitrary, a random initial configuration of the rest of the particles doesn't present a problem.

In the Appendix, we obtain an explicit expression of the distribution of the marginal process (i.e. the distribution of the position of the most left particle at a time $t$) using \cite{rakos2006bethe} or equivalently the conditional probability expression in chapter 3, and show manifestly its symmetry with regards to $\alpha_{1}$ and $\alpha_{2}$.
\qed

\begin{lem}\label{lemma2}
	Consider two systems with initial configurations: $ (\eta_{0},x_{1}^{\alpha_{1}}(0),x_{2}^{\alpha_{2}}(0)) = (\tilde{\eta}_{0},\tilde{x}_{1}^{\alpha_{2}}(0),\tilde{x}_{2}^{\alpha_{1}}(0))$, where $x_{2}^{\alpha_{2}}(0) = x_{1}^{\alpha_{1}}(0) + 1$, then it is possible to couple the two systems so that for all $t \geq 0$ we have $ x_{1}^{\alpha_{1}}(t) = \tilde{x}_{1}^{\alpha_{2}}(t)$.	
\end{lem}

\subsubsection*{Remark}
It is possible to use the hole-particle symmetry to generate a dual lemma of lemma \ref{lemma2} . For that purpose, we tag two holes instead of two particles, we set their rates to $\beta_{1}$ and $\beta_{2}$. (That obviously means that jumping over these holes would be determined now by the clocks attached to them). We denote the configuration at time $t$: $(\eta_{t},y_{1}^{\beta_{1}}(t),y_{2}^{\beta_{2}}(t))$.
Assume that $ y_{2}^{\beta_{2}} (0) = y_{1}^{\beta_{1}} (0) - 1 $, then the process $y_{1}^{\beta_{1}}(t)$ is invariant under the exchange of $ \beta_{1} $ and $ \beta_{2} $. It is as well possible, as in the previous lemma, to couple the movement of the two holes $y_{1}^{\beta_{1}}$ and $\tilde{y}_{1}^{\beta_{2}}$ belonging to the two systems with the same initial configuration: $(\eta_{0},y_{1}^{\beta_{1}}(0),y_{2}^{\beta_{2}}(0)) = (\tilde{\eta}_{0},\tilde{y}_{1}^{\beta_{1}}(0),\tilde{y}_{2}^{\beta_{2}}(0))$, providing that at $t = 0$ the two holes are consecutive. 

Note that the clocks involved in this coupling are all located to the left of $y_{1}$, $(\tilde{y}_{1})$ and are all attached to holes.

\begin{lem}\label{lemma3}
	
	Let's consider an initial configuration with a particle at the origin, a hole at the site -1 and a Bernoulli product measure on the other sites with a parameter $\mu$ for the positive sites and $\nu$ for the negative sites starting from -2. Let's call this initial configuration: $(\eta_{0}, y^{1}(0) ,x^{1}(0)) = (\eta_{0}, -1 ,0)$ Then:
	\newlist{alphalist}{enumerate}{1}
	\setlist[alphalist,1]{label=\textbf{\alph*.}}
	\begin{itemize}
		
		\item [$(a)$] $x^{1}(t)$ will have the same distribution as the position of a free particle of rate $1 - \mu$ starting at the origin.
		\item [$(b)$]  $y^{1}(t)$ will have the same distribution as the position of a free hole of rate $\nu$ starting at the site -1.
	\end{itemize}
\end{lem}

\textit{Sketch Proof}.
Noticing that $x^{1}(t)$ obviously depends only on the movement of the particles located on the positive sites, (a) becomes nothing but a restatement of example 3.2 of Spitzer (1970) \cite{spitzer1991interaction}. (b) is obviously obtained from (a) using the hole-particle symmetry.
\qed

\subsection{Probability distribution of a second class particle of arbitrary rates in a step initial configuration}
Let $ 0 \leq \alpha, \beta \leq 1$ and $\alpha+\beta \geq 1 $. We consider a second-class particle of rates $\beta,\alpha$ located initially at the origin with no particle to its right and no hole to its left. We denote this system: $(\eta_{t}, {}^{\beta}z^{\alpha}(t))$

with ${}^{\beta}z^{\alpha}(t)$ being the position of the second class particle at time $t$.

Let's consider as well the configuration with a second-class particle of rates equal to 1 located at the origin in a Bernoulli product measure initial configuration with a parameter $1-\alpha$ for the positive sites and a parameter $\beta$ for the negative sites. We denote this system $(\eta_{t}^{R},z^{R}(t))$, we call it the reference system.

Our main result here is:

$$ z^{R}(t)| \eta_{0}^{R} \stackrel{d}{=} {}^{\beta}z^{\alpha}(t)| \eta_{0} $$

\subsection*{Proof:}

Let us define the configuration $\xi^{R}$ as follows:

\begin{equation}
\xi^{R}_{0}(0) := 1  \quad \xi^{R}_{0}(-1) := 0 \quad \xi^{R}_{0}(k) = \left\{
\begin{array}{ll}

\eta_{0}^{R}(k) & k > 0 \\
\eta_{0}^{R}(k+1) & k < -1

\end{array}
\right.
\end{equation}

It is a classical procedure to see the second-class particle of rates equal to one as a couple of a hole followed by a first-class particle. We will track the position of this $(0,1)$ couple, (defined as the position of the particle component of it) using the variable $z^{R}$, this would require a suitable coupling between $\eta^{R}$ and $\xi^{R}$ that allows this identification.
Let's as well tag the particle (the hole) located initially at the origin (at -1) with the variable $x^{R}$, ($y^{R}$).
Note that $x^{R}$ and $y^{R}$ coincide only initially with the couple $(0,1)$ constituting the second class particle.

Let's define the configuration: $$(\xi_{0}^{(0)},x_{(0)}^{\alpha}(0),y_{(0)}^{\beta}(0)) := (1_{\{k \leq 0\}}, -1 , 0) $$
In words, this is a free tagged hole followed by a free tagged particle of rates $\beta$ and $\alpha$ respectively. We choose a clock for the tagged particle that rings each time $x^{R}$ makes a jump (we will sometimes call a tagged particle by the name of its position variable) While the clock of the tagged hole rings each time the $y^{R}$ is jumped over. We track the position of the couple $(0,1)$ at the origin using a variable $z^{(0)}(t)$. Again, $z^{(0)}$ coincides only initially with $x^{\alpha}_{(0)}$.

We set $t_{0} = 0$ and we define a real decreasing sequence of intervals $([t_{n}, \infty))_{n\in \mathbb{N}}$ and a sequence of initial configurations $((\xi_{t_{n}}^{(n)},x_{(n)}^{\alpha}(t_{n}),y_{(n)}^{\beta}(t_{n})))_{n \in \mathbb{N}}$ by induction, where the system $\xi^{(n)}$ is defined on the time interval $[t_{n},\infty($, and for $n \geq 1$:
$$ t_{n} := min(\min_{t}\{t \geq t_{n-1} \; and \;  x^{\alpha}_{(n-1)}(t) > x^{\alpha}_{(n-1)}(t_{n-1}) \}, \min_{t}\{t \geq t_{n-1} \; and \;  y^{\beta}_{(n-1)}(t) < y^{\beta}_{(n-1)}(t_{n-1}) \} ) $$

In words, $t_{n}$ is the moment when either the tagged particle or the tagged hole of the configuration $\xi^{(n-1)}$ first decides to move.

We define now the configuration $\xi_{t}^{(n)}$ over the interval $[t_{n}, \infty)$ as one starting with the initial configuration:

$$ (\xi_{t_{n}}^{(n)},x_{(n)}^{\alpha}(t_{n}),y_{(n)}^{\beta}(t_{n})) := \left\{
\begin{array}{ll}
(\xi_{t_{n}}^{(n-1)},x_{(n-1)}^{\alpha}(t_{n}) - 1,y_{(n-1)}^{\beta}(t_{n}))&if \; y_{(n-1)}^{\alpha}(t_{n}) = y_{(n-1)}^{\alpha}(t_{n-1}) - 1 \\
(\xi_{t_{n}}^{(n-1)},x_{(n-1)}^{\alpha}(t_{n}),y_{(n-1)}^{\beta}(t_{n})+1)&if \; x_{(n-1)}^{\alpha}(t_{n}) = x_{(n-1)}^{\alpha}(t_{n-1}) + 1
\end{array}
\right.$$

To avoid possible confusion, we always assume the trajectories of the particles to be Càdlàg functions.

Note that the definition of $\xi_{t}^{(n)}$ would allow its tagged hole and its tagged particle to be consecutive in the interval $[t_{n},t_{n+1})$ and only in this interval.

As we defined $z^{(0)}$, we can define $z^{(n)}$ on the interval $[t_{n}, \infty($ as the position of the second component of the couple $(0,1)$ that coincides initially with $ (y^{\beta}_{(n)},x^{\alpha}_{(n)})$.

So, one of two possible events for $\xi^{(n-1)}$ illustrated in the table can cause the creation of $\xi^{(n)}$.

\begin{center}
\begin{tikzpicture}

\draw [<->]  (-0.5,0.15) -- (-0.5,-0.15);
\draw [<->]  (-0.9,0.15) -- (-0.9,-0.15);
\draw [<->]  (1.9,0.15) -- (1.9,-0.15);
\draw [<->]  (2.3,0.15) -- (2.3,-0.15);

\node (tab1) {%
	\renewcommand{\arraystretch}{1.5}
	\begin{tabular}{l|c|r}
	$\xi_{t_{n}}^{(n-1)}$ & $ ...y^{\beta} x^{1}  x^{\alpha}... $  & $ ...y^{\beta}  y^{1}  x^{\alpha}... $\\
	\hline
	$\xi^{(n)}_{t_{n}}$ & $ ...y^{\beta} x^{\alpha}  x^{1} ...$ & $... y^{1}  y^{\beta}  x^{\alpha}... $\\

	\end{tabular}};
\end{tikzpicture}
\end{center}

\begin{itemize}
\item The first event is the jumping of $x^{\alpha}_{(n-1)}$, in this case, we know that there is a hole between the tagged hole and the tagged particle at time $t_{n}$. Let’s tag this hole in the middle with the variable $y^{1}_{(n-1)}$.We have $y^{1}_{(n-1)}(t_{n}) = y_{(n-1)}^{\beta}(t_{n})+1$. We are now in a position that allows us to use the the dual lemma of lemma \ref{lemma2} to couple the movements of $ y^{\beta}_{(n)}(t) $ and $ y^{1}_{(n-1)}(t) $ for $t \geq t_{n}$, We will as well couple all the particles to the right of $x^{\alpha}_{(n-1)}$ with all the particles to the right of $x^{\alpha}_{(n)}$ inclusive for $t \geq t_{n}$, this is obviously possible since this two segments coincides at $t_{n}$. So we have:
$$ \xi_{t}^{(n-1)}[k] = \xi_{t}^{(n)}[k]  \; for \; k \geq x_{n-1}^{\alpha}(t) \; for \; t \geq t_{n} $$
That means as well that segments: $y^{1}_{(n-1)}$ to $x^{\alpha}_{(n-1)}$ and $y^{\beta}_{(n)}$ to $x^{\alpha}_{(n)}$  will be identical to the one between
:

$$ \xi_{t}^{(n-1)}[k] = \xi_{t}^{(n)}[k]  \; for \; y_{(n-1)}^{1} \leq k \leq x_{n-1}^{\alpha}(t) \; for \; t \geq t_{n} $$

From the previous statement it becomes obvious that:
\begin{equation}\label{zz}
z^{(n)}(t) = z^{(n-1)}(t) \; for \; t \geq t_{n}
\end{equation}

\item The second event is the backward jumping of $y^{\beta}_{(n-1)}$. One can show in a similar fashion as previously by using the Lemma \ref{lemma3} and by coupling the particles left to $y^{\beta}_{(n-1)}$ with the ones left to $y^{\beta}_{(n)}$. We show again that \ref{zz} holds in this case too.
\end{itemize}

By induction on \ref{zz} we get:
\begin{equation}\label{zzR}
z^{(n)}(t) = z^{(0)}(t) = z^{(R)}(t)  \; for \; t \geq t_{n}
\end{equation}

The final step of our proof is to define the system $\zeta$ as follows:

$$ \zeta_{t} = \xi_{t}^{(n)} \; for \; t \in [t_{n}, t_{n-1})  $$

In this system, the tagged particle of rate $\alpha$ and the tagged hole of rate $\beta$ are always consecutive, so $\zeta$ can be coupled with $\eta$, and the proof is completed using \ref{zzR}.
\qed

\subsection{Appendix}
We obtain here an explicit formula for the marginals of the process described in lemma \ref{lemma1} and show explicitly its symmetry with respect to $\alpha_{1}$ and $\alpha_{2}$. Consider a finite system of $N$ particles of initial positions $\mathbf{y} = \{y_{1} < y_{2}<  ... < y_{N}\}$ with forward hopping rates \{$\alpha_{1}$,...,$\alpha_{N}$\} respectively. The conditional probability of having these particles at positions 
$\mathbf{x} = \{x_{1} < x_{2}<  ... < x_{N}\}$ has been found in \cite{rakos2006bethe}, and is given by 

\begin{equation}
P(\mathbf{x} | \mathbf{y} ; t) = \prod_{i=1}^{N} (e^{-t \alpha_{i}} \alpha_{i}^{x_{i} - y_{i}} )
\begin{vmatrix}
F_{1,1}(x_{1}-y_{1},t) & F_{1,2}(x_{2}-y_{1},t) & \ldots & F_{1,N}(x_{N}-y_{1},t) \\ 
F_{2,1}(x_{1}-y_{2},t) & F_{2,2}(x_{2}-y_{2},t) & \ldots & F_{2,N}(x_{N}-y_{2},t) \\ 
\vdots & \vdots  &   & \vdots  \\
F_{N,1}(x_{1}-y_{N},t) & F_{N,2}(x_{2}-y_{N},t) & \ldots & F_{N,N}(x_{N}-y_{N},t) 
\end{vmatrix}
\end{equation}
where:
\begin{equation}
F_{k,l}(x,t) := \frac{1}{2 \pi i} \oint e^{\frac{t}{z}} z^{x-1} \prod_{i=1}^{l-1} (1-\alpha_{i} z)^{-1} \prod_{i=1}^{k-1} (1- \alpha_{i} z) dz
\end{equation}
For our case, we have: $y_{2} = y_{1}+1$ and $\alpha_{i} = 1$ for $i>2$. We are interested in the probability distribution of the first particle regardless of the final position of the rest of the particles, this amounts to the sum over all their possible final positions:
\begin{equation}
P(x_{1} | \mathbf{y}; t) =
\sum_{x_{n} = x_{1}+n-1}^{\infty}...
\sum_{x_{3} = x_{1}+2}^{x_{4}-1}
\sum_{x_{2} = x_{1}+1}^{x_{3}-1}
P(x_{1},...,x_{n} | \mathbf{y})
\end{equation}\label{summation}
One can show that the functions $F_{k,l}(x,t)$ verify the following properties:

For $l \geq 3$
\begin{equation}
\sum_{n = n_{1}}^{n_{2}}
F_{k,l}(n,t) =
F_{k,l+1}(n_{1},t) - F_{k,l+1}(n_{2}+1,t)
\quad \text{for} \quad l \geq 3
\end{equation}

\begin{equation}
\sum_{n = n_{1}}^{\infty}
F_{k,l}(n,t) =
F_{k,l+1}(n_{1},t) \quad l \geq 3
\end{equation}

And for $l=2$
\begin{equation}
\sum_{n = n_{1}}^{n_{2}}
\alpha_{2}^{n}F_{k,2}(n,t) =
\alpha_{2}^{n_{1}}F_{k,3}(n_{1},t)-
\alpha_{2}^{n_{2}+1}F_{k,3}(n_{2}+1,t)
\end{equation}

Note that the function $F_{k,l}(x,t)$ are symmetric with respect to $\alpha_{1}$ and $\alpha_{2}$ only when $k \neq 2$ and $l \neq 2$. Now we can perform the summation in \ref{summation} column by column starting from the second one and getting rid each time of the term that is proportional to the next column: we get as a result:

\begin{equation}
P(\mathbf{x} | \mathbf{y} ; t) = \prod_{i=1}^{N} (e^{-t \alpha_{i}})
(\alpha_{1}\alpha_{2})^{-y_{1}}
\alpha_{1}^{x_{1} }
\alpha_{2}^{x_{2}  - 1} 
\end{equation}

\begin{equation}
\begin{split}
&P(x_{1} | \mathbf{y}; t) =
\prod_{i=1}^{N} (e^{-t \alpha_{i}})
(\alpha_{1}\alpha_{2})^{-y_{1}}
\alpha_{1}^{x_{1} }
\alpha_{2}^{- 1}\\
\times &
\begin{vmatrix}
F_{1,1}(x_{1}-y_{1},t) & \alpha_{2}^{x_{1}+1} F_{1,3}(x_{1}+1-y_{1},t) &F_{1,4}(x_{1}+2-y_{1},t) &  \ldots & F_{1,N}(x_{1}+N-1-y_{1},t) \\ 
F_{2,1}(x_{1}-y_{2},t) &  \alpha_{2}^{x_{1}+1} F_{2,3}(x_{1}+1-y_{2},t) &F_{2,4}(x_{1}+2-y_{2},t) & \ldots & F_{2,N}(x_{1}+N-1-y_{2},t) \\ 
\vdots & \vdots  &\vdots  &   & \vdots  \\
F_{N,1}(x_{1}-y_{N},t) & \alpha_{2}^{x_{1}+1} F_{N,3}(x_{1}+1-y_{N},t) &F_{N,4}(x_{1}+3-y_{N},t) & \ldots & F_{N,N}(x_{1}+N-1-y_{N},t)  
\end{vmatrix}
\end{split}
\end{equation}

The final step is to manipulate the second line where we have $k=2$. We notice that:

\begin{equation}
F_{2,l}(x,t) = F_{1,l}(x,t) - \alpha_{i} F_{1,l}(x+1,t)
\end{equation}
If we apply this to the second line, the second term will be proportional to the first line (remember that  $y_{2}= y_{1}+1$ ), and thus we get the final formula:

\begin{equation}
\begin{split}
&P(x_{1} | \mathbf{y}; t) =
\prod_{i=1}^{N} (e^{-t \alpha_{i}})
(\alpha_{1}\alpha_{2})^{-y_{1}}
(\alpha_{1}\alpha_{2})^{x_{1}}
\\
\times &
\begin{vmatrix}
F_{1,1}(x_{1}-y_{1},t) &  F_{1,3}(x_{1}+1-y_{1},t) &F_{1,4}(x_{1}+2-y_{1},t) &  \ldots & F_{1,N}(x_{1}+N-1-y_{1},t) \\ 
F_{1,1}(x_{1}-y_{1}-1,t) &   F_{1,3}(x_{1}-y_{1},t) &F_{1,4}(x_{1}-y_{1}+1,t) & \ldots & F_{1,N}(x_{1}-y_{1}+N,t) \\
F_{3,1}(x_{1}-y_{3},t) &  F_{3,3}(x_{1}+1-y_{3},t) &F_{3,4}(x_{1}+2-y_{3},t) &  \ldots & F_{3,N}(x_{1}+N-1-y_{3},t) \\
\vdots & \vdots  &\vdots  &   & \vdots  \\
F_{N,1}(x_{1}-y_{N},t) &  F_{N,3}(x_{1}+1-y_{N},t) &F_{2,4}(x_{1}+3-y_{N},t) & \ldots & F_{N,N}(x_{1}+N-1-y_{N},t)  
\end{vmatrix}
\end{split}
\end{equation}

This formula is manifestly symmetric with regards to $\alpha_{1}$ and $\alpha_{2}$.

This result is still valid for an infinite system $N \rightarrow \infty$ since at each instant $t$, with a probability $1$ there exists a particle that didn't try to jump in the interval $[0,t]$, and so only the finite number of particles behind it will be involved.

\typeout{}
\bibliography{f}
\bibliographystyle{ieeetr}

\end{document}